# Learning Self-Awareness Models for Physical Layer Security in Cognitive and AI-enabled Radios

**ALI KRAYANI**

Department of Electrical, Electronic, Telecommunications Engineering and Naval Architecture (DITEN)
University of Genoa

School of Electronic Engineering and Computer Science (EECS)
Queen Mary University of London

This dissertation is submitted for the degree of
*Doctor of Philosophy*

Joint Doctorate in Interactive and
Cognitive Environments - Cycle 34

April 2022

**Learning Self-Awareness Models for Physical Layer Security in Cognitive and AI-enabled Radios**

Ali Krayani

Joint Doctorate in Interactive and Cognitive Environments
JD-ICE

XXXIV cycle

## Acknowledgements

This PhD Thesis has been developed in the framework of, and according to, the rules of the Joint Doctorate in Interactive and Cognitive Environments JD-ICE with the cooperation of the following Universities:

Università degli Studi di Genova (UNIGE)
DITEN - Dept. of Electrical, Electronic, Telecommunications Engineering and Naval Architecture
ISIP40 - Information and Signal Processing for Cognitive Telecommunications

Primary Supervisor: Prof. Carlo Regazzoni
Secondary Supervisor: Prof. Lucio Marcenaro

Queen Mary University of London (QMUL)
EECS - School of Electronic Engineering and Computer Science
CSI - Centre for Intelligent Sensing

Primary Supervisor: Asst. Prof. Atm Shafiul Alam
Secondary Supervisor: Prof. Arumugam Nallanthan

I would like to dedicate this thesis to my loving family and friends …

*“When wireless is fully applied the earth will be converted into a huge brain, capable of response in every one of its parts.”*

- Nikola Tesla

# Acknowledgements

Three years that it is what it took me to complete this thesis. My entire PhD journey was full of joy and passion because I enjoyed searching and learning new things every moment. This thesis could not have been produced without the support and guidance from my fantastic supervisors and mentors.

Foremost, I would like to express my sincere gratitude to Prof. Carlo Regazzoni for his wise guidance, patience and continuous support of my PhD research. With his profound knowledge on signal processing, cognitive radios and Artificial intelligence, he offered many constructive advice and suggestions on my PhD research and the directions for my future work. My sincere gratitude also goes to Prof. Atm S. Alam for his priceless technical advice, availability, guidance, attention to detail and encouragement. I would also have to thank Prof. Lucio Marcenaro for his invaluable suggestions and comments for my research and scientific writing. I would also like to express my appreciation to Prof. Yue Gao with whom I worked during my first year for his effective instructions, discerning suggestions and insightful technical advice. I also wish to thank Prof. Arumugam Nallanathan and Prof. James Kelly for their invaluable feedback and constructive comments. Finally, I could not have imagined having better supervisors and mentors for my PhD research. From them, I learned the technical knowledge and the crucial aspects of being a good researcher, specifically, meticulousness, passion, intelligence, and hard work.

Special thanks to all of my colleagues at UNIGE and QMUL, who provided me with a friendly research environment.

Last but not least, I would like to thank all my friends and family members for their encouraging supports and love.

# Abstract

Cognitive Radio (CR) is a paradigm shift in wireless communications to resolve the spectrum scarcity issue with the ability to self-organize, self-plan and self-regulate. On the other hand, wireless devices that can learn from their environment can also be taught things by malicious elements of their environment, and hence, malicious attacks are a great concern in the CR, especially for physical layer security. This thesis introduces a data-driven Self-Awareness (SA) module in CR that can support the system to establish secure networks against various attacks from malicious users. Such users can manipulate the radio spectrum to make the CR learn wrong behaviours and take mistaken actions. The SA module consists of several functionalities that allow the radio to learn a hierarchical representation of the environment and grow its long-term memory incrementally. Therefore, this novel SA module is a way forward towards realizing the original vision of CR (i.e. Mitola's Radio) and AI-enabled radios. This thesis starts with a basic SA module implemented in two applications, namely the CR-based IoT and CR-based mmWave. The two applications differ in the data dimensionality (high and low) and the PHY-layer level at which the SA module is implemented. Choosing an appropriate learning algorithm for each application is crucial to achieving good performance. To this purpose, several generative models such as Generative Adversarial Networks, Variational AutoEncoders and Dynamic Bayesian Networks, and unsupervised machine learning algorithms such as Self Organizing Maps Growing Neural Gas with different configurations are proposed, and their performances are analysed. In addition, we studied the integration of CR and UAVs from the physical layer security perspective. It is shown how the acquired knowledge from previous experience within the Bayesian Filtering facilitates the radio spectrum perception and allows the UAV to detect any jamming attacks immediately. Moreover, exploiting the generalized errors during abnormal situations permits characterizing and identifying the jammer at multiple levels and learning a dynamic model that embeds its dynamic behaviour. Besides, a proactive consequence can be drawn after estimating the jammer's signal to act efficiently by mitigating its effects on the received stimuli or by designing an efficient resource allocation for anti-jamming using Active Inference. Experimental results show that introducing the novel SA functionalities provides the high accuracy of characterizing, detecting, classifying and predicting the jammer's activities and outperforms conventional detection methods such as Energy detectors and advanced classification methods such as Long Short-Term Memory (LSTM), Convolutional Neural Network (CNN) and Stacked Autoencoder (SAE). It also verifies that the proposed approach achieves a higher degree of explainability than deep learning techniques and verifies the capability to learn an efficient strategy to avoid future attacks with higher convergence speed compared to conventional Frequency Hopping and Q-learning.

# Table of contents

# List of figures

# List of tables

# Nomenclature

**Acronyms / Abbreviations**

3GPP Third Generation Partnership Project

5G Fifth-generation

6G Sixth-Generation

AAE Adversarial AutoEncoder

ABD Anomaly Based Detection

ABSF Absorptive Band Stop filter

AI Artificial Intelligence

AIn Active Inference

AMC Automatic Modulation Classification

AMCC Automatic Modulation Conversion and Classification

BS Base Station

CFD Cyclostationary Feature Detector

CNN Convolutional Neural Network

CR Cognitive Radio

DCNN Deep Convolutional Neural Network

DGDBN Double Generalized Dynamic Bayesian Network

DGM Deep Generative Model

DL Deep Learning

DNN Deep Neural Network

DoS Denial of Service

DRL Deep Reinforcement Learning

DSSS Direct Sequence Spread Spectrum

ED Energy detector

EKF Extended Kalman Filter

FAA Federal Aviation Administration

FFT Fast Fourier Transform

FHSS Frequency Hopping Spread Spectrum

GAN Generative Adversarial Network

GM Generative Model

GPS Global Positioning System

HDBN Hierarchical Dynamic Bayesian Network

HMM Hidden Markov Model

IF Intermediate Frequency

INF Instantaneous Frequency

IoT Internet of Things

JSR Jammer to Signal Ratio

KF Kalman Filter

LOS Line Of Sight

LSTM Long Short Term Memory

LTE Long Term Evolution

MDP Markov Decision Process

MFD Match Filter Detector

ML Machine Learning

mMTC massive Machine Type Communication

mmWave Millimeter Wave

MTC Machine-Type Communication

OFDM Orthogonal Frequency-Division Multiplexing

PDR Packet Delivery Ratio

PDA Personal Digital Assistant

PF Particle Filter

PGM Probabilistic Graphical Model

PHY Physical Layer

PN Pseudo Noise

PSD Power Spectral Density

PUE Primary User Emulation

PU Primary User

QoI Quality of Information

QoS Quality of Service

RAN Radio Access Network

RFAL Radio Frequency Adversarial Learning

RF Radio Frequency

RL Reinforcement Learning

SAE Stacked Autoencoder

SA Self-Awareness

SDR Software Defined Radio

SNR Signal to Noise Ratio

SS Signal Strength

SVM Support Vector Machine

TVWS TV White Spaces

UAV Unmanned Aerial Vehicle

UKF Unscented Kalman Filter

V2X Vehicle to Everything

VAE Variational AutoEncoder

WLAN Wireless Local Area Network

# Chapter 1

# Introduction

## 1.1 Motivation and Scope

The rapid and exploding growth of global wireless data traffic generated by massive Internet of Things (IoT) devices and high-density terrestrial/aerial vehicles over the last years have brought new challenges for utilizing resources intelligently to meet the need of growing diversity in services and user behaviour [1]. In addition, the newly emerging wireless communications scenarios under the umbrella of the fifth-generation (5G), including the massive Machine Type Communications (mMTCs), Unmanned Aerial Vehicles (UAVs), Vehicle to Everything (V2X) and IoT, have opened new challenges and research opportunities for the development of Artificial Intelligence (AI) based solutions to meet the stringent performance requirements on delay, reliability, accuracy and security. 5G has provided different service opportunities to the new emerging ecosystems, and it is still evolving to reach the comfort of having things, vehicles, and people fully connected anytime everywhere [2]. However, 5G cannot completely meet the rising technical criteria such as autonomous, highly dynamic and fully intelligent services [3]. Building upon the 5G vision, studies from academia and industry (Nokia, Ericsson, Huawei, Samsung, LG, Apple, etc.) have already started to discuss the sixth generation (6G), which is expected to be a transformative technology revolutionizing the wireless evolution from *connecting everything* to *connecting intelligence* and pave the road towards intelligent radios [4, 5]. Nevertheless, Intelligent Radios are not a new concept. Indeed, Joseph Mitola formalised this concept in 2000 and coined it Cognitive Radio (also known as Mitola cognitive radio by many) [6]. Thus, it is expected that 6G will likely be a self-contained ecosystem of AI making the original Cognitive Radio (CR) a reality after more than 20 years of effort [7].

The rapid development of the IoT has attracted much research and industrial interests [8]. With the explosive growth in the number of connected IoT devices in a variety of applications, the scarcity of spectrum resources has become a serious problem [9]. In addition, most of the studies on IoT are focused on the communication, computing and connectivity aspects which are of great concern. However, IoT cannot fulfill its potentials and deal with growing challenges without comprehensive cognitive capabilities and empowering IoT with high-level intelligence [10]. It is envisioned that future IoT networks should be equipped with cognitive abilities to think, learn and understand the demands of both the physical and social world [11]. Moreover, reliable communications realize the key to allow IoT devices to connect, interact and exchange information securely anytime, anywhere.

On the other hand, UAVs are expected to play a significant role in future wireless communications (5G and beyond) serving as flying base stations, mobile relays or Aerial users [12]. UAVs can provide reliable and high capacity communications with an improved coverage and quality of service (QoS) owing to their fascinating features like mobility, dominant Line of Sight (LoS) channels, and fast deployment [13, 14]. However, UAVs must have the capability to maintain wireless connectivity in dynamic radio environments and to adapt to varying traffic loads, network failure and radio frequency interference [15]. Besides, the unprecedented growth of UAVs in various applications such as smart cities, surveillance and IoT [16] will also overcrowd the spectrum resources causing spectrum insufficiency [17].

Millimeter Wave (mmWave) band can be a solution to serve the vast number of connected IoT and UAV devices and mitigate spectrum scarcity problem with its abundant spectrum resources at high frequencies (from 30 to 300 gigahertz), offering low latency and high-speed data connection [18–21]. However, in heterogeneous networks, spectral resources of mmWave band are managed by different operators, which imposes tight restrictions regarding the spectrum sharing process and results in an inefficient resource utilization [22, 23]. Hence, a suitable hybrid access scheme is needed in the mmWave band to efficiently manage the licensed and unlicensed spectrum.

CR has been proposed to be integrated with different radio applications to tackle the aforementioned issues. The integration of CR and IoT (known as CR-IoT) could relieve the spectrum scarcity problem by enhancing spectral efficiency and improving the network performance [24]. Furthermore, CR endows the IoT devices with the cognitive facility to take smart decisions and perform intelligent operation by analyzing network conditions [25]. It can also provide intelligent services by processing different types of data generated by connected devices in a variety of applications such as smart manufacturing, smart homes and smart cities and handle intelligent tasks. Also, CR can be a promising solution to make UAVs capable of reaching and maintaining connectivity with a high degree of autonomy as well as managing the dynamic spectrum access in mmWave communications [14].

CR is envisaged as a computational intelligent radio that observes, thinks and acts to achieve a high level of competence in radio-related domains [26]. It can autonomously learn by sensing the radio environment, infers the signals' dynamic behaviours to plan, decide and act accordingly, providing a promising solution for wireless devices to achieve flexible connectivity and adjusting link parameters dynamically under high autonomy. Learning the radio environment and adapting dynamically based on observation and previous experiences are main characteristics of CR. The cognition cycle by which CR interacts with the external environment was first described by Mitola in [27] and includes the following capabilities: *1)* collect data by sensing the radio environment; *2)* learn a representation of the collected data; *3)* take a decision based on such representation to act on; *4)* observe the environmental feedback in response to the action and update the acquired knowledge (autonomous incremental knowledge acquisition) to improve the future decision-action process, consequently.

The cognition cycle realizes the core of CR, and it is the fundamental building block of radio's cognition. AI-empowered CR (known as AI-enabled radios) allows reaching a high level of intelligence. Although AI systems can perform smart activities, none can understand why they do what they do, nor how [28]. Thus, a sort of *Self-Awareness (SA)* is needed to achieve the required cognition. This was confirmed by Mitola's declaration: "*A radio which should be termed as Cognitive must be Self-Aware*". SA as defined in [29] is the "ability of the radio to understand its own capabilities, i.e., to understand what it does and does not know, as well as the limits of its capabilities". SA is concerned with a radio's knowledge about itself and its prevalent environment. It is the ability of a radio to interpret the surrounding environment according to the knowledge encoded in its internal models and to adapt its behaviour according to the detected environmental changes to reach the dynamic equilibrium. *An important research question that needs to be addressed is: which SA representation should be used to realize computationally the cognition cycle of the original CR?* To answer this question, we propose introducing a novel SA module in CR that aims to organize the main functionalities of the cognition cycle by learning autonomously and incrementally emergent dynamic representations.

On the other side, another main challenge in designing CR is related to the physical layer security. As all wireless communications, CR is vulnerable to malicious attacks (e.g. jamming attacks) because of the radio propagation and broadcasting nature [30, 31]. Moreover, the situation could be worse in CRs, where a smart jammer with cognitive abilities can estimate system parameters and manipulate radio spectrum, thus forcing CR to learn wrong behaviours and take non-optimal actions. Additionally, the dominant LOS channel links in UAV communications are more prone to terrestrial jammer threats that might affect the system's performance drastically causing increased latency and transmission delays [32–36]. Furthermore, the dynamic environment that involves multiple signal transmissions due to the shared wideband spectrum in mmWave and the coexistence in tight integration with different wireless systems as well as heterogeneity in IoT poses additional security threats in CR-related applications (such as, CR-IoT and CR-mmWave) [16, 25, 30, 37–39]. All these motivations indicate the necessity that the security threats need to be investigated in CR-based applications with more comprehensive solutions in detecting, characterizing and mitigating the jamming attacks at the physical layer. Thereupon, enhancing the physical layer security is still an open issue and of great concern to guarantee successful deployment of CR. Thus, the proposed SA module aims to enhance the physical layer security against jamming attacks. Nevertheless, the functionalities of the SA module will be useful for realizing other goals targeted by the radio and pave the road towards full Self-Aware Radio (i.e. original CR).

The proposed SA module allows the radio to perform the following functionalities: *i) Learning Generative Models* autonomously by observing radios' states and the occurring environmental changes; *ii) Radio Spectrum Perception* by predicting the future environmental states according to the rules encoded in the Generative model and estimating current states of the observed stimuli received from the environment; *iii) Abnormality Detection* which refers to the process of deciding whether communications in the spectrum occur according to the expected behaviour and noticing any deviation from the normal situation (i.e. similar to what was learned from previous experience); *iv) Abnormality Characterization* by analyzing the new behaviour (the detected differences which are not seen before) and characterizing it to draw up the dynamic rules of the new emergent force, explaining how the new situation is evolving; *v) Incremental Learning* which is the process of acquiring new representations that encode the occurred environmental changes associated with the detected signals representing new behaviour; *vi) Internal Action* by removing the abnormal signal caused by a malicious user from the received stimuli; *vii) Abnormality Classification* by using multiple models learned so far and performing multiple predictions to discriminatively select the model that better fits to current experience. This functionality is crucial to understand when the radio should learn a new model and which communication policy should apply (e.g., how to change the defence policy after being able to identify the detected jammer). *viii) Learning Interactive Models* by observing multiple signals related to different sources (e.g. RF and GPS signals) and learning the cross-correlation between them, or by learning the causality among different users (e.g. CR device and jammer) interacting inside the radio spectrum to design a defense strategy and change the policies of communication (changing the configuration: power, modulation, frequency, etc.).

The proposed SA module endows CR with the capability of maintaining a dynamic equilibrium with the external environment under the free-energy principle (FEP) (proposed by Karl Friston in [40]). The latter is one of the basic principles that underwrite the self-organization of biological systems like brains. A biological system exposed to random and unpredictable variations in its surrounding environment can learn to restrict itself to occupy limited and unsurprising states to survive by minimizing the free energy (FE) [41]. In this perspective, a radio can start operating in the field without prior knowledge and tends to distil structural regularities from the environmental variations through sensory signals and embody them in its internal dynamic models. These models explain the causal structure in the wireless environment and enable the radio to predict what will happen next and register surprising violations (prediction errors) of those predictions. This allows the radio to maintain a certain homeostasis by having steady states it finds itself in and resists the tendency to disorder. It is of great interest to explain the states occupied by the radio and their temporal evolution in a probabilistic manner and connect it with the Bayesian inference. This forms some sort of statistical dependency between environmental variations (external states) and radios' states (internal states) in terms of a probabilistic graphical model. The notion of statistical dependency is central to the existence of the Markov blanket that separates states into external and internal building by that a circular causality explaining the perception-action cycle in the brain where active and internal states minimize a FE functional of sensory states. Inspired by how the brain infers the causes of its sensory inputs, the proposed SA module allows the radio to learn Generative Models of how data were generated given some observations and infers its cause through the Bayesian Filtering. SA also provides a powerful account of optimal control and decision-making through the Active Inference (*AIn*) (i.e., the FEP for action and perception) using the active states that control how environmental variations are sampled by sensory states (i.e., by changing the physical configuration as power, frequency, etc.).

## 1.2 Summary of Contributions

Firstly, this thesis starts with a general overview to put CR in its historical context to understand the interests that led to its advent at the design and utilization phases. It highlights the physical layer security issues regarding jamming attacks by reviewing the proposed solutions from the literature and shedding light on the gap related to the development of a framework that studies jammer's detection, characterization, mitigation, classification and anti-jamming dependently through an unsupervised incremental approach. We will show how the thesis can fill this gap by proposing a logical and coherent framework.

Secondly, this thesis discusses various Generative Models to learn an efficient representation of the wireless environment which realizes the first essential step to equip CR with SA by exploring different CR applications (CR-based IoT and CR-based mmWave) [42–45]. These applications differ in the data dimensionality (high- and low-dimensional signals) and the physical layer level at which the CR performs sensing and perception operations (near or far from the antenna). A framework for the joint spectrum representation and abnormality detection is proposed to learn SA representations encoded in generative models learned from Generalized States GSs [46] (i.e., pattern vectors consisting of the signals' features and the corresponding temporal derivatives). The fundamental advantage of generalized states is capturing temporal correlations on random variations in hidden environmental states, facilitating the predictions and estimations of those states. In the high-dimensional application, the method is employed just after the receiving antenna and the down-conversion block, where multi-signals representations are extracted from a mmWave wideband spectrum with a high sampling rate. Consequently, a high-dimensional GS is built from which the dynamic models can be learned. Three deep generative models are compared, namely, Conditional Generative Adversarial Network (C-GAN), Auxiliary Classifier GAN (AC-GAN) and Variational AutoEncoder (VAE), by evaluating their performance in capturing the dynamic behaviour of multi-transmissions and their ability in detecting any anomalous signals injected by malicious users (i.e., jammers) inside the spectrum. In the low-dimensional application, the method is employed after down-conversion, cyclic prefix removal and fast Fourier transform block at which the CR-IoT extracts signals with low dimensionality and low sampling rate from Orthogonal Frequency Division Multiplexing (OFDM) sub-carriers to build the GS and learn a probabilistic graphical model, i.e., the Dynamic Bayesian Network (DBN). Additionally, a possible way of bridging high-dimensional and low-dimensional data through the DBN model is presented [47]. The advantage of this approach is that, representing sufficient features extracted from high-dimensional multi-signals in a probabilistic graphical model as the DBN endows the radio with the ability to produce explainable and interpretable predictions allowing to understand surprising events. The exciting peculiarities of the DBN in modelling the environment as dynamic processes describing the signals' temporal evolution at hierarchical levels. As well as providing a graphical structure representing hidden and observed states in terms of random variables encoding the conditional dependencies among them explicitly that specify a compact parameterization of the model and its dynamic evolution makes it worthy for deep investigation as we will see throughout the thesis. Numerical results reveal that the proposed method effectively detects malicious or jammer attacks after learning the corresponding dynamic models and outperforms conventional detection methods. Validation is performed on a real mmWave dataset in the high-dimensional CR application and simulated OFDM data in the low-dimensional application.

Thirdly, this thesis proposes introducing a SA module at the physical layer, which endows the CR with high-level intelligence equipped with explainability and interpretability [48]. It presents the SA functionalities logically and coherently because they are dependent, while one incrementally leads to the other, allowing the radio to build up the knowledge of its own memories (brain) based on a hierarchical layered representation. This representation permits the radio to predict the spectrum's future states accurately using Bayesian filters (e.g. Particle Filter [49] and Kalman Filter [50]) which permits the radio to anticipate the received (or sensed) signal through prediction (top-down inference) and then matching it with the actual measurement (bottom-up inference) when a stimulus is received from the environment and leading by that to a better perception and understanding of the surrounding environment. Initially, the radio starts with an empty memory without any prior knowledge about the radio environment. Then it makes an initial guess about the probability distribution of the environmental states. After that, by observing the environment, it begins to change the initial guess based on the distribution in question using Bayes theory. Indeed, the radio assumes (at the beginning of the learning process) that the rules of how the signals are changing inside the radio spectrum are fixed. After that, the generated errors (i.e. the difference between predictions and observations) allow the radio to extract the dynamic rules of the spectrum and construct the initial dynamic model. This thesis will focus on learning hierarchical DBN models due to their ability in studying the causality between random hidden variables at different hierarchical levels explicitly and represent the corresponding dynamics over time following probabilistic reasoning [51]. Observing new signals might follow the same dynamic rules learned by the radio from previous experience or might deviate due to a new situation (e.g. during jamming attacks). The radio can recognize the new signals where their dynamic rules can not be explained by its memories (abnormality detection). The generalized errors during the new situation allow the radio to study and analyse the dynamics of such abnormal behaviour in order to characterize its nature. The results of Abnormality Characterization can be used later on to incrementally learn new dynamic models or to choose an appropriate action (abnormality mitigation or adjusting the operating parameters like power, frequency, modulation, etc.). Also, learning new dynamic models incrementally allows the radio to identify between different jammers attacking the system and predict their activities which allows the radio to avoid their threat in the future. Moreover, the radio can learn interactive models between different entities (user-jammer or primary-secondary users) interacting inside the radio spectrum or between different signals received by the cognitive device (e.g. the RF and GPS signal) which support the implementation of an efficient anti-jamming technique, dynamic spectrum access strategy and autonomous navigation.

Finally, the thesis also studies the integration of CR and UAV from the physical layer security perspective [52, 53]. Additionally, this thesis explains how SA enrich the radio with discriminative capability by comparing the current experience to multiple predictions generated from a set of models learnt so far describing different wireless experiences (e.g., multiple jammers with different modulation schemes) and selecting the model that better fits actual experience (i.e., jammer identification and automatic modulation classification) [54, 55]. For the first time, it also introduces the joint signal modulation conversion and classification, which is a prospective candidate technology in future wireless communications (6G) based on generalized filtering integrated with transport planning [56]. Numerical results (based on real datasets and simulations) demonstrate: 1) that the proposed method for signal modulation classification outperforms both Long Short-Term Memory (LSTM), Convolutional Neural Network (CNN) and Stacked Autoencoder (SAE) in terms of classification accuracy and achieves a higher degree of explainability of its own decisions by interpreting causes and effects at hierarchical levels under the Bayesian learning and reasoning processes; 2) the effective performance of our novel approach on jointly converting and classifying multiple modulation formats. The thesis concludes the journey by proposing a novel resource allocation strategy for anti-jamming using the Active Inference approach [57]. Under the Active inference framework, radio can jointly learn a joint internal representation of the external environment encoding: the physical signals that the radio is supposed to receive through sensory inputs under normal circumstances and the available spectral resources. Such representation explains the dynamic interaction between CR-user and jammer in the spectrum, which drives the CR-user to learn the best set of actions that leads to the minimum surprise (i.e., the radio learns to restrict itself to occupy unsurprising states). Simulation results verify the effectiveness of the proposed AIn approach in minimizing abnormalities (maximizing rewards) and has a high convergence speed by comparing it with conventional Frequency Hopping and Q-learning.

To summarize, the main contributions of this thesis are the following:

- it is recalling the original vision of CR (Mitola CR) that goes beyond spectrum sensing and proposes a SA module to empower CR with a brain for high-level intelligence.
- the proposed SA module can be implemented at different sides, at the Base Station side or at the mobile device side. The physical layer (PHY) level at which the SA module can be installed depends on the scenario, but in any case, it must be installed before demodulator/decoder blocks. The advantage of implementing the module at this PHY level is to provide flexibility in deploying the module at the CR device regardless its role, it could be a transmitter, a receiver or even a device that is only sensing and collecting data. Such installation and implementation flexibility ensure the generalizability of the proposed approach.

- it is following a data-driven unsupervised approach by allowing the radio to build up knowledge about the radio spectrum from null memory. Hence, radio's autobiographical memories are grown up incrementally by observing real-time data and learning autonomously from the cognition cycle.

- it shows that the proposed SA module has a higher degree of explainability of its own decisions: hidden variables used in the generalized model make it possible to draw explicit causal dynamic probabilistic relationships among continuous signals and their symbolic higher-level counterparts. Therefore, CR can analyse the dynamic evolution of the situation by designing more straightforwardly proactive mitigation strategies thanks to the underlying Bayesian learning and reasoning processes. On the contrary, Deep learning approaches where the hidden variables are at a sub-symbolic level and hidden variables are considered as a black box cannot provide a similar level of explainability of their decisions and create results that are hard to understand.

- it highlights how the proposed module is relying on raw *IQ* data which are easy to extract. Also, using *IQ* data provides flexibility in implementing the proposed approach in different systems and environments.

- the modulation classification problem is formulated in terms of an objective function that aims to minimize the surprise (i.e. abnormality) by testing different models learned by the radio and weighting them to select the model that causes the minimum surprise and thus that better explains the modulation scheme of the detected jamming signals.

- extensive simulations verify that the proposed functionality for the automatic jamming signal classification performs with superiority classification accuracy than LSTM, CNN and SAE and can achieve higher interpretability than Deep Learning-based models since they can explain the predictions explicitly at hierarchical levels and use the abnormality measurements and Generalized Errors as self-information to keep learning by understanding incrementally.

- it introduces the automatic modulation conversion in wireless communications, which allows an AI-enabled node to predict signals' dynamics of different modulation schemes and explain how it can be transported (converted) with minimal effort and forwarded with higher spectral efficiency.

- it proposes a novel resource allocation strategy for anti-jamming using *Active Inference* (it is the first time that *Active Inference* is adopted in wireless communications). Operating in a pure belief-based setting is one of the main features of *Active Inference* that enables speeding up the learning process and resolving uncertainty in a Bayesian optimal fashion. It also ensures a dynamic balance between exploration and exploitation. Moreover, in *Active Inference* the reliance on an explicit reward signal coming from the environment is not necessary; the reward is substituted by GEs that can be treated as self-information to avoid surprising states (i.e., states under attack) and reach the equilibrium.

## 1.3 Outline of the Thesis

- **Chapter 2** provides a general overview of CR and its related applications. Additionally, it highlights the physical security issues regarding jamming attacks.
- **Chapter 3** proposes a reasonable framework to learn SA representations of the wireless environment in various CR applications.
- **Chapter 4** introduces a data-driven SA module at the PHY of CR and discusses the corresponding SA functionalities in detail.
- **Chapter 5** designs a joint automatic jamming detection and classification framework and introduces a novel modulation conversion method.
- **Chapter 6** investigates the Active Inference theory by developing a novel resource allocation strategy for anti-jamming.
- **Chapter 7** draws the conclusions and a plan for future directions and work.

# Chapter 2

# Literature Review

In this chapter, we provided a general overview to put Cognitive Radio (CR) in its historical context in order to fully understand the interests that led to its advent at both the design and utilization phases. In addition, a general overview has been provided to highlight the physical layer security issue in CR regarding jamming attacks. Different jamming strategies (types) are discussed, along with the proposed detection, characterization and classification methods in literature. Moreover, proposed solutions are reviewed, including jammer suppression and anti-jamming.

## 2.1 Cognitive Radios

### 2.1.1 History

The marvellous evolution in digital technologies during the 1980s and '90s has enabled certain base-band processing techniques to be performed within digital integrated circuits and led to the term *Software Defined Radio (SDR)* [58]. SDRs are re-configurable transceivers whose communication functions are running as programs on suitable processors using computer language (i.e, Software) [59]. From a pure architectural point of view, SDRs differs from traditional radios in the point where the digitization can take place which is also a distinguishing factor for different types and levels of SDR itself. SDR can refer to Radio Frequency (RF) digitization in some cases, or intermediate frequency (IF) digitization in other cases or even to base-band digitization. A pure SDR should be able to modulate (at transmission) and demodulate (at reception) any communication standards on any platform by means of software. A traditional radio performs in an analog fashion all the functions of RF transceiver (e.g., channel selection, interference suppression, amplification, and baseband transposition) using dedicated hardware. In SDRs the digital processing unit perform several radio operations (e.g., frequency transposition, channel selection, demodulation, etc.) using software processing directly after sampling and filtering the RF broadband signal.

The emergence of SDRs motivates the functional objectives of CR. SDRs have considerable computational capacity but little cognitive ability [6]. A step ahead from SDRs is CR, introduced by Joseph Mitola in 1999 as an example of agent technology in telecommunications to fill this gap by combining radio transceivers with computerized intelligence to automate coordination of devices, networks, and services for improved functionality, interoperation, and spectrum utilization [60]. The term "cognitive" originates from the way biological organisms interact with their environment using a regime of goals, sensory inputs and reactive behaviour to sustain themselves.

### 2.1.2 Original Vision and Evolution

CR has attracted intensive attention from academia and industry since its introduction in 1999 [6]. While there is a rich literature on CR focusing on spectrum sensing and dynamic spectrum access, there have been few studies on the original vision of CR in being Self-Aware, which is nowadays also known as Intelligent Wireless Communications. As claimed in [29], [61] and [62], the original vision of CR goes beyond the spectrum sensing (which is, of course, one of the main components of the cognition cycle) and aims to improve Quality of Service (QoS) [63], Quality of Information (QoI) [64], optimizing the wireless users' configuration [1] (i.e., the device's ability to change and adjust automatically the operating parameters such as frequency, modulation, power output, resource allocation, antenna orientation/beamwidth, transmitter bandwidth etc.). Several recent studies in the literature tried to highlight this fact by spotting the light on the original vision of CR and focusing on new functionalities rather than spectrum sensing, encouraged by the recent advances in Artificial Intelligence (AI) methods and their effectiveness in achieving detection, classification and prediction tasks that can empower the CR realization and support it to effectuate the desired functionalities. However, many of these studies mixed between CR and intelligent wireless communications (as in [1] and [65]). We believe that they are grasping the main broader perspective of providing the radio with the capability of SA (in our definition) or the smartness and intelligence capabilities (in their definition). Nevertheless, the state of the art is still lack of studies that help to realize the original Mitola's CR and to systematize the functionalities of the cognition cycle and learn incrementally through its consecutive iterations.

After 10 years of CR, a special issue [62] has been organized to provide an overview of the achievements, developments and challenges in this technology. The special issue composed of different papers containing a lot of references (in which we suggest to refer to) dealing with three functions of the cognition cycle: sensing, learning and decision making/action. After 20 years of CR, a special issue addressing the evolution of the CR to intelligent radio and the role that AI and machine learning can play in this evolution is introduced in [66]. A brief overview of the evolution of CR and the developments on intelligent radio during the last 20 years is provided in [67]. The authors started from the perception-action cycle which represents the process of intelligent decision-making. Then they listed the main functionalities (Perception, Learning from radio environment and action) that the CR device must perform to achieve the cognition. The spectrum sensing and sharing approaches have been reviewed along with the recent achievements on the AI-enabled intelligent radio to improve the perception and action capabilities as well as the main challenges to be addressed. Authors in [1] provided a comprehensive overview of CR technology and Machine Learning (ML) focusing on their roles and relationship in achieving intelligent wireless communications. The article discussed the importance of cognition and reviewed how learning, perception and reconfiguration or adaptation (components of the cognition cycle) can be achieved by using ML techniques. The article provides a detailed overview on spectrum sensing methods which are the key for better radio perception, including traditional approaches (e.g energy detection, matched filter, etc.), local/cooperative (i.e. sensing performed at the CR-user side/sensing performed by multiple CR-users) scenarios and narrow-band/wide-band spectrum. The article detailed a plethora of ML techniques used for different tasks such as spectrum sensing, channel estimation, signal classification, power optimization, channel selection, dynamic spectrum access etc., to enhance the perception capability and reconfigurability (by specifying what to use and what to learn) and then to achieve the intelligent wireless communications. In addition, the article addresses the mission of updating the radio's knowledge based on reinforcement learning, thus the radio learns to take optimal actions. A pathway to intelligent radio is presented in [65], the authors started by reviewing the CR and the latest progress in ML and deep learning for wireless communications. Where they asked an important question of how to make CR more intelligent (which in our understanding is how to achieve the original CR). To this purpose, an intelligent radio structure is proposed based on learning and reasoning which give a radio the perception capability with reconfigurable functions. In addition, they introduce an example of how to make the spectrum sensing more intelligent and then to access and share the spectrum efficiently (intelligent actions). However, these articles did not propose an explicit framework to reach the required functionalities incrementally and did not study the link between them to make the radio grows its memories from experience to achieve the real intelligence or cognition. The work in [68] discussed Game theory and reinforcement learning as learning mechanisms that provide the radio with the capability to learn from its past actions and those of others. In addition, an insight into reasoning frameworks in CR is provided and open research issues are viewed. The article concluded by addressing several challenges that may face the introduction of learning and reasoning in CR (e.g. complexity, reasoning time, etc.). Therefore, an efficient reasoning approach must select carefully the required knowledge within the cognition cycle and intertwines reasoning with learning. In addition to the mentioned learning and reasoning frameworks, we are proposing a combination of probabilistic reasoning and learning while the radio is observing the cognition cycle. Another interesting research presented in [69] on the opportunities, challenges and future vision for the realization of a fully intelligent radio. An ML-based architecture with three hierarchical levels is proposed which enables the cognitive user to autonomously perceive, understand, and reason about the unknown environment. However, this work studied these levels independently without introducing a link among them to achieve an incremental perception, understanding and reasoning of the surrounding radio environment. In addition, using the feature as the Primary User's (PU's) transmit power might affect the performance of this method in practice (different users might have the same power level and will not allow the radio any more to identify between them).

Understanding the wireless environment is a crucial step in CR to achieve spectrum awareness and then SA [70]. This includes the ability of the radio to identify different signals (e.g. modulation classification) inside the spectrum and detect any unexpected behaviour while monitoring it. Such tasks support the radio to analyse and reason the radio environment and allow it to learn efficiently. For this purpose a deep learning method based on Long Short Term Memory (LSTM) for automatic modulation classification (AMC) and spectrum anomaly detection based on an adversarial autoencoder (AAE) are proposed in [71] and [72], respectively. Authors in [73] proposed an unsupervised anomaly detection method for the CR using LSTM mixture density networks applied to time series data. A deep predictive coding neural network for radio-frequency anomaly detection in wireless systems has been proposed in [74] where image sequences generated from the spectrum by monitoring real-time wireless signals. In [75], classification on RF spectrum modulations and detection of radio frequency anomalies in radar systems is implemented by using Convolutional Neural Networks (CNNs) trained on waveform images. They proposed two techniques that use the activations of the last hidden layer of the CNNs to detect anomalies. While in [76], autonomous detection of electromagnetic spectrum anomalies based on spectrum amplitude probability and Hidden Markov Model (HMM) has been proposed. The training process estimates the HMM parameters for different models while the testing process decides which abnormality pattern the testing data belongs to. In [77], authors apply the deep-structure auto-encoder neural networks to detect signal anomalies. Signal time-frequency features are used to train the auto-encoders network, which acts as a one-class classifier, relying on the reconstruction error of the network to decide whether the signal is anomalous or not. An interference mitigation algorithm based on neural networks and spectral correlation for wide-band radios to detect and classify signals with different modulation schemes is presented in [78].

### 2.1.3 Cognition Cycle

The cognition cycle depicted in Fig. 2.1 by which CR might interact with external radio environment was first introduced by Joseph Mitola. CR as an example of agent technology in telecommunications can observe the radio environment continually, orient itself, create plans, decide, and then act [79]. According to Mitola's vision, the CR including the wireless personal digital assistants (PDAs) and the related networks are sufficiently computationally intelligent about radio resources and related computer-to-computer communications to detect user communications needs as a function of use context, and to provide radio resources and wireless services most appropriate to those needs [80].

On the other hand, Simon Haykin defined the CR as an intelligent wireless communication system built on a software-defined radio for complete reconfigurability, that is aware of its environment and uses the methodology of understanding-by-building to learn from the environment around it and adapt to statistical variations in the input stimuli [81]. Thus, a simplified cognition cycle is introduced by Haykin as represented in Fig. 2.2.

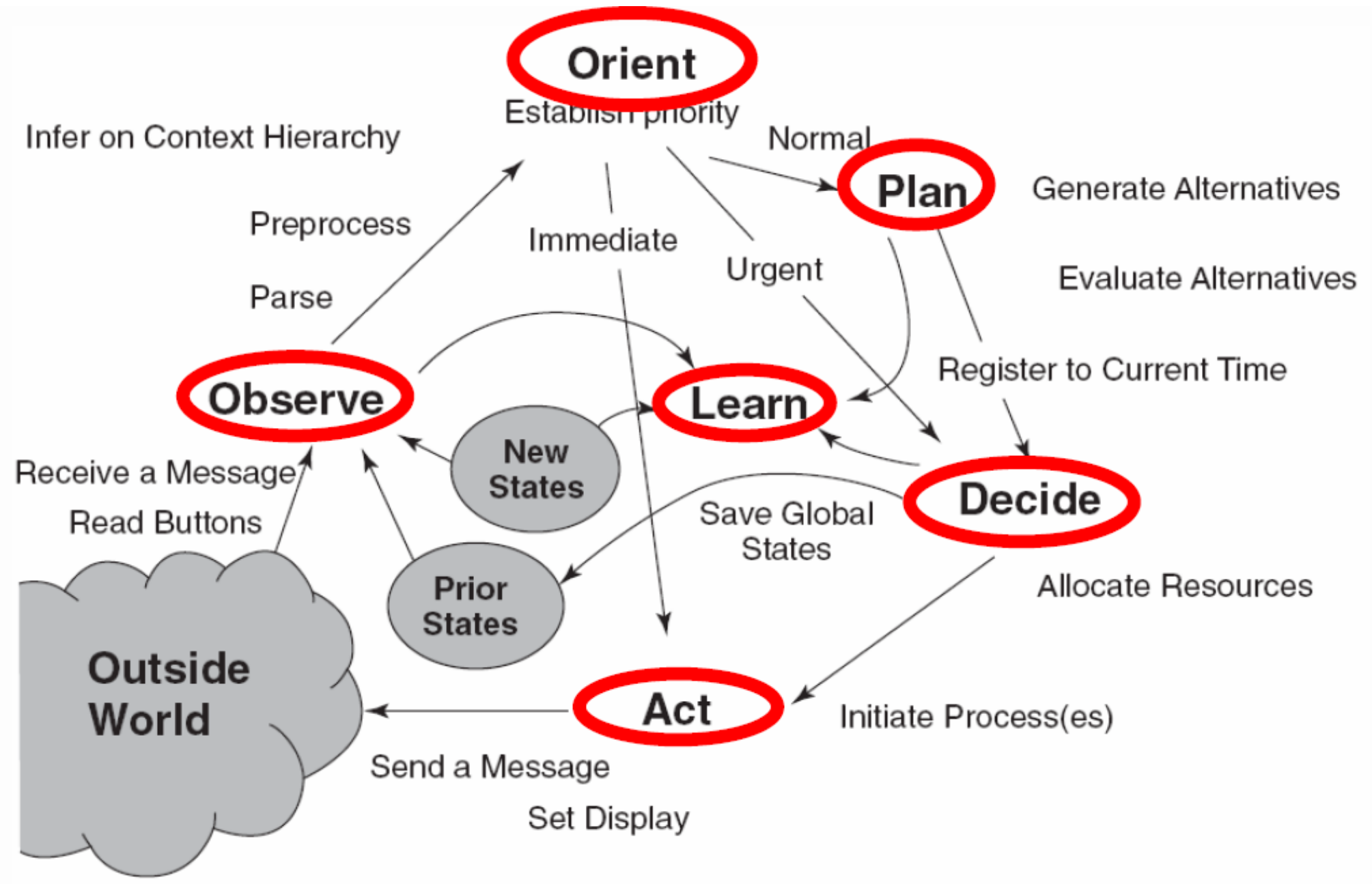

Fig. 2.1 Mitola's cognition cycle.

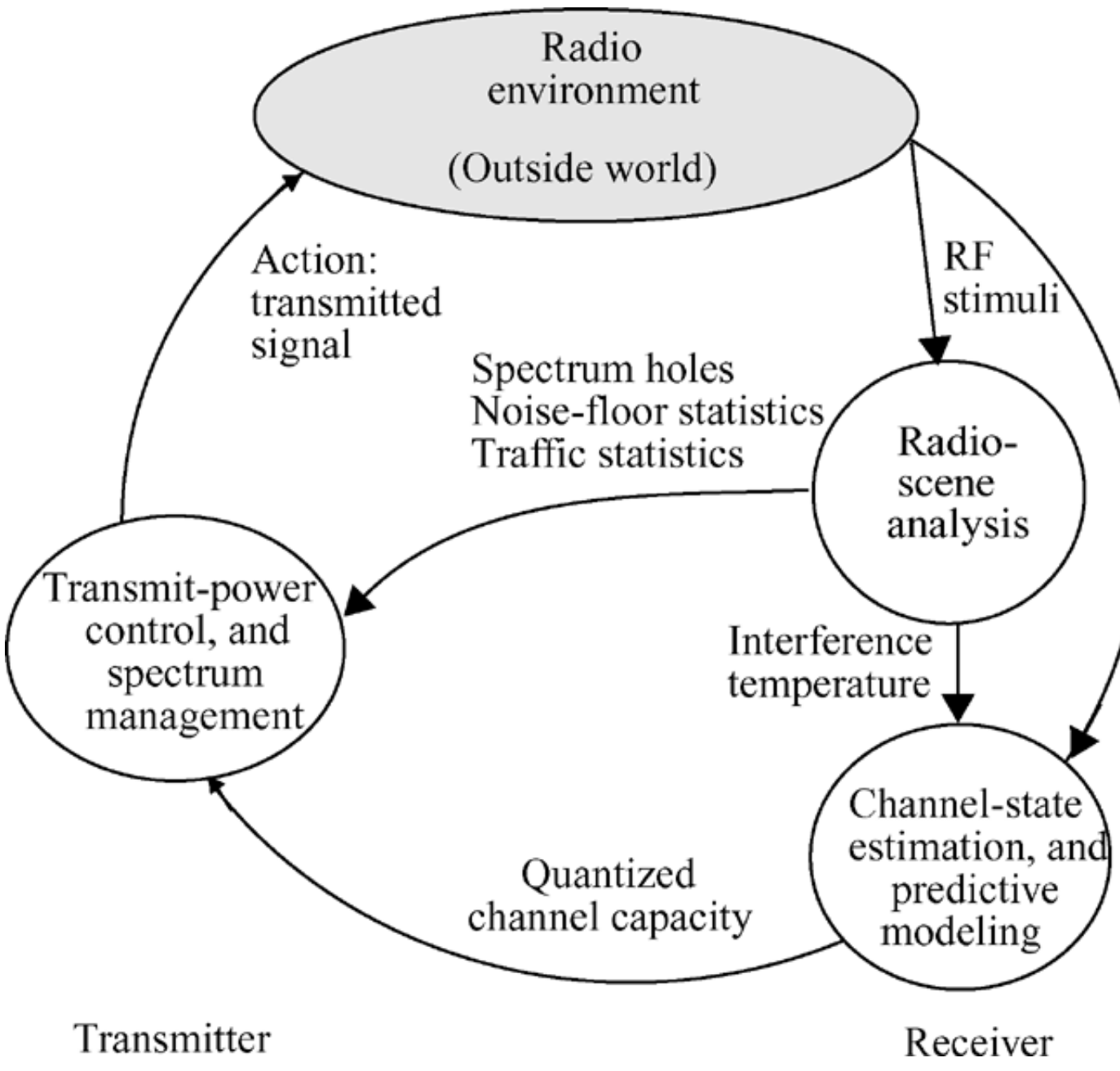

Fig. 2.2 Haykin's Cognition Cycle.

According to Mitola, the CR sequentially observes the radio environment by sensing and perceiving the RF stimuli dispatched to the cognition cycle for a response. The CR starts by orienting itself to determine the significance of observation through binding the observation to a previously known set of stimuli. After that, the CR creates plans by analysing and reasoning over time. Later on, the CR can make decisions by selecting among the candidate plans which depend on the task the radio is already oriented to achieve and finally to act by initiating the selected plans using the actuators that access the external world or the CR's internal states. This loop (observe, orient, plan, decide and act) revolves around learning (the most critical capability in CR) and updating the acquired knowledge by precisely understanding the radio environment.

On the other side, Haykin focused on three on-line cognitive tasks: 1) Radio-scene analysis, by estimating the interference temperature of the radio environment to detect the opportunities (i.e. spectrum holes); 2) Channel identification, by estimating the channel-state information and predicting the channel's capacity; 3) transmit power control and dynamic spectrum management. The cognition process starts with the passive sensing of RF stimuli and culminates with action. The tasks mentioned earlier 1) and 2) are carried out at the receiver side, while task 3) is carried out by the transmitter. Moreover, the cognitive module in the transmitter must work harmoniously with the cognitive modules in the receiver.

To sum up, Haykin defines the behavior of the CR system through a simplified Cognition Cycle, similar to Mitola's one but much more clustered in macro-phases. Mitola is much more interested in the impact of cognitive capabilities on the communications market. Haykin faces the problem from a more general point of view. Conversely, both types of research agree on the fact that the SDR systems are the natural platform for the implementation of CR devices. While in Mitola's vision, the CR is suited to realize the user's preferences, in Haykin's one, it is well explained as a cognitive communication between a transmitter and a receiver. In both of the previous visions, it is clear the effort to model the CR as an entity that can reason about and analyze the external world in order to modify its internal configuration to reach the best solution.

### 2.1.4 General Comparison with existing work

Compared to some selected papers from the literature introduced and discussed in Section 2.1.2, in this thesis we are proposing a Self-Awareness (SA) module which organizes the main functionalities (Table 2.1) of the cognition cycle and ensures an effective relationship among them allowing by that the radio to learn in an incremental fashion and from experience (by observing the cognition cycle). In this work, the radio is focusing on the goal of enhancing the physical layer security. However, the proposed module can be adapted to reach any goal oriented by the radio ensuring its generalizability to different radio applications.

| **Functionalities** | [1] | [65] | [67] | [68] | [69] | [71] | [72] | [73] | [74] | [75] | **This Work** |
|---|---|---|---|---|---|---|---|---|---|---|---|
| Learning from Radio Environment | ✓ | ✓ | ✓ | ✓ | ✓ | ✓ | ✓ | ✓ | ✓ | ✓ | ✓ |
| Radio Spectrum Perception | ✓ | ✓ | ✓ | ✓ | ✓ | ✓ | ✓ | ✓ | ✓ | ✓ | ✓ |
| Abnormality Detection | - | - | - | - | - | ✓ | ✓ | ✓ | ✓ | ✓ | ✓ |
| Abnormality Characterization | - | - | - | - | - | - | - | - | - | - | ✓ |
| Action Selection | ✓ | ✓ | ✓ | ✓ | ✓ | - | - | - | - | - | ✓ |
| Incremental Learning | - | - | - | - | - | - | - | - | - | - | ✓ |

Table 2.1 General comparison with the state-of-the-art. Publications that investigate the original vision of CR and recommend the ideas to achieve such vision without proposing an explicit framework are shaded.

### 2.1.5 The role of CR in 6G

The Fifth-generation (5G) technology was envisaged to meet the necessitated requirements supporting both Human Type Communications and Massive Machine Type Communications [82]. The deployment of 5G realizes the vision of *connecting everything* and evolves towards enhancing its capabilities to stilt this lofty vision. Building upon the 5G vision, the Sixth-generation (6G) will realize the concept of *connecting every intelligent thing* by integrating AI in wireless operations from the core to the edge optimizing system performance and enabling AI radios [83, 84], thus, making the ideal Cognitive Radio (CR) a reality after 20 years of effort. Indeed, some of the glimpses of 6G were rendered in [80] by Joseph Mitola, who formalised the concept of CR, and it is believed that Mitola's CR will reach full potential in the 6G era [7]. According to [85], 6G will be established on ubiquitous AI to achieve data-driven Machine Learning solutions in heterogeneous and massive scale networks. Thus, it is expected that CR and AI-enabled radios will play a significant role in the future 6G.

## 2.2 CR related wireless applications

As an intelligent and hybrid technology involving smart transceivers and SDRs, CR can be integrated with different radio applications and inter-operate with different wireless communication standards. In this section, we present different radio applications with which the CR can coexist and operate, including Internet of Things (IoT), millimeter Wave (mmWave) and Unmanned Aerial Vehicles (UAVs), where we highlight at the end of each subsection the physical layer security issue regarding jamming attacks.

### 2.2.1 CR-based Internet of Things

Recent developments in information technologies and machine to machine communications (Known as Machine-Type Communications MTCs), brought a new technology which is deployed to intelligently connect different objects in a network, (i.e. IoT). IoT is a pervasive paradigm in the information and communication field connecting a wide variety of ubiquitous objects [86]. Typical objects in IoT networks are sensors, actuators, mobile phones and other devices, which are equipped with powerful data capabilities and use standard protocols to access any service at any time by using ideally any available path, service and network. IoT is an emerging technology which is penetrating into many fields such as industrial manufacturing, logistics process, transportation, health care, automation, and many more [87]. It is shaping real-world assets by providing unique features to remain interconnected with anything at any place and time by using any available network [88]. In IoT, objects are either connected through wired or wireless networks, however, wireless networks provide cost-effective and remote access solution as compared to the wired network [89].

IoT objects might generate massive data as many devices intend to utilize the spectrum and exchange information in order to remain connected and access services. This increases spectrum resources requirements and creates a bottleneck in the IoT network and gives rise to the problem of spectrum scarcity especially if a static spectrum allocation policy is adopted [90]. Hence, technological trends are shifting to integrate CR into IoT to eradicate the spectrum scarcity problem [91]. In this scenario (CR-IoT), IoT objects act as secondary users to opportunistically access the primary user's spectrum whenever this is free [92]. It is expected that after incorporating CR in the IoT network, IoT devices will be supplied with cognitive capabilities to deal with spectrum scarcity problem and to refrain from insufficient utilization of spectrum [32, 93]. Spectrum measurements show that most of the time, the radio spectrum is not being used by the licensed users and remains vacant. CR exploits vacant spaces to enhance spectrum utilization by assigning unused spectrum to other unlicensed users in the network. CR has been implemented in many applications such as mobile communication, wireless sensor networks, and radio-based smart grids [94].

CR makes IoT devices intelligent by allowing them to sense, learn and adopt the best possible transmission strategy in a given operating environment. However, for CR to achieve the aforementioned objectives and execute the required tasks, the physical layer needs to be highly flexible and adaptable through an adaptive physical layer modulation technique. To this purpose, Orthogonal Frequency-Division Multiplexing (OFDM) has been recognized as an attractive transmission modulation technique for CR systems [95]. OFDM, due to its unique features, is a broadly used technology and has become dominant in many existing wireless and broadcasting standards such as Wireless Local Area Network (WLAN) standards (IEEE 802.11a, IEEE 802.11g, IEEE 802.11n, IEEE 802.11ac, IEEE 802.11ad), IEEE 802.22 Wireless Regional Area Network based on CR in TV White Spaces (TVWS), Long Term Evolution (LTE) mobile network, LTE-Advanced and it is also considered as one of the key technologies in 5G Radio Access Network (RAN) [96]. Moreover, to provide larger coverage area with the low-cost operation for thousands of connected objects in IoT, IEEE 802.11ah has been proposed in recent years which employs OFDM [97]. In OFDM, all the elements of the time-frequency grid can be scanned without any extra hardware by exploiting the Fast Fourier Transform (FFT), which simplifies the sensing process of the operating spectrum. Furthermore, OFDM can be adapted to different transmission environments by simply changing some parameters as FFT size, modulation and transmission power, achieving good adaptation and scalability. Moreover, since OFDM modulation is successfully implemented in various wireless standards as mentioned above, it will be easy for CR employing OFDM to inter-operate with such technologies [95].

Although the use of CR has already been advocated for many IoT applications as an imminent solution to overcome spectrum scarcity and reach smartness. However, CR-IoT network suffers from various kinds of malicious attacks. Such kind of attacks aims to disrupt communication, deplete the bandwidth and seize transmission. Therefore, to achieve CR-IoT network objectives, providing secure communication under jamming attacks is a basic yet challenging task [98].

### 2.2.2 CR-based mmWave

The high congestion levels in the lower parts of the electromagnetic spectrum caused by the exponentially increasing traffic demand and the tremendous wireless connectivity in IoT and Vehicle to Everything (V2X) applications are forcing the telecommunications industry to explore new frequencies within the mmWave band [99–101]. The mmWave provides sizeable available bandwidth at high frequencies which operate in the range of 30 to 300 GHz, offering low latency and high-speed data connection [102, 103]. Such frequencies impose several limitations due to the fact that the signal will suffer from high propagation loss and get distorted due to raindrops and humidity absorption as well as its sensitivity to blockages, making the implementation of the mmWave communications possible to a few kilometres in small cells and heterogeneous networks which are efficient to serve the IoT and V2X scenarios [104, 105]. Besides, the 5G technology will provide a system structure for these emerging V2X and IoT applications that require high reliability and strict delay for secure message delivery between transmitters and receivers which impose the need of an efficient hybrid access scheme for licensed and unlicensed spectrum in mmWave band. Thus, CR has been proposed to manage the dynamic spectrum access in mmWave communications allowing secondary users to sense frequently and access opportunistically the mmWave channels which are not in use by the primary licensed users and without damaging the quality of service [14], [106–109].

However, since the propagation characteristics of mmWave are different from those of lower frequencies (e.g., sub-6GHz), analyzing the statistical characteristics of the received/sensed signals is a bit challenging and requires advanced spectrum sensing methods [110]. In addition, physical layer security in Cognitive mmWave Radios has attracted broad interest to achieve secured communications that involve multiple signal transmissions due to the shared wideband spectrum and the coexistence in tight integration with different wireless systems [16, 37]. Such open access medium and dynamic environment makes the system vulnerable to malicious users that aim to manipulate the radio spectrum by injecting anomalous signals and enforce the system to learn wrong behaviours that lead the radio to take mistaken actions [38, 30]. Autonomous learning is a crucial component in CR system to adapt to the perceived wireless environment and potentially maximize the utility of the available spectrum resources and allow the radio to take an optimal decision and act efficiently [1]. Therefore, precise detection of spectrum anomalies is crucial to enhance the physical layer security and improve the system's performance.

### 2.2.3 CR-based Unmanned Aerial Vehicles

Recently, UAVs have attracted extensive attention to the telecommunication community and industry due to its remarkable features as deployment flexibility, high mobility and dominant Line Of Sight (LOS) links facilitating wireless broadcast and supporting high data rate transmissions [111–113]. UAVs are already being studied for 4G LTE [114] and they are expected to play a vital role in the upcoming 5G technology as mentioned in the content approval of Release 17 produced by the Third Generation Partnership Project (3GPP) [32, 115]. UAVs can be deployed as flying Base Stations (BSs) for improving reliability, coverage and capacity of terrestrial networks [116], or flying gateways/relays to support V2X services [117] and IoT [118] devices or even as aerial user equipment by connecting them to the cellular system [119]. The explosive number of wirelessly connected UAVs adopted in a wide range of applications including the package delivery, environment/traffic monitoring, surveillance and sanitization during COVID-19 will overcrowd the radio spectrum resources and lead to spectrum scarcity [120–122]. According to the Federal Aviation Administration (FAA) report, the fleet of connected UAVs will be more than doubled from an estimated 1.1 million in 2017 to 2.4 million units by 2022 and further growth is expected by enabling a myriad of applications in the coming years [123]. Furthermore, since UAVs are battery-powered, they will face another challenge regarding the energy efficiency [124]. Therefore, for a successful deployment of UAV-based wireless communications, different issues regarding power consumption, physical layer security, communication capability and path planning must be addressed [119, 125]. Incorporating CR in UAV communications which we refer to as Cognitive-UAV-Radios has been proposed to increase the spectrum efficiency [121, 126], improve the UAV's communication capability [127], optimize the energy and power consumption [17, 128, 124], maximize the signal-to-interference plus noise ratio [129] and for cognitive channel modelling [125]. In this thesis, the integration of CR and UAV communications is studied from the physical layer security perspective.

Due to the radio propagation and broadcasting nature, CRs are vulnerable to jamming attacks [31]. Moreover, the situation could be worse in CRs, where a smart jammer with cognitive abilities can estimate system parameters and manipulate radio spectrum, thus forcing CR to learn wrong behaviours and take non-optimal actions. In addition, due to the dominant LOS communication links, UAVs are more vulnerable to terrestrial jammers [32]. Manual and autonomous UAVs can communicate with ground control or satellite system to complete a specific mission. A jammer can intercept to launch malicious signals reducing the SNR at the RF receiver and preventing the UAV from responding to the received commands from the ground control and thus neutralizes its operation (i.e., it might be turned off remotely, hijacked, flown away or stolen) [122]. Jamming attacks on CRs and UAV communications have been widely explored in literature [30, 31, 130–132] to enhance the physical layer security and improve the performance. However, the state of the art is still lack of advanced methods for accurate and timely detection of the jammer and its anomalous behavior as well as the development of anti-jamming strategies to protect the Cognitive-UAV-Radios effectively.

## 2.3 Physical Layer Security

Security has always been a common problem in wireless communications because of the open-access nature of the wireless medium. In this broadcasting nature, unauthorized entities such as eavesdroppers can easily intercept the transmitted signals, decode them and extract sensitive information in order to damage the ongoing communication between legitimate terminals. A plethora of works have been done in the literature employing traditional cryptographic techniques at the upper layer of the network stack to address the security needs [133]. However, cryptographic techniques may not be sufficient to protect transmitted signals of all network scenarios due to the stringent constraints of computational power, memory and communication rates. Physical Layer (PHY) security has emerged as a practical solution that well suits dynamic networks and distributed processing techniques by exploiting the dynamic channel characteristics and randomness of noise to limit the amount of information that unauthorized users can evoke [134]. On the other hand, wireless networks are susceptible to a severe class of security threats known as jamming attacks at the PHY that needs to be adequately addressed via advanced methods allowing to detect and understand the type of jammer present in the wireless environment and evaluate its effect on the network [135]. Due to the fact that jammers can simply ignore any medium access protocol and wireless standard requirements to launch intentional radio interference continually, it is not sufficient to address such attacks through conventional security mechanisms. To better understand this problem, we need to analyse and discuss various jamming strategies, jammer detection and classification methods, as well as anti-jamming solutions.

### 2.3.1 Jammer Types

Launching RF attacks by a jammer can be accomplished in different ways. Consequently, various levels of effectiveness can be distinguished to select the proper detection tactic for each attacking model. The following are the most commonly jamming strategies studied in literature that have been proved to be effective in damaging wireless systems [122, 136–140].

- *Constant jammer:* The constant jammer follows a predefined attacking strategy by injecting continually electromagnetic radio signals generated from a random sequence of bits that interfere with the legal transmission on one or more particular channels. A drawback of this type is the high power consumption due to the continuous emission of signals.

- *Deceptive jammer:* The deceptive jammer uses a legitimate bit sequence to inject regular signals continually and deceive legitimate users into believing that regular transmissions are propagating through the channel.

- *Random jammer:* unlike deceptive and constant jammer, random jammer saves energy by alternating between sleeping and jamming states. It will attack for $t_a$ time and enters the sleeping mode by turning off its radio for $t_s$ after which it can resume jamming. During sleep mode, the random jammer does not use energy which reduces the power consumption. During jamming mode, it can either behave like a deceptive or constant jammer. Adjusting the sleeping and jamming times allows to reach the trade-off between energy efficiency and jamming effectiveness.

- *Reactive jammer:* A reactive jammer can conserve energy by avoiding attacking the channels continuously like the random jammer and by avoiding transmitting interference signals whether or not there is communication activities inside the spectrum as Proactive jammers (i.e., Constant, Deceptive and Random) do. However, it senses the spectrum and reacts by injecting disturbing signals as soon as it detects active transmissions by legitimate users. Reactive jammers require more power for jamming than sensing the spectrum (or listening to the channels). Reactive jamming strategy is harder to detect compared to constant, deceptive and random strategies.

- *Smart jammer:* The smart jammer is power efficient and effective due to its ability to analyse the transmitted signals, extract their properties, and identify critical points to deny successful communication. Thus, smart jammers are protocol aware that can be developed using SDRs and then generating jamming signals similar to the targeted signals in terms of hopping patterns, Pseudo Noise (PN) code data rate and modulation techniques.

### 2.3.2 Jammer and spectrum anomaly detection

CR uses some of the characteristics of the radio network to address security challenges; however, traditional radio networks may differ from one another with respect to the different strategies that are being adopted by each network to mitigate malicious attacks. This variation in using different strategies comes from the fact that each network is exposed to a dynamic environment, and the CR network is more vulnerable to security threats as compared with other radio networks due to its unique features. CR network attacks include Primary User Emulation (PUE), Spectrum sensing data falsification, Denial of Service (DoS) and spoofing attacks. Jamming attacks are one of the most frequent and effective forms of DoS and are considered to be the most menacing that disrupt the communication and reduce the bandwidth of the CR network [141].

Detecting the existence of a jammer and its anomalous behaviour in the radio spectrum realize the first essential stage towards defending the network. Jammer detection is a challenging task due to the different attacking models that the jammer can follow to interrupt the wireless communication and due to the difficulty to discriminate jamming situations from network conditions as congestion, similar situations as low signal to noise ratio, battery running out of power, or a receiver moving out of transmission range [137]. In this section we review various detection methods based on conventional and advanced approaches.

Conventional jammer detection methods are mainly based on Energy detector (ED), Match Filter Detector (MFD) and Cyclostationary Feature Detector (CFD) [142]. ED is one of the most straightforward approaches for deciding between two hypothesis, i.e., the presence or absence of the jammer inside the spectrum based on the estimated energy of received siganls [143]. MFD can improve the detection performance compared to ED by maximizing the Signal to Noise Ratio (SNR) under additive noise if the jamming signal is deterministic and known a priori. It is based on correlating the known signal with the unknown received signal to compare the output with a predefined threshold and decide whether the received signal is corrupted by jamming interference or not. MF is not suitable for jammer detection in some conditions (e.g., low SNR conditions) due to the difficulty in achieving the synchronization. CFD exploits the inherent periodicity of the jamming signals and uses a second-order cyclostationary process to characterize the modulated signal, and the detection task boils down to a binary hypothesis testing as in ED and MFD. CFD has the advantage of performing signal classification and distinguishing co-channel interference compared to ED. Since CR has limited knowledge about legitimate and non-legitimate users present in the spectrum, ED is more favourable while MFD and CFD are not well suited. However, due to its simplicity and sensibility to noise, ED might not perform well in complex and dynamic environments. Thus, advanced methods based on machine learning are needed to achieve better performances.

In CR, it is imperative to learn legitimate and non-legitimate user's behaviours to make inferences about their states inside the spectrum by exploiting certain statistical properties related to the dynamic behavior of each user through a learning process [144]. In this way, user's activities in the spectrum as time evolves can be predicted and jammer's behaviour can also be exploited in a much more vivid way. In this perspective, several methods have been studied and proposed to detect such attacks in CR-IoT network including signature-based method and Anomaly-Based Detection (ABD) [109]. An ABD method has been studied and proposed to detect malicious attacks by using machine learning techniques in [145]. For ABD system, CR network uses various features of the signals such as SNR, traffic flow, signal modulation, Packet Delivery Ratio (PDR), sensing threshold and Signal Strength (SS) to learn the behavior of the users under normal and jamming conditions [92]. In [141], authors used SS and PDR to implement ABD technique. Learning of the network is accomplished by using SS and PDR under no jamming conditions in learning phase. During the testing phase, jammer detection is done by comparing the normal and abnormal situations by using the baseline profile. The work in [146] provides more comprehensive solution to detect jammer in the CR network. However, it is not applicable for IoT applications because it employs many nodes to perform ON/OFF line monitoring and anomaly behavior analysis. In [147], authors provided anomaly detection method to countermeasure the effects of a jammer in CR network. The proposed method is not well suited for the CR-IoT network since it involves many nodes in data processing to perform anomaly behaviour analysis. In [148], authors proposed anomaly-based detection method which uses K-mean clustering for a mobile network. In [149], deep auto-encoders are deployed to identify abnormal signals in wireless spectrum whereas, machine learning based anomaly detection method is formulated in [74]. For practical IoT systems, machine learning based security techniques are explored and discussed in detail in [105].

Spectrum anomaly detection has been explored in literature. However, it does not provide exhaustive work based on ML and DL techniques, making it still a challenging task. Authors in [73] proposed an unsupervised anomaly detection method for the CR using long-short-term memory mixture density networks applied to time series data by considering the In-Phase (I) components and discarding the Quadrature (Q) components of digital radio transmissions. Such networks are able to learn rapidly from the training set and produce probability distribution functions for the expected signal as a function of time where the loss function metric allows anomalous signals (as jammers) to be detected. However, discarding the Q components will impose some limitations on analysing how the signal dynamics are changing with time at both I and Q channels and cause confusion in identifying one sample from the other by looking only at the I component (e.g., two samples might have the same I with different Q values). In addition to the manual setting of the threshold which may not exhibit perfect performance. The use of an adversarial auto-encoder model in wireless spectrum data anomaly detection for dynamic spectrum access in CR is analysed in [71] by using power spectral density data. Where authors explored the model's capabilities to learn interpretable features for compression and signal classification purposes, this model achieves perfect detection and classification accuracy by working on features as Power Spectral Density (PSD), signal bandwidth and center frequency even though the extraction of such features may require an additional effort that could be inconvenient in the IoT and UAV scenarios. A deep predictive coding neural network for radio-frequency anomaly detection in wireless systems based on video frame predictor has been proposed in [74] where image sequences generated from the spectrum by monitoring real-time wireless signals. Additionally, the CNN trained on waveform images is investigated for Cognitive Radar System in [75] to detect anomalies in the RF spectrum by using the activation of the last hidden layer of the CNNs. The methods proposed in [74, 75] are based on video frame predictor and learning from images, respectively; this requires the generation of video frames and waveforms images that can be unfeasible in specific scenarios like IoT and UAV (i.e., to elaborate and process such type of data), because of the battery and power consumption limitations. In [76], autonomous detection of electromagnetic spectrum anomalies based on spectrum amplitude probability and Hidden Markov Model (HMM) has been proposed. The training process estimates the HMM parameters for different models while the testing process decides which abnormality pattern the testing data belongs to. In [77], authors apply the deep-structure auto-encoder neural networks to detect signal anomalies. Signal time-frequency features are used to train the auto-encoders network, which acts as a one-class classifier, relying on the reconstruction error of the network to decide whether the signal is anomalous or not. An interference mitigation algorithm based on neural networks and spectral correlation for wide-band radios to detect and classify signals with different modulation schemes is presented in [78]. In [111], scaling deep learning models are built to capture spectrum usage patterns and use them as baselines to detect LTE spectrum usage anomalies resulting from faults and misuse. An adversarial auto-encoder using interpretable features as power spectral density (PSD) data is proposed in [72] for wireless spectrum anomaly detection. In [111], and [72], the data is relative to narrow ranges of frequencies and is represented by bidimensional spectrograms. Besides, anomalies are not related to changes in the dynamics of the signals. Indeed, in the abnormal spectrum, there is either an additional signal or the signal is corrupted concerning the normal situation. The aforementioned citations [74, 76, 72, 77, 78, 75] played an important role in implementing AI-based techniques in CR to detect abnormalities inside the radio spectrum. However, these approaches consider only a certain level at which the AI method is applied at the physical layer and deal with a certain type of data dimensionality. In this thesis, we will propose a framework that can be implemented at different physical layer levels and deal with different data dimensionalities, providing by that a flexible and generalizable framework which mimic the real-world applications and the dynamic nature of CRs.

### 2.3.3 Jamming signal classification

In complex electromagnetic environments as in CR, communication links are prone to different kinds of network interference and jamming attacks. It is significant for a CR to sense the spectrum and identify the type of attack it has already detected to change the transmission parameters adaptively.

The traditional jamming signal classification methods include Likelihood-based and Feature-based methods. The former requires building a mathematical model based on prior knowledge and then calculating the jamming signal's likelihood function which can be compared with a certain threshold to recognize the jamming type [150]. However, this approach is computationally complex due to the required mathematical modeling and it is inapplicable in real radio applications such as CR since it needs prior information about the jammer. The latter approach requires extracting jamming signal features to design the classifiers based on supervised or/and unsupervised machine learning techniques. Usually, jamming signal features are extracted from the time domain, frequency domain, time-frequency domain, polarization domain etc. [151, 152]. In particular scenarios it is necessary to characterize the jammer in order to extract the proper features and consequently learn the jamming models. Artificial features are hard to extract and require some effort from external experts and the use of additional material resources. Thus, deep learning has been used and implemented to learn discriminative and invariant features automatically from the sensed or received data. The work in [153] focuses on designing a CNN to classify audio, narrowband, sweep, and spread spectrum jamming modes and using different jamming signal parameters as carrier frequencies and bandwidth which makes the proposed approach limited to specific radio cases. In [154], the problem of unknown interference (such as jamming interference) classification is solved by using the Hilbert transform on the received signal and projection approximation to estimate the interference. Then, the classifier is learnt in an unsupervised way by using the approximated interference on the Self Organizing Map neural network. However, Hilbert transform and projection estimation increase computational complexity of the classifier. The Radio Frequency Adversarial Learning (RFAL) framework has been proposed in [155] to build robust system identifying rogue RF transmitters by implementing the generative adversarial net. First, RFAL detects the trusted transmitters from the rogue ones then a neural network is used to differentiate among multiple trusted transmitters. In [156], a jamming identification scheme has been proposed using Naive Bayes classifiers.

However, the proposed approach requires labelled features during training which may not be feasible since it is expensive and needs domain expertise, and this realizes a common problem in supervised ML methods. In [150], a method is proposed to classify jamming signals using three sub-networks: 1D-CNN to extract deep features from time-frequency spectrograms, 2D-CNN to extract time-frequency deep features and a fusion network to fuse the extracted features from 1D-CNN and 2D-CNN. In addition, a soft label smoothing is proposed to alleviate the problem of over-fitting and improve the generalization ability. A jammer identification methodology is proposed in [157] and includes a pre-processing step to obtain multi-resolution images that can be used as input to the classifier. The Support Vector Machine (SVM) is used for classifying spectrograms images and compared with the Deep CNN (DCNN), which is used to extract features of the transformed signals automatically and classify them consequently. A novel approach for global navigation satellite system jamming classification is proposed in [158] utilizing transfer learning from the imagery domain and considering the jamming signal power spectral density, spectrogram, raw constellation and histogram signal representations as images. Transfer learning is applied to avoid training the neural networks from scratch and various ML classifiers are evaluated including SVM, Logistic Regression and Random Forest. The methods in [150, 158] and [157] rely on spectrograms images which are difficult to extract and might not be suitable for IoT, UAV and V2X applications.

### 2.3.4 Jammer Characterization and Suppression

Jammer detection and jammer classification are essential steps towards mitigating jammer's impact on system performance. The former allows to detect the presence of the jammer and to locate it inside the spectrum, while the latter allows identifying jammer's type. Moreover, jammer characterization is an important stage to protect the communication system and design an anti-jamming or interference mitigation strategies after extracting crucial features about the jammer. It allows extracting jammer's characteristics as position in the ground (for terrestrial jammers), jamming power, sensing time (i.e., the time required by the jammer to sense the spectrum and design the jamming strategy), modulation scheme, and the type of information (e.g., data sequence of random bits or legitimate bits) from which the jamming signal has been generated. In this section, we review several related methods on jammer characterization and jammer suppression from literature.

A jammer characterization algorithm is proposed in [159] for in car jammer with high jammer-to-signal ratio (JSR) using time-frequency analysis. The short-term time-frequency Renyi entropy is used to compute jammer model parameters from which the Instantaneous Frequency (INF) estimates are generated. The improved INF estimates are then used by the notch filter for jammer removal. In [160], a multi-scale model is employed to characterize jamming signals and design a new class of adaptive notch filters. The multi-scale notch filter is able to cope with state transitions of the jamming signal by tracking the fast frequency variations of the jammer and mitigating jamming interference. The work in [161] proposes a framework for estimating the modulation characteristics of multiple jammers (by exploiting the sawtooth patterns) and there respective locations using the S-transform (i.e., a specific type of time-frequency transforms). The work in [162] addresses the suppression of frequency modulated jammers considering the case of sparsely signals and compressed observations. The proposed method is based on estimating the INF of jamming signals using structure-aware Bayesian compressive sensing to exploit the contiguous nature of jammer time-frequency signature and enhance the suppression. Different applications of time-frequency signal representations were briefly described in [163]. The role of time-frequency analysis techniques for the separation of desired and undesired signal components was studied, and consequently, the jammer suppression is evaluated.

In general, basic features as magnitude and frequency of the jamming signals are unknown and dynamic which require tunable filters capable of tuning the frequency to suppress the dynamic interference consequently. Authors in [164] proposed a fully autonomous jammer suppressor by using an absorptive band-stop filter (ABSF) and an open-loop tuner to tune the ABSF. However, [164] is based on designing a filter to mitigate jammer's interference in a pure hardware manner which is not suitable in CR scenarios. In [165], the randomization of preamble technique in OFDM based 802.11 system is presented to mitigate jammer attacks. The work in [166] proposes a jammer interference mitigation scheme to avoid the jammer effects in OFDM communication system. The implementation of additional FFT module in parallel of OFDM block with wider window to reduce jamming effects in IEEE 802.11 Wi-Fi system is introduced in [167]. The effectiveness of interference mitigation techniques using adaptive notch filters while dealing with frequency hopped tick jammers is analysed in [168]. In [169], a modified Q-learning is proposed for jammer mitigation in the CR networks. Moreover, [169] engages many tiny nodes in learning jammer behaviour which raises energy-constrained issues.

### 2.3.5 Anti-Jamming

All wireless communications are extremely prone to jamming attacks due to the openness of wireless channels and the inherent broadcast nature [170]. Physical layer security has received extensive attention to fight against jamming attacks [171]. Spectrum spreading technology is one of the conventional anti-jamming protocols used widely, among which frequency hopping spread spectrum (FHSS) [172] and direct sequence spread spectrum (DSSS) [173] are proposed to reduce the effectiveness of jamming attacks. However, the inevitable reliance on the pre-defined and pre-shared secrets such as the hopping sequences in FHSS and spreading codes in DSSS between the communicating node pairs prior to communication make these techniques significantly limited [174, 175]. In addition, FHSS and DSSS are considered to be spectrally inefficent since they require a wideband spectrum [176]. Thus, developing anti-jamming schemes that are spectrally efficient through the optimal resource allocation is of great importance to obtain effective use of resources under jamming scenarios. Furthermore, the development of AI and SDR technology has raised new trends related to diversity, dynamics and jammers' intelligence that requires advanced methods for communication anti-jamming [177]. Resource allocation for anti-jamming and waveform reconfiguration can be accomplished in multi domains, namely, power domain, frequency domain (or spectrum domain) and space domain. In the power domain, jamming attacks can be avoided by adjusting the transmission power. Nevertheless, increasing the transmission power leads to high power consumption and deteriorate the linearity of power amplifier affecting the performance of some modulation formats that are sensitive to that linearity. Channel switching can be adopted in frequency domain to beat the jammer. But, switching among different channel may cause performance loss due to the time needed to re-settle the communication link in addition to the different propagation characteristics of diverse frequencies that might cause other problems as well such as throughput reduction. Therefore, a trad-off between the adopted strategy and its side-effects on power consumption, delay and other signal processing aspects is necessary during the development phase.

Recent advancements in ML and DL have pointed out a new research direction for anti-jamming design. Particularly the Reinforcement Learning (RL) methods which is a major branch of ML allowing to deal with uncertain and unknown jamming pattern models in dynamic wireless communication systems [178]. A multi-agent RL framework was introduced in [177] to make online channel selection to tackle malicious external jamming and avoid internal mutual interference among sensor nodes in the network. A time-domain anti-pulse jamming method based on RL is proposed in [179]. The method is based on learning continuously from the dynamic interaction with the jamming environment to achieve an optimal timing strategy enabling the transmitter to keep silent in time slots with high probability of pulse jamming and active in other time slots thus maximizing the system throughput. However, RL is inefficient and hard to converge if the state or/and action spaces are large. Thus, Deep Neural Network is proposed to be combined with RL (i.e., Deep RL) to overcome such an issue. A deep RL (DRL) based anti-jamming scheme is developed in [180]. The scheme is based on using a double DRL to model the confrontation between CR and the jammer and a transformer encoder Q network to estimate action-value from raw spectrum data. An anti-intelligent UAV jamming strategy through deep Q-networks is proposed in [181]. The work in [182] proposed a sequential DRL algorithm to identify jamming patterns and then to make on-line channel selection based on the recognized jamming patterns. An optimal defence policy is obtained by the transmitter in [183] through the Markov Decision Process (MDP). Also, Q-learning, deep Q-learning and deep dueling have been developed to maximize the long term throughput and minimize the packet loss. Other techniques leveraged game theory to optimize power control policy of transmitters against jammer. Game theory provides a powerful approach to study the interactions among jammers and legitimate users [184, 185]. The optimal power strategy under jamming attacks can be formulated in Stackelberg Game from which the Stackelberg equilibrium of the anti-jamming game can derived as in [186–189]. In [190], the power interaction among secondary users (SUs) and a jammer is formulated as an anti-jamming game. Stackelberg equilibrium has been presented and RL has been applied by the SUs to determine their transmission powers in the dynamic environment without knowing a priori the underlying game model. The design of an anti-jamming defence mechanism allowing SUs adapting their strategy on how to switch among control and data channels based on the observed spectrum availability, channel quality and attackers' actions was studied in [191]. The interactions between SUs and cognitive attackers who are able to use an adaptive strategy in time-varying environment was modeled as a stochastic zero-sum game integrated with minimax-Q learning. A spectrum agile CR dealing with a sophisticated jammer has been proposed in [192]. The spectrum co-existance of the radio and the jammer is modeled as a non-cooperative stochastic game where the radio uses RL to make better future channel selection.

## 2.4 Conclusion

This chapter provides a comprehensive survey on CR and the PHY security issue regarding jamming attacks. We first give the historical motivation of CR and its evolution in the last 20 years by explaining its original vision and the proposed cognitive cycles. Also, we briefly mention the expected role of CR in 6G. We then provide three potential areas of wireless application with which the CR can coexist and operate including IoT, mmWave and UAV. Later, we spot the light on the PHY security issue in CR where we studied different types of jammers and reviewed the proposed methods for detecting, classifying, characterizing and mitigating jamming signals as well as anti-jamming. The key lessons learned are: 1) Although, after 20 years of CR, state of the art is still lack extensive studies on its original vision and the techniques that can be used to achieve that vision; 2) Enhancing PHY security is of great concern to ensure the successful deployment of CR in various wireless applications; 3) jammer detection is the first essential step towards protecting the network. Jammer detection and spectrum abnormality detection has been a significant concern in CR realizing the first essential step towards protecting the network. Several techniques based on AI have been implemented to detect abnormal jamming signals. Nevertheless, the literature does not provide exhaustive work on that task. Therefore, abnormal jamming signal detection at the PHY of the CR is still a challenging task. 4) jammer characterization is significant to classify multiple jammers attacking the network and consequently mitigating its interfering effect and designing an anti-jamming strategy. However, most of the work in the literature studied these tasks independently. Thus, it is of great importance to develop a framework allowing the radio to detect the attack, characterize its nature and classify its identity and act accordingly via an unsupervised incremental approach, which will be the contribution of this thesis.

# Chapter 3

# Self-Awareness Representations for Cognitive Radio Applications

This chapter will focus on Self-Aware Representation Learning based on generalized random states allowing the radio to organize sensorial experiences while observing the wireless environment into a Generative Model (GM) representing emergent knowledge. A novel framework is presented, including various GMs to handle both high- and low-dimensional observations in real-world practical applications (CR-based mmWave and CR-based IoT), and a coherent inference mechanism for spectrum jamming anomaly detection.

## 3.1 Introduction

CR has been defined to exhibit three integral attributes which are observations, reconfiguration, and cognition [193]. In the observation process, CR gathers information about the radio environment. In the reconfiguration step, radio parameters are adjusted or changed. Whereas, cognition is related to understanding the radio environment, taking decisions on gathered information and learning the implications of such decisions on radio performance [194]. Learning and reasoning are fundamental aspects of cognition that may be achieved if the CR network subsumes a certain degree of *Self-Awareness (SA)* which can be developed by implementing Artificial Intelligence (AI) techniques. SA has been defined in the literature as the ability of a system to generate knowledge about itself and its environment and determine actions to be executed based on that knowledge [195–197]. The minimum requirements to consider a CR as Self-Aware are the following: *i)* the ability of the CR device to autonomously learn *Generative Models* by observing its states and the occurring environmental changes simultaneously; *ii)* the ability of a CR device to decide whether communications inside the radio spectrum occur according to a normal behaviour among the device itself and other devices (*Abnormality Detection*); *iii)* the detected abnormalities can be used either by the control system to apply *Abnormality Mitigation* strategies or by the SA module itself to *Incrementally Learn* new models that describe different dynamic situations not included in previous experiences.

Learning a good representation of the wireless environment (i.e., RF representation) indicates the core component of Self-Aware and AI-enabled radios whose performance heavily depends on that representation. In this perspective, it is crucial to answer the following question: what makes one representation better than another? Principally, a good representation is the one that makes subsequent tasks (it can be a learning or testing task) easier while dealing with different dimensional data (i.e., high- or low-dimensional data). Thus, the representation's choice is usually dependent on the targeted subsequent tasks and the data dimensionality with which the radio is dealing.

There are various ways of learning representations to understand the surrounding radio environment and build effective predictive models and classifiers. Understanding the surrounding environment can be achieved only after identifying and disentangling underlying explanatory factors hidden in the observed stimuli of low-level sensory signals [198]. Generative Models (GMs) are successful at learning spectrum representations and generating realistic data samples from complex underlying distributions. Probabilistic Graphical Models (PGMs) are a specific class of GMs providing stochastic behavior modelling of interacting variables whose relationships are represented in a graphical structure. An extension of GMs called deep generative models (e.g., Generative Adversarial Networks GANs and Variational Autoencoders VAEs) form a combination of GMs and deep neural networks which are suitable to deal with high dimensional signals [199].

We propose a framework for the joint spectrum representation and Abnormality detection that can be adopted in different radio applications. The two applications differ in the data dimensionality and the PHY-layer level at which the AI-method is implemented. In the first application, the method is applied just after the receiving antenna and the down-conversion process where multi-signals representations are extracted from a wideband spectrum with a high sampling rate and consequent high dimensionality data. While in the second application, the method is employed after down-conversion, cyclic prefix removal and Fast Fourier Transform (FFT) block at the CR device side where signals with low dimensionality and low sampling rate are extracted. The proposed framework is depicted in Fig. 3.1 illustrating two different approaches. The first approach, on top of the diagram, is thought for a high dimensional data scenario and investigates a wide-band spectrum with multiple signals. The second approach, at the bottom of the diagram, is suitable for low dimensional data such as sub-carriers in Orthogonal Frequency Division Multiplexing (OFDM) signals.

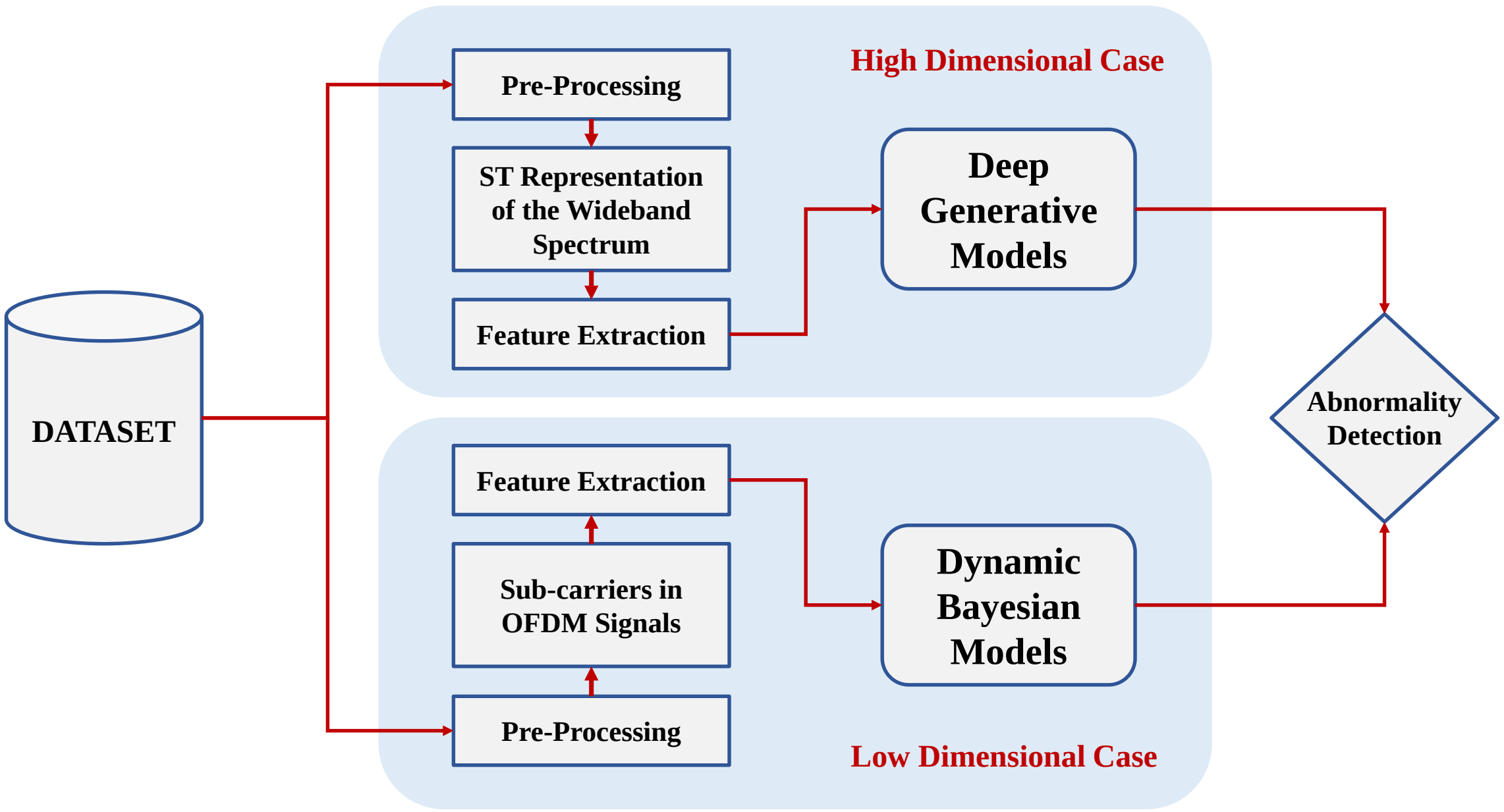

Fig. 3.1 The Proposed Framework for Joint Spectrum Representation and Abnormality Detection in different radio applications: (Top) Application to High-Dimensional Data, (Bottom) Application to Low-Dimensional Data.

## 3.2 Applications to High-Dimensional Data (Part I)

The general scheme of the proposed research is depicted in Fig. 3.2. The radio environment represents wireless communication in which transmissions are involved in the mmWave band. A CR system observes and gathers information about the spectrum occupancy where multiple signals dynamically occupy the available channels. However, processing and sensing such dynamic spectrum in the considered scenario requires suitable techniques. To this end, the Stockwell Transform (ST)-based dual-resolution proposed in [200] is used here to extract the time-frequency representation of the spectrum following the process explained in [200]. Note that the ST-based dual-resolution reduces the computational complexity significantly compared to the conventional ST [200].

In the discrete model of the ST distribution, $r[p]$, $p = 0, 1, \cdots, P-1$ denotes the discrete time series corresponding to a continuous signal $r(t)$ with a time sampling interval $T$. The

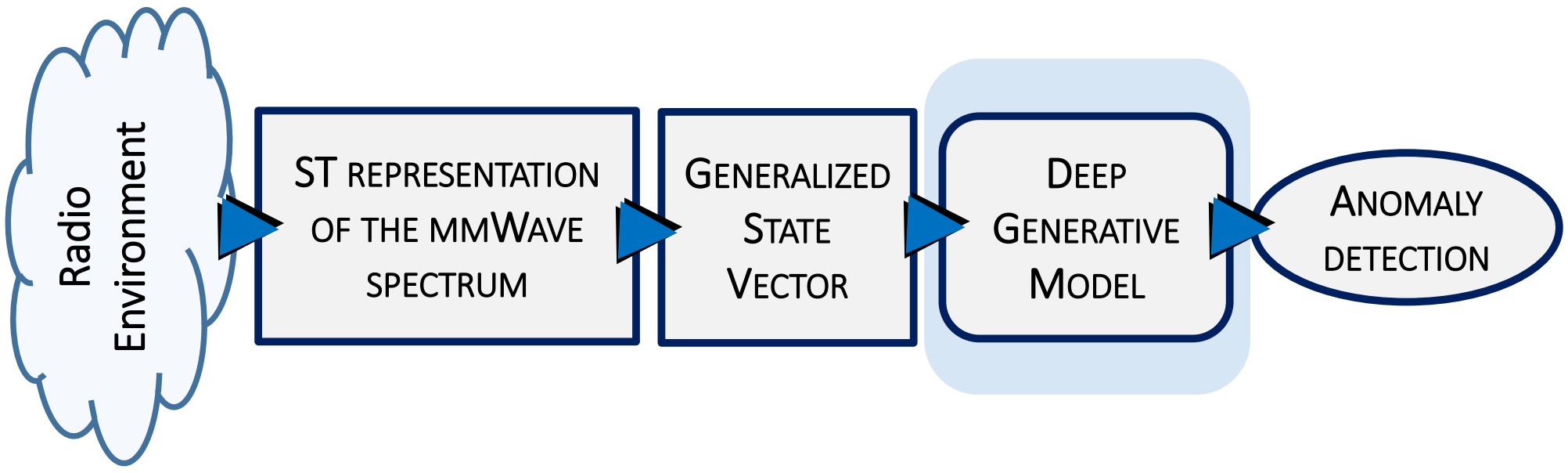

Fig. 3.2 The deep model-based anomaly detection scheme for CR.

discrete ST of $r[p]$ is then given by [201]:

$$S_T[m,n] = \sum_{p=0}^{P-1} r[p] \frac{|n|}{\sqrt{2\pi}kN} e^{-\left(n^2(m-p)^2/2k^2N^2 + j2\pi pn/N\right)}, \tag{3.1}$$

when $m = 0, \cdots, M-1$; $n = 1, \cdots, N-1$; and by

$$S_T[m,0] = \frac{1}{P} \sum_{p=0}^{P-1} r[p], \tag{3.2}$$

when $n = 0$; $m$ is the time delay of the sliding window, $n$ denotes the index of frequency range, $p$ denotes the time index, and $k$ is a scaling factor that controls the time-frequency resolution. When $k$ increases, the frequency resolution increases, with a corresponding loss of time resolution. The discrete ST can be modeled as a linear vector equation $\mathbf{s} = \mathbf{Tr}$ where $\mathbf{T}$ is the ST matrix (refer to [200] for more details about $\mathbf{T}$).

From such representation, a generalized state vector ($\mathbf{x}$) is formed, as defined in [46]. It consists of the state in terms of amplitude ($A_{n,k}$) and its first-order derivative ($\dot{A}_{n,k}$[1]):

$$\mathbf{x} = [A_{n,k},\ \dot{A}_{n,k}]; \quad \mathrm{n} \in \{1,\ldots,N\}, \tag{3.3}$$

where $A_{n,k} = abs(S_T[m,n])$ as the ST output is complex, $k$ is the time instant at which each value $A_{n,k}$ related to the $n$-th channel is extracted from ST and $N$ is the total number of channels.

In Fig. 3.2, the deep generative model block represents the deep generative models (i.e., C-GAN, AC-GAN and VAE) that are employed to learn the dynamics of the radio environment from the generalized state vector by analysing how the signals are evolving with time. In addition, the anomaly detection block realizes the process of evaluating whether the observations support the predictions performed by the deep generative models (i.e., received signals match predicted signals) using a proper indicator.

[1] The first order derivative is obtained by calculating the difference between the amplitude at time instant $k$ and that at time instant $k-1$, such that $\dot{A}_k = A_k - A_{k-1}$ if $k > 1$ and $\dot{A}_k = 0$ if $k = 1$.

### 3.2.1 Generative Adversarial Network (GAN)

Generative Adversarial Network (GAN) is one of the most crucial research avenues in the field of AI as an unsupervised learning technique and its outstanding data generation capacity has received extensive attention [202]. The adversarial nets consist of both a Generative model $G$ and a Discriminative model $D$. The generator $G$ defines a model distribution $p_g$ that captures the data distribution $p_{data}(\mathbf{x})$. The discriminator $D$, learns to determine whether a sample is from $p_g$ or $p_{data}(\mathbf{x})$ by estimating the probability $D(\mathbf{x})$ that a sample comes from the data distribution rather than the model distribution. The training data is denoted as $\mathbf{x}$ which is defined in (3.3). To learn the generator's distribution $p_g$ over data $\mathbf{x}$, a prior on input random noise $\mathbf{z}$ is defined as $p_{\mathbf{z}}(\mathbf{z})$, then the mapping to data space is represented by $G(\mathbf{z};\theta_{\mathbf{g}})$ where $G$ is a differential function represented by a multilayer perceptron with parameters $\theta_{\mathbf{g}}$. In addition, a second multilayer perceptor $D(\mathbf{x};\theta_{\mathbf{d}})$ is defined with parameters $\theta_{\mathbf{d}}$ that outputs a single scalar and $D(\mathbf{x})$ represents the probability that $\mathbf{x}$ came from the data rather than $p_g$. Both $G$ and $D$ can be represented by a non-linear mapping function such as a multilayer perceptron. The two models are simultaneously trained. The training procedure for $G$ is to minimize the $log(1-D(G(\mathbf{x})))$ that $D$ makes correct decision. While $D$ is trained to maximize $log(D(\mathbf{x}))$ of correctly differentiating the training samples from generated samples. This framework corresponds to a two-player min-max game with cost function $V(G,D)$ given in [203]. In the space of arbitrary functions $G$ and $D$, a unique solution exists with $G$ recovering the training data distribution and $D$ equal to $1/2$ everywhere. In an unconditioned generative model, there is no control on modes of the data being generated. However, by conditioning the model on additional information, it is possible to direct the data generation process as we will discuss in the following section.

### 3.2.2 Conditional Generative Adversarial Network (C-GAN)

By conditioning the basic GAN model [203] on additional information $\mathbf{y}$ (i.e., class labels), it is possible to direct the data generation process. This model is called Conditional-GAN (C-GAN) [204] shown in Fig. 3.3.

The auxiliar information is denoted by $\mathbf{y}$ and used to perform the conditioning when $\mathbf{y}$ (i.e., class labels) form the additional input layer at both $G$ and $D$ sides as in Fig. 3.3. In the generator, $p_{\mathbf{z}}(\mathbf{z})$ and $\mathbf{y}$ are combined in a joint hidden representation, while in the discriminator $\mathbf{x}$ and $\mathbf{y}$ form the inputs of the discriminative function. In C-GANs, the objective

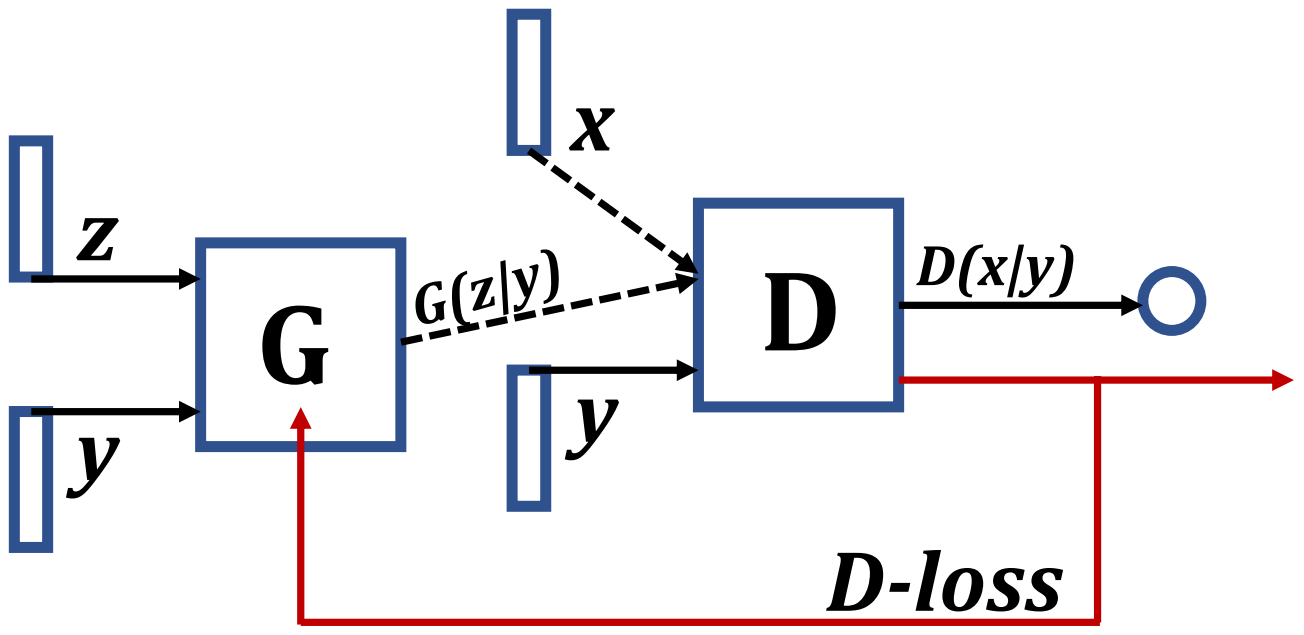

Fig. 3.3 Diagram of the Conditional Generative Adversarial Networks (C-GAN).

function of the two-player game is given by:

$$\min_{G}\max_{D} V(D,G) = \mathrm{E}_{\mathbf{x}\sim p_{data}(\mathbf{x})}\left[\log D(\mathbf{x}|\mathbf{y})\right] + \mathrm{E}_{\mathbf{z}\sim p_{z}(\mathbf{z})}\left[\log\left(1-D(G(\mathbf{z}|\mathbf{y}))\right)\right], \quad (3.4)$$

In the proposed method, $G$ consists of 14 layers based on neural networks, while $D$ consists of 9 layers according to [205]. Due to the conditioning variable $\mathbf{y}$, C-GANs belong to the supervised learning family. Nevertheless, in this work, no actual labels are assigned to $\mathbf{y}$ since a null value is constantly given regardless of the input data (normality or abnormal). Considering that, the C-GAN has been employed as **unsupervised learning** method because no priory information is required for each kind of input data which consists of samples extracted from ST representation with different patterns and the amplitude is used as a feature as described in Sec. 3.3.2. The D-loss is computed as the difference between the adversarial losses obtained when the input data at the discriminator is real data and fake data, respectively, as detailed in the testing phase section.

**Training Phase:** the C-GAN consists of both a generative model $G$ that captures the data distribution and a discriminative model $D$ that estimates the probability of a sample comes from that data distribution. Both $G$ and $D$ can be represented by a non-linear mapping function that is learnt during the training phase. $G$ maps a random noise $\mathbf{z}$ to data space $\mathbf{x}$. This mapping is represented by $G(\mathbf{z}|\mathbf{y})$. While $D$ acts as a binary classifier and outputs a single scalar represented by $D(\mathbf{x}|\mathbf{y})$. The training procedure for $G$ is to minimize the probability that $D$ makes the correct decision. While $D$ is trained to maximize the probability of correctly differentiating the training samples from generated samples. This framework corresponds to a two-player min-max game. The corresponding cost function is given by:

$$\min_{G}\max_{D} V(D,G) = \mathrm{E}_{\mathbf{x}\sim p_{data}(\mathbf{x})}\left[\log D(\mathbf{x}|\mathbf{y})\right] + +\mathrm{E}_{\mathbf{z}\sim p_{z}(\mathbf{z})}\left[\log\left(1-D(G(\mathbf{z}|\mathbf{y}))\right)\right], \quad (3.5)$$

where $p_{data}(\mathbf{x})$ is the data distribution and $p_z(\mathbf{z})$ is the prior.

**Testing Phase:** the parameters of both $G$ and $D$ networks are not updated through the optimization of the cost function which is only utilized to detect deviations between prediction and observation, based on the following anomaly measurement:

$$\text{db0} = |l_{real} - l_{fake}|, \tag{3.6}$$

where $l_{real}$ is the loss computed at the discriminator when the input is the data $\mathbf{x}$ while $l_{fake}$ is the loss when the input is the one generated by the generator from G($\mathbf{z}|\mathbf{y}$), respectively, and $|\cdot|$ represents the absolute value function.

Let $\mathbf{u}$ be the output of the discriminator, $\mathbf{v}$ the corresponding adversarial ground truth and $N$ the batch size. Then, the adversarial loss representing the Mean Square Error (MSE) between the two vectors $\mathbf{u}$ and $\mathbf{v}$ can be expressed as:

$$l(\mathbf{u},\mathbf{v}) = \frac{1}{N}\sum_{n=1}^{N}(u_n - v_n)^2. \tag{3.7}$$

Consequently, $l_{real}$ in (3.6) can be defined as:

$$l_{real} = \frac{1}{N}\sum_{n=1}^{N}(u_{n,real} - v_{n,real})^2, \tag{3.8}$$

that depicts the MSE loss computed at the discriminator when the input is $\mathbf{x}$. Whereas, $l_{fake}$ in (3.6) can be expressed as:

$$l_{fake} = \frac{1}{N}\sum_{n=1}^{N}(u_{n,fake} - v_{n,fake})^2, \tag{3.9}$$

depicting the MSE loss when the input is the one generated by the generator from $G(\mathbf{z}|\mathbf{y})$.

### 3.2.3 Auxiliary Classifier GAN (AC-GAN)

Alternatively, the discriminator can be modified with reconstructing the class information $\hat{\mathbf{y}}$. In this way, the discriminator will contain an auxiliary decoder network that outputs the class label for the training data. This variant of the GAN architecture is called auxiliary classifier GAN (or **AC-GAN**) [206] and shown in Fig. 3.4.

**Training Phase:** $G$ uses both the class labels $\mathbf{y}$ and the noise $\mathbf{z}$ to generate data samples (fake data), $\mathbf{x}_{fake}$. While, the discriminator computes both the probability distribution of the sources, $p(\mathbf{s}|\mathbf{x})$, and of the class labels, $p(\mathbf{y}|\mathbf{x})$ such that $D(\mathbf{x}) = (p(\mathbf{s}|\mathbf{x}), p(\mathbf{y}|\mathbf{x}))$. The source of the data, $\mathbf{s}$, refers to the decision of the discriminator, namely either real data, $\mathbf{s}_{real}$, or fake data, $\mathbf{s}_{fake}$. Consequently, the objective function consists of both the log-likelihood of the correct source, $L_{\mathbf{s}}$, and the log-likelihood of the correct class, $L_{\mathbf{y}}$, as follows:

$$L_{\mathbf{s}} = \mathrm{E}\left[\log p\left(\mathbf{s}_{real}|\mathbf{x}_{real}\right)\right] + \mathrm{E}\left[\log p\left(\mathbf{s}_{fake}|\mathbf{x}_{fake}\right)\right], \quad (3.10)$$

$$L_{\mathbf{y}} = \mathrm{E}\left[\log p\left(\hat{\mathbf{y}}|\mathbf{x}_{real}\right)\right] + \mathrm{E}\left[\log p\left(\hat{\mathbf{y}}|\mathbf{x}_{fake}\right)\right]. \quad (3.11)$$

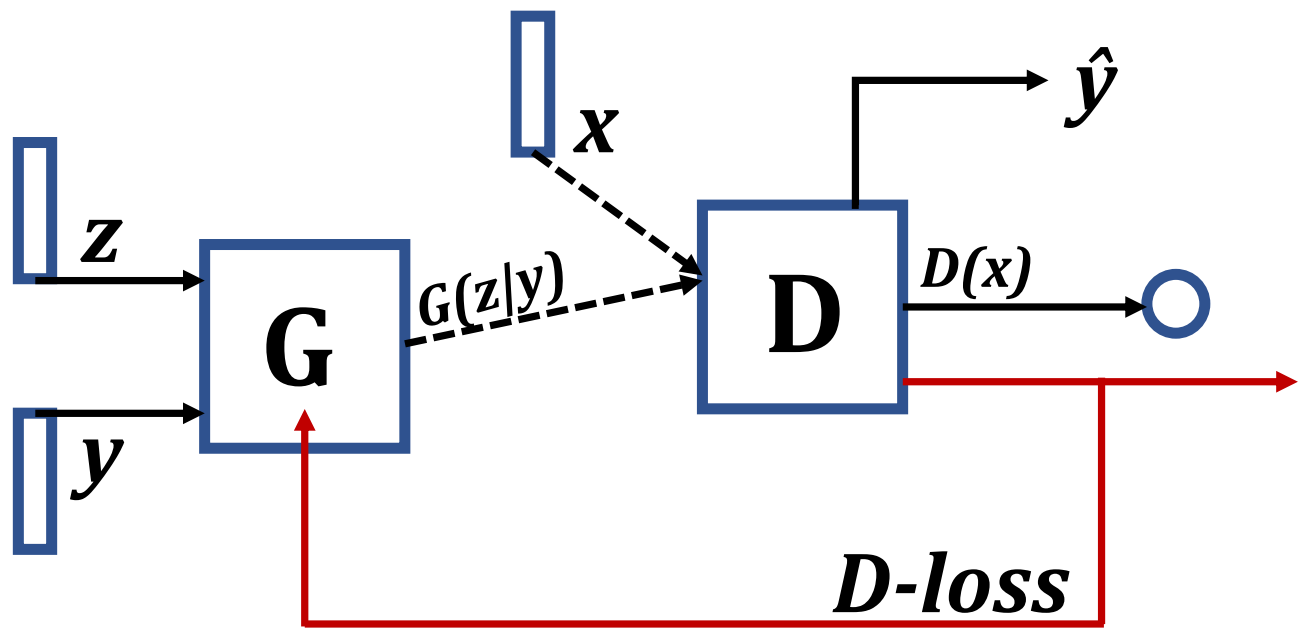

Fig. 3.4 Diagram of the Auxiliary Classifier Generative Adversarial Networks (AC-GAN).

$D$ maximizes the probability of correctly classifying real and fake samples ($L_{\mathbf{s}}$) and correctly predicting the class label ($L_{\mathbf{y}}$) of a real or fake sample ($L_{\mathbf{s}} + L_{\mathbf{y}}$). $G$ minimizes the ability of the discriminator to discriminate real and fake samples while also maximizing the ability of the discriminator in predicting the class label of real and fake samples ($L_{\mathbf{y}} - L_{\mathbf{s}}$).

**Testing Phase:** as in C-GAN, in this phase the parameters of both $G$ and $D$ networks are not updated, and the anomaly measurement defined in (3.6) is utilized to detect deviations where, in addition to the loss computed on data, both $l_{real}$ and $l_{fake}$ take also into account an auxiliary loss term from real and fake class labels, respectively.

### 3.2.4 Variational Auto Encoder (VAE)

VAE in its vanilla version, is composed of an encoder $q_\theta(\mathbf{z}|\mathbf{x})$ and a decoder $p_\phi(\mathbf{x}|\mathbf{z})$, which can be estimated through the learning process of the VAE on training data. Through $\theta$ and $\phi$, we define the parameters of the encoder and decoder, respectively. The encoder $q_\theta(\mathbf{z}|\mathbf{x})$ allows to represent each sample $\mathbf{x}$ inputted to it through two bottleneck features, i.e., a mean $\mu$ and a variance $\sigma^2$. On the other hand, the decoder $p_\phi(\mathbf{x}|\mathbf{z})$ synthesizes an observation $\mathbf{x}$ from a latent state $\mathbf{z}$ sampled from $\mathcal{N}(\mu, \sigma^2)$. Therefore, the former allows reduction of dimensionality of the observations (i.e., wide-band signals) and thus pass to low-dimensional data while the latter allows to obtain again high-dimensional data through reconstruction starting from latent space in low-dimensional data.

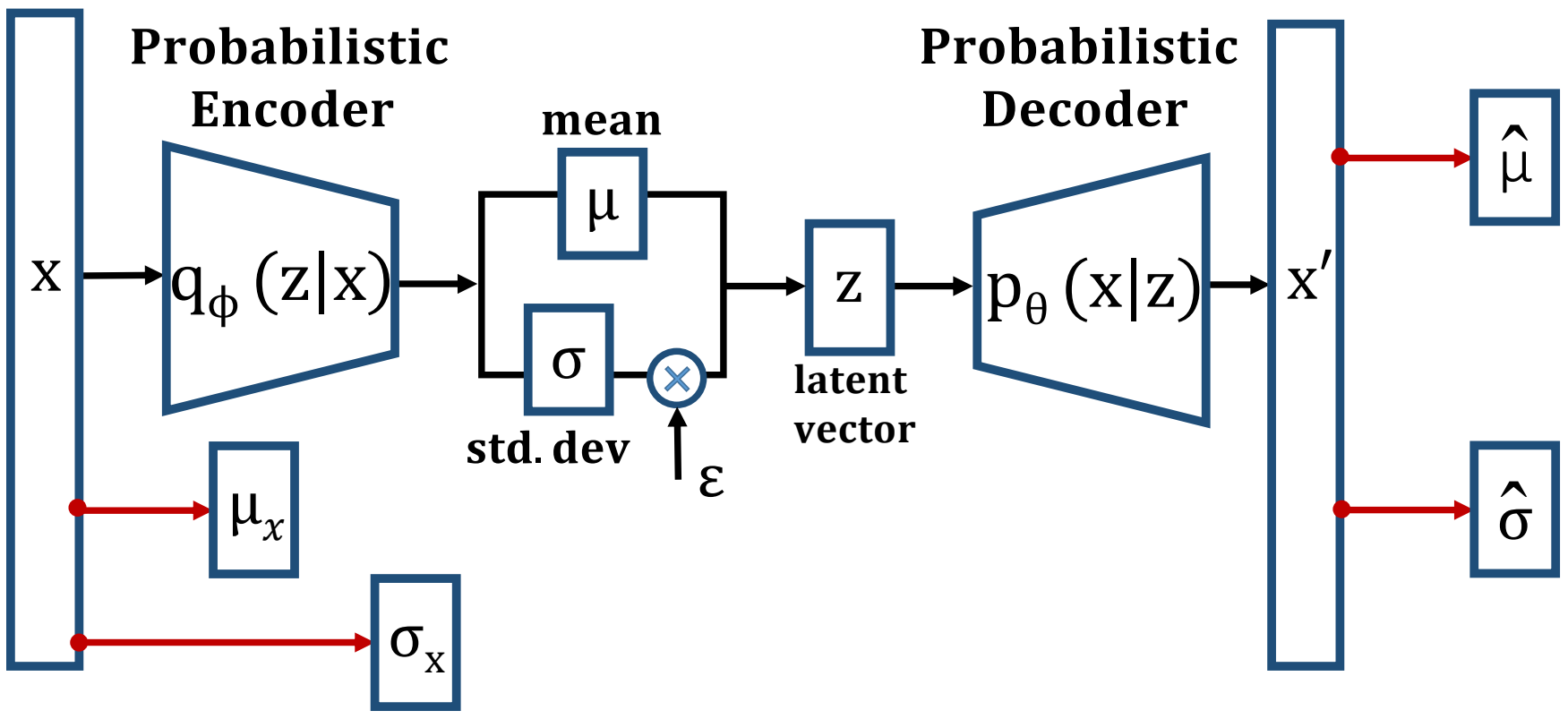

Fig. 3.5 Diagram of the Variational Auto-Encoder (VAE).

**Training Phase:** VAEs learn a stochastic mapping between an observed data space $\mathbf{x}$, whose empirical distribution is typically complicated, and a latent space $\mathbf{z}$, whose distribution can be relatively simple [207]. $\mathbf{z}$ represents a compressed low dimensional representation of the input $\mathbf{x}$. VAEs consist of two models, the encoder or inference model, and the decoder or generative model (refer to Fig. 3.5). The generative model (decoder) learns the joint distribution $p_\theta(\mathbf{x},\mathbf{z})$. The inference model (encoder) $q_\phi(\mathbf{z}|\mathbf{x})$, approximates the true but intractable posterior $p_\theta(\mathbf{z}|\mathbf{x})$ of the generative model. The model parameters of the decoder and encoder are denoted by $\theta$ and $\phi$, respectively. While, $\mu$ and $\sigma$ are the mean and standard deviation of the multivariate distribution $q_\phi(\mathbf{z}|\mathbf{x})$. $\varepsilon \sim \mathcal{N}(\mathbf{0},\mathbf{I})$ is a noise random variable. Ideally, the reconstructed input $\mathbf{x}'$ is approximately identical to $\mathbf{x}$, $\mathbf{x} \approx \mathbf{x}'$.
VAEs provide a computationally efficient way for optimizing the generative model jointly with the corresponding inference model. The model parameters ($\phi$), also called variational parameters, are optimized such that:

$$q_\phi(\mathbf{z}|\mathbf{x}) \approx p_\theta(\mathbf{z}|\mathbf{x}), \tag{3.12}$$

by using the Evidence Lower Bound (ELBO) which is the variational lower bound on the log-likelihood of the data. It includes the Kullback-Leibler (KL) divergence between $q_\phi(\mathbf{z}|\mathbf{x})$ and $p_\theta(\mathbf{x},\mathbf{z})$. Maximization of the ELBO w.r.t. the parameters $\theta$ and $\phi$, will approximately maximize the marginal likelihood $p_\theta(\mathbf{x})$ and minimize the KL divergence of the approximation $q_\phi(\mathbf{z}|\mathbf{x})$ from the true posterior $p_\theta(\mathbf{z}|\mathbf{x})$.

Thus, the learning of the VAE is conducted to optimize the parameters $\theta$ and $\phi$, maximizing the sum of the lower bound on the marginal likelihood of each observation $\mathbf{x}$ of the dataset $D$, as described in [208, 209]:

$$\mathcal{L}_{\phi,\theta}(D) = \sum_{\mathbf{x}\in D} \mathcal{L}_{\phi,\theta}(\mathbf{x}), \tag{3.13}$$

where $\mathcal{L}_{\phi,\theta}(\mathbf{x})$ is defined as:

$$\mathcal{L}_{\phi,\theta}(\mathbf{x}) = -D_{KL}(q_{\phi}(\mathbf{z}|\mathbf{x})||p_{\theta}(\mathbf{z})) + E_{q_{\phi}(\mathbf{z}|\mathbf{x})}[\log p_{\theta}(\mathbf{x}|\mathbf{z})], \tag{3.14}$$

where the term $D_{KL}$ is the KL divergence. Therefore, the first term measures the difference between the encoder's distribution $q_{\phi}(\mathbf{z}|\mathbf{x})$ and the prior $p_{\theta}(\mathbf{z})$; the prior typically being a standard normal distribution $\mathcal{N}(0,1)$. The second term is the expected log-likelihood of the observation $\mathbf{x}$ and forces the VAE to reconstruct the input data.

**Testing Phase:** the parameters $\theta$ and $\phi$ are not updated so that the encoder and decoder are the ones learned during training. In this phase, a way of measuring the similarity between the observation and prediction is related to the reconstruction error which gives the anomaly measurement $db0$ (refer to Fig. 3.5) computed as follows:

$$\text{db0} = \left( (\mu_x - \widehat{\mu})^T C_{\widehat{\sigma}^2}^{-1} (\mu_x - \widehat{\mu}) \right)^{\alpha}, \tag{3.15}$$

where $\mu_x$ is the mean vector from the input data with dimension $d$ (for the sake of completeness, $\sigma_x$ is the standard deviation vector from the input data), and $\widehat{\mu}$ and $\widehat{\sigma}$ are the mean and standard deviation vectors from the reconstructed data vector with the same dimension $d$. The variable $\alpha$ is a tuning parameter that we selected to be equal to 8 after numerous trials using different integer values. These quantities are the output of neural networks whose input is $\mathbf{x}$ and $\mathbf{x}'$, respectively. $C_{\widehat{\sigma}^2}^{-1}$ is a covariance matrix given by $\text{diag}(\widehat{\sigma}_1^2, \ldots, \widehat{\sigma}_d^2)$.

# 3.3 Simulation and Performance Evaluation (Part I)

The following experiments have been performed on the generative models described in the previous section to demonstrate the practical feasibility of the proposed approach for spectrum anomaly detection. First, the mmWave testbed and the real dataset are described, then results are presented.

## 3.3.1 mmWave Communication Testbed

An enormous amount of under-utilized bandwidth lies in the millimeter-wave bands. The significant advantages offered by the propagation characteristics in terms of frequency re-usability and large channel bandwidths make millimeter-wave suitable for the very high capacities required by the fifth-generation (5G) wireless communication systems [210, 211]. Indeed, mmWave can achieve Giga-bits/sec data rate and large-data-capacity. Consequently, high interest in this part of the electromagnetic spectrum has risen in the recent years and on October 22$^{nd}$ 2015, the Federal Communications Commission (FCC) proposed new rules (FCC 15138) for wireless broadband frequencies of 28 GHz, 37 GHz, 39 GHz and 64 - 71 GHz bands. Line of sight communication is a limit of mmWave technology which greatly limits the communication range to a few meters.

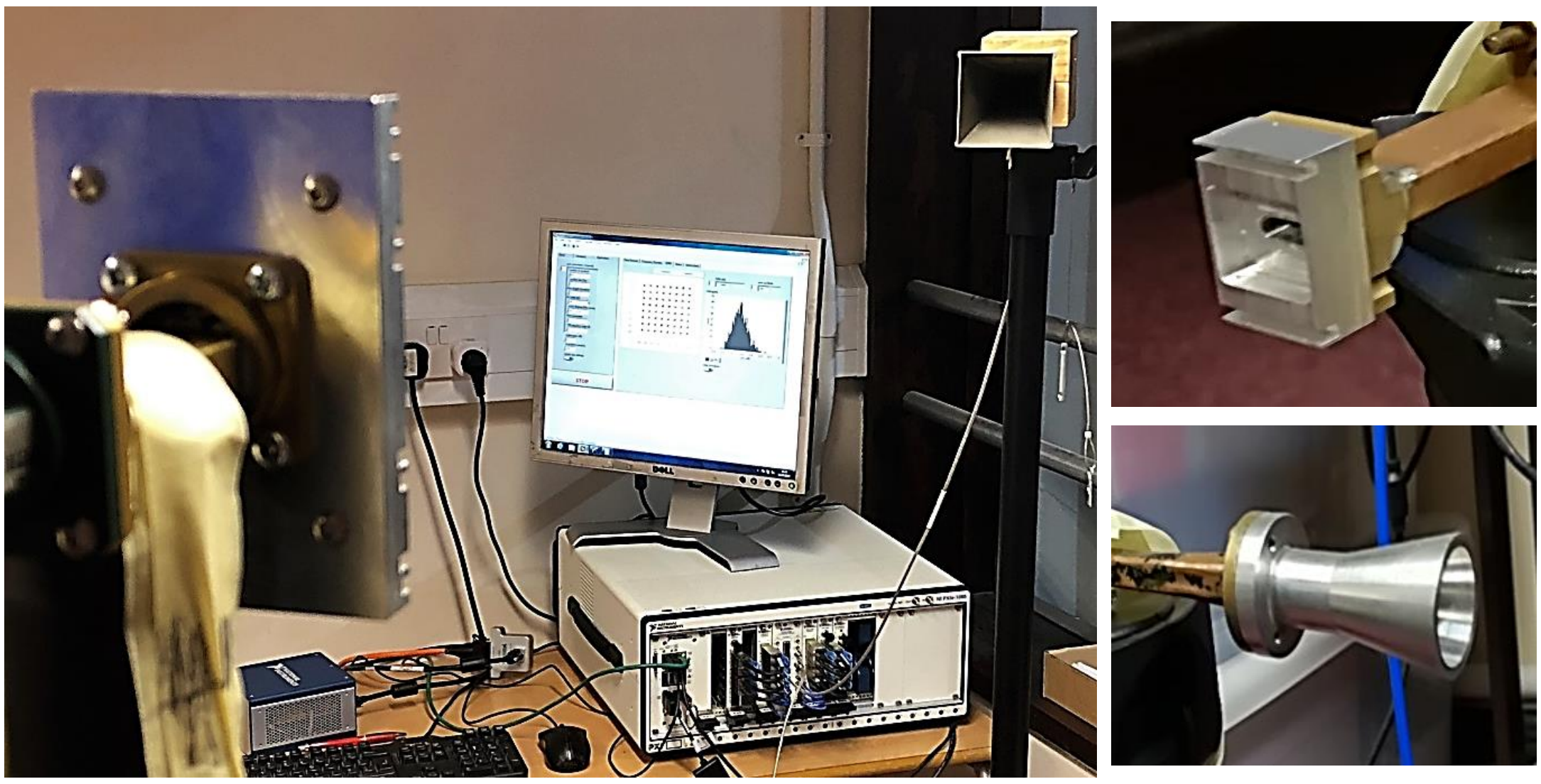

Fig. 3.6 The mmWave testbed setup.

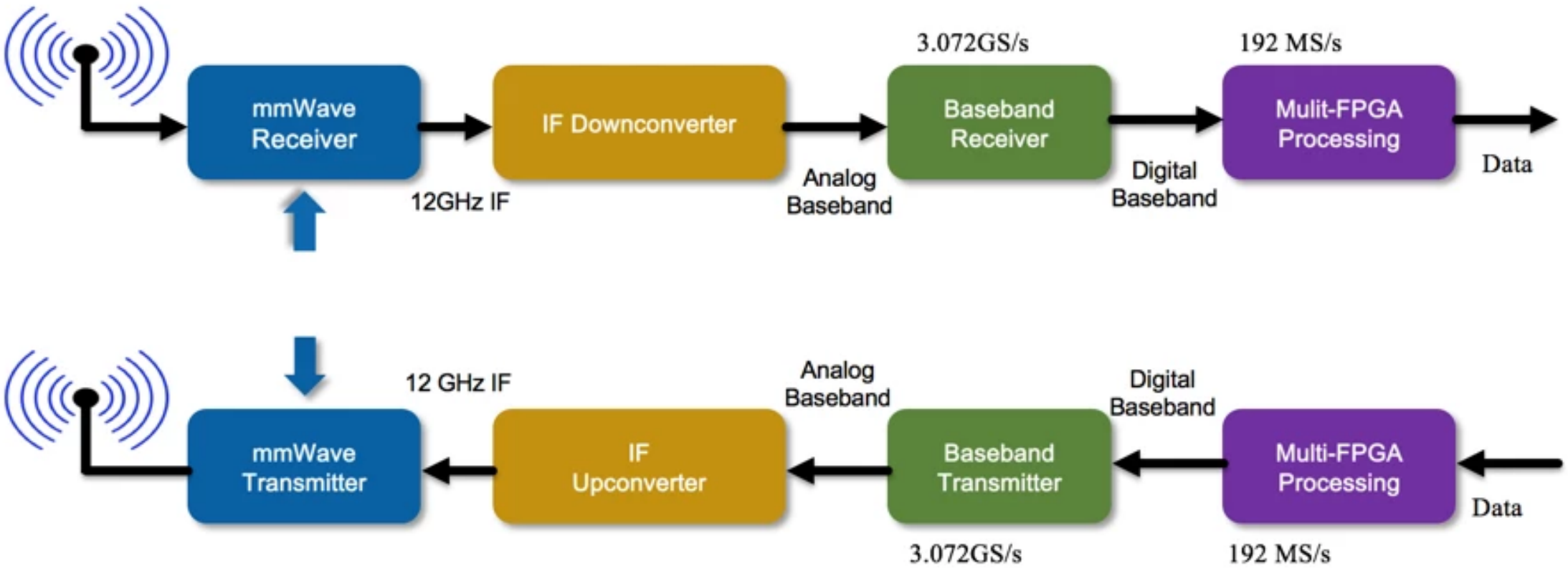

Fig. 3.7 The mmWave System Diagram.

The National Instruments (NIs) mmWave Transceiver System at Queen Mary University of London, Fig. 3.6, used to collect the dataset with the objective of validating the proposed approach in learning the dynamics of the signals transmitted over the mmWave channels and detecting deviations from what has been learned by the radio so far. The NI communication system is a Software Defined Radio (SDR) platform consisting of hardware equipment and application software that enables real-time over the air mmWave communications. The transceiver system illustrated in Fig. 3.7 is comprised of chassis, controllers, a clock distribution module, 192 MS/s Field-Programmable Gate Array (FPGA) modules, high-speed Digital-to-Analog Converters (DACs) and Analog-to-Digital Converters (ADCs) (3.072 GS/s), Local Oscillator (LO) and Intermediate Frequency (IF) modules, and mmWave radio heads (24.25 - 33.4 GHz) for up-conversion from 12 GHz IF to mmWave band and down-conversion from mmWave band to 12 GHz IF. A detailed description can be found in [212].

The radio heads are connected to a Ka-band circular horn transmitting antenna (26-40 GHz) and a slot antenna at 28.5 GHz for receiving the signal [213], respectively. The mmWave transceiver operates at 28 GHz (central carrier frequency), and the analysed spectrum consists of 8 $\times$ 100 MHz channels with 800 MHz total bandwidth, as shown in Fig. 3.8. The 8 channels are respectively found at the corresponding offset frequencies: -350, -250, -150, -50, +50, +150, +250, and +350 MHz with respect to the central frequency of the mmWave band and complex I/Q data is collected at base-band after the down-conversion process (see Fig. 3.9). Cyclic-Prefix Orthogonal Frequency Division Multiplexing (CP-OFDM) signals with 1200 sub-carriers are transmitted inside the mmWave band with 75 kHz sub-carrier spacing and 2048 FFT size. Different modulation schemes are supported (BPSK, QPSK, 16-QAM, and 64-QAM). The sampling frequency is 3.072 GS/s (12-14 bits).

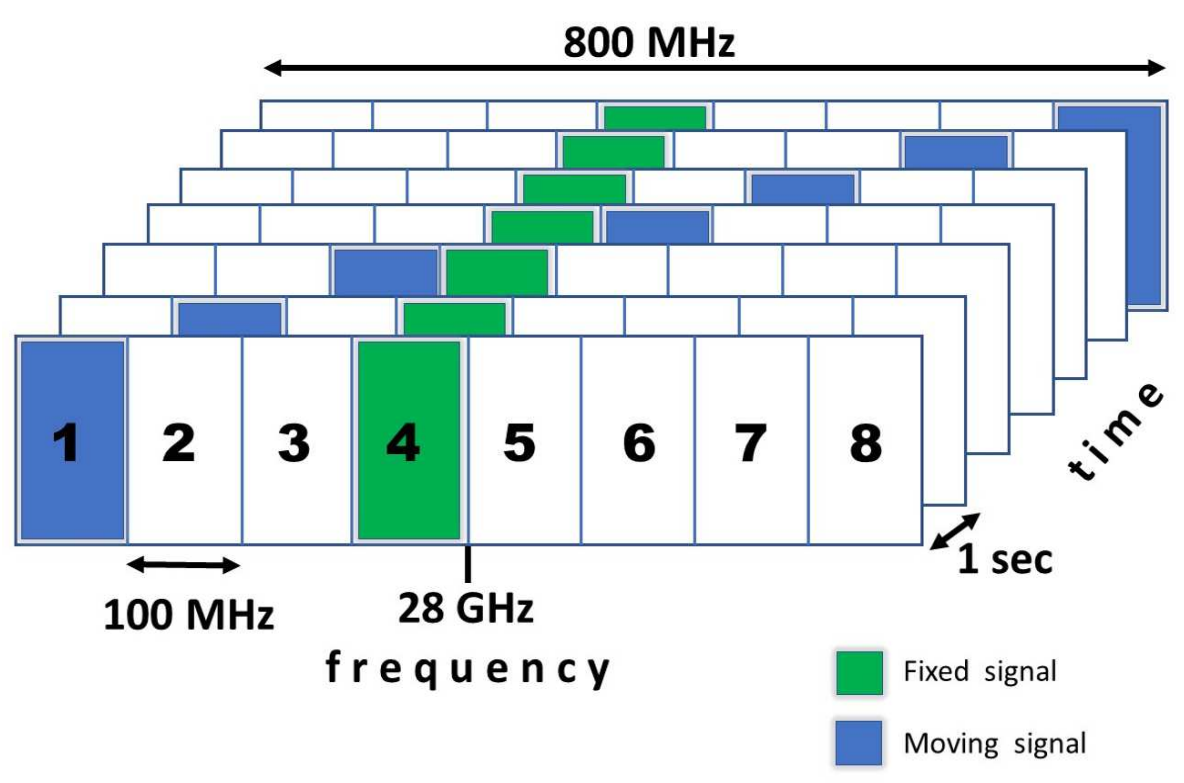

(a)

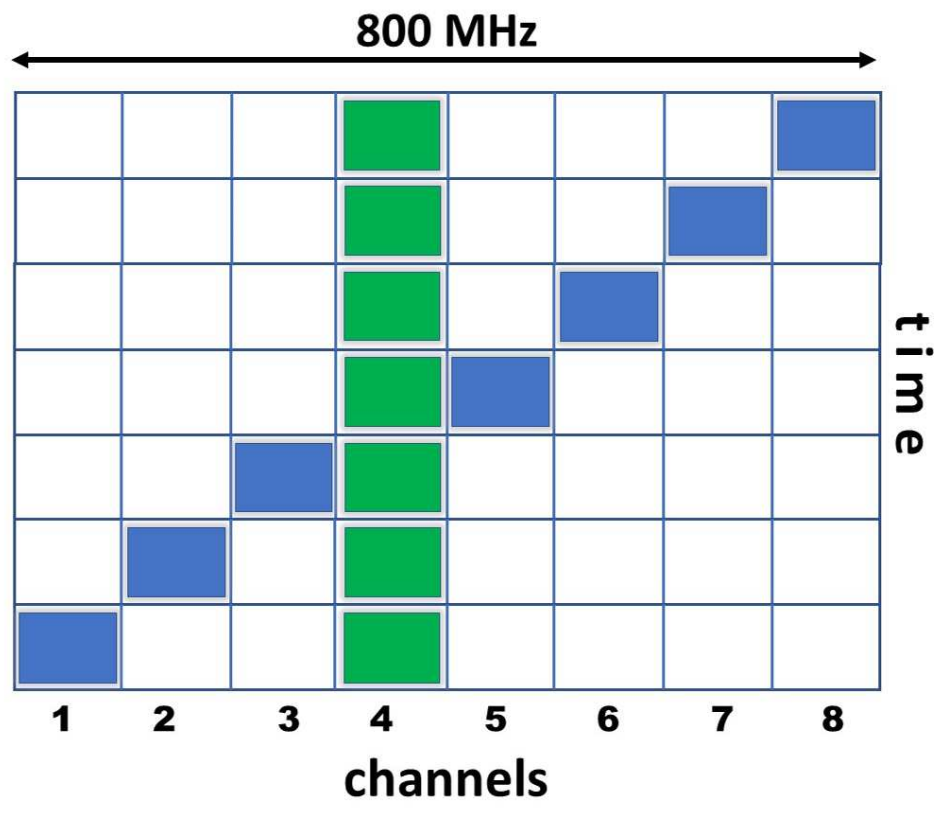

(b)

Fig. 3.8 Representation of a dynamic pattern in the 800 MHz mmWave band with a fixed OFDM signal (the green rectangles) at channel 4 and a moving OFDM signal (the blue rectangles). (a) 3D frequency-time diagram and (b) the corresponding 2D time-channels version.

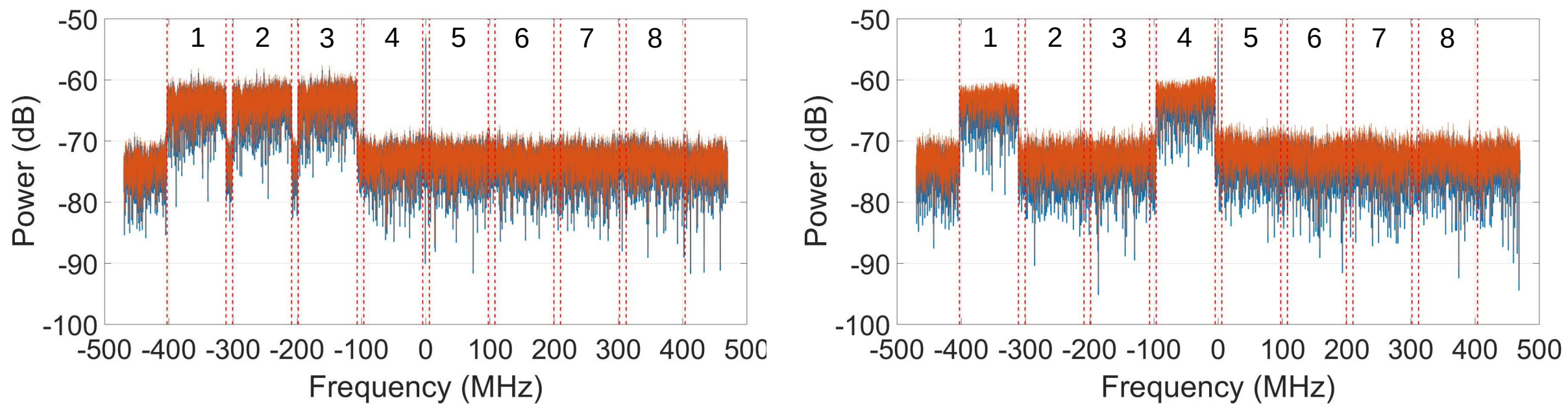

(a) (b)

Fig. 3.9 The mmWave spectrum at base-band: 8 $\times$ 100 MHz bandwidth. Occupied channels: (a) 1$^{st}$, 2$^{nd}$, 3$^{rd}$, (b) 1$^{st}$ and 4$^{th}$.

The system specifications relative to the observed spectrum and the signal pre-processing resulting in the images of Fig. 3.10 are summarized in details in Table 3.1 and Table 3.2.

Table 3.1 Data from the system specification datasheet

| | |
|---|---|
| **Maximum bandwidth** | 2 GHz |
| **Central frequency** | 28 GHz in the mmWave band |
| **Sampling Rate** | 3.072 GS/s, resampled to/from 153.6 MS/s |
| | 3.072 GS/s ÷ 153.6 MS/s = 20 |
| **Spectrum of interest** | 800 MHz bandwidth (27.6 GHz - 28.4 GHz) |
| **Channels (width)** | 8 (100 MHz) |
| | 192 MS/s x 2 (I/Q) x 8 channels = 3.072 GS/s |
| **OFDM signal** | 1200 subcarriers/channel with 75 kHz spacing for each subcarrier (75 kHz x 1200 = 90 MHz) |
| **Symbol Rate** | 153.6 MS/s (oversampling on each channel, I/Q data) |
| **FFT for OFDM** | 2048 points (153.6 MHz / 75 kHz) |

### 3.3.2 Real Dataset

*Feature Extraction through ST*: an example of raw ST representation of a mmWave signal in a dynamic scenario is shown in Fig. 3.11 (left image) where a fixed signal is found at channel number 4. While a moving signal jumps in the spectrum at different time instants (sequentially, ch 7, ch 2, and ch 6). The delay index is the shift number of the sliding window of the ST. Signals can be detected from the ST representation to obtain occupied areas in both the frequency and time domains, as shown in Fig. 3.11 (middle image). Afterwards, occupancy (considered as energy locations) and corresponding contours for each occupied area can be extracted as in Fig. 3.11 (right image). A detailed description of the theoretical aspects of the ST employed in this work can be found in [200]. This kind of analysis allows for the extraction of features such as bandwidth, central frequency, received power or amplitude, and variance. The extracted values can be mapped into the time-frequency location of the corresponding signals and, then, represented as bi-dimensional localized dynamic features. Derivatives can be computed to obtain the generalized state vector for each of the considered features. In this case, since the generalized state vector is extracted from columns from the ST representation and the corresponding derivatives, its dimensionality is high and mostly related to the frequency resolution in the ST.

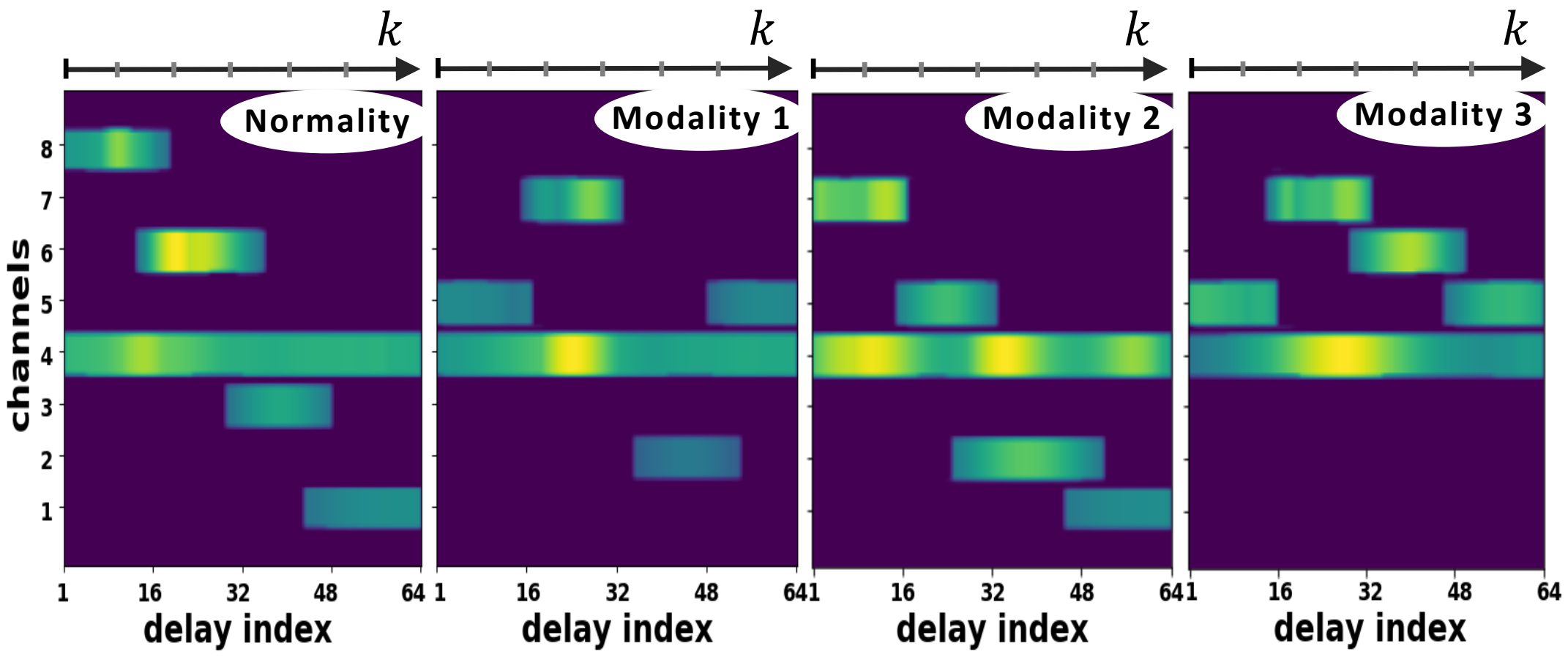

(a) Amplitudes

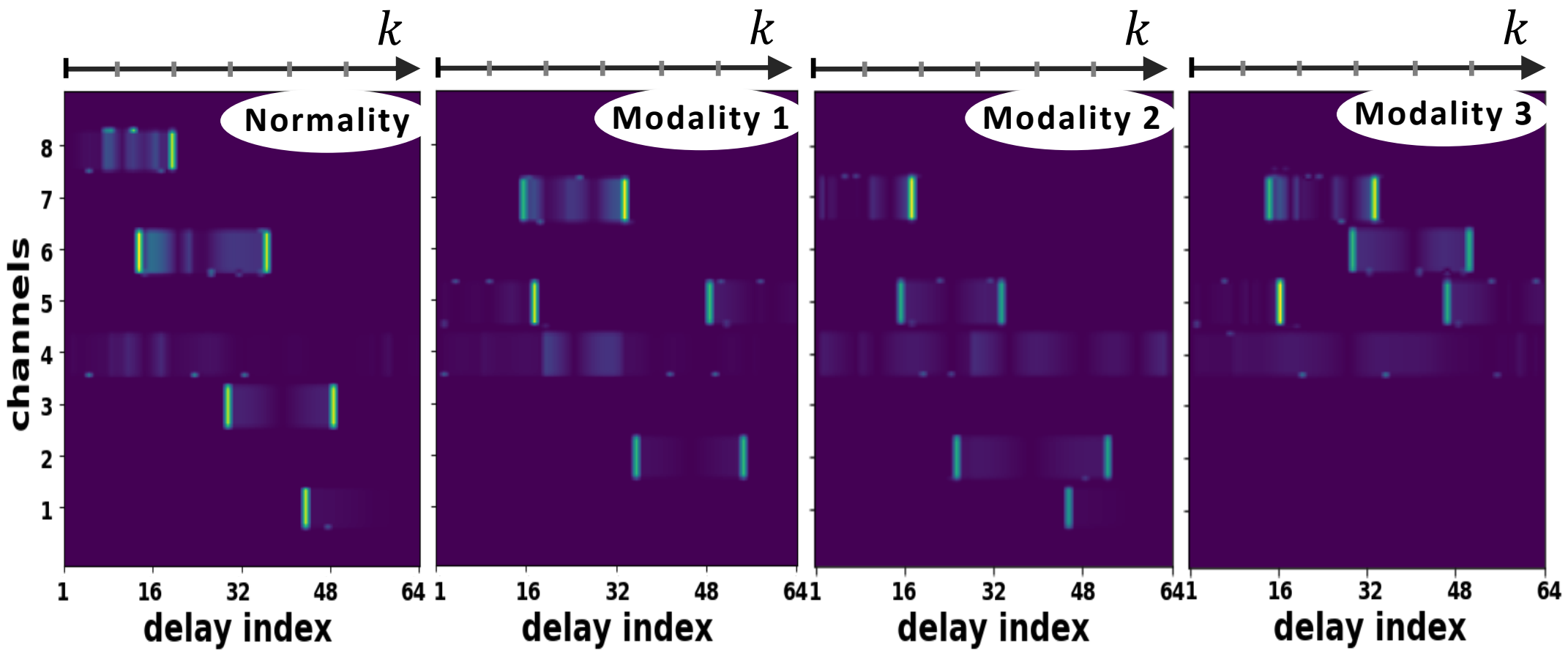

(b) Derivatives

Fig. 3.10 Four patterns performed by the signals inside the spectrum.

Table 3.2 Signal pre-processing specifications

| | |
|---|---|
| **Samples for each burst (I/Q signals)** | 4096 |
| **Stockwell transform size** | 512 |
| **Dual-resolution ST (frequency-time index)** | 512 (f) x 64 (k) |
| **Sub-channel division (frequency-time index)** | 128 (f) x 64 (k) |

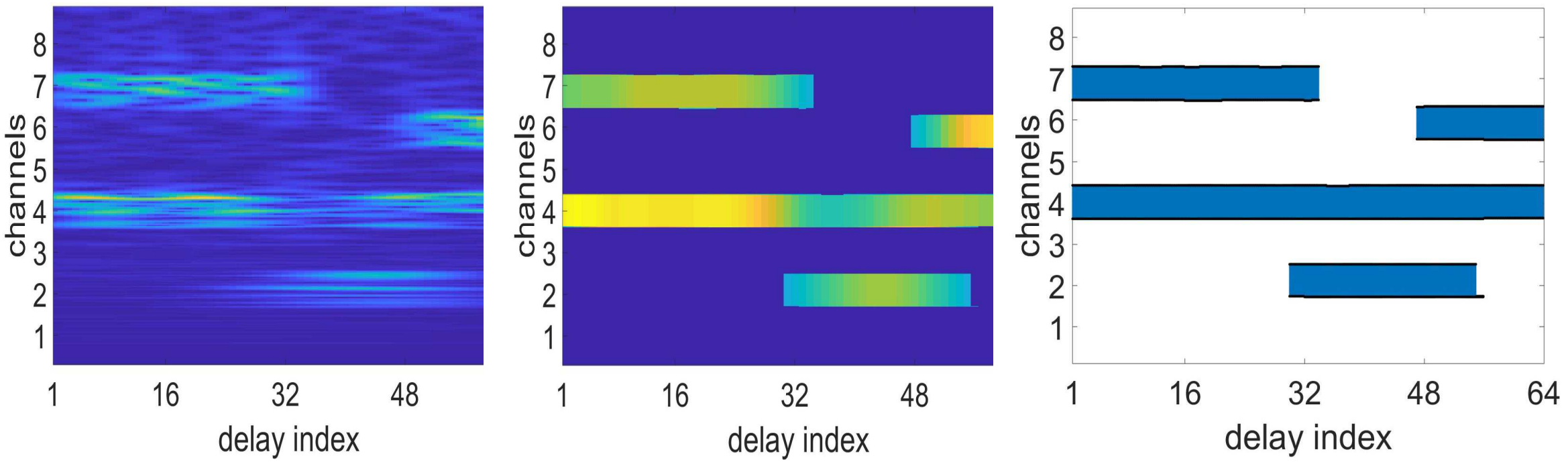

Fig. 3.11 ST representation of a wideband dynamic signal: raw representation (left), occupied areas (centre) and occupancy represented by blue areas with contours represented as black lines (right).

The dataset is divided into two sets: one for the training phase which represents the normal behaviour (no malicious behaviour) of the signals inside the spectrum and the second is used during the testing phase including three different anomaly modalities in which the behaviour of the signal is different from the normal one. Fig.3.10(a) shows the time-frequency representation of the dynamic spectrum obtained by ST in terms of amplitude (only one snapshot for each modality is displayed due to space limitations). The $k$ axis represents the time domain in terms of 64 shifts of the sliding window denoted as delay index. While the vertical axis represents the frequency domain consisting of 8 channels divided into 128 sub-channels. And the corresponding derivatives are shown in Fig.3.10(b). The generalized state vector is formed by inserting the values relative to each vertical line from ST representation and concatenated with the corresponding vertical line of the derivative. The state vector is thus composed of 256 elements (128 for amplitudes and 128 for the derivatives) at each time instant $k$.

***Normality data:*** the normal behaviour consists of a fixed signal which occupies channel *ch* number 4 and a moving signal jumping at four different channels: sequentially 8, 6, 3, and 1 as shown in the first pattern of Fig. 3.10(a).
***Testing data:*** testing patterns with 3 different behavior modalities are also shown in Fig. 3.10(a) and described as follows.

- Modality 1: a fixed signal is occupying *ch-4* and a moving signal jumps between *ch-5*, *ch-7*, *ch-2*, *ch-5*.
- Modality 2: a fixed signal is occupying *ch-4* and a moving signal jumps between *ch-7*, *ch-5*, *ch-2*, *ch-1*.
- Modality 3: a fixed signal is occupying *ch-4* and a moving signal jumps between *ch-5*, *ch-7*, *ch-6*, *ch-5*.

Specifically, in the observed spectrum, a signal is anomalous when its behaviour (or dynamics) is different from the one previously seen during the training phase. Namely, the strategy by which the signal jumps in the spectrum changes with respect to the normal behaviour. In this case, an anomaly is said to have happened. This could be due to a new device in the network or to a malicious user. In particular, our approach is capable of learning the dynamics of signals and, each time a change in the dynamics happens, the generative models produce predictions that deviate from the observations that are classified as anomalies, as demonstrated in next section.

### 3.3.3 Performance Metrics

Distance metrics are used to provide an abnormality indicator (i.e., $db0$) as described previously. In order to evaluate the performance of the proposed methods, we used a range of confidence thresholds to build a corresponding Receiver Operating Characteristic (ROC) curve. We applied these confidence thresholds to the abnormality signals provided in the testing phase by $db0$. The ROC curve represents the probability of detection, $P_d$, over the probability of false alarm, $P_{fa}$. Where, $P_d$ is the number of times when abnormalities are correctly identified, and $P_{fa}$ refers to the times where abnormalities are classified incorrectly (when normal behaviour is identified as abnormal behaviours).

### 3.3.4 Results

#### Training of the generative models

The training data consists of 59520 *k-samples* in the time domain of the generalized state vector and 256 in the frequency domain for C-GAN, while 164480 *k-samples* and 256 for AC-GAN and VAE. By providing the normality data, *the generative models are learnt in an unsupervised way*. Indeed, in this work, the conditioning information, **y** consists of a fictitious input label because it is assigned the same value regardless of the input data **x** whether normal or anomalous. The training data is used to train neurons in the latent space of the networks, and it has been shown that in generative models, each neuron learns to detect specific types of features from the input data. Intrinsic clustering on input data is also obtained thanks to neurons that learn to detect similarity characteristics of groups of input samples. The Adam optimizer is used to train $G$ and $D$ of C-GAN and AC-GAN as well as the encoder and decoder of VAE. MSE loss is used as adversarial loss in C-GAN, while $L^p$ loss (with p = 8) in AC-GAN which also includes a Cross-Entropy loss as an auxiliary loss. By setting p = 8, anomaly peaks (when an anomaly happens) and fluctuations (when signals in the spectrum follow a normal behaviour) in the indicator signal are optimized. The KL divergence is included in the loss function in VAE. Experiments have been performed on 'NVIDIA® GeForce® GTX 1080 Ti' GPU.

#### Testing of the generative models

In this phase 25280 *k-samples* in time domain and 256 in frequency domain forming the generalized state vector are tested and anomaly measurement is obtained (Figs. 3.12-3.14-3.16) for each of the 3 models and modalities.

It can be seen that, when the deep model is given a generalized state vector as input, it is capable of detecting abnormal patterns, when they happen, in which malicious behaviour produces deviations of predictions from observations. This would be a novel approach by applying a generalized state vector to a deep model. These results can be analyzed by considering that groups of samples in the testing data could exhibit different types of features (anomalous situation) from the ones observed during the training of the generative network with normality data. In other words, since no neuron in the latent space was trained to detect these features, abnormality data cannot activate any neuron in the neural networks and the consequent deviation of prediction from observation produces high values of the abnormality measurements.

Additionally, to evaluate the performance of the models, ROC curves are also shown in Figs. 3.13-3.15-3.17 that confirm that each of the deep model can provide high detection probability with low $P_{fa}$. In addition, the $P_d$ can be optimized through a sensible choice of the threshold in the binary testing. Indeed, Area Under Curve (AUC) and Accuracy (ACC) values are extracted and listed in the Table 3.3 where the AC-GAN seems to provide better performance than C-GAN and VAE models. From another point of view, when GAN-based models are compared to VAE, it can be noticed that: in the first case, since the generator is trained to learn a mapping between a random noise vector, $\mathbf{z}$ in Figs. 3.3-3.4, and the generated data (by learning hidden, complex structure in the real data $\mathbf{x}$), then $G$ is able to capture the dynamics in the real data. In the second case, a VAE model returns the posterior probability that an observation belongs to a specific cluster by learning the latent vector, **z** in Fig. 3.5. In this way, observations **x** from different clusters will correspond to different **z** vectors and the dynamics of **x** is captured according to the way and the time instants the vector **z** changes. In effect, learning from dynamic data as in the first case should provide better performance as confirmed by the results. Alternatively, an important advantage of the VAE, with respect to GAN, is the possibility to exploit the encoder's output latent variables ($\mu$ and $\sigma$) that represent probabilistic distributions. Indeed, such variables can be clustered to learn temporal dependencies among them and draw a probabilistic graphical representation. The latent variables can also be used to reduce the complexity due to high dimensionality data in wideband RF spectrum. Finally, Table 3.4 gives an idea about the time required to train and test the models under investigation. Among the 3 analysed models, VAE required less computational time to perform both training and testing processes, since KL is faster than MSE and $L^p$ methods.

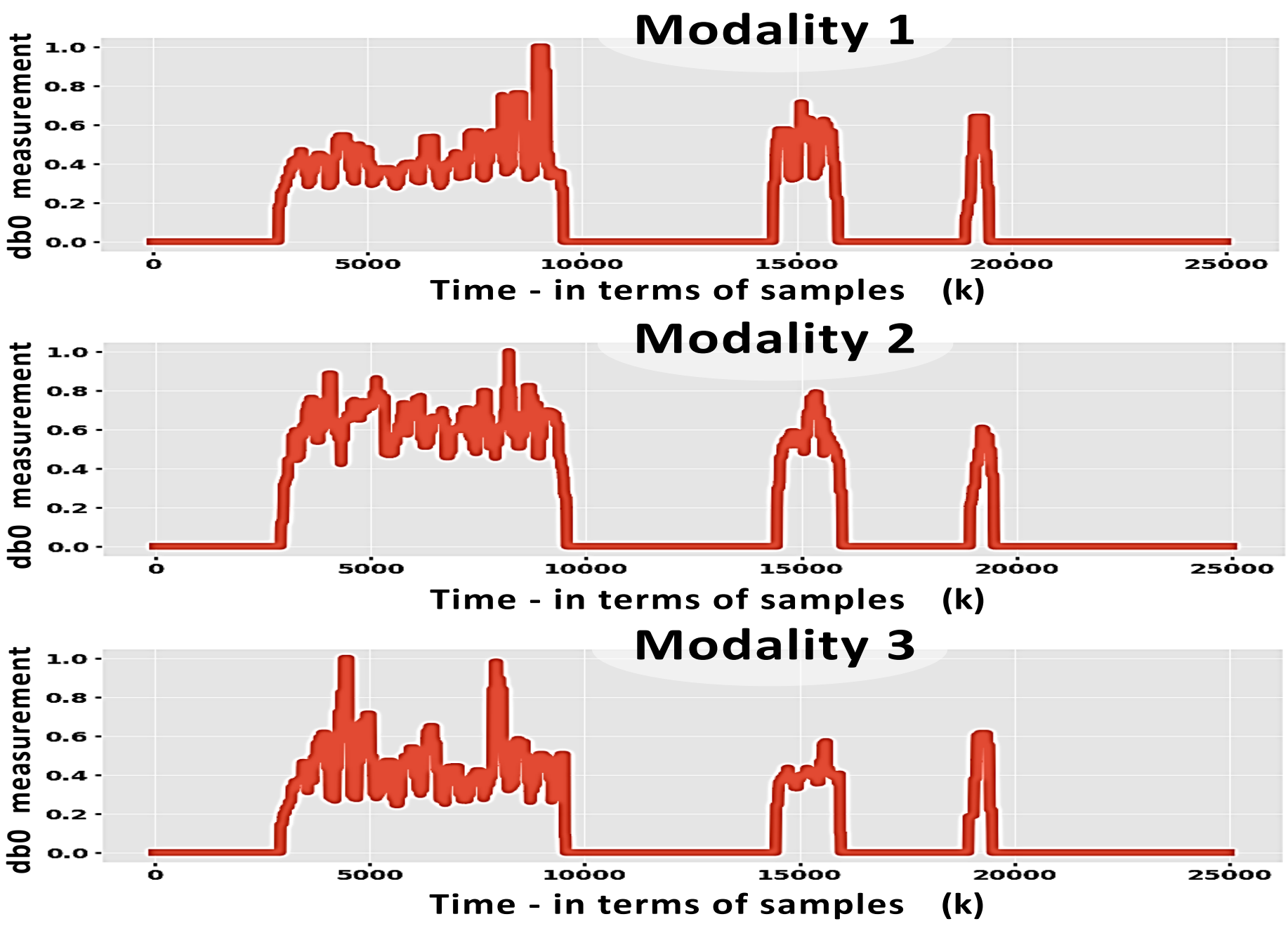

Fig. 3.12 Anomaly indicator (C-GAN model).

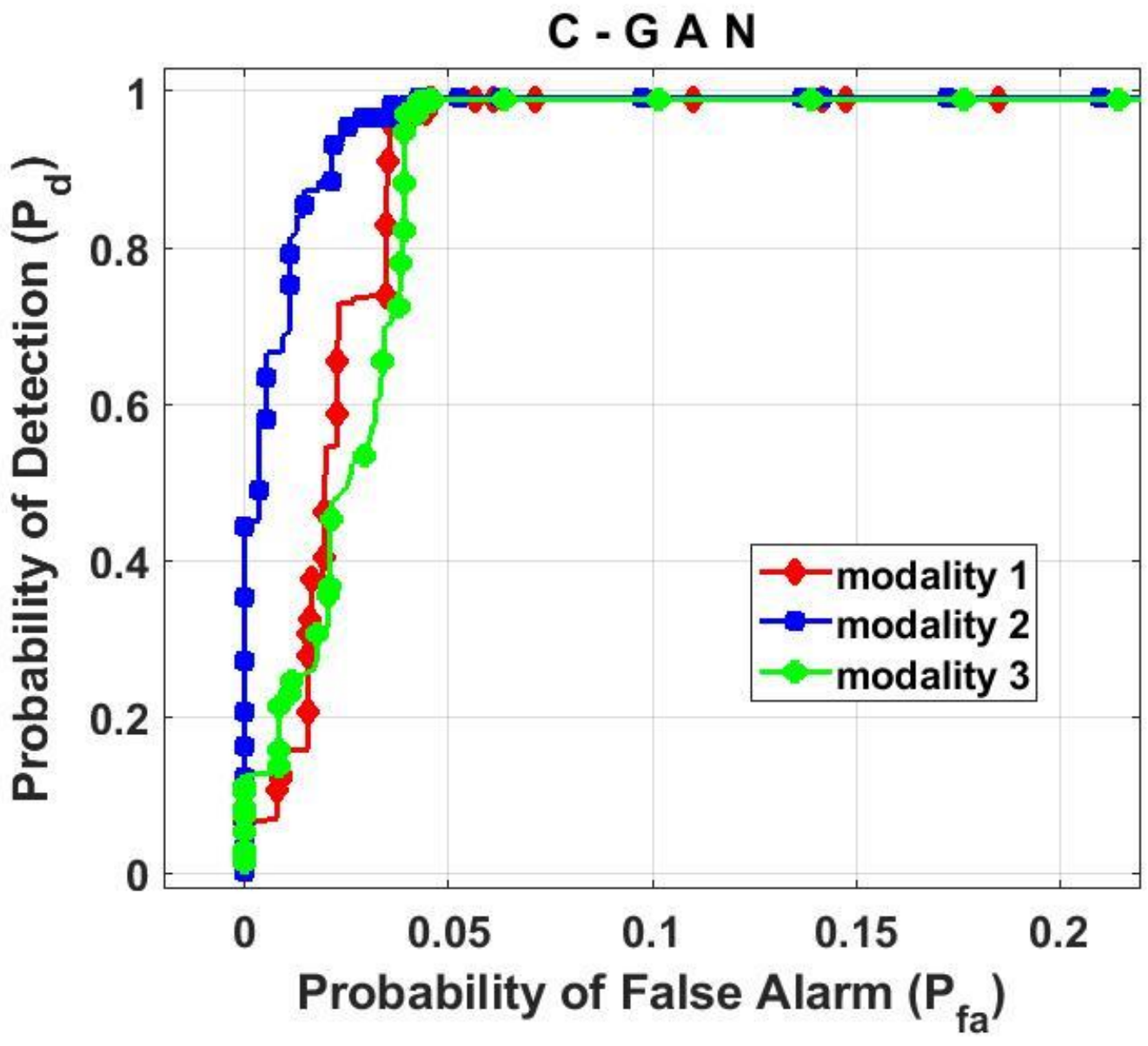

Fig. 3.13 ROC curves (C-GAN model).

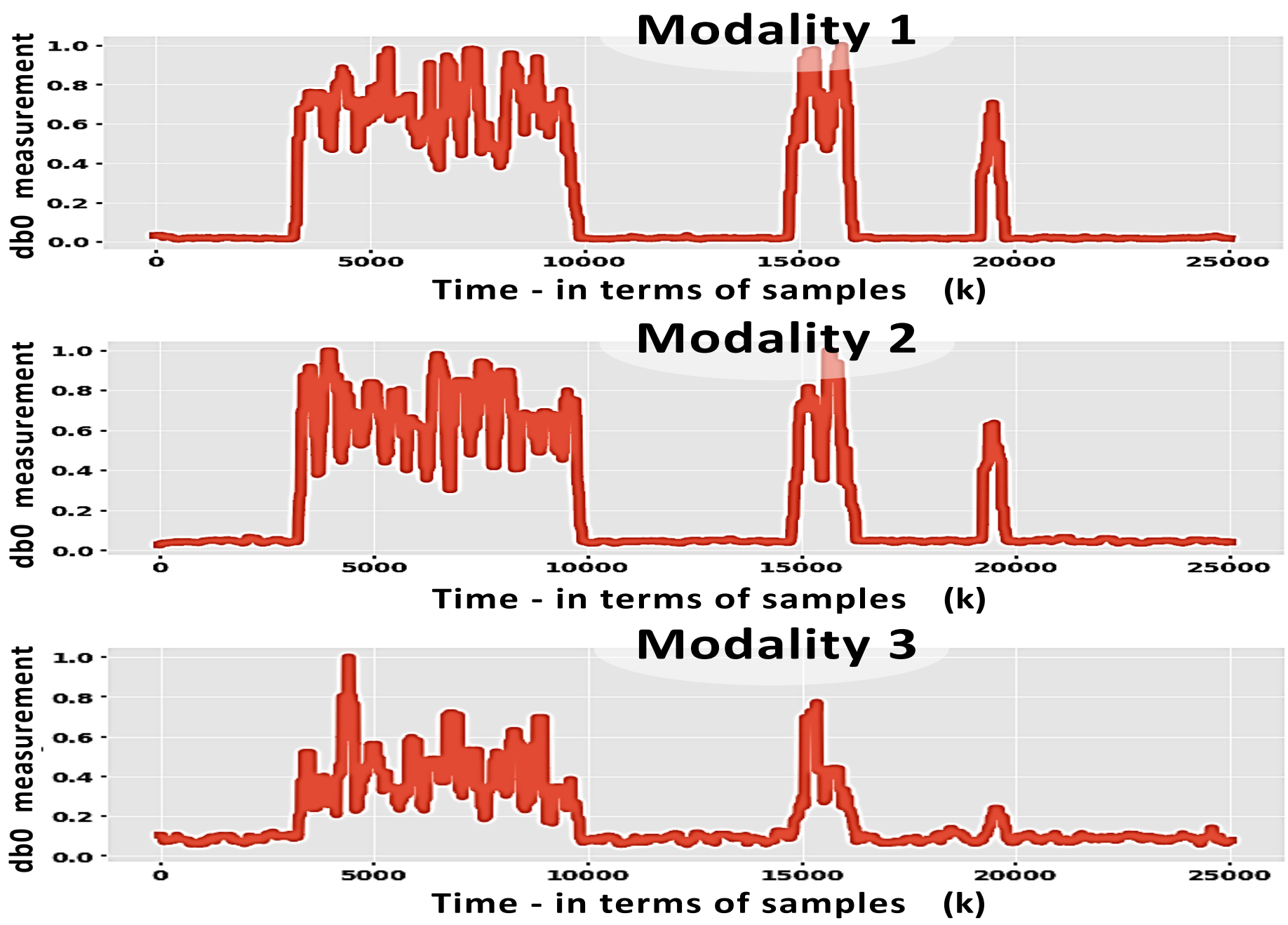

Fig. 3.14 Anomaly indicator (AC-GAN model).

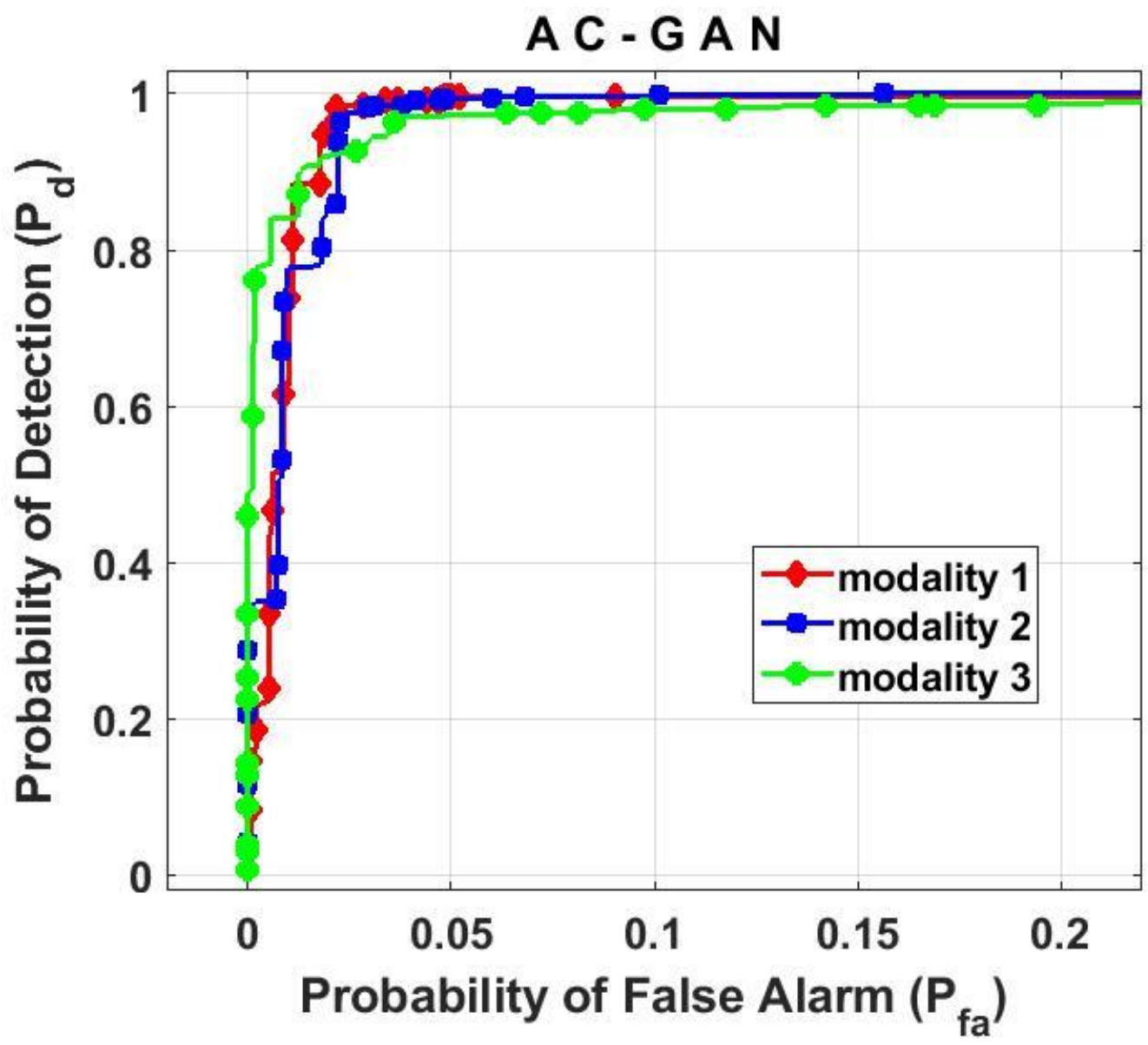

Fig. 3.15 ROC curves (AC-GAN model).

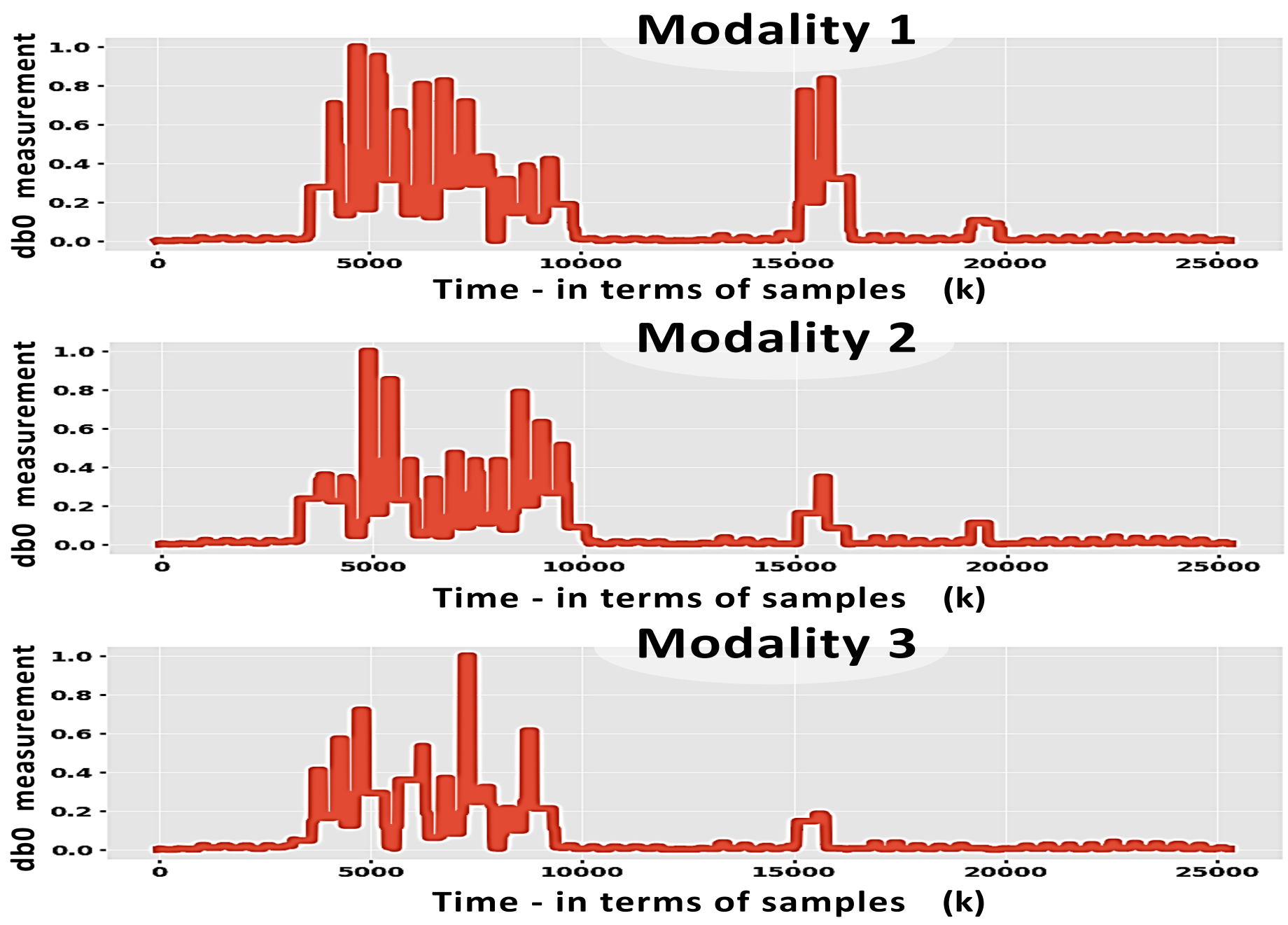

Fig. 3.16 Anomaly indicator (VAE model).

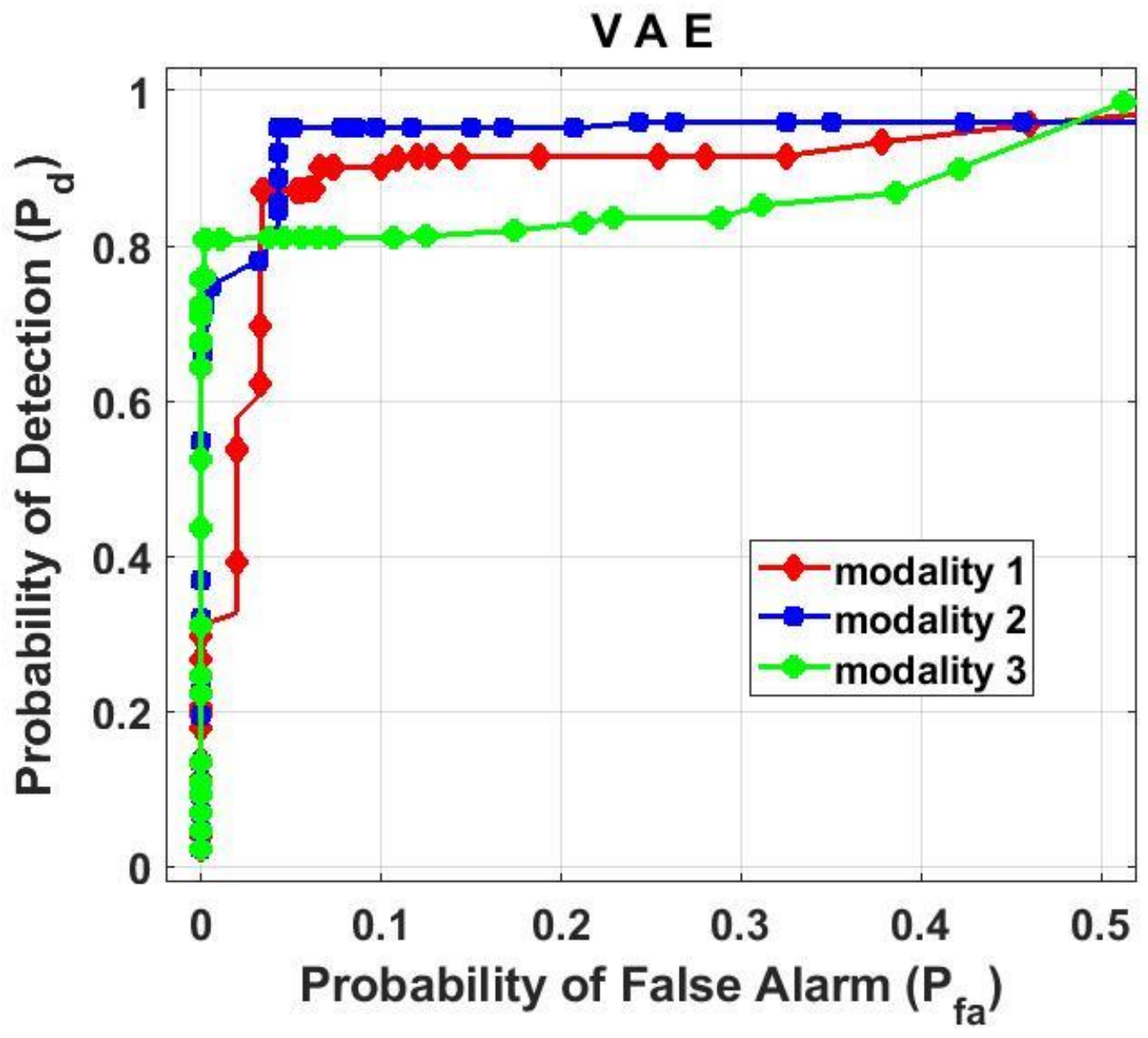

Fig. 3.17 ROC curves (VAE model).

Table 3.3 AUC and ACC values for the 3 Deep Learning Models

| | | AUC | ACC |
|---|---|---|---|
| C-GAN | modality 1 | 0.9566 | 0.9657 |
| | modality 2 | 0.9737 | 0.9696 |
| | modality 3 | 0.9545 | 0.9668 |
| AC-GAN | modality 1 | **0.9741** | **0.9804** |
| | modality 2 | **0.9751** | **0.9757** |
| | modality 3 | **0.9742** | **0.9660** |
| VAE | modality 1 | 0.9365 | 0.9356 |
| | modality 2 | 0.9577 | 0.9551 |
| | modality 3 | 0.9232 | 0.9382 |

Table 3.4 Computational times for the 3 Deep Learning Models

| Deep Learning Models | Training time [mm:ss] | Testing time [mm:ss] |
|---|---|---|
| *C-GAN* | 15:16 | 01:36 |
| *AC-GAN* | 30:42 | 03:16 |
| *VAE* | **15:09** | **01:00** |

### 3.3.5 Summary

This part (Part I) has demonstrated the effective implementation of C-GAN, AC-GAN, and VAE models to detect mmWave spectrum anomalies in a CR system. A comparison is made between deep generative models learned from the generalized state vector which incorporates the signals amplitude and the corresponding derivative extracted from ST representation of the dynamic spectrum. Extensive experiments have been conducted on a real dataset collected by using a mmWave testbed. In all the tested modalities, anomaly measurements showed good performance for the three models, particularly the AC-GAN. ROC curves confirmed that the probability of detection is high with a low false alarm probability. However, from computational time analysis, the VAE resulted in being faster than the other two networks. Moreover, in a VAE, the encoder's output latent variables could be clustered to learn temporal dependencies among them and draw a probabilistic graphical representation. These latent variables can also be used to reduce the complexity due to high dimensionality data.

## 3.4 Applications to Low-Dimensional Data (Part II)

In the first part (Sec. 3.2), we presented a framework for the joint spectrum representation and abnormality detection which is intended for huge cognitive devices dealing with multi wide-band signals (i.e., high-dimensional data) such as Base stations. This part is intended for cognitive devices dealing with narrow band signals (i.e., low-dimensional data) such as IoT sensors, aerial/terrestrial vehicles and user equipment. The proposed framework for the joint spectrum representation and abnormality detection dealing with low-dimensional data consists of extracting features, clustering the extracted features and using Bayesian Filters to predict and detect abnormal behaviours in the spectrum.

An OFDM based model has considered in this approach in which malicious signals corrupt any of the given sub-carriers (as shown in Fig. 3.18). Let us suppose that $N$ x $M$ OFDM signal grid is transmitted where $N$ and $M$ are the number of sub-carriers and OFDM symbols, respectively, and at least one of the sub-carrier suffers from malicious attacks. Upon reception, FFT output is picked up at the CR device to form a state vector, as shown in Fig. 3.18. There are three following dominant reasons due to which FFT output is being exploited. First of all, FFT gives amplitude and phase information, and the signal is easily analyzed statistically by using this information. The second reason stems with the fact that an anti-jamming technique can be implemented before demodulation of the signal to mitigate any abnormal signal effects at this level. In this way, the burden on the CR device processing can be potentially reduced. Third, exploiting FFT output simplifies the spectrum sensing process and reduce the complexity where the CR device can scan the entire time-frequency grid without any extra hardware.

For any given sub-carrier consisting of $M$ symbols, there is a temporal evolution between consecutive symbols that allows describing how amplitude and phase values are dynamically changing in a specific sub-carrier '$sc$'. Therefore, we can define the generalized state vector

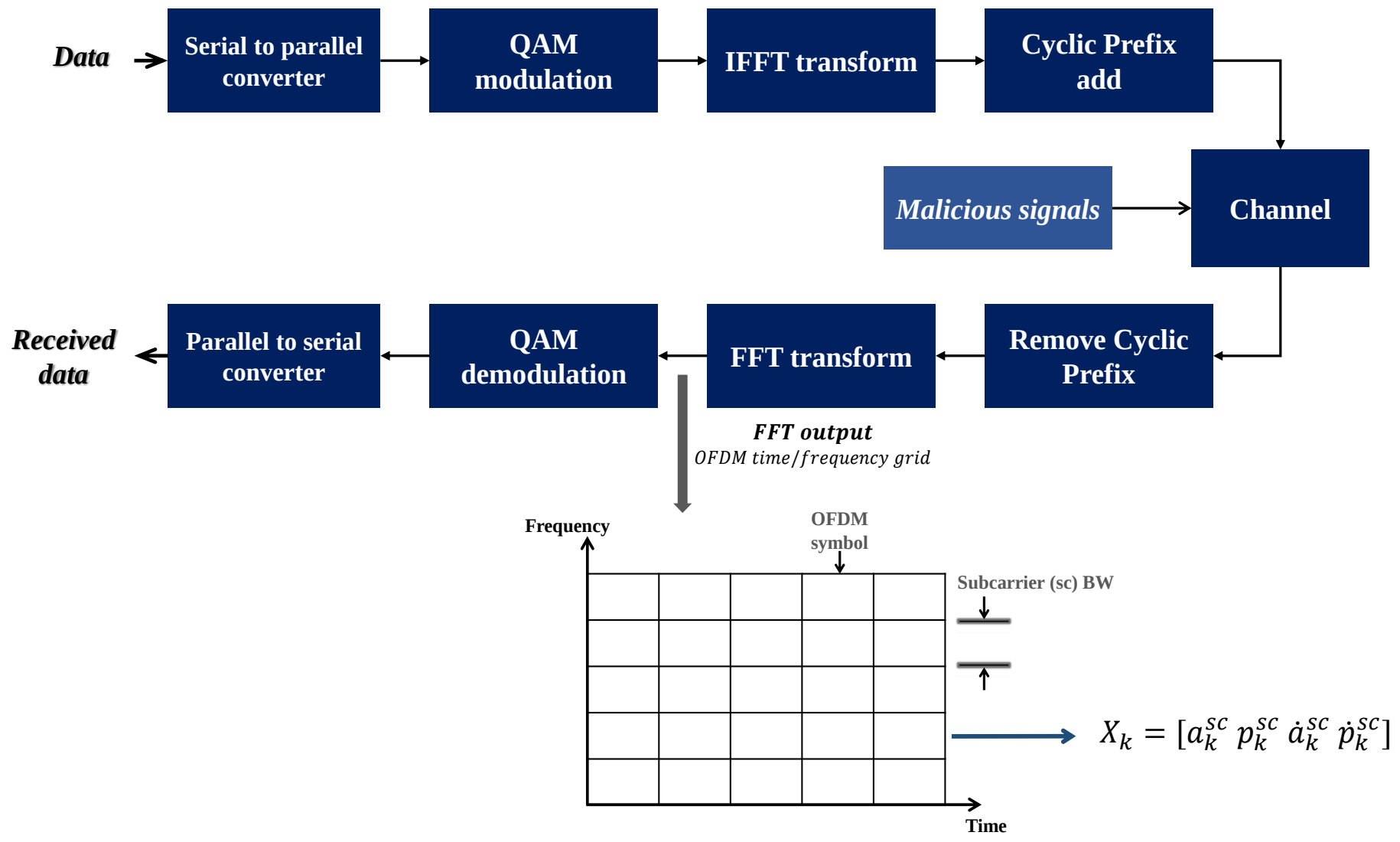

Fig. 3.18 Extracting FFT output to form the state vector by the CR device.

at each time instant $k \in \{1, 2, \ldots, M\}$ as,

$$X_k^{sc} = [\, a_k^{sc}\ p_k^{sc}\ \dot{a}_k^{sc}\ \dot{p}_k^{sc}\, ], \tag{3.16}$$

where $X_k^{sc}$ represents a low dimensional generalized state vector from which the radio learned a representation structured in a Dynamic Bayesian Network (DBN) encoding the signals' dynamics and used as a GM. $a_k^{sc}$, $p_k^{sc}$ are the amplitude and phase while $\dot{a}_k^{sc}$, $\dot{p}_k^{sc}$ are the corresponding derivatives. The 1st order temporal derivatives are obtained by calculating the difference between the amplitude (phase) at time instant $k$ and amplitude (phase) at time instant $k-1$ according to:

$$\dot{a}_k^{sc} = \begin{cases} 0 \ \text{ if } \ k = 1, \\ a_k^{sc} - a_{k-1}^{sc} \ \text{ if } \ k > 1, \end{cases} \tag{3.17}$$

and

$$\dot{p}_k^{sc} = \begin{cases} 0 \ \text{ if } \ k = 1, \\ p_k^{sc} - p_{k-1}^{sc} \ \text{ if } \ k > 1. \end{cases} \tag{3.18}$$

### 3.4.1 Dynamic Bayesian Networks

Bayesian Networks (BNs), also known as Probabilistic Networks (PNs), are directed acyclic graphical models where nodes in the graph represent a set of random variables and edges encode a particular factorization of the joint distribution of that set from a single point in time. However, in many real-world applications, most of the events are not detected based on a single point in time but on multiple observations that yield to a certain event [214]. Thus, for time-dependent applications, BNs do not provide temporal dependencies among random variables. Hence they cannot be used to make inference for such applications.

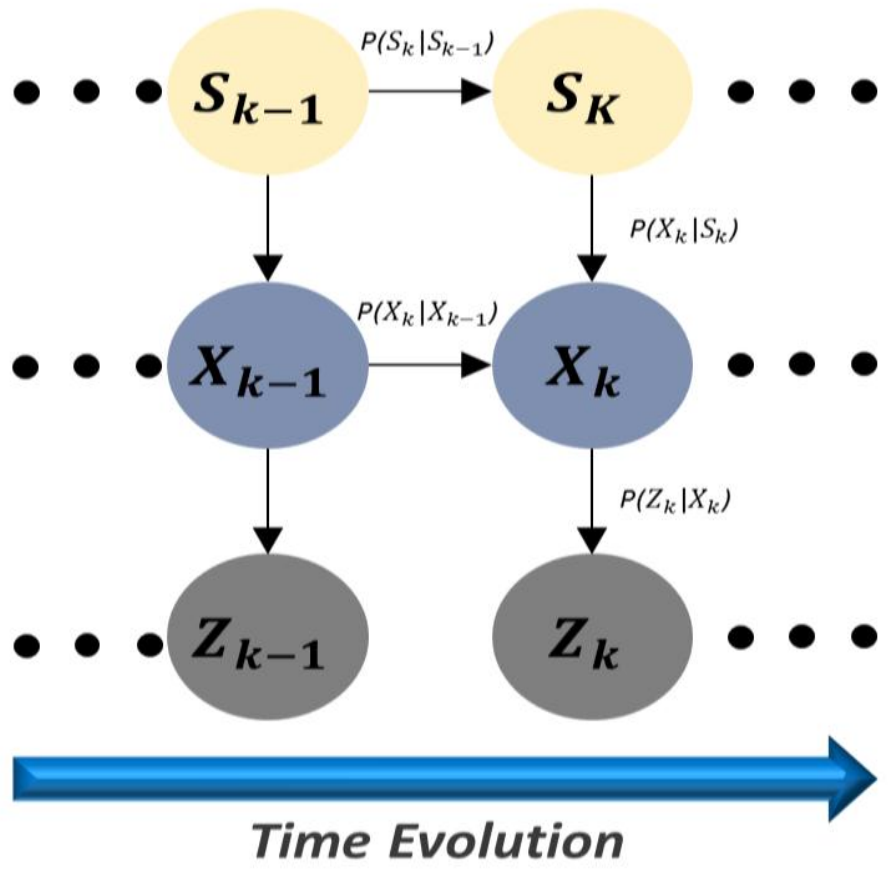

Fig. 3.19 DBN model comprises of two parts (discrete and continuous) representing the dynamics of the signals in spectrum at multiple levels.

A Dynamic Bayesian Network (DBN) is an extension of the BNs that can model dynamic processes and describes a system's evolution with time at hierarchical levels. DBN allows encoding probabilistic dependencies and feedbacks between random variables over different time slices. Such relationships provide a way of making inferences about the system and draw some conclusions. A DBN is typically represented by two sets of parameters. The first includes the number of nodes in each time slice and the corresponding topology while the second set consists of the conditional probability distributions (CPDs) described by the edges of the network. DBN generalizes linear dynamical systems by representing the hidden and observed states in terms of random state variables providing a graphical structure that specifies the corresponding conditional dependencies and a compact parameterization of the model. DBN can decompose data with complex and non-linear dynamics into segments that are explainable by simpler dynamical units. A specific class of DBN models known as Switching linear dynamic systems (SLDS) [215] can be used to discover the dynamical units and explain their switching behaviour and their dependency on both observations and continuous hidden states.

In the low-dimensional application, the behaviour of signals is investigated using DBN model which provides a probabilistic switching model and comprises of coupled hidden layers for each temporal slice for modeling and predicting the dynamics of the signal over time. This facilitates the detection of malicious signals inside the spectrum. The proposed DBN model is a generative model where continuous states are encoded inside discrete regions. DBN makes spectrum inferences at two levels, discrete and continuous, through Bayesian Filtering.

### 3.4.2 Bayesian Filtering

Bayesian theory is a branch of mathematical probability theory that incorporates prior knowledge and evidence to model uncertainty about the world. In Bayesian inference, states and parameters describing uncertainties are treated as random variables which are either time-varying or deterministic [216]. The objective is to use causal and prior knowledge to infer conditional probabilities at hierarchical levels given only observations. Graphical models as DBNs allows constructing hierarchical statistical models and illustrate the Bayesian inference. Bayesian filtering can be employed using two assumptions: states ($\mathbf{x}$) are following a first-order Markov process i.e., $p(\mathbf{x}_k|\mathbf{x}_{1:k-1}) = p(\mathbf{x}_k|\mathbf{x}_{k-1})$, and observations are independent of the states. Filtering is the process of extracting information about a quantity of interest at time $k$ and using observed data up to and including $k$. The purpose of Bayesian Filtering is to compute the marginal posterior distribution $p(\mathbf{x}_k|\mathbf{z}_{1:k})$ of the state $\mathbf{x}_k$ at each time step $k$ given the observations up to $k$ [217]. The fundamental steps of Bayesian filtering are the following:

- *Initialization*, the filtering starts from a prior distribution $p(\mathbf{x}_k)$.
- *Prediction step*, given the dynamic model, predictive posterior can be computed using the Chapman-Kolmogorov equation:

$$p(\mathbf{x}_k|\mathbf{z}_{k-1}) = \int p(\mathbf{x}_k|\mathbf{x}_{k-1})p(\mathbf{x}_{k-1}|\mathbf{z}_{k-1})d\mathbf{x}_{k-1}, \tag{3.19}$$

  It is to note that if $\mathbf{x}_{k-1}$ is a discrete distribution, then the above integral is replaced with summation over $\mathbf{x}_{k-1}$.

- *Update step*, given the observation $\mathbf{z}_k$ at time step $k$, posterior of state $\mathbf{x_k}$ can be updated using the Bayes rule:

$$p(\mathbf{x}_k|\mathbf{z}_k) = \frac{1}{\mathbf{Z}_k}p(\mathbf{z}_k|\mathbf{x}_k)p(\mathbf{x}_k|\mathbf{z}_{k-1}), \tag{3.20}$$

where $\mathbf{Z}_k$ is the normalization constant given by:

$$\mathbf{Z}_k = \int p(\mathbf{z}_k|\mathbf{x}_k)p(\mathbf{x}_k|\mathbf{z}_{k-1})d\mathbf{x}_k. \quad (3.21)$$

Kalman Filter (KF) [218] is the closed form solution to the Bayesian filtering equations under the assumption that the dynamic system is linear and the noise is Gaussian. Under non-linearity hypothesis, an extension of KFs was proposed and known as the Extended KF (EKF) [219] which perform local linearizations of the state transition model and the observation model. In addition, due to the fact that some systems cannot be linearized (their first order derivatives are null) a second order extension to KF has been proposed and called Unscented KF (UKF) [220] which also requires the hypothesis of Gaussian noise. On the other hand Monte Carlo methods denoted as Particle Filters (PF) have been developed to represent the posterior density in terms of random and weighted samples which can be applied to non-linear and non-Gaussian environments [221]. Sampling in high-dimensional spaces is the main drawback of PF degrading its performance and makes it inefficient. To overcome such an issue, marginalization can be carried out using Rao-Blackwellisation PF to reduce the space size over which the sampling is performed [222].

A combination of PF and KF can be applied in DBNs providing probabilistic inference about the spectrum at multiple levels (discrete and continuous). The combined approach is called Markov Jump Particle Filter (MJPF) [223], it is sharing the same form of state-space models but with a jump Markov dynamics at multiple models which can be linear/non-linear Gaussian/non-Gaussian.

### 3.4.3 Unsupervised Clustering Methods

The DBN's vocabulary can be learned from time-series data (i.e., a data-driven approach). The vocabulary includes a set of discrete switching variables at the discrete level along with the associated dynamic models at the continuous level. Discrete variables can be thought as discrete regions of the spectrum which can be learned by clustering the received stimuli while sensing the spectrum. The clustering procedure can be performed in two different ways, supervised or unsupervised [224]. The main difference between the two ways is that supervised learning requires prior information of what the output should be given the time-series data. Thus, the supervised method aims to learn the best relationship among the observed data in input and the given desired output [225]. Besides, unsupervised learning aims to learn the inherent structure of the data without the explicit use of known labels. We are interested to cluster the incoming data following an unsupervised manner due to the fact that CR usually starts operating in the field without any prior knowledge. The most popular unsupervised learning methods include K-means, Self-Organizing Maps (SOM) and Growing Neural Gas (GAS). K-means is an iterative procedure aiming to partition observations into k clusters based on the sum of squared euclidean distance between each observation and its nearest cluster mean. However, it does not specify how the initial centroids should be selected during the initialization procedure and this might lead to many local optima of the solution space before reaching the convergence. SOM was proposed as a possible subtitute for K-means which is less prone to local optima than k-means [226]. SOM is a type of Artificial Neural Network (ANN) that is trained in an unsupervised manner with a simple structure and computational form to provide low-dimensional discretized representation (map) of the input data samples [227]. GNG is an incremental neural network that learns the relation between a given set of input patterns and adapts the topological structure based on nearest neighbour relationships and local error measurements [228]. GNG shows higher flexibility in representing the input data if compared to fixed size networks like the Self Organizing Maps (SOMs). This chapter uses the SOM method, while the GNG will be implemented in the following chapters.

## 3.5 Simulation and Performance Evaluation (Part II)

### 3.5.1 OFDM signal generation

A random signal is generated by using a random bit generator, with a 16-QAM modulation scheme, and then mapped onto $N$ sub-carriers by using Inverse Fast Fourier Transform (IFFT) followed by cyclic prefix augmentation. The produced OFDM signal is assumed to follow mmWave waveform specifications and to be affected by only additive white Gaussian noise (AWGN) noise. There is no frequency offset between transmitter and receiver since a perfect synchronization is assumed. The OFDM symbols $X(k)$, which are mapped to $N$ sub-carriers, can be represented in the time domain as

$$x(t) = \sum_{k=1}^{N} X(k) e^{j2\pi kt/N}, \tag{3.22}$$

and the received OFDM signal can be written as

$$r(t) = h(t) \otimes x(t) + w(t), \tag{3.23}$$

where $h(t)$ is the channel response, $x(t)$ is the transmitted signal and $w(t)$ is AWGN with zero mean and power spectral density $\sigma_w^2$. For simulation purposes, we consider an OFDM signal consisting of $N = 64$ subcarriers and $M = 3000$ symbols. CP removal and FFT implementation come under post-processing after the OFDM signal is captured at the CR terminal. The output data of the FFT block is extracted to form the state vector. The data is divided into two sets: one carries only the normal data (clean data) for the training phase, and the second set contains malevolent data (affected data) for the testing phase. The extracted data corresponds to only one sub-carrier, which can be any of the available $N$ sub-carriers.

### 3.5.2 Training Phase

The DBN model, as depicted in Fig. 3.19, can establish the dependency among random variables with the evolution of time graphically. The network consists of three levels; the lowest one corresponds to the observation $Z_k$, $X_k$ shows the medium level that encodes continuous information and $S_k$ stands for the higher level represented by super-states which discretize the continuous states. Arrows show conditional probabilities between the involved variables. Vertical arrows describe causalities between both continuous and discrete levels of inference and observed measurements. Horizontal arrows represent temporal causalities between hidden variables.

The DBN model can be learned by implementing the following steps:

**Learning superstates**. As initial step of learning super-states, SOM is employed to obtain, from the state vector $X_k$, a set of learned super-states $S$ where similar information (quasi-constant derivatives) is valid, such that:

$$S = \{S_1, S_2, \dots, S_L\}, \tag{3.24}$$

where $S_k \in S$ and $L$ is the total number of super-states.

**Learning Transition Model**. It is feasible to estimate the transition matrix which encodes the probabilities $P(S_k|S_{k-1})$ of passing from a current superstate to another one by observing activated superstates over the training time.

**Learning Region Properties**. There are certain variables which describe the region $S_k$. Such variables include $\xi_{S_k}$ , $Q_{S_k}$ and $\psi_{S_k}$ which show mean, covariance and boundary, respectively. The boundary is defined as:

$$\psi_{S_k} = \xi_{S_k} + 3(\sigma(\xi_{S_k})) \tag{3.25}$$

Such variables are used to perform predictions during the testing phase.

**Learning Dynamic Model**. The dynamic system can be analyzed and inferred in the context of two models: the measurement model that captures observation into states and the dynamic model that explicates state estimation with time and it can be written as:

$$X_k = AX_{k-1} + BU_{S_{k-1}} + w_k, \tag{3.26}$$

where $A = [A_1 \;\; A_2]$ is a dynamic model matrix: $A_1 = [I_2 \;\; 0_{2,2}]^\intercal$ and $A_2 = 0_{4,2}$. $I_n$ represents a square identity matrix of size $n$ and $0_{l,m}$ is a $l \times m$ null matrix. $B = [I_2\Delta k \;\; I_2]^\intercal$ is a control input model and $\Delta k$ represents the sampling time. $w_k$ represents the prediction noise. The variable $U_{S_{k-1}}$ is a control vector that encodes the spectrum's action when it is inside a superstate $S_k$, such that:

$$U_{S_k} = [\dot{a}_{S_k} \;\; \dot{p}_{S_k}]^\intercal \tag{3.27}$$

Accordingly, it is possible to estimate the probability of obtaining a future spectrum's state given its present state $P(X_k|X_{k-1}, S_{k-1})$ for each superstate $S_{k-1}$.

### 3.5.3 Testing Phase

After acquiring a set of state vector during the training phase, that describes the behaviour of the signal inside the spectrum under no malicious attacks (normal behaviour), we apply the MJPF in the testing phase that performs predictions at different levels and provides abnormal signals based on the Bhattacharyya distance between prediction $p(X_k^*|X_{k-1}^*(S_k^*))$ and

- probability of being inside the predicted superstate of particle $p(X_k^*|S_k^*)$:

$$db1 = -\ln \int \sqrt{p(X_k^*|X_{k-1}^*(S_k^*))p(X_k^*|S_k^*)}dX_k^*; \tag{3.28}$$

- evidence $p(z_k|X_k^*)$ to have solutions near the measurement:

$$db2 = -\ln \int \sqrt{p(X_k^*|X_{k-1}^*(S_k^*))p(Z_k|X_k^*)}dX_k^*; \tag{3.29}$$

where $(.)^*$ indicates the considered particle and $S_k^*$ means that the prediction depends on the superstate. $Z_k$ are the measurements/observations, $X_k$ represent states, and $S_k$ are the superstates. The $db1$ indicator corresponds to the discrete level of the DBN and its value is related to the similarity between the prediction of the state and the likelihood to be in the predicted superstate. If the predicted state is out of the learned model (outside the clusters), $db1$ will provide a high abnormality signal; otherwise, it will provide low abnormality signal. While the $db2$ indicator corresponds to the continuous level of the DBN and its value is related to the similarity between the state prediction and the continuous state evidence corresponding to the new observation in each superstate.

### 3.5.4 Results

The proposed model is first trained with normal data (clean data mentioned in Sec. 3.5.1) and, then, it is tested with affected data (Sec. 3.5.1), by considering the following two distinct scenarios.

- **Scenario 0** (Reference Situation): represents the normal behaviour of the signal related to the clean data (Fig.3.20) which is used during the training phase to learn the DBN model. The MJPF is applied to the learned DBN to test the other scenario with attacks and detect any abnormal behaviour.
- **Scenario I** (Multiple Jammer attacks): a jammer dynamically attacks multiple OFDM symbols randomly in a specific sub-carrier with different power values ($P_J$ = 5 dB, $P_J$ = 4 dB, $P_J$ = 3 dB, $P_J$ = 2 dB), as shown in Figures (3.21, 3.22, 3.23, 3.24).

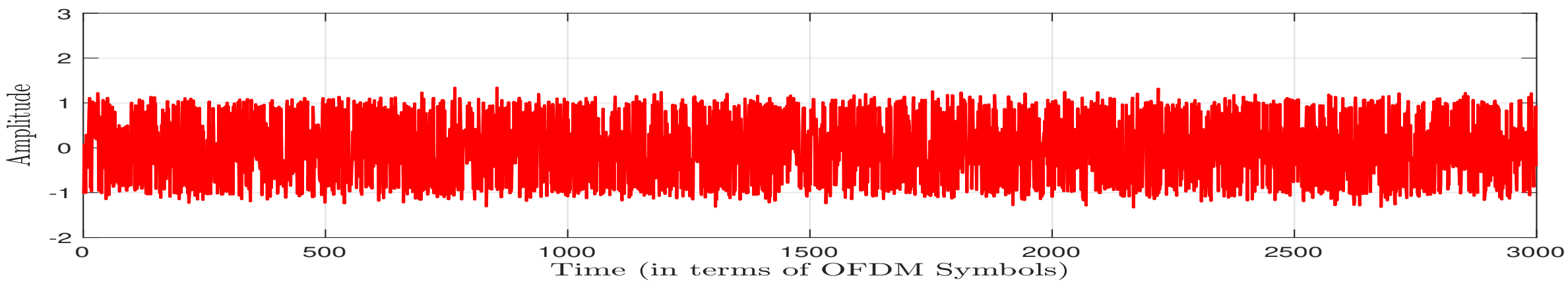

Fig. 3.20 Reference Situation (scenario 0): OFDM Signal which does not contain any abnormal signal.

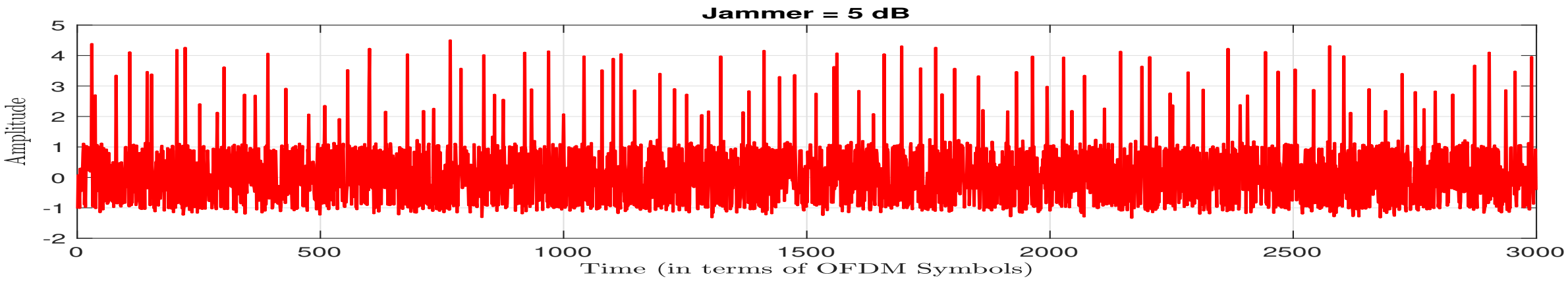

Fig. 3.21 Scenario I: $P_J = 5dB$.

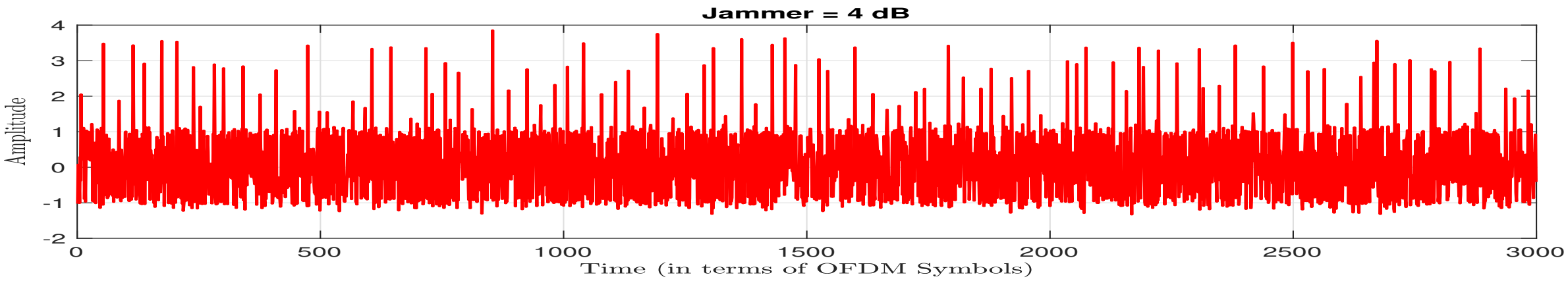

Fig. 3.22 Scenario I: $P_J = 4dB$.

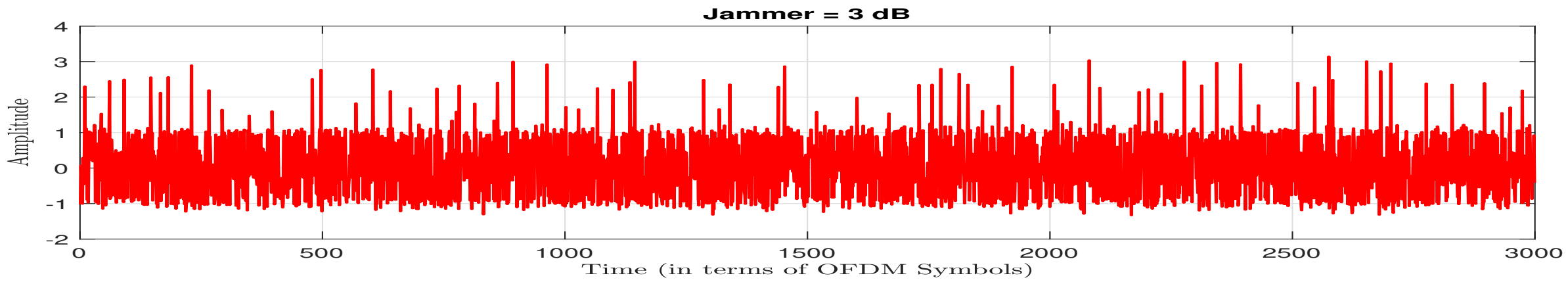

Fig. 3.23 Scenario I: $P_J = 3dB$.

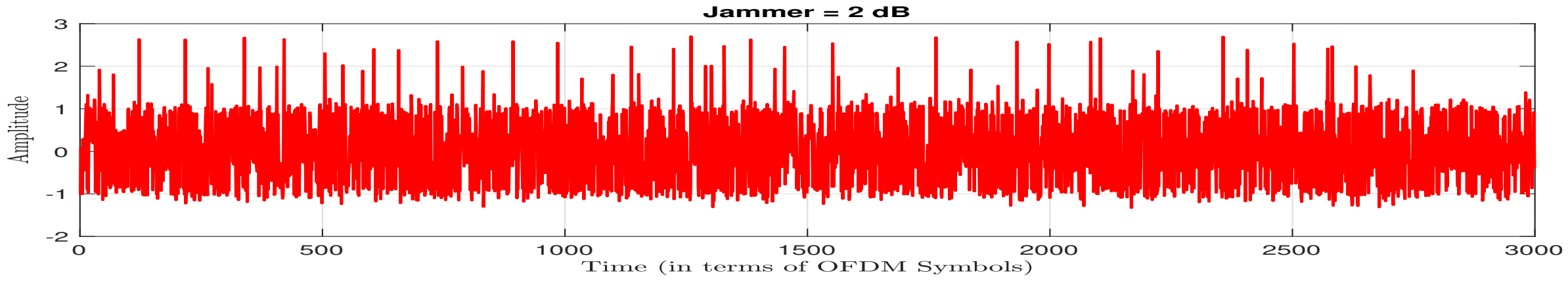

Fig. 3.24 Scenario I: $P_J = 2dB$.

Testing new signals could follow the same rules with which the dynamic model has been learned (reference situation), but they could deviate due to the presence of malicious attacks in the signal. In order to detect any abnormal situation during the online phase, two abnormality measurements are defined for discerning the abnormal signals based on the indicators defined in Eqs. 3.28 and 3.29.

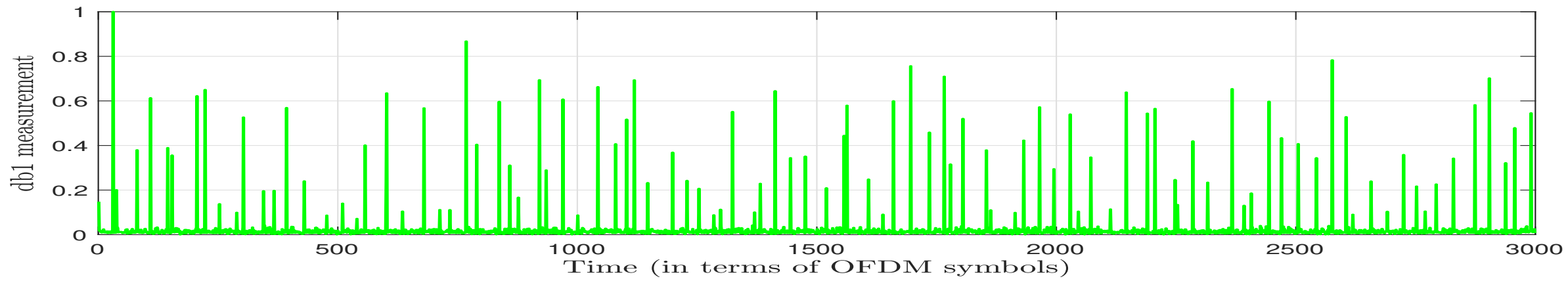

(a) discrete level.

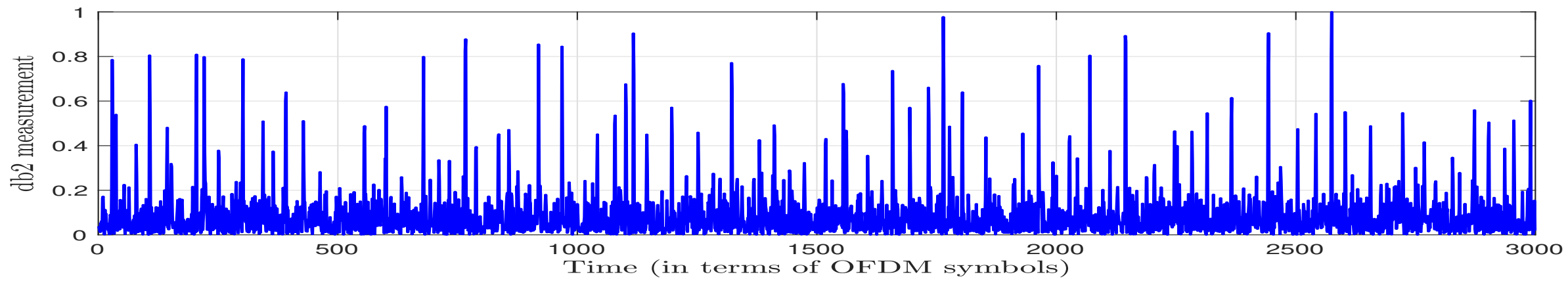

(b) continuous level.

Fig. 3.25 Scenario I ($P_J = 5dB$): Abnormality indicators

Figures (3.21, 3.22, 3.23, 3.24) contain multiple jammer attacks (with different jammer power values) in the OFDM signal. It is evident from the corresponding abnormality

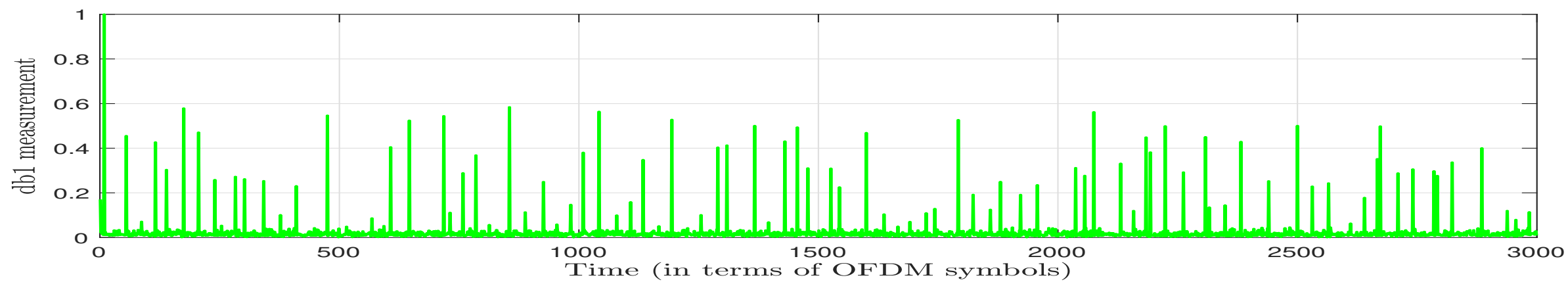

(a) discrete level.

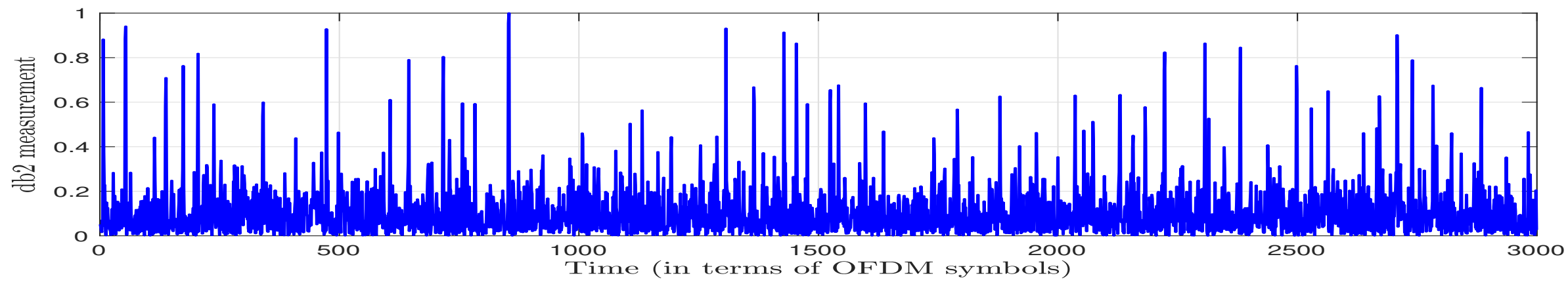

(b) continuous level.

Fig. 3.26 Scenario I ($P_J = 4dB$): Abnormality indicators

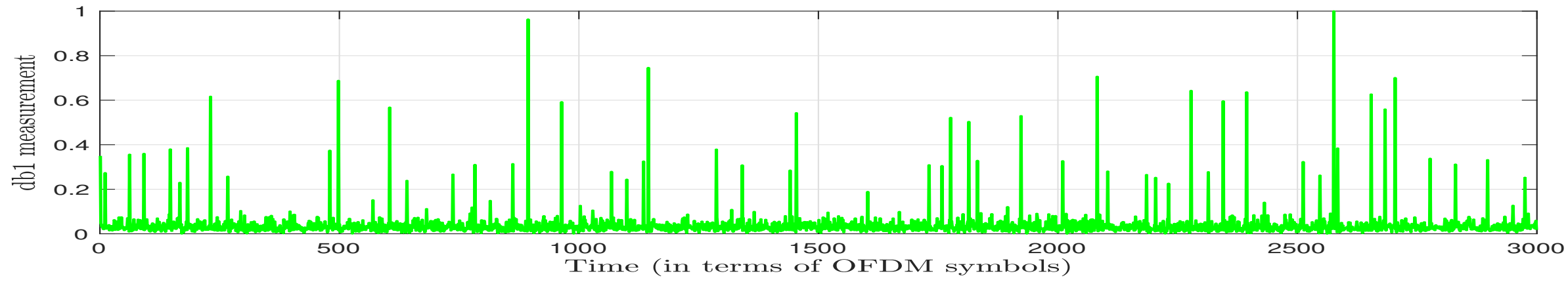

(a) discrete level.

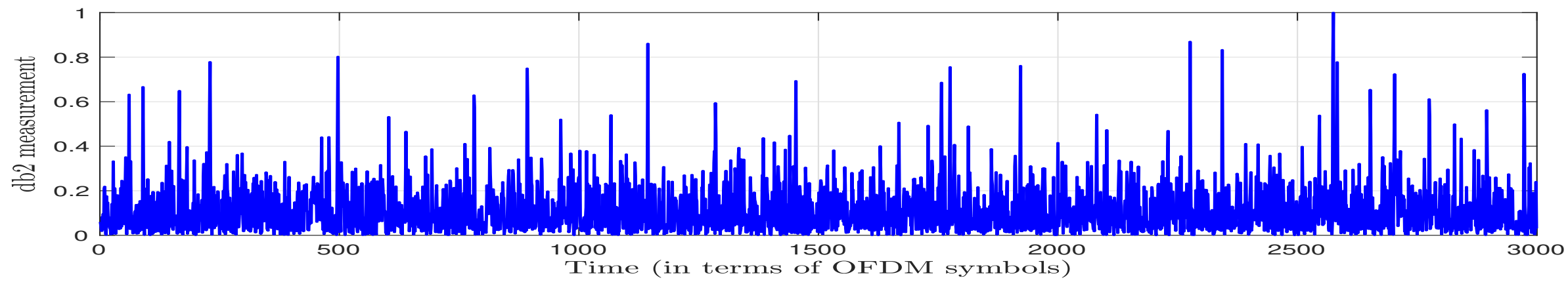

(b) continuous level.

Fig. 3.27 Scenario I ($P_J = 3dB$): Abnormality indicators

measurements in Figures (3.25, 3.26, 3.27, 3.28) that DBN model is able to detect abnormal signals at both the continuous level, in which the prediction based on the learned model is different from observation, and the discrete level in which the predicted state is outside the learned discrete regions (superstates).

Even in this application, where the dimensionality of the data is low, the ROC curves in Fig. 3.29 (a) and Fig. 3.29 (b) show that the DNB can provide high detection probability with low $P_{fa}$ at both continuous and discrete levels when the jammer's power is medium (5 dB and 4 dB), respectively. The $P_d$ can be optimized through a sensible choice of the threshold in the binary testing. After observing the ROCs in the second application, we can see how the performance degrades as the jammer decreases its power. At a certain point, we have a high probability of False Alarm which makes it important to learn new models that are able to detect such a new situation (Jammer with new behaviour).

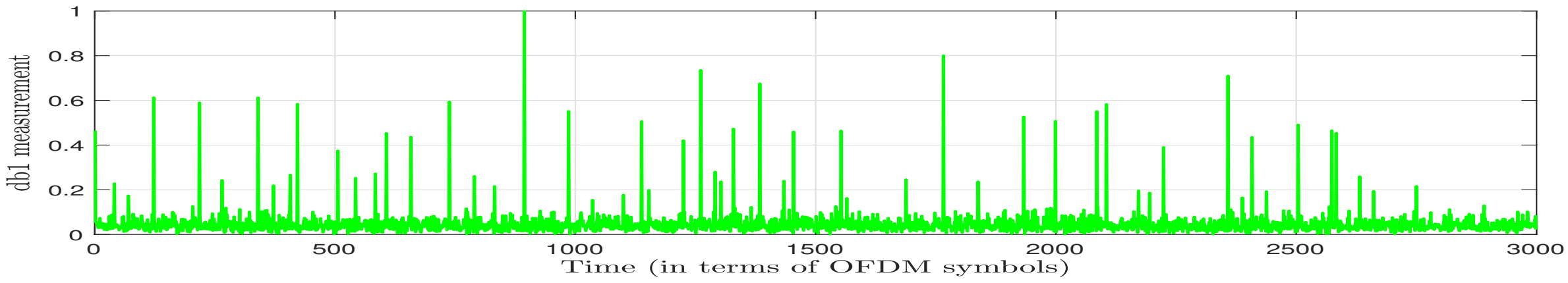

(a) discrete level.

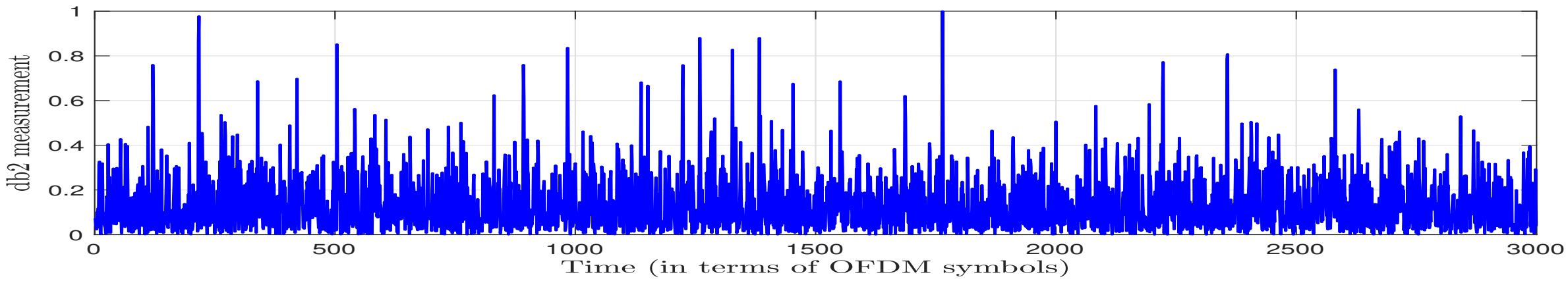

(b) continuous level.

Fig. 3.28 Scenario I ($P_J = 2dB$): Abnormality indicators

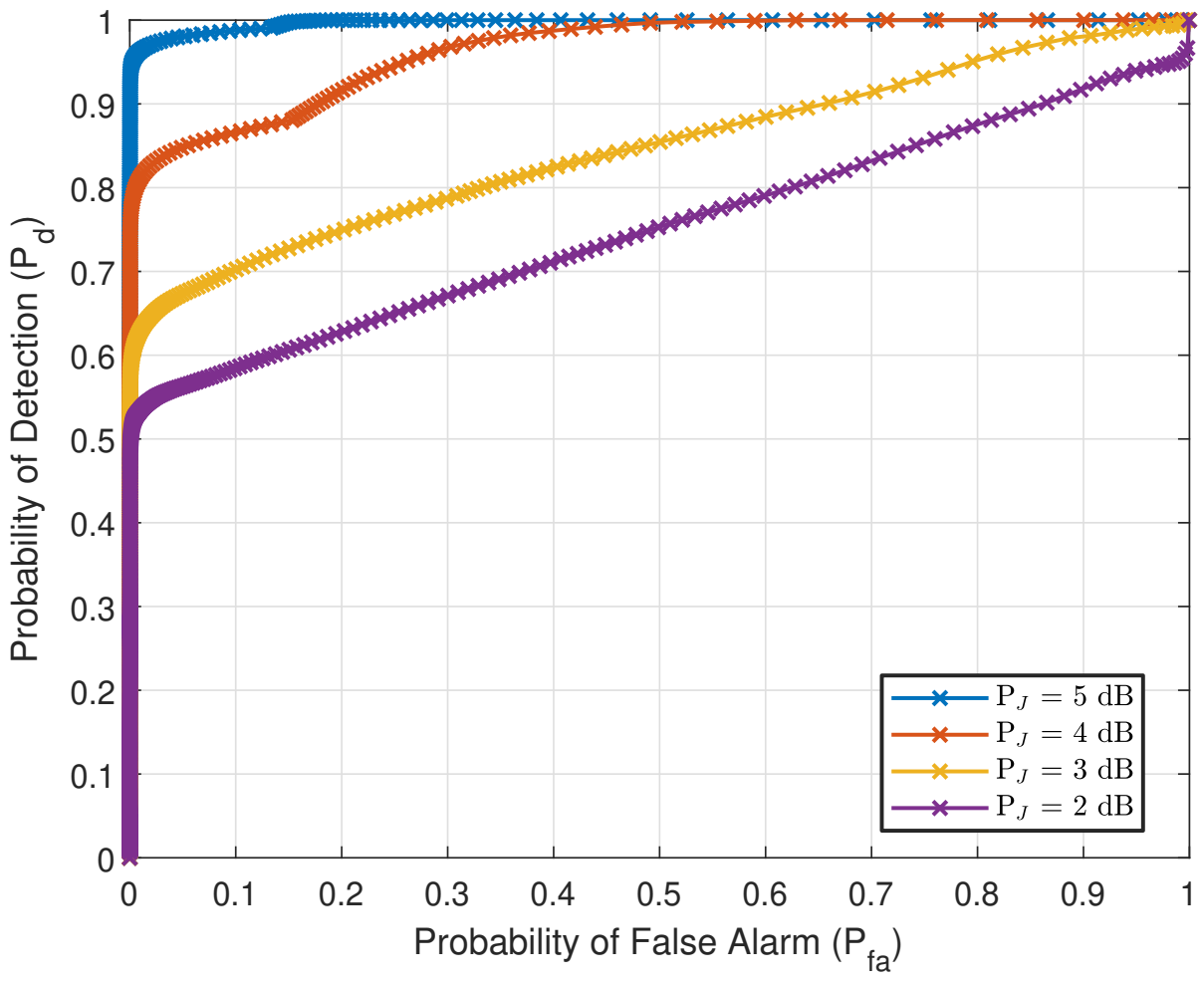

(a) $db1$.

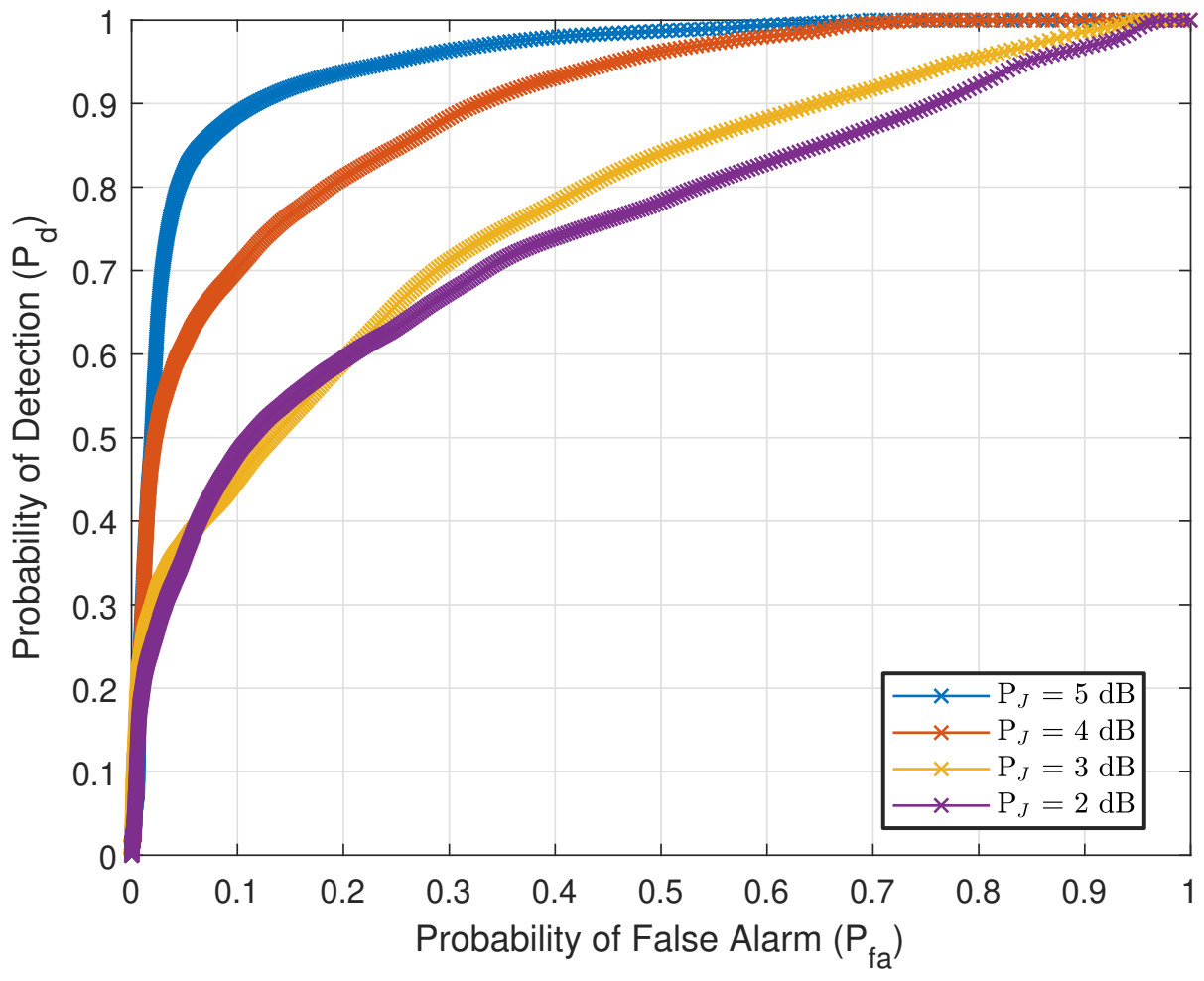

(b) $db2$.

Fig. 3.29 Scenario I: ROC

Before concluding, it should be summarized that the two performance metrics, abnormality indicators and ROC curves described in Sec. 3.3.3, can help to detect abnormalities and evaluate the performance of the model. According to ROC curves, the SA module can decide whether to incrementally learn new generative models whenever it performs poorly (in case of low $P_d$ and high $P_{fa}$) or to act accordingly to mitigate malicious attacks.

### 3.5.5 Single and Bank-Parallel DBN for multiple sub-carries

#### System Model

We consider a CR-IoT network consisting of a group of Cognitive Radio Users (CRUs) and a jammer trying to disrupt the communication as shown in Fig. 3.30. CRUs sense the spectrum continuously and try to detect the abnormal situation. The radio spectrum contains OFDM waveforms based on IEEE 802.11ah standard, which is adopted in this work. OFDM divides the band channel into many narrower sub-carriers allowing different users to transmit simultaneously with different orthogonal frequencies. The OFDM modulated signal consists of a set of $N$ sub-carriers:

$$C = \{C_1, C_2, \dots, C_N\}, \tag{3.30}$$

each sub-carrier is divided into $Q$ symbols in the time domain, forming a $N \times Q$ time-frequency grid. In the previous section, only one sub-carrier is picked to employ the proposed method, supposing that OFDM use 16-QAM for all the sub-carriers in the set (3.30). Instead, here we consider multiple sub-carriers modulated with different QAM (4, 16, 64, and 256-QAM according to the standard IEEE 802.11ah). Exploiting FFT output which consists of amplitude and phase of each symbol makes the spectrum sensing easier and less complex where CRUs can scan the entire grid. Moreover, by using the Amplitude and Phase information at this level, permits to implement a jammer detection technique before demodulation of the signal which reduces the receiver complexity. The jammer attacks at different time instants by jumping from one frequency into another. We assume that there is perfect synchronization between the transmitter and receiver. To evaluate the dynamics of the amplitudes and phases related to consecutive symbols and how they are evolving with time, we consider the derivatives ($\dot{a}$, $\dot{p}$) of both amplitudes ($a$) and phases ($p$), and the generalized state vector can be defined at each time instant $k$ for a specific sub-carrier as,

$$X_{k,C_n} = [\, a \; p \; \dot{a} \; \dot{p} \,] \;\; n = \{1, 2, \dots, N\}, \;\; C_n \in C \tag{3.31}$$

A set of generalized state vectors corresponding to each sub-carrier is defined as:

$$X = \{X_{k,C_1}, X_{k,C_2}, \dots, X_{k,C_N}\}, \tag{3.32}$$

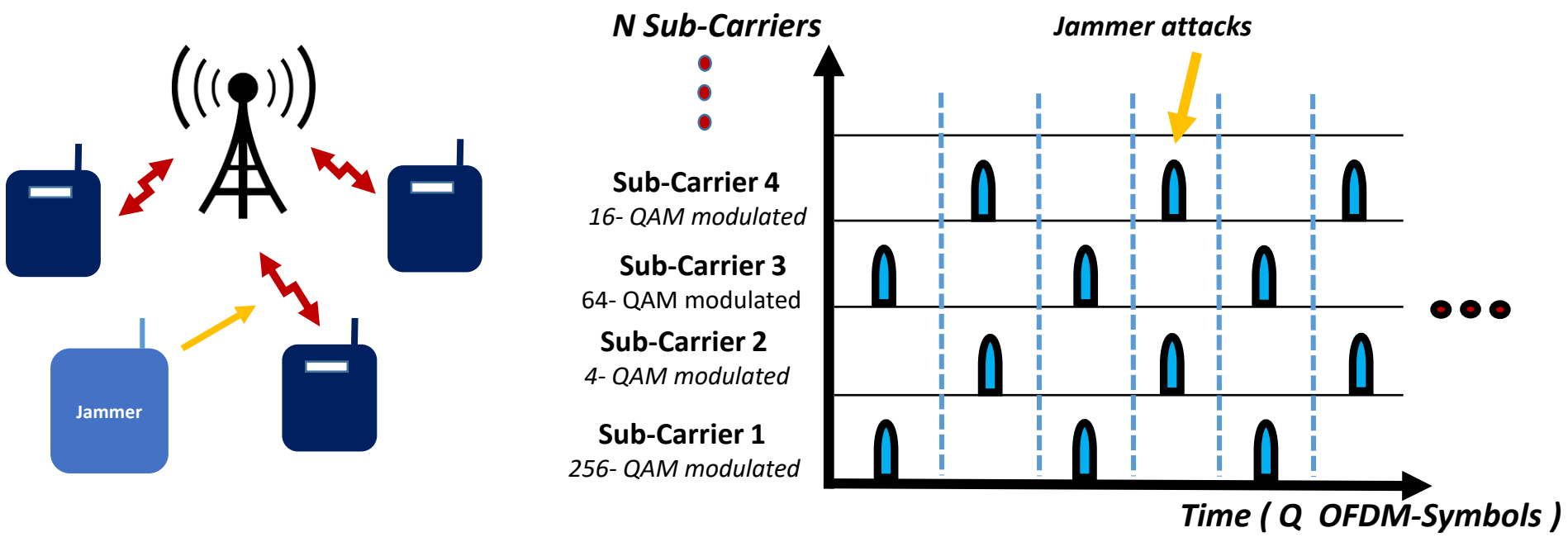

Fig. 3.30 Spectrum of the M-ary QAM modulated OFDM users in the CR-IoT Network under jammer attacks.

### Single Dynamic Bayesian Network

As shown in figure 3.31, we use the set of state vectors corresponding to each sub-carrier in (3.32) to learn a single DBN. During the Offline Learning Process, $X$ is considered as input of the SOM which outputs a set of neuron $S$. In this approach, $S$ consists of the discretization of the entire spectrum. However, single DBN keeps a memory of the spectrum's behaviour in time and frequency domain. Additionally, a single abnormality indicator is provided during the online Process.

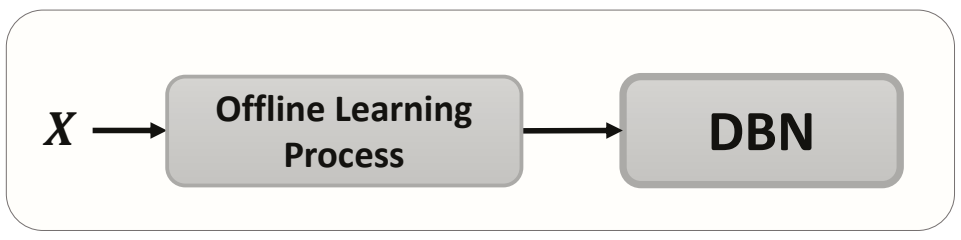

Fig. 3.31 Single DBN system.

### Bank-Parallel Dynamic Bayesian Network

In this approach, we don't have any correlation between the sub-carriers, where the spectrum's behaviour at each sub-carrier $X_{k,C_n}$ is processed individually (Fig 3.32). Accordingly, for each $X_{k,C_n}$ we learn a DBN, such as:

$$DBN = \{DBN_1, DBN_2, \dots, DBN_N\}, \tag{3.33}$$

In the online process, a MJPF is applied on each $DBN_n$ providing an Abnormality signal, such as:

$$db1 = \{db1_1, db1_2, \dots, db1_N\}, \tag{3.34}$$

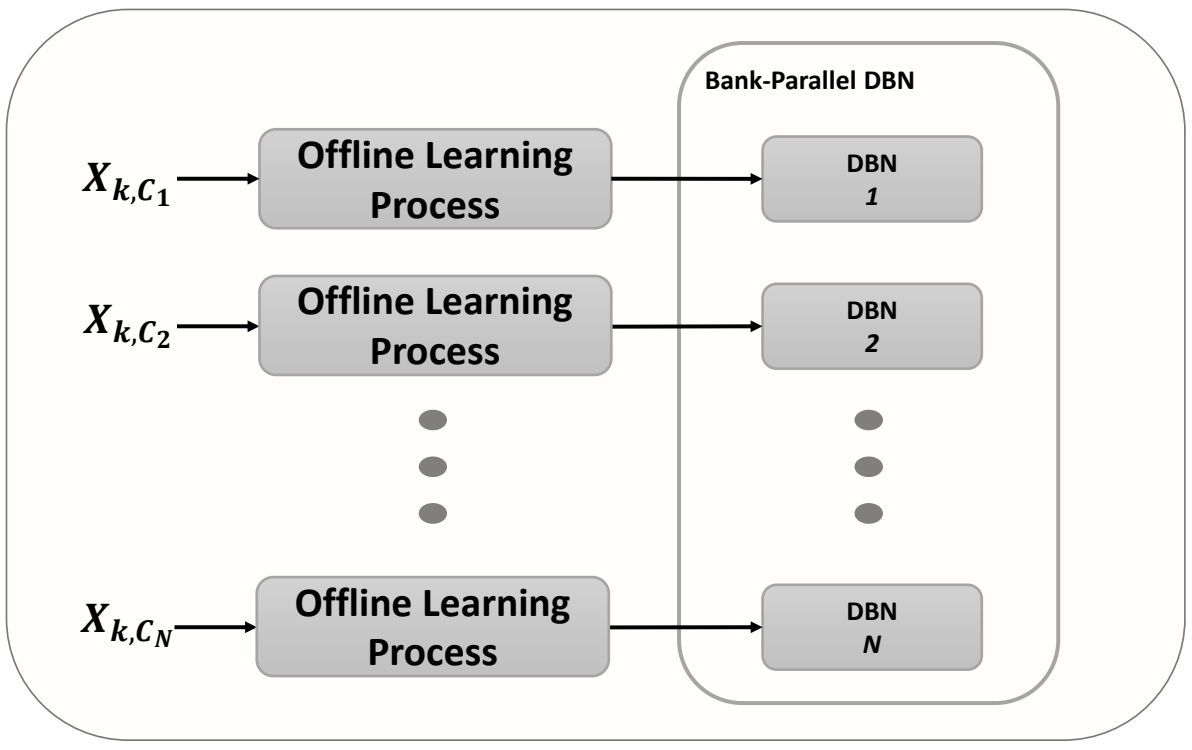

Fig. 3.32 Bank-Parallel DBN system.

### Data Source

We use the OFDM system based on the IEEE 802.11ah standard. We use a simulated OFDM signal consists of $N = 64$ sub-carriers and $Q = 1000$ symbols. The source generates random independent data. Each sub-carrier of the OFDM signal is modulated with different QAM modulation. For our experiments, we pick four sub-carriers with different QAM modulation (4, 16, 64, 256). The received signal is assumed to be affected by additive white Gaussian noise (AWGN) with zero mean and power spectral density $\sigma_w^2$. Data is cleaved into two data sets: first set contains clean data (without jammer attacks) which is used during the training phase and the second one includes jammer's attacks which is used during testing, immediately after the cyclic prefix (CP) is removed and FFT is performed on received data. We consider that jammer launches attacks into multiple sub-carriers with equal power.

### Performance evaluation of M-ary QAM with SOM size

The performance of Single and Bank-Parallel DBN models are evaluated under multiple attacks and results are shown in terms of ROC curves which consist of Probability of Detection ($P_d$) and Probability of False Alarm ($P_f$), and Area Under Curve (AUC). The abnormality measurement ($db1$) is used to calculate the ($P_d$) and ($P_f$) respectively. ($P_d$) is the number of times where abnormalities (related to jammer attacks) are correctly identified, while ($P_f$) are the times where anomalies are wrongly assigned to normal symbols. Fig. 3.33 illustrates the ROC curve obtained from Single DBN when a different number of neurons is selected. It is evident from Fig. 3.33 and Table 3.5 that 1024 neurons are the most appropriate for a Single DBN. Whereas, Fig. 3.34a, 3.34b, 3.34c and 3.34d present Bank-Parallel DBN ROC curves. For every ROC curve, each DBN deploys different QAM and optimum SOM size is analyzed. In case of 4-QAM, the optimum SOM size is 4 (see Fig. 3.34a and Table 3.6). In 16-QAM is 4 (refer to Fig. 3.34b and Table 3.6). For 64-QAM, is 8 (see Fig. 3.34c and Table 3.6), and for 256-QAM is 8 (refer to Fig. 3.34d and Table 3.6). We believe that the optimum number of neurons depend on the data and the number of symbols. For the simulated data used in our experiments and from the obtained results, we can notice that the Bank-Parallel system performs well for a small number of neurons, where the Single system performs well for a large number of neurons. This is due to the fact that Single-DBN uses the generalized state vector consisting of a large number of samples ($4Q$ symbols), which is 4 times the number of symbols used in Bank-Parallel system.

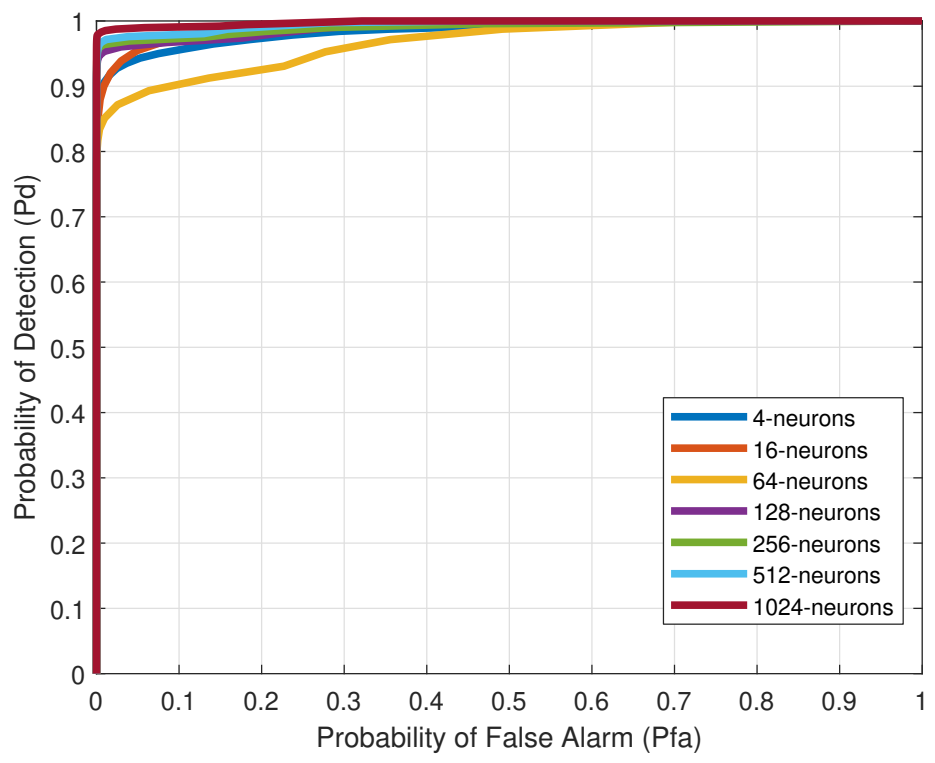

Fig. 3.33 ROC for a Single-DBN under attacks while varying SOM size.

Table 3.5 Precision measurements for a Single-DBN

| SOM size | 4 | 16 | 64 | 128 | 256 | 512 | 1024 |
|---|---|---|---|---|---|---|---|
| *AUC* (%) | 98.95 | 99.75 | 99.05 | 99.89 | 99.71 | 99.93 | **99.95** |

Table 3.6 Precision measurements for a Bank-Parallel-DBN

| | SOM size | | | | | | |
|---|---|---|---|---|---|---|---|
| | 4 | 8 | 16 | 32 | 64 | 128 | 256 |
| AUC (%) 4-*QAM* | **99.99** | 99.89 | 96.65 | 97.82 | 96.64 | 98.23 | 96.38 |
| AUC (%) 16-*QAM* | **99.86** | 99.79 | 99.51 | 99.7 | 97.46 | 96.51 | 91.76 |
| AUC (%) 64-*QAM* | 99.16 | **99.89** | 99.61 | 99.67 | 96.31 | 95.11 | 97.13 |
| AUC (%) 256-*QAM* | 98.55 | **99.83** | 98.87 | 99.2 | 96.1 | 91.36 | 95.35 |

### Comparison between Single and Parallel DBN

After using the optimum number of neurons obtained previously to make a fair comparison between the two systems. The performance of both systems is somehow similar as shown in Fig. 3.35. We can deploy either of the proposed methods depending on the receiver complexity and specific task. For instant, Single DBN learns single vocabulary for all sub-carriers, whereas, Bank-Parallel DBN learns multiple vocabularies corresponds to each sub-carrier which increases complexity. Subsequently, implementing bank parallel DBN is suitable for the source characterization tasks. Tracking the jammer and keeping its profile history in the entire spectrum is much more convenient in Single DBN.

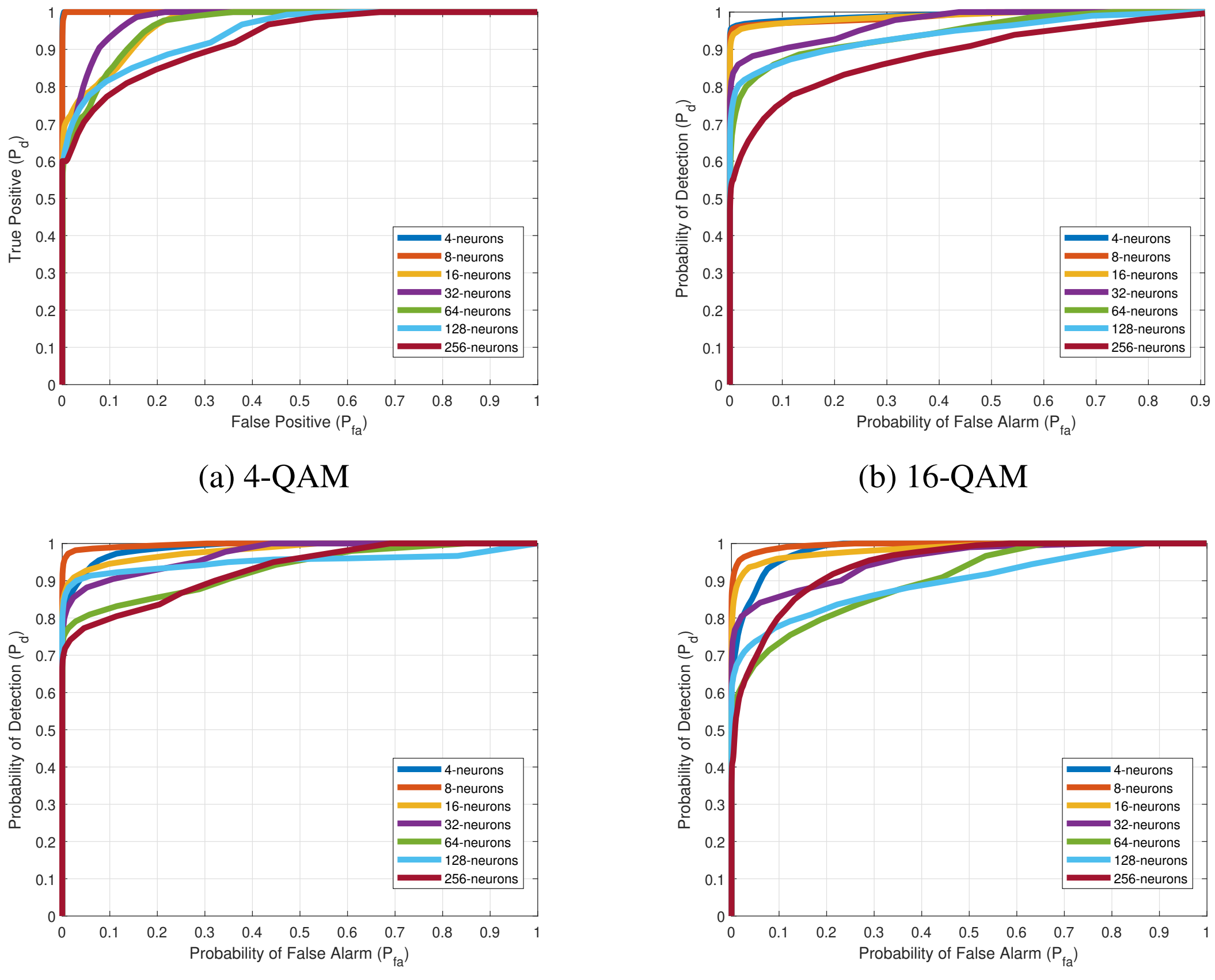

(a) 4-QAM (b) 16-QAM

(c) 64-QAM (d) 256-QAM

Fig. 3.34 ROC for an individual DBN in Bank-Parallel-DBN employs different M-QAM under attacks while varying SOM size.

## Summary

In this part (Part II), we present a jammer detection method in multiple OFDM sub-carriers by using two different systems. Sub-carriers are modulated with different M-QAM and optimum SOM size is selected for each QAM modulation based on the probability of detecting multiple attacks. As a conclusion we have learned that using the two systems, Single DBN and Bank-parallel DBN, exhibits similar performance under multiple attacks. The results presented in this part provide an understanding to further investigate the dynamic behaviour of the jammer in order to track its activity inside the spectrum and characterize it.

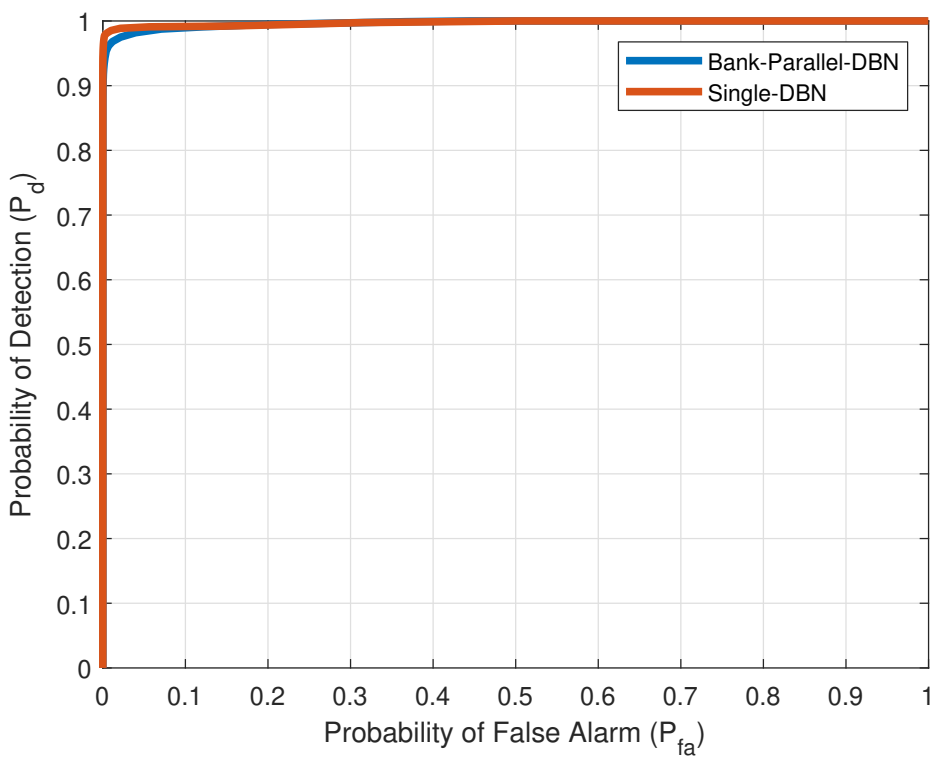

Fig. 3.35 Performance comparison between Single-DBN and Bank-Parallel-DBN in terms of ROC.

## 3.6 Comparison with existing methods

### 3.6.1 Comparison with conventional methods

Among the others, two conventional approaches that can be employed to detect abnormality signals inside the radio spectrum in CR are Energy Detector (ED) and Cyclostationary Feature Detection (CFD). ED is one of the simplest and most popular sensing techniques which compares the energy of the received signal with a certain threshold to decide whether the signal is present or absent. However, it suffers from the noise uncertainty [229]. In the CFD, the $\alpha$-profile is extracted from the Spectral Correlation Function (SCF) of the observed signals to be used as a feature. This method is computationally complex and requires knowledge of prior information of the signal as the cyclic frequency which is not always possible in real applications [230].

In this section, the performance of the C-GAN is compared with the conventional CFD and shown in Fig. 3.39, while DBN is compared with the conventional ED whose performance is shown in Fig. 3.40. Specifically, Fig. 3.39 contains the ROC curves corresponding to the 4 abnormal modalities depicted in Fig. 3.37 and Fig. 3.38, for the two methods (C-GAN vs. CFD). 'C-GAN mean' represents the mean ROC curve over the 4 modalities from Fig. 3.39 (the mean is plotted since the ROC curves related to the 4 modalities are somehow similar). It is obvious from the Figure how stable the performance of CGAN and unstable that of CFD are.

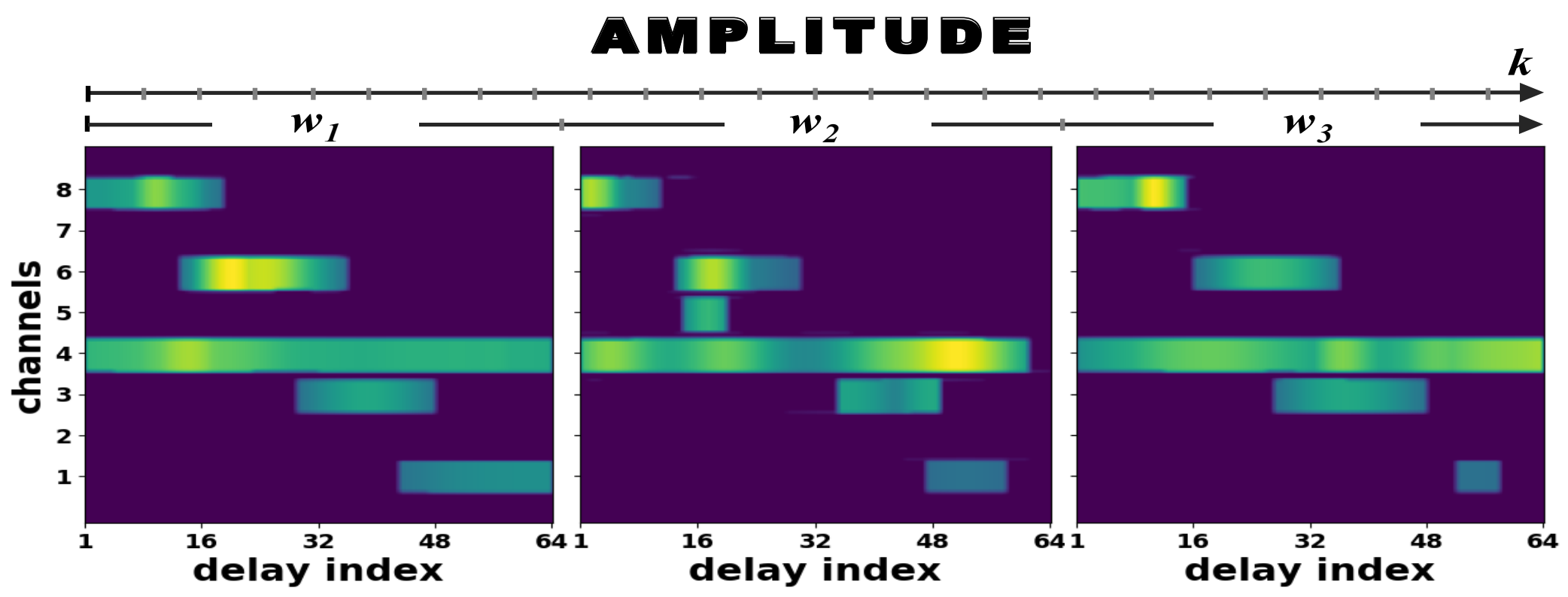

(a) Amplitude

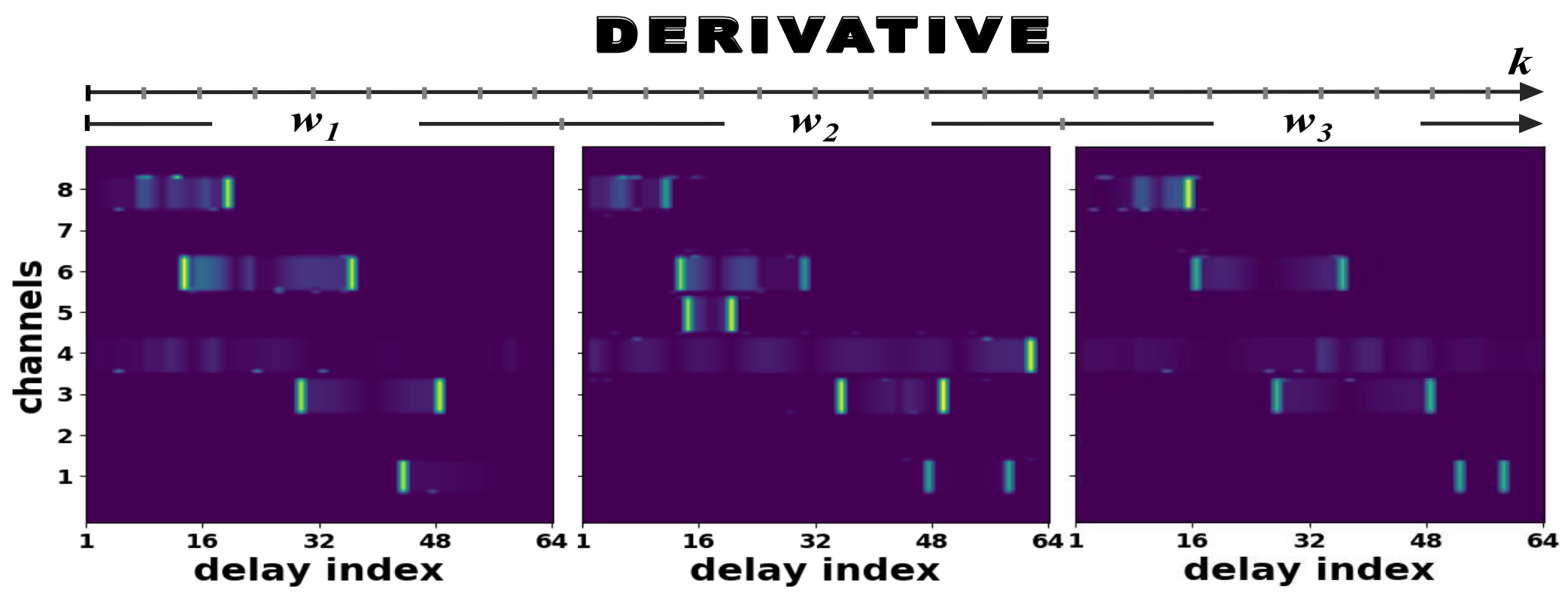

(b) Derivative

Fig. 3.36 Independent data samples from bi-dimensional localized dynamic amplitude. Pattern used as normality data (in this example: $w = 1,2,3$; 64 shifts in the ST; and $k = 1,\cdots,192$).

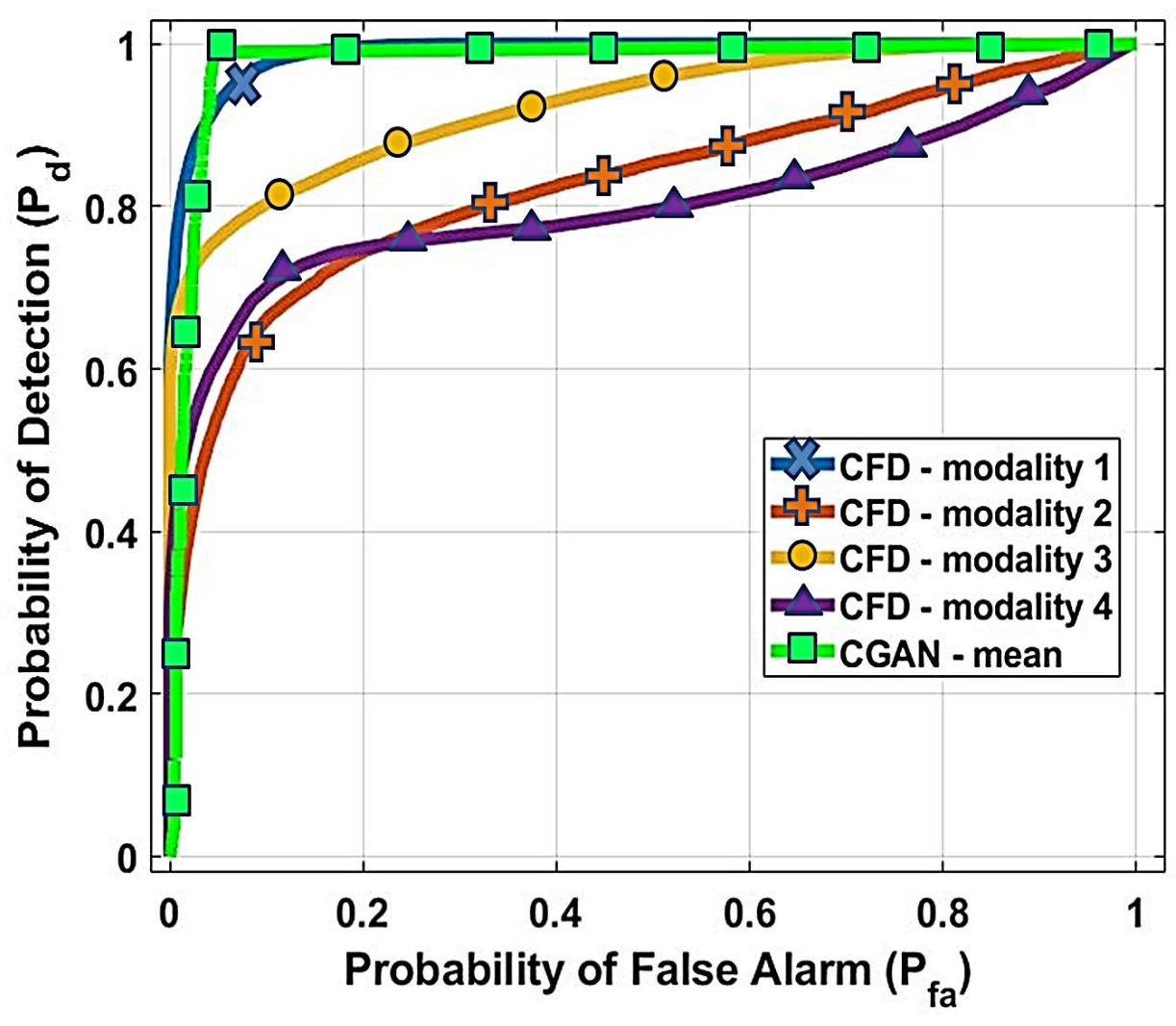

Fig. 3.39 C-GAN vs. CFD.

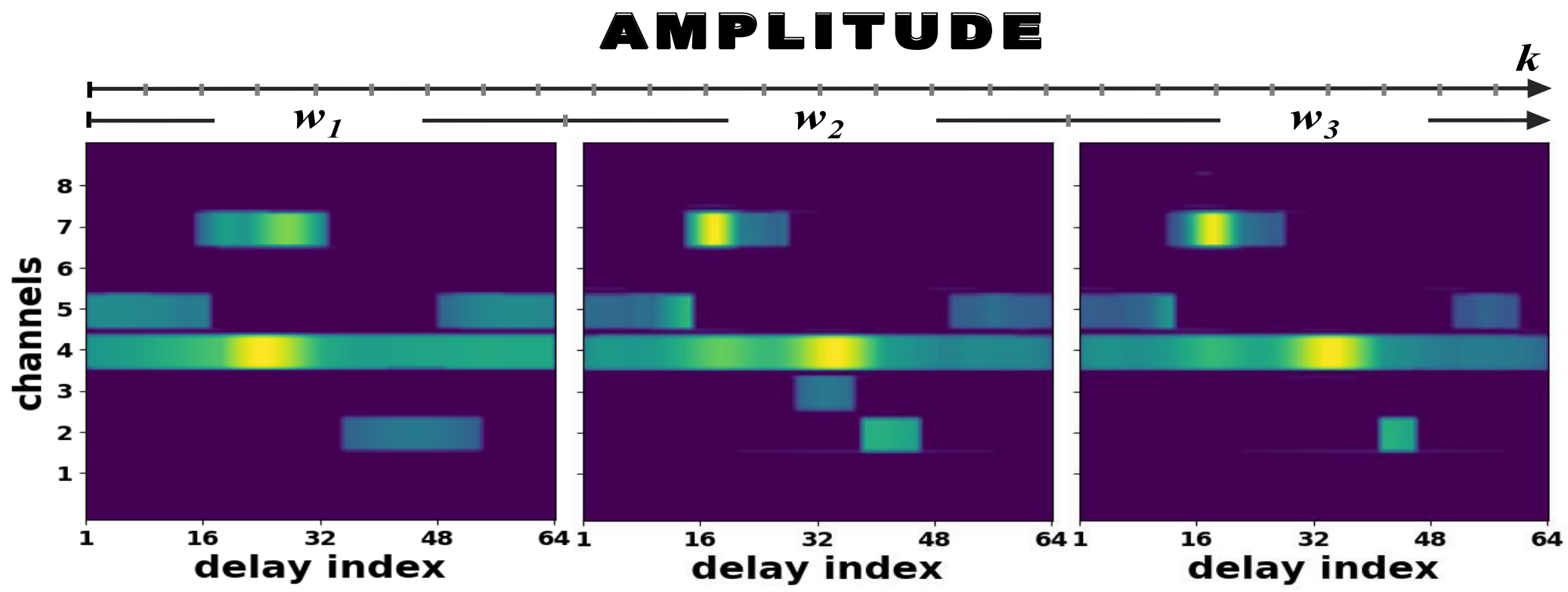

(a) Amplitude

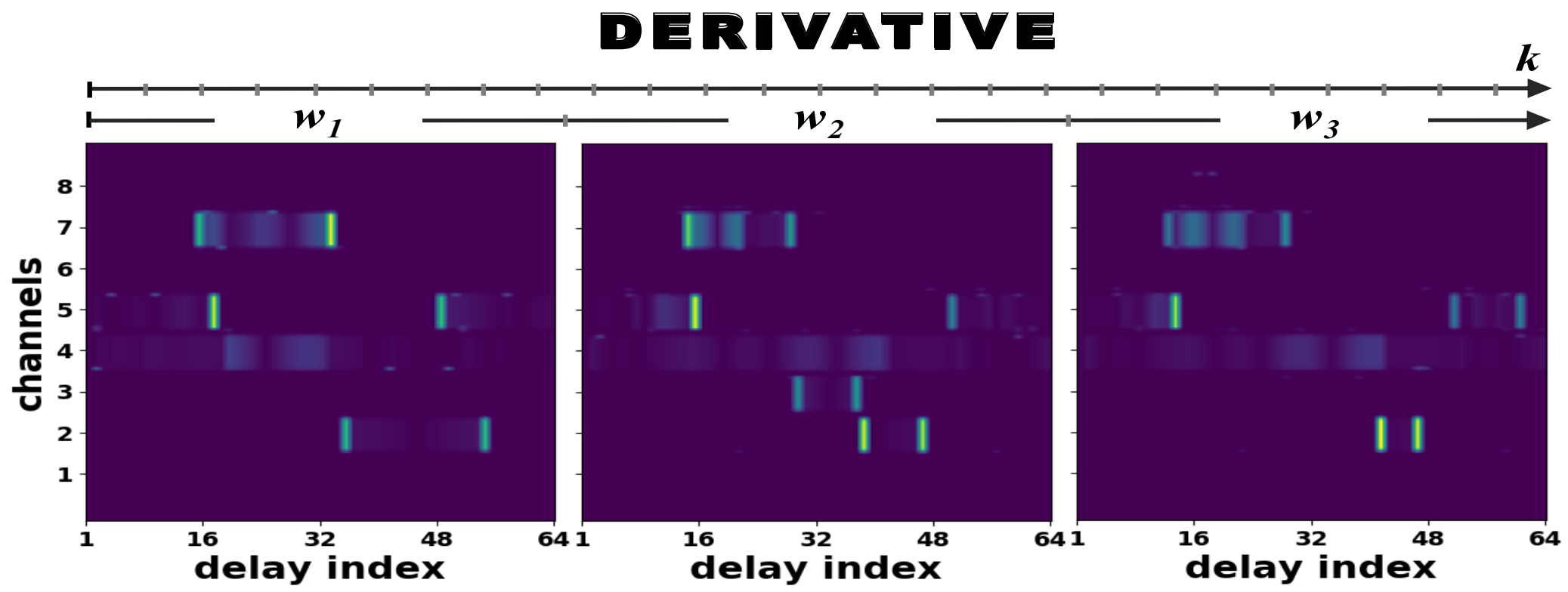

(b) Derivative

Fig. 3.37 Independent data samples from bi-dimensional localized dynamic amplitude. Pattern used as abnormal data during testing phase ($w = 1,2,3$; 64 shifts in the ST; and $k = 1,\cdots,192$) and corresponding to modality 1.

The corresponding Accuracy (ACC) values are listed in Table 3.7.

In Fig. 3.40, ROC curves represent the performance at 4 different jammer's powers in both continuous and discrete levels, as in Sec. 3.5.4, for the two methods (DBN vs. ED). The corresponding ACC values are listed in Table 3.4. It is worth noticing that the same set of thresholds is used to make a fair comparison between DBN-ED and CGAN-CDF. Experiments show that the proposed framework outperforms the conventional detection methods due to its twofold ability to predict the signal states (not only to detect them as in the conventional approach) and, consequently, to detect abnormalities resulting in a reduced amount of appearances of false alarms.

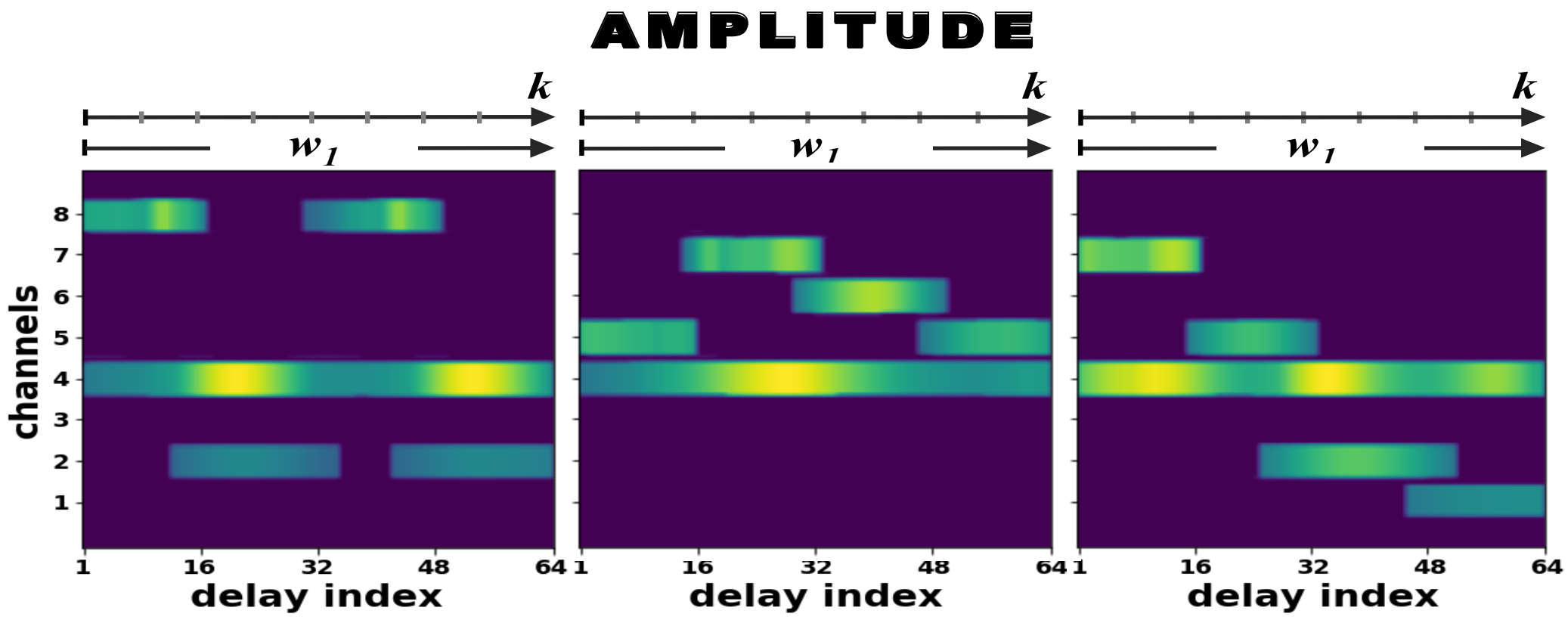

(a) Modality 2 (b) Modality 3 (c) Modality 4

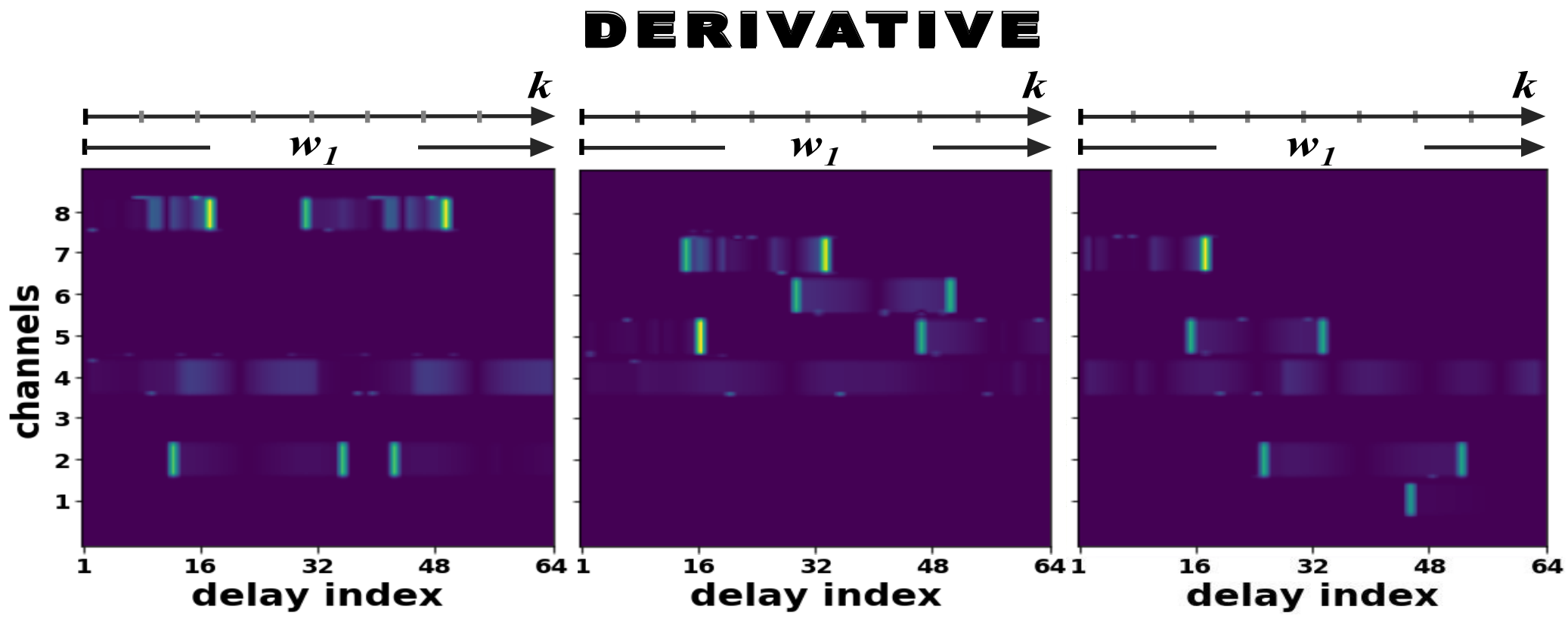

(a) Modality 2 (b) Modality 3 (c) Modality 4

Fig. 3.38 Three patterns used as abnormal data during testing phase and corresponding to modalities 2, 3, and 4, respectively, where only one sample is shown ($w_1$) for each modality.

| | $C-GAN(db0)$ | $CFD$ |
|---|---|---|
| $ACC$ (modality 1) | **0.9938** | 0.9434 |
| $ACC$ (modality 2) | **0.9862** | 0.8262 |
| $ACC$ (modality 3) | **0.9951** | 0.8950 |
| $ACC$ (modality 4) | **0.9960** | 0.8424 |

Table 3.7 Comparison in terms of accuracy measures between C-GAN and CFD

| | DBN ($db1$) | DBN ($db2$) | $ED$ |
|---|---|---|---|
| $ACC$ ($P_J$=5dB) | **0.9997** | 0.9677 | 0.9667 |
| $ACC$ ($P_J$=4dB) | **0.9967** | 0.9687 | 0.9670 |
| $ACC$ ($P_J$=3dB) | **0.9883** | 0.9687 | 0.9667 |
| $ACC$ ($P_J$=2dB) | **0.9840** | 0.9670 | 0.9667 |

Table 3.8 Comparison in terms of accuracy measures between DBN and ED.

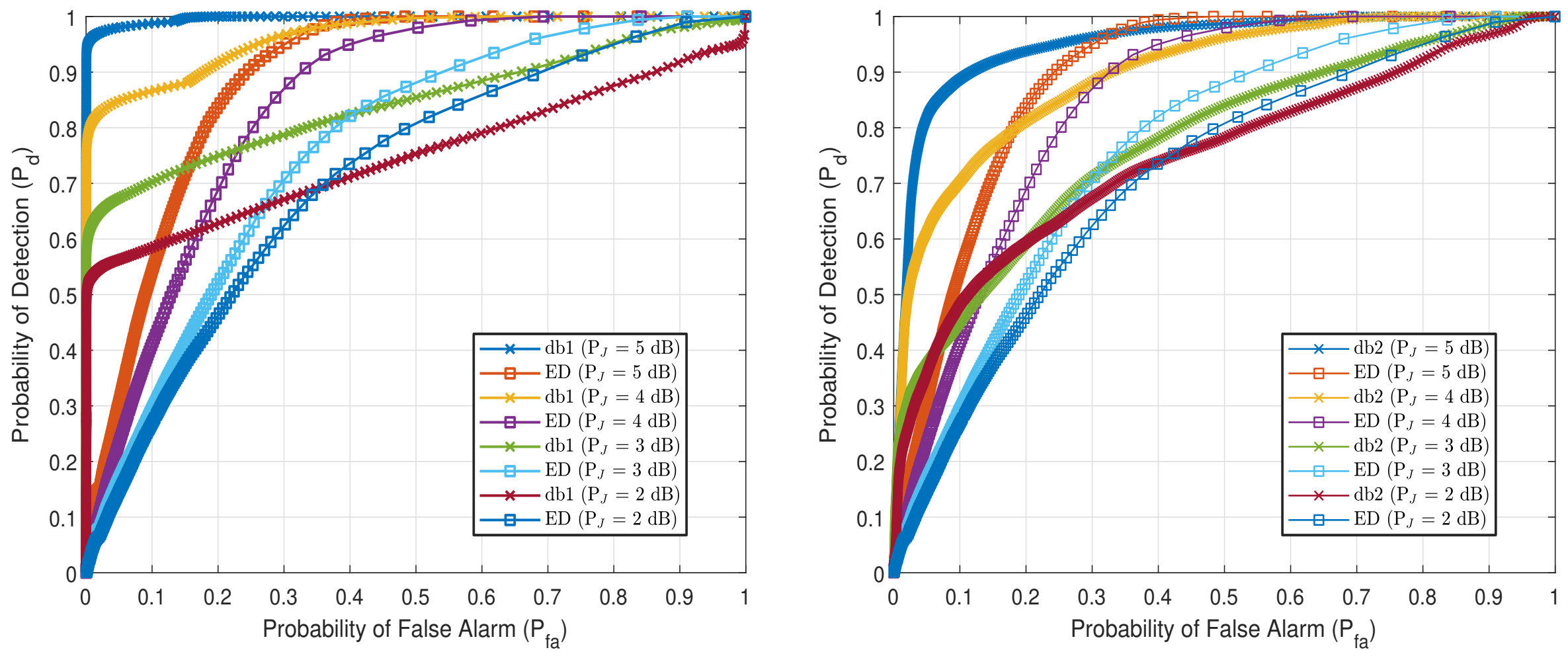

(a) $db1$. (b) $db2$.

Fig. 3.40 Scenario I: DBN vs. ED.

### 3.6.2 Comparison with AI-based methods

The main differences with respect to the some AI-based techniques ([72, 74–78]) selected from Chapter 2 are that: *i)* our framework is based on learning generative dynamic models by using the generalized state vector to incorporate both the state and its dynamics (namely the corresponding derivatives), which facilitates the prediction of the spectrum's future state and allows to detect any abnormality. Forming the generalized state vector is more efficient and less complicated by relying on raw I/Q data compared to other methods that deal with more complex features. *ii)* the proposed framework covers two applications and can be implemented at different sides, at the base station-side which deals with multi wideband signals and at the cellular device-side dealing with narrowband signals. This makes it possible to flexibly select both the signal dimensionality and, consequently, the receiver stage at which abnormality detection has to be applied. *iii)* As demonstrated with the considered OFDM case, the approach can be applied to generic wideband modulations at different and eventually multiple post- and pre- detection stages, providing in this way a flexible security tool for enabling CR-receivers with AI capabilities.

## 3.7 DBN: bridging High- and Low-dimensional Data

This section presents a possible future direction of bridging high-dimensional and low-dimensional data through a DBN model. The bridging task must be preceded by dimensionality reduction, which plays a vital role in data processing, making the processing of high-dimensional data more efficient [231]. Dimensionality reduction aims to extract from high-dimensional data low-dimensional features that can be represented in a probabilistic graphical model as in [232–236]. VAE is one of the most effective techniques to obtain low-dimensional probabilistic latent representations from data, unlike the GAN-based models, which do not approximate explicit probability distributions but seek to learn it implicitly [237]. Fig. 3.41 depicts the necessary steps to pass from high-dimensional multi-signals received by the CR device (e.g., BS) to low-dimensional semantic representations structured in a DBN. The DBN model can be learned by clustering the extracted low-dimensional features (i.e., the output of VAE's encoder on high-dimensional signals) in an unsupervised manner. The advantage of this approach is that representing useful features extracted from high-dimensional multi-signals in a probabilistic graphical model endows the radio with the ability to produce explainable and interpretable predictions allowing to understand surprising events and link them to the causal interpretation that provides the motivation supporting actions as we will show in the coming chapters.

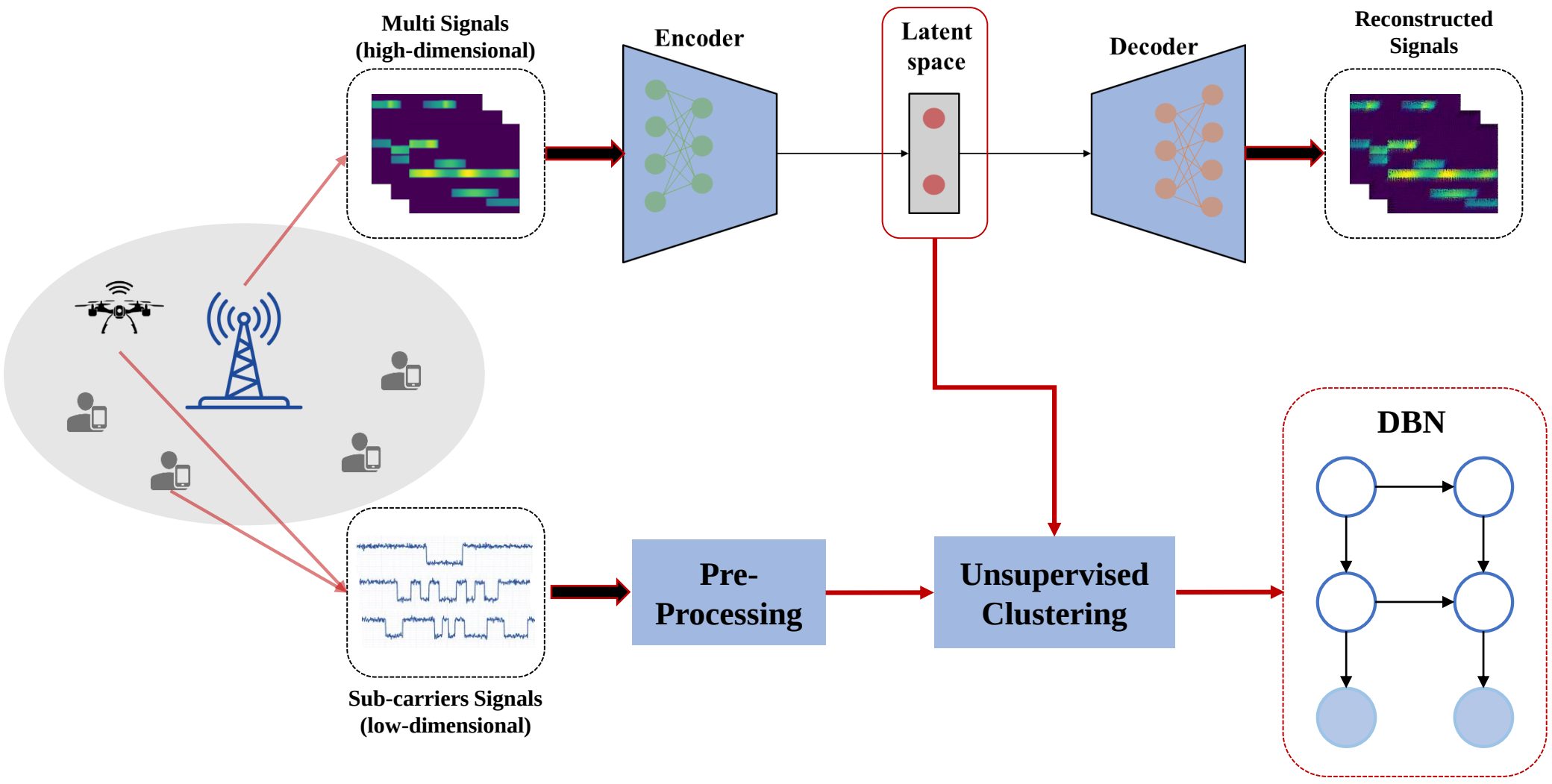

Fig. 3.41 Block diagram illustrating the process of bridging high-dimensional and low-dimensional signals through the DBN.

## 3.8 Conclusion

This chapter presents a framework which is foreseen to enhance the PHY-layer security in CR. A basic Self-Awareness (SA) module is introduced that includes the capability to learn dynamic generative models and consequently detect abnormal signals inside the wireless spectrum. The potential of the proposed methods lies in a fact that they can be incorporated in CR systems and give a facility to adopt either of them depending on the application (data dimensionality, sampling rate). C-GAN, AC-GAN and VAE are employed in the high-dimensional data application due to its ability to deal with high data samples extracted from a wideband signal while the DBN model is implemented in low-dimensional data application due to its strength in representing in a probabilistic way low data samples extracted from sub-carriers of an OFDM signal at two levels. The learned dynamic models in both methods are generative models which have been investigated when the input spectrum data is a generalized state vector (high dimensional generalized state vector in the first method while low dimensional generalized state vector in the second method). GMs are capable of both generating synthesized data and providing a distance metric to measure the deviation of the predicted data from the observed one. Indeed, results reveal that both of the methods can effectively detect malicious or jammer attacks after learning the corresponding dynamic models. Validation is performed on real mmWave dataset in Part I and simulated OFDM data in Part II. The techniques and results presented in this chapter allow for achieving further SA functionalities as incremental learning and abnormality mitigation which will be investigated in the following chapters.

# Chapter 4

# An Emergent Self-Awareness Module for Physical Layer Security

This chapter picks up where the previous chapter ended by introducing a data-driven Self-Awareness (SA) module in CR at the PHY. The corresponding functionalities are presented in Section 4.2. In Section 4.3, the system model is described. Section 2.1.4 provides a brief comparison with selected related works. Section 4.4 gives the simulation results and Section 4.5 concludes this chapter.

## 4.1 Introduction

As claimed in Chapter 3, learning by understanding and reasoning realize the main aspects of cognition which could not be achieved until and unless CR subsumes a certain degree of Self-Awareness (SA). AI-empowered CR (i.e., AI-enabled radios) is a paradigm shift to achieve the highest level of SA in future wireless communications.

SA endows the radio with the capability of maintaining a dynamic equilibrium with the external environment guided by the Free-Energy (FE) minimization. Originally, the FE principle (FEP) was presented as an account of biological Self-Organization [238, 239]. The existence of a living organism and its all possible organismal states are explained by a set of random variables, which can be represented as a state-space model [240]. The organism tends to occupy certain states (small volume) in that space to stay alive, forming the so-called organism's phenotype. For example, the typical human's body temperature is around $36.6^{\circ}$, hence the body must occupy points clustered around that value to continue living. The FEP conceives phenotypes as an attracting set of states (i.e., random dynamical attractor, which is assumed to be ergodic) constraining the organism's path through its state-space. Thus, to reach the equilibrium (e.g., stay alive), an organism must constantly revisit the states that belong to its phenotype, allowing to interpret the average amount of time a state is occupied as a probability of the system being in that state when observed at random [241]. This allows defining a probability function over the state-space where phenotypic states are highly probable and non-phenotypic states are improbable. In this sense, the probability distribution will be sharply peaked around phenotypic states and very flat on the remaining states. Moreover, that probability distribution has low entropy, which means that the organism tends to occupy states with low surprise. The surprise quantifies how far is a given state from the phenotypic states. Therefore, an organism must avoid surprising states (minimize FE) to survive and reach equilibrium.

The proposed SA representation allows quantifying the FE in terms of abnormality measurements (or surprise) at multiple levels. The abnormality measurements involve two probability densities: a generative density encoding the joint probability of environmental and sensory states given a generative model of how sensory signals are generated, and a recognition density on the causes of sensory states. Various types of generative models can be used (as we discussed in Chapter 3) to predict spectrum dynamics and consequently detect anomalies. However, this chapter and the remaining ones will focus on DBN models due to their pecularity in modeling uncertaintes and representing a set of random variables and their conditional dependencies at multiple levels through directed acyclic graphs. The notion of dependency is central to the existance of a Markov blanket that induces a partition of states into internal and external. The external states are hidden from the internal states by the Markov blanket comprising sensory and active states. External states are states of the surrounding environment that cause sensory states and depend on action. Sensory states are the radio's sensations that influence (but are not influenced by) internal states and constitutes a probabilistic mapping from actions and external states. Internal states are the states of the radio that depend on sensory states and cause active states (actions) that influence (but are not influenced by) external states. Thus, the Markov blanket build a circular causality explaining the perception-action cycle in the brain where active and internal states minimize a FE functional of sensory states. This chapter and chapter 5 will focus on the perception process while chapter 6 will focus on both perception-action cycle to jointly update internal states based on the coming sensory signals and the active states that change sensory signals to avoid surprises [242].

Building upon the above discussion, we propose a data-driven SA module involving several functionalities allowing to build up knowledge about the wireless environment and reach the goal oriented by the *radio*. A radio equipped with SA is capable to perform the following functionalities:

1. *Learning Generative Models* autonomously by observing the occurring environmental changes;
2. *Radio Spectrum Perception* by predicting the future states of the spectrum and estimating the current states of the observed stimuli received from the environment;
3. *Abnormality Detection* which refers to the process of noticing any deviation from the normal situation (i.e., similar to what it was learned from previous experience);
4. *Abnormality Characterization* by analyzing the new behaviour (the detected differences which are not seen before) and characterizing it to draw up the dynamic rules of how the new situation is evolving;
5. *Incremental Learning* of new representations of the occurring environmental changes related to the detected signals which represent new behaviour;
6. *Internal Action* by removing the detected abnormal signal caused by a malicious user from the received stimuli;
7. *Abnormality Classification* by using multiple models learned so far and performing multiple predictions to discriminatively select the model that better fits to current experience. This functionality is crucial to understand when the radio should learn a new model and which communication policy should apply (e.g., how to change the defence policy after being able to identify the detected jammer).
8. *Learning Interactive Models* by observing multiple signals related to different sources (e.g. LTE and GPS signals) and learning the cross-correlation between them, or by learning the causality among different users (e.g. CR device and jammer) interacting inside the radio spectrum to draw a deference strategy by changing the policies of communication (changing the configuration: power, modulation, frequency, etc.);

These functionalities are dependent, the one leads to the other in an incremental way and without any external supervision. The radio augmented with the SA module has the capability of **Learning Generative Dynamic Models** that defines a hierarchical representation of the radio environment and explain the dynamic changes of the sensed signals inside it. Based on the knowledge acquired, the radio is capable to predict and precept the future states of the environment (**Radio Spectrum Perception**) by switching between the learned dynamic models and/or by activating the best models that fit the received stimulus. Accordingly, the radio can identify whether those stimuli are normal or abnormal (**Abnormality Detection**) and then to characterize the abnormal behaviour (**Abnormality Characterization**) and learn a new representation that encodes such behaviour (**Incremental Learning**) or to mitigate its effect from the received data (**Internal Action**). Also, SA endows the radio with the discriminative capability by comparing the current experience to multiple predictions generated from a set of models describing different radio experiences (**Abnormality Classification**). Recognizing the current experience allows the radio to know when it should learn incrementally a new model and the set of actions to follow according to the identified external force that caused the current abnormality (**Learning Interactive Models**). The set of actions can be learned by the radio while interacting in the environment to avoiding surprising states (e.g., states under jamming attacks). The SA module depicted in Fig. 4.1 revolves around learning and updating the acquired knowledge by understanding precisely the radio environmental circumstances. Introducing a SA module in CR can support the system to improve the decision and action cycles after detecting abnormal behaviours such as jammer attacks that can manipulate the radio spectrum and teach the CR malicious behaviours. Thus, the SA module can lead to establishing secure networks against various attacks.

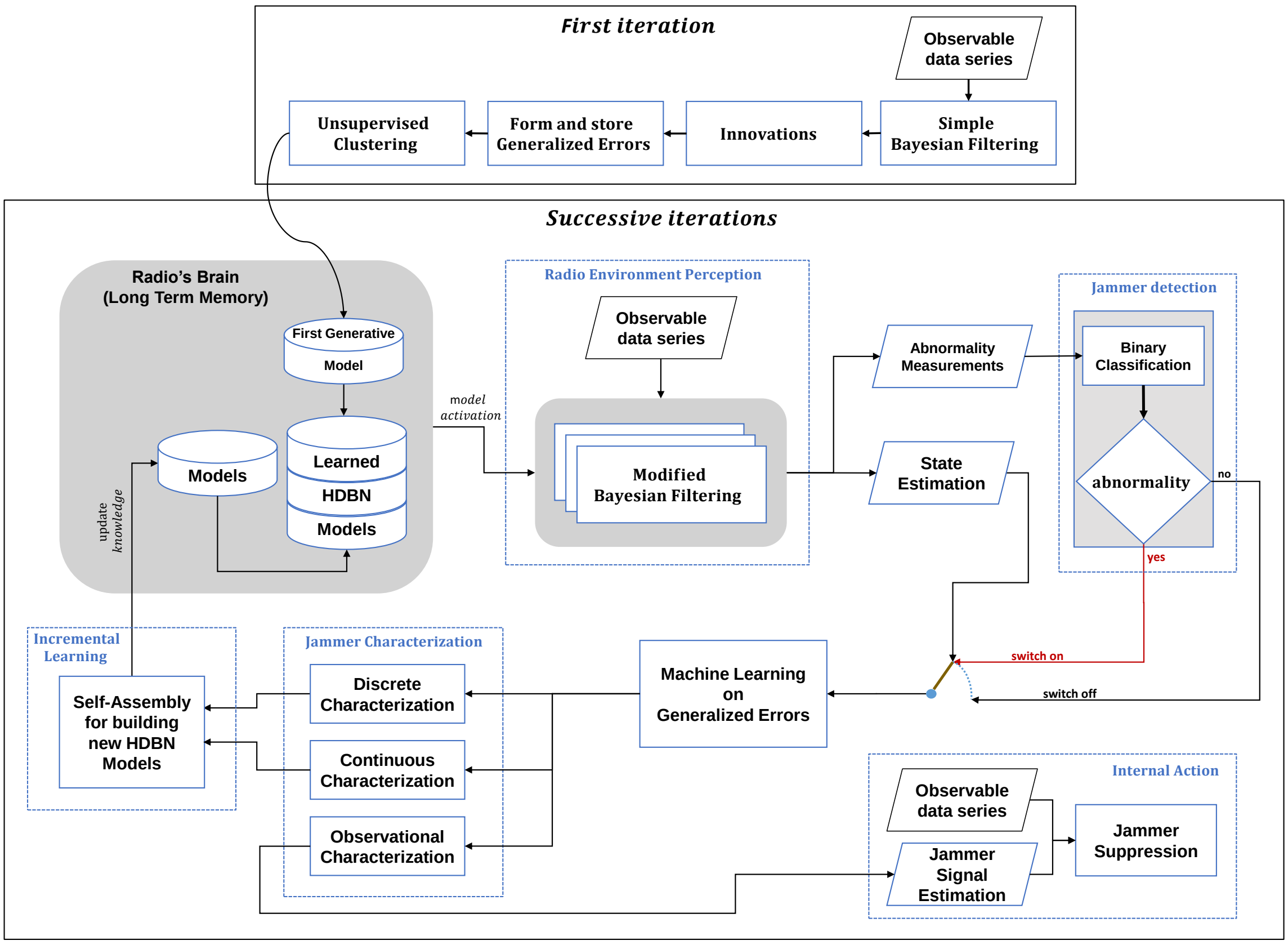

Fig. 4.1 The Proposed Self-Awareness (SA) Module.

## 4.2 Proposed Self-Awareness functionalities

### 4.2.1 Learning Generative Dynamic Models

One of the most promising approaches towards developing algorithms that can analyze and understand data are Generative Models [243]. Generative Models are statistical models that aim to learn the joint distribution over a set of random variables from which data samples can be generated from that distribution. In this work, we propose to learn a generative model consisting of a Hierarchical Dynamic Bayesian Network (HDBN) [51], due to its ability to represent joint distributions of random variables at different abstraction and temporal levels. HDBN extends the standard Bayesian Networks by introducing the temporal slices where relationships among nodes at the same time instant provide causal links between different abstraction levels and links between consecutive slices represent causal temporal probabilities. The relationship among variables associated with nodes encoded through conditional probabilities is represented by directed edges. HDBN models provide also graphically decomposition of the joint probability directly related to causal relationships among different hidden and observable variables at consequent temporal instants. Therefore, the capability of learning such models from data is equivalent to be capable of generating data sequences at various abstraction levels (hierarchical inference) and temporal levels (temporal inference) coherent with the joint model. So, this makes it possible to individuate in learned models conditional distribution describing the dynamics (both linear and non-linear) over time as well as the semantic rules to associate hidden discrete symbols to variables describing continuous signals. The proposed HDBN whose graphical representation is depicted in Fig. 4.2-a can be associated with inference mechanisms that work simultaneously at different hierarchical levels, like switching models [244]. A main aspect of the proposed HDBNs is that the variables they include at all levels can be considered as Generalized random variables (i.e., random variables represented in Generalized coordinates of motion) as allowing both advantages in representation and inference with dynamic systems. The concept of Generalized coordinates of motion proposed by Friston in [245] and used here, implies to represent a pattern vector composed of the random variable per se and its temporal derivative allowing the radio to represent the dynamic rules of how the signal is evolving with time in terms of forces[1] and to facilitate the prediction process. For example, the sensed signals associated to a certain radio situation are represented by nodes at the lower level of the hierarchy that can be defined as Generalized Observations ($\tilde{\mathbf{Z}}$), such that:

[1]Forces are dynamic rules explaining how the signal evolves with time, including the signal data samples' transitions and the power of those transitions.

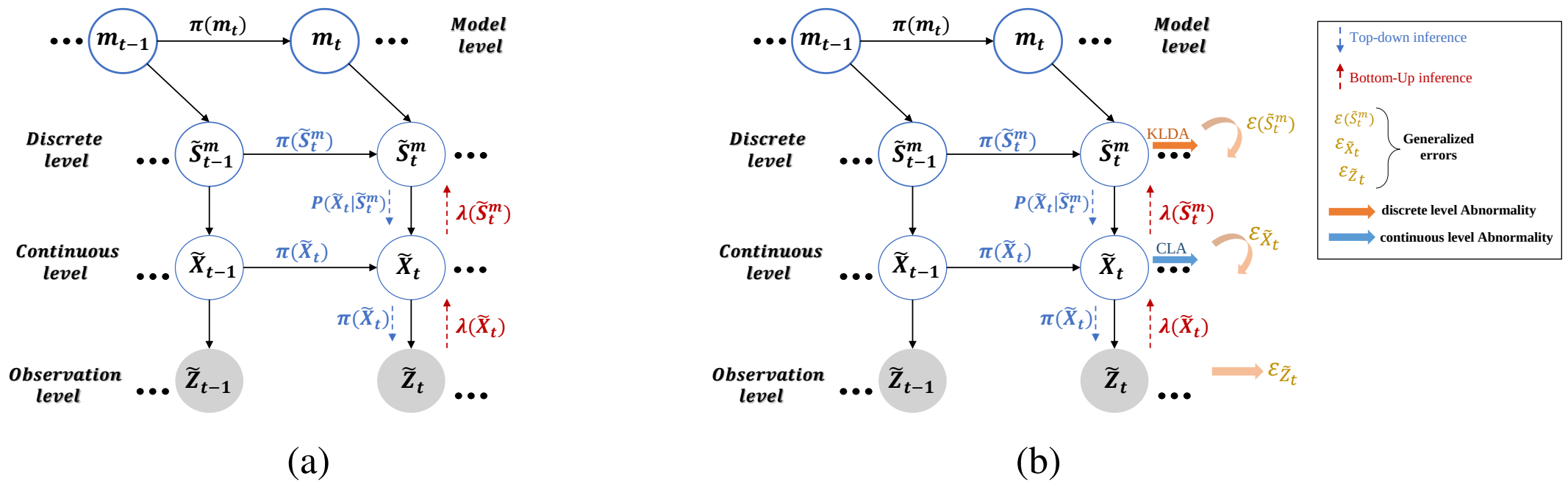

Fig. 4.2 (a) The Proposed Hierarchical Dynamic Bayesian Network (HDBN). (b) Abnormality measurements and generalized errors at hierarchical levels estimated during the real-time inference.

$\tilde{Z}_t = \Big[\underbrace{I_{t,f_1},\ldots,I_{t,f_d},Q_{t,f_1},\ldots,Q_{t,f_d}}_{Z_t},\overbrace{\dot{I}_{t,f_1},\ldots,\dot{I}_{t,f_d},\dot{Q}_{t,f_1},\ldots,\dot{Q}_{t,f_d}}^{\dot{Z}_t}\Big]$, where $\tilde{Z}_t \in \tilde{\mathbf{Z}}$, $d$ is the number of the sensed sub-carriers, and $I_{t,f_\eta}$, $Q_{t,f_\eta}$ are in-phase and quadrature components of the sensed signal at different sub-carriers and $\dot{I}_{t,f_\eta}$, $\dot{Q}_{t,f_\eta}$ are the corresponding derivatives[2], such that $\eta = \{1,2,\ldots,d\}$. Such variables are connected with hidden generalized states that are considered as the causes that generate observations and their changes.

At the very beginning of the learning process (first iteration in Fig. 4.1), the radio's brain is empty ($m = 0$). The radio is assumed to start as a quasi white blackboard where only knowledge describing a static condition assumption is available. Such initial generative model consists of a simplified HDBN of only two levels where the linear dynamic model assumes that generalized continuous variables of the environment (i.e. the state and its derivative) are not changing. The inference mechanism associated with such an HDBN uses a simpler Bayesian Filtering based on an Unmotivated Kalman Filter (UKF) [246] (i.e. a KF that assumes that changes are only due to Gaussian noise). In UKF, predictions can be done using the following dynamic model:

$$X_t = X_{t-1} + v_t, \tag{4.1}$$

where $X_t$ are the predicted states and $v_t$ is a Gaussian noise. The temporal innovations (deviations from the predicted derivatives) of a signal that violates the UKF assumption of

[2] In this work, we consider the first order temporal derivative which is obtained by calculating the difference between the $I$ ($Q$) at time instant $t$ and $I$ ($Q$) at time instant $t-1$, such that $\dot{I}_t = I_t - I_{t-1}$ if $t > 1$ and $\dot{I}_t = 0$ if $t = 1$. Likewise, $\dot{Q}_t = Q_t - Q_{t-1}$ if $t > 1$ and $\dot{Q}_t = 0$ if $t = 1$.

static sequences (i.e. null derivative) can be calculated as:

$$\dot{X}_t = H^{-1}\left(\frac{Z_t - HX_{t-1}}{\Delta t}\right), \tag{4.2}$$

where $\Delta t$ is the sampling time and $H = [H_1\ H_2]$ is the observation that maps hidden states to observations where $H_1 = [I_{2d}\ 0_{2d}]$, $H_2 = [0_{2d}\ 0_{2d}]$. The estimated elements $X_t$ and $\dot{X}_t$ can be considered equivalent to filtered observations derived from $I$ and $Q$ radio signal features and they can be seen as a data series describing forces that caused violations of static hypothesis defined as Generalized errors ($\tilde{\mathbf{X}}$) expressed as:

$$\tilde{X}_t = \begin{bmatrix} X_t & \dot{X}_t \end{bmatrix}, \tag{4.3}$$

where $\tilde{X}_t \in \tilde{\mathbf{X}}$. The HDBN model ($m = 1$) can be trained by using unsupervised clustering techniques (e.g. Growing Neural Gas (GNG), Self Organizing Maps (SOM)) to group those errors. GNG is an incremental neural network that learns the relation between a given set of input patterns and adapts the topological structure based on nearest neighbour relationships and local error measurements. GNG shows higher flexibility in representing the input data if compared to fixed size networks like SOMs. Here, we employ the GNG algorithm that receives the Generalized errors $\tilde{\mathbf{X}}$ provided by the UKF model defined in (4.3) and produces a set of superstates (or clusters) $\mathbf{S}$ in which $\tilde{\mathbf{X}}$ are encoded, such that: $S^m = \{S_1^m, S_2^m, \ldots, S_L^m\}$, where $S_k^m \in S^m$ and $L$ is the total number of superstates associated to the first model ($m = 1$). Additionally, each superstate is associated with mean value ($\mu_{S_k^m}$) and covariance matrix ($\Sigma_{S_k^m}$) of the set of hidden filtered $\tilde{X}_t$ samples clustered inside $S_k^m$, so providing the $P(\tilde{X}_t|S_t^m)$ link in the HDBN.

Within a slice learned, vertical links describe causal relationships between $m_t$, $S_t^m$, $\tilde{X}_t$ and $\tilde{Z}_t$ at a given time instant $t$. Besides, links between variables at consecutive time instants allow representing conditional temporal probabilities among generalized variables starting from obtained superstates related to model $m$, i.e. dynamic causality that drives changes in the signal. In particular, the $L$x$L$ transition matrix $\Pi$ defined as:

$$\Pi = \begin{bmatrix} \pi(S_t^m = S_1^m) \\ \pi(S_t^m = S_2^m) \\ \vdots \\ \pi(S_t^m = S_L^m) \end{bmatrix} = \begin{bmatrix} \pi_{11} & \pi_{12} & \dots & \pi_{1L} \\ \pi_{21} & \pi_{22} & \dots & \pi_{2L} \\ \vdots & \vdots & \ddots & \vdots \\ \pi_{L1} & \pi_{L2} & \dots & \pi_{LL} \end{bmatrix}, \tag{4.4}$$

that embeds the dynamic causal models at the discrete level is learned by estimating the transition probabilities $\pi_{ij} = P(S_t^m = j|S_{t-1}^m = i),\ i, j \in S^m$ over a period of time. Such a probability allows the HDBN to embed knowledge describing the discrete dynamics of the signal, namely transitions from superstate $i$ to superstate $j$ as time evolves. To extend the concept of generalized variables to also include higher HDBN levels, generalized superstates $\tilde{\mathbf{S}}$ can be defined as follows: $\tilde{S}_t^m = [S_t^m \, \dot{S}_t^m] = [S_t^m \, E(S_t^m|S_{t-1}^m)]$, where, $S_t^m$ stands for the current superstate and $E(S_t^m|S_{t-1}^m)$ represents the event of transiting to $S_t^m \in \mathbf{S^m}$ and conditioned to be in $S_{t-1}^m$ in the previous time instant.

The dynamic causal models represented in the HDBN and formulated in generalized superstates $\tilde{\mathbf{S}}^m$ (hidden discrete variables) and generalized states $\tilde{\mathbf{X}}$ (continuous hidden variables) have the following forms:

$$\tilde{S}_t^m = f(\tilde{S}_{t-1}^m, m_{t-1}) + w_t = f(\pi_{ij}^m) + w_t, \tag{4.5}$$

$$\tilde{X}_t = g(\tilde{X}_{t-1}, \tilde{S}_{t-1}^m, m_{t-1}) + w_t = A\tilde{X}_{t-1} + BU_{\tilde{S}_{t-1}^m} + w_t. \tag{4.6}$$

The discrete non-linear function $f(.)$ determines the superstates temporal evolution based on the learned $\Pi$ at the discrete level. On the other hand, the continuous linear function $g(.)$ determines the state temporal evolution at the continuous level of the HDBN. Both $f(.)$ and $g(.)$ are subject to a process noise $w_t$ which is assumed to be drawn from a zero multivariate normal distribution with covariance $\sigma_t$ such that $w_t \sim \mathcal{N}(0, \sigma_t)$. In (4.6), $A = [A_1 \, A_2]^\intercal$ is a dynamic model matrix where $A_1 = [I_{2d} \, 0_{2d}]$ and $A_2 = [0_{2d} \, 0_{2d}]$. $I_{2d}$ and $0_{2d}$ are the identity and zero matrices, respectively. In addition, $B = [I_{2d} \, 0_{2d}]^\intercal$ is the control model and $U_{\tilde{S}_t^m}$ is the vector parameter describing the dynamics that the signal is expected to follow, when its states belong to a given cluster $\tilde{S}_t^m$ and it is defined as the mean of generalized states' derivative in the cluster, i.e. $\mu(\tilde{S}_t^m)$. So, $\Pi$ encodes not only transitions between clusters but also jumps between different linear models (different $U_{\tilde{S}_t^m}$ vectors) at the continuous level.

The observation model stands for the bottom level of the HDBN and it is defined as:

$$\tilde{Z}_t = h(\tilde{X}_t) + v_t = H\tilde{X}_t + v_t, \tag{4.7}$$

where $\tilde{Z}_t$ is the observable signal, $h(.)$ is a linear function that maps the hidden generalized state to the observed data and $v_t$ is the measurement noise which is assumed to be zero mean Gaussian noise with covariance $R_t$ such that $v_t \sim \mathcal{N}(0, R_t)$. In (4.7), $H = [H_1 \, H_2]$ is the observation matrix where $H_1 = [I_{2d} \, 0_{2d}]$, $H_2 = [0_{2d} \, 0_{2d}]$.

The model level in the HDBN showed in Fig. 4.2-a links the set of models ($m > 1$) stored in the radio's long term memory that are learned in previous experiences related to different radio situations ($m > 1$ depicts the successive iterations in Fig. 4.1). Such hierarchical representation provides the radio with the capability of predicting not only the dynamics of the continuous states (the direct cause of the observation/bottom level) but also to maintain coherent knowledge between discrete and continuous variables as well as to predict the dynamics at higher levels of abstractions enabling by that an anticipatory process in explaining the behaviour inside the radio spectrum deeply (at higher levels).

### 4.2.2 Radio Spectrum Perception

Classical probabilistic inference approaches (e.g. Belief propagation [247]) make it possible to use Probabilistic Graphical Models (PGMs) [248] like an HDBN for online predictions and estimations. However, here we employ a Modified Bayesian Filtering to integrate within such inference operation also multilevel abnormality measurements. It is known that prediction and estimation related inference in multilevel HDBNs like switching models [244] can be based on a combination of Particle Filter (PF) and Kalman Filter (KF). This allows inferring the radio environmental states at different hierarchical levels. Markov Jump Particle Filter (MJPF) firstly proposed in [249] is the type of inference methods that can be here applicable and from which a Modified Bayesian filtering block can be derived and used as shown in Fig. 4.1. The inference process concerning prediction and estimation within an MJPF considers the generative model of the HDBN by using its specific components, i.e. dynamic and likelihood inter-level models to parametrize the PF and the KF based on local generative properties required in their algorithmic steps (i.e. the knowledge acquired and stored in the radio's memory is used to drive the inference). In fact, the switching variables of the MJPF can be considered as a finite number of discrete superstates variables associated with the activated model and learned as described in the previous section. Moreover, the corresponding dynamics for each superstate is related to the linear model at the continuous level where the velocity (the average derivative component of the superstate in question) is encoded. PF is employed by the radio to perform superstate predictions due to its ability in dealing with non-linearity, while KF is employed to perform state predictions due to the linear relationship between the continuous variables of the dynamic model at the continuous level.

However, to allow Modified MJPF model to perform additional functionality, namely multilevel abnormality prediction, the local information flow of conditional generative predictions in the HDBN must be enriched by additional inference operations than those included in a classical MJPF. To understand the extension here proposed, it can be useful to recall that in hierarchical Bayesian models as MJPF, an inference can be described as a distributed message passing between nodes where top-down component of generative nature can be associated with both predictivity on temporal links (from slice to slice) and on hierarchical abstraction messages (within a slice or intra-slice). Generative nature in HDBNs also allows generating data series in a bottom-up way, i.e. messages from lower nodes to higher-level nodes. Thus, depending on the direction of message passing, a prognostic or predictive (top-down) component of inference can be distinguished from a diagnostic (bottom-up) one.

Temporal and semantic switching predictive top-down messages ($\pi(\tilde{X}_t)$, $\pi(\tilde{S}_t^m)$) depend on the knowledge learned in dynamical models within an HDBN. The bottom-up inference is based on likelihood models and consists of passing messages ($\lambda(\tilde{X}_t)$, $\lambda(\tilde{S}_t^m)$) in a feed-backward manner to adjust the expectations (predictions provided by top-down messages) given a sequence of observations. Thus, the availability of synchronous top-down messages and bottom-up messages in classical Bayesian inference allows changing the belief (i.e., what radio is expecting) in hidden variables (i.e., $\tilde{X}_t$ and $\tilde{S}_t^m$ that are not directly observed) in a distributed way by updating the belief using messages coming from connected variables. MJPF consists of two steps, prediction and update at each time instant $t$. In the prediction step, PF relies on the transition matrix ($\Pi$) as a proposal to predict future superstates ($\tilde{S}_t^{m*}$) as mentioned in (4.5), by propagating a set of particles each being associated with a specific superstate associated ($\tilde{S}_{t,n}^m$) and a weight $W_{t,n} = 1/N$, where $N$ is the total number of particles at each $t$. Then for each particle ($n$), a KF is used to predict the continuous state as pointed out in (4.6). This prediction depends on the hypothesized superstate as (4.6) can be written as a conditional probability $P(\tilde{X}_t|\tilde{X}_{t-1}, \tilde{S}_{t,n}^m)$. Accordingly, the posterior probability associated with the predicted state is:

$$\pi(\tilde{X}_t) = P(\tilde{X}_t, \tilde{S}_{t,n}^m|\tilde{Z}_{t-1}) = \int P(\tilde{X}_t|\tilde{X}_{t-1}, \tilde{S}_{t,n}^m) \overbrace{\lambda(\tilde{X}_{t-1})}^{P(\tilde{Z}_{t-1}|\tilde{X}_{t-1})} d\tilde{X}_{t-1}, \tag{4.8}$$

In the update step, the posterior probability $P(\tilde{X}_t, \tilde{S}_{t,n}^m|\tilde{Z}_t)$ is corrected by using the message ($\lambda(\tilde{X}_t)$) backward propagated from the current observation $\tilde{Z}_t$, in the following way: $P(\tilde{X}_t, \tilde{S}_{t,n}^m|\tilde{Z}_t) = \pi(\tilde{X}_t)\lambda(\tilde{X}_t)$. After that, the message $\lambda(\tilde{S}_t^m)$ backward propagated from the bottom level towards the discrete level can be used to update the belief in $\tilde{S}_t^m$, and it can be calculated as:

$$\lambda(\tilde{S}_t^m) = \lambda(\tilde{X}_t)P(\tilde{X}_t|\tilde{S}_t^m) = P(\tilde{Z}_t|\tilde{X}_t)P(\tilde{X}_t|\tilde{S}_t^m), \tag{4.9}$$

where $P(\tilde{X}_t|\tilde{S}_t^m) \sim \mathcal{N}(\mu_{\tilde{S}_k^m}, \Sigma_{\tilde{S}_k^m})$ denotes a Gaussian distribution with mean $\mu_{\tilde{S}_k^m}$ and covariance $\Sigma_{\tilde{S}_k^m}$. While, $\lambda(\tilde{X}_t) \sim \mathcal{N}(\mu_{\tilde{Z}_t}, R)$ denotes a Gaussian distribution with mean $\mu_{\tilde{Z}_t}$ and covariance $R$. The multiplication between $\lambda(\tilde{X}_t)$ and $P(\tilde{X}_t|\tilde{S}_t^m)$ can be obtained by calculating the Battacharyya distance ($D_B$) [250, 251] as follows:

$$D_B\big(\lambda(\tilde{X}_t), P(\tilde{X}_t|\tilde{S}_t^m = \tilde{S}_k^m)\big) = -ln \int \sqrt{\lambda(\tilde{X}_t)P(\tilde{X}_t|\tilde{S}_t^m = \tilde{S}_k^m)} d\tilde{X}_t, \tag{4.10}$$

where $\tilde{S}_k^m \in \tilde{S}^m$. The vector $D_\lambda$ containing all the $D_B$ values between $\lambda(\tilde{X}_t)$ and all the superstates in the set $\tilde{S}^m$ is here estimated as:

$$D_\lambda = \left[D_B(\lambda(\tilde{X}_t), P(\tilde{X}_t|\tilde{S}_t^m = \tilde{S}_1^m)), \ldots, D_B(\lambda(\tilde{X}_t), P(\tilde{X}_t|\tilde{S}_t^m = \tilde{S}_L^m))\right]. \quad (4.11)$$

So, $\lambda(\tilde{S}_t^m)$ can be rewritten in the following form:

$$\lambda(\tilde{S}_t^m) = D_\lambda = \left[\overbrace{D_B(\lambda(\tilde{X}_t), P(\tilde{X}_t|\tilde{S}_t^m = \tilde{S}_1^m))}^{D_\lambda(1)}, \ldots, \overbrace{D_B(\lambda(\tilde{X}_t), P(\tilde{X}_t|\tilde{S}_t^m = \tilde{S}_L^m))}^{D_\lambda(L)}\right]. \quad (4.12)$$

Thus, $\lambda(\tilde{S}_t^m)$ is a vector containing $L$ elements (i.e., $\lambda(\tilde{S}_t^m) \in \mathbb{R}^{1\times L}$). However, since $0 \leq D_B \leq \infty$, $\lambda(\tilde{S}_t^m)$ includes values ranging from 0 to $\infty$. Hence, we need to convert the elements of $\lambda(\tilde{S}_t^m)$ into probabilities. We know that $\lambda(\tilde{S}_t^m)$ is a discrete probability distribution where $0 \leq \lambda(\tilde{S}_l^m) \leq 1$ and $\sum_l^L \lambda(\tilde{S}_l^m) = 1$ such that $l \in L$. Therefore, $\lambda(\tilde{S}_t^m)$ in terms of probability can be achieved by performing normalization as follows:

$$\lambda(\tilde{S}_t^m) = \left[\frac{1/D_\lambda(1)}{1/\sum_{l=1}^{L} D_\lambda(l)}, \ldots, \frac{1/D_\lambda(L)}{1/\sum_{l=1}^{L} D_\lambda(l)}\right]. \quad (4.13)$$

After calculating $\lambda(\tilde{S}_t^m)$ and unlike [249], herein the weight $W_{t,n}$ of the particle $\tilde{S}_{t,n}^m$ is updated using $W_{t,n} = W_{t,n}\lambda(\tilde{S}_{t,n}^m)$ and then normalized by considering the Sequential Importance Resampling (SIR) technique.

In classical MJPF, predictive and diagnostic messages are at the basis of Bayesian updating, i.e. are used to estimate an updated joint posterior at different levels. However, some information is lost in this process, namely an evaluation according to some probabilistic metric of the differences between two messages arriving at a given node. In fact, in such difference, i.e. the surprise or abnormality can be estimated based on the difference of expectation w.r.t evidence coming from data on a given variable. The following section defines how this is here obtained to provide Self-Aware radio with a general basis for abnormality detection.

### 4.2.3 Abnormality Detection

In the modified approach here proposed, classical Bayesian inference is enriched by another capability correlated to exploit and describe differences between top-down and bottom-up messages incoming into a generic node of the HDBN that provide hierarchical abnormality signals as shown in Fig. 4.2-b.

**Kullback-Leibler-Divergence Abnormality (*KLDA*) at Discrete Level**

in this case the abnormality is defined as a distance between the messages entering to node $\tilde{S}_t^m$, namely the predictive ($\pi(\tilde{S}_t^m)$) and the diagnostic ($\lambda(\tilde{S}_t^m)$) messages. Differences between the probability profiles of predictive support and evidence indicate that involved components of the generative model used to predict radio environment dynamics do not fit current observations, i.e. provide to the radio an awareness signal indicating whether and how much the current surrounding environment is behaving in a way different to the rules learned in the generalized model. Since these two messages represent discrete probability distributions, Kullback-Leibler-Divergence (KLD) [250, 252] is here proposed as probability distance measurement to calculate the difference between them as follows:

$$KLDA = D_{KL}\big(\pi(\tilde{S}_t^m)||\lambda(\tilde{S}_t^m)\big) + D_{KL}\big(\lambda(\tilde{S}_t^m)||\pi(\tilde{S}_t^m)\big), \tag{4.14}$$

where "||" operator indicates the "divergence". To computationally apply this distance in the run time of the MJPF, at each time instant $t$ and after the updated process (KF) and resampling (PF) the histogram of the particles is extracted and the relative frequency of each particle is used to approximate the local new prior over $\tilde{S}_t^m$ as follows:

$$p(\tilde{S}_t^m = i) = \frac{y(\tilde{S}_t^m = i)}{N} \quad i \in S, \tag{4.15}$$

where $y(.)$ is the frequency (or number of occurrences of superstate $i$ in the histogram) and $N$ is the total number of particles. It is worth to note that $\lambda(\tilde{S}_t^m)$ is the same for all the particles propagated by PF at time instant $t$. As $\pi(\tilde{S}_t^m)$ is available as a row vector related to $\tilde{S}_t^m$ picked from the transition matrix, the predictive message can be approximated by selecting the row to be used in the *KLDA* depending on the updated prior histogram previously introduced. Lets define $\mathbb{S}$ as the set of the winning particles (whose entries in the histogram is greater than zero), where:

$$\mathbb{S} = \{i|p(\tilde{S}_t^m = i) > 0\} \quad i \in S. \tag{4.16}$$

$D_{KL}$ is calculated between $\lambda(\tilde{S}_t^m)$ and the rows of the transition matrix $\pi(\tilde{S}_t^m)$ related to the winning particles $\mathbb{S}$ (that are the most probable before update). Therefore, (4.14) becomes:

$$KLDA = \sum_{i\in\mathbb{S}} \big[p(i) \sum_{j=1}^{L} \pi_{ij} \log(\frac{\pi_{ij}}{\lambda_j})\big] + \sum_{i\in\mathbb{S}} \big[p(i) \sum_{j=1}^{L} \lambda_j \log(\frac{\lambda_j}{\pi_{ij}})\big]. \tag{4.17}$$

The space of the predicted discrete variables can be divided into two sub-sets. One corresponding to the normality region $\Omega_N$, and the other to its complement ($\Omega_A$), that can be interpreted as the subset of unexpected super-states, i.e. outside the model's trusted prediction area. $\Omega_N$ can be defined as the sub-set that contains the most probable particles (with high frequency of occurrence in the histogram) and satisfy the following condition:

$$\sum_{i \in \Omega_N} p(\tilde{s}_t = i) \geq \zeta, \tag{4.18}$$

while, $\Omega_A$ is the sub-set that contains the less probable particles (with low frequency of occurrences) and satisfy the following condition:

$$\sum_{k \in \Omega_A} p(\tilde{s}_t = k) \leq 1 - \zeta, \tag{4.19}$$

where, $k \neq i$ and $\Omega_N \subset \mathbb{S}$, $\Omega_A \subset \mathbb{S}$ while $p(.)$ and $\mathbb{S}$ are defined in (4.15) and (4.16), respectively. $\zeta$ is the acceptance ratio that varies from 0 to 1. It is possible evaluating to what extent the probability mass function of the evidence message falls in the region $\Omega_N$ and its complement to 1 in $\Omega_A$. A value $\alpha$ can be used to indicate the area (i.e., the probability mass) of $\lambda(\tilde{S}_t^m)$ in the normal region. The KLDA defined in (4.14) covers the global discrete state space, however by dividing the discrete space into two regions we can rewrite KLDA as:

$$\begin{aligned} KLDA = \mathcal{KLD}_N + \mathcal{KLD}_A = \\ \sum_{i \in \Omega_N} p(\tilde{s}_t = i) \times \Big[ D_{KL}\big(\pi(\tilde{S}_t^m = i) || \lambda(\tilde{S}_t^m)\big) + D_{KL}\big(\lambda(\tilde{S}_t^m) || \pi(\tilde{S}_t^m = i)\big) \Big] \\ + \sum_{i \in \Omega_A} p(\tilde{s}_t = i) \times \Big[ D_{KL}\big(\pi(\tilde{S}_t^m = i) || \lambda(\tilde{S}_t^m)\big) + D_{KL}\big(\lambda(\tilde{S}_t^m) || \pi(\tilde{S}_t^m = i)\big) \Big]. \end{aligned} \tag{4.20}$$

$\mathcal{KLD}_N$ measures the similarity between the set of particles with the highest prior probability and the probabilistic evidence (i.e., likelihood characterized by $\lambda(\tilde{S}_t^m)$), while $\mathcal{KLD}_A$ measures the similarity with low probability predictions. The abnormal situation can so be defined as:

$$\mathcal{KLD}_N > \mathcal{KLD}_A, \tag{4.21}$$

From (4.20),

$$\mathcal{KLD}_A = KLDA - \mathcal{KLD}_N. \tag{4.22}$$

After substituting (4.22) in (4.21) we have: $\mathcal{KLD}_N > KLDA - \mathcal{KLD}_N$. Thus, $KLDA < 2(\mathcal{KLD}_N)$. In addition, during a normal situation $\mathcal{KLD}_N$ is supposed to be small due to the fact that the observation often votes for the most probable particles, thus:

$$\mathcal{KLD}_N \leq (1-\zeta)\alpha, \tag{4.23}$$

where $\alpha$ is the support mass expected by $\lambda(\tilde{S}_t^m)$ to consider a prediction as not associated with abnormalities. Therefore the normal situation occurred if:

$$KLDA < (1-\zeta)\alpha. \tag{4.24}$$

**Abnormality at Continuous Level**

At this level, the abnormality indicator can be defined as a distance between different probabilistic messages incoming to node $\tilde{X}_t$. It is based on the Bhattacharyya distance ($D_B$) between the predictive message $\pi(\tilde{X}_t)$ and the diagnostic message $\lambda(\tilde{X}_t)$ incoming from the observation level. Thus, the continuous level abnormality (CLA) is defined as:

$$CLA = D_B(\pi(\tilde{X}_t), \lambda(\tilde{X}_t)) = -\ln \int \sqrt{P(\tilde{X}_t, \tilde{S}_t^{m*}|\tilde{Z}_{t-1})P(\tilde{Z}_t|\tilde{X}_t)}d\tilde{X}_t. \tag{4.25}$$

$\pi(\tilde{X}_t)$ is distributed according to a multivariate Gaussian and it has the following probability density function (PDF):

$$P(\pi(\tilde{X}_t)) = \frac{1}{\sqrt{(2\pi)^d|\Sigma_\pi|}} \exp\big[-\frac{1}{2}(\tilde{x}_t - \mu_\pi)^T \Sigma_\pi^{-1}(\tilde{x}_t - \mu_\pi)\big]. \tag{4.26}$$

The shape of $\pi(\tilde{X}_t)$ is characterized by the covariance matrix $\Sigma_\pi$ and the Mahalanobis distance ($D_M$) of each data sample $\tilde{x}_t$ from the distribution's centroid. Thus, a good generative model can generate data samples $\tilde{x}_t$ that lie in the confidence region almost of the time. The confidence region is often represented as an ellipsoid around the data samples that satisfy the following condition:

$$(\tilde{x}_t - \mu_\pi)^T \Sigma_\pi^{-1}(\tilde{x}_t - \mu_\pi) \leq \chi_d^2, \tag{4.27}$$

where $(\tilde{x}_t - \mu_\pi)^T \Sigma_\pi^{-1}(\tilde{x}_t - \mu_\pi)$ is the Squared Mahalanobis distance and $\chi_d^2$ is the quantile function of the chi-squared distribution with $d$ degrees of freedom. Then,

$$\begin{cases} \tilde{x}_t \in \mathcal{R}_N, \text{ if } (\tilde{x}_t - \mu_\pi)^T \Sigma_\pi^{-1}(\tilde{x}_t - \mu_\pi) \leq \chi_d^2, \\ \tilde{x}_t \in \mathcal{R}_A, \text{ if } (\tilde{x}_t - \mu_\pi)^T \Sigma_\pi^{-1}(\tilde{x}_t - \mu_\pi) > \chi_d^2, \end{cases} \tag{4.28}$$

where $\mathcal{R}_N$ is the normal region and $\mathcal{R}_A$ is the abnormal region. From the Mahalanobis distances we can obtain the upper ($\tilde{x}_t^{UB}$) and lower ($\tilde{x}_t^{LB}$) bounds of $\pi(\tilde{X}_t)$. The area of each region can be written as:

$$\begin{aligned}\mathcal{R}_A = &\int_{-\infty}^{\tilde{x}_t^{LB}} \frac{1}{\sqrt{(2\pi)^d|\Sigma_\pi|}} \exp\big[-\frac{1}{2}(\tilde{x}_t-\mu_\pi)^T\Sigma_\pi^{-1}(\tilde{x}_t-\mu_\pi)\big]d\tilde{x}_t + \\ &\int_{\tilde{x}_t^{UB}}^{+\infty} \frac{1}{\sqrt{(2\pi)^d|\Sigma_\pi|}} \exp\big[-\frac{1}{2}(\tilde{x}_t-\mu_\pi)^T\Sigma_\pi^{-1}(\tilde{x}_t-\mu_\pi)\big]d\tilde{x}_t,\end{aligned} \tag{4.29}$$

$$\mathcal{R}_N = \int_{\tilde{x}_t^{LB}}^{\tilde{x}_t^{UB}} \frac{1}{\sqrt{(2\pi)^d|\Sigma_\pi|}} \exp\big[-\frac{1}{2}(\tilde{x}_t-\mu_\pi)^T\Sigma_\pi^{-1}(\tilde{x}_t-\mu_\pi)\big]d\tilde{x}_t, \tag{4.30}$$

where the performance of the generative model depends upon these regions. A good generative model can generate more data samples that lie in $\mathcal{R}_N$, such that, $\mathcal{R}_N > \mathcal{R}_A$.

On the other side, the message $(\lambda(\tilde{X}_t))$ tells if the real signal matches the predicted signal. Thus, we aim to study how much the diagnostic message supports the predictive message generated by the generative model. This can be done by calculating the product between the two messages and then integrating it to obtain the corresponding area. After that we must see if the calculated area belongs to the normal or the abnormal region defined in (4.29) and (4.30) respectively, to make the correct decision. We propose to employ the $D_B$ to calculate the similarity between the two messages. The analytical advantage of using $D_B$ is due to its ability in measuring the overlap area between the two distributions through the Bhattacharyya coefficient ($\mathcal{BC}$) and then converting it to a distance metric. The $\mathcal{BC}$ is divided into 3 regions as follows:

$$\begin{aligned}\mathcal{BC} = \int_{-\infty}^{+\infty}\sqrt{\pi(\tilde{X}_t)\lambda(\tilde{X}_t)}d\tilde{X}_t = &\overbrace{\int_{-\infty}^{\tilde{x}_t^{LB}}\sqrt{\pi(\tilde{X}_t)\lambda(\tilde{X}_t)}d\tilde{X}_t}^{\mathcal{BC}_A^1} + \overbrace{\int_{\tilde{x}_t^{LB}}^{\tilde{x}_t^{UB}}\sqrt{\pi(\tilde{X}_t)\lambda(\tilde{X}_t)}d\tilde{X}_t}^{\mathcal{BC}_N} + \\ &\underbrace{\int_{\tilde{x}_t^{UB}}^{+\infty}\sqrt{\pi(\tilde{X}_t)\lambda(\tilde{X}_t)}d\tilde{X}_t}_{\mathcal{BC}_A^2},\end{aligned} \tag{4.31}$$

where $\mathcal{BC}_A$ is the overlap that falls inside the abnormal region $\mathcal{R}_A$ (i.e. $\mathcal{BC}_A \subset \mathcal{R}_A$) and $\mathcal{BC}_N$ is the overlap that falls in the normal region $\mathcal{R}_N$ ($\mathcal{BC}_N \subset \mathcal{R}_N$). The normal situation occurred if:

$$\mathcal{BC}_N > \mathcal{BC}_A, \tag{4.32}$$

$\mathcal{BC}_A$ is the overlap between $\pi(\tilde{X}_t)$ and $\lambda(\tilde{X}_t)$ inside the region $\mathcal{R}_A$, which can be written as:

$$\mathcal{BC}_A = \int_{-\infty}^{\tilde{x}_t^{LB}} \sqrt{\pi(\tilde{X}_t)\lambda(\tilde{X}_t)} d\tilde{X}_t + \int_{\tilde{x}_t^{UB}}^{+\infty} \sqrt{\pi(\tilde{X}_t)\lambda(\tilde{X}_t)} d\tilde{X}_t. \tag{4.33}$$

According to the product rule:

$$\mathcal{BC}_A = \int_{-\infty}^{\tilde{x}_t^{LB}} \sqrt{\pi(\tilde{X}_t)} d\tilde{X}_t \,.\, \int_{-\infty}^{\tilde{x}_t^{LB}} \sqrt{\lambda(\tilde{X}_t)} d\tilde{X}_t + \int_{\tilde{x}_t^{UB}}^{+\infty} \sqrt{\pi(\tilde{X}_t)} d\tilde{X}_t \,.\, \int_{\tilde{x}_t^{UB}}^{+\infty} \sqrt{\lambda(\tilde{X}_t)} d\tilde{X}_t. \tag{4.34}$$

Then, according to the Cauchy–Schwarz inequality:

$$\begin{aligned}(\mathcal{BC}_A)^2 &\leq \int_{-\infty}^{\tilde{x}_t^{LB}} \pi(\tilde{X}_t) d\tilde{X}_t \,.\, \int_{-\infty}^{\tilde{x}_t^{LB}} \lambda(\tilde{X}_t) d\tilde{X}_t + \int_{\tilde{x}_t^{UB}}^{+\infty} \pi(\tilde{X}_t) d\tilde{X}_t \,.\, \int_{\tilde{x}_t^{UB}}^{+\infty} \lambda(\tilde{X}_t) d\tilde{X}_t \\ &\leq Q_{\tilde{x}_t^{LB}}(\tilde{X}_t) \,.\, [\beta_1(1-\alpha)] + Q_{\tilde{x}_t^{UB}}(\tilde{X}_t) \,.\, [\beta_2(1-\alpha)],\end{aligned} \tag{4.35}$$

where $Q$ denotes the Q-function. Since $\pi(\tilde{X}_t)$ is a symmetric multivariate Gaussian we have: $\tilde{x}_t^{LB} = \tilde{x}_t^{UB} = \mathcal{T}$. Therefore,

$$\mathcal{BC}_A \leq \sqrt{Q_{\mathcal{T}}(\tilde{X}_t) \,.\, [\beta_1(1-\alpha)] + Q_{\mathcal{T}}(\tilde{X}_t) \,.\, [\beta_2(1-\alpha)]}, \tag{4.36}$$

where $\alpha$ is the area of $\lambda(\tilde{X}_t)$ falling inside the normal region $\mathcal{R}_N$, $\beta_1$ and $\beta_2$ indicate the amount of $(1-\alpha)$ falling in $(-\infty, \tilde{x}_t^{LB})$ and $(\tilde{x}_t^{UB}, +\infty)$, respectively, such that $(\beta_1 + \beta_2) = 1$. From (4.31):

$$\mathcal{BC}_N = \mathcal{BC} - \mathcal{BC}_A. \tag{4.37}$$

After substituting in (4.32), we have: $\mathcal{BC} > 2(\mathcal{BC}_A)$. Hence, the normal situation occurred if:

$$\mathcal{BC} > \sqrt{Q_{\mathcal{T}}(\tilde{X}_t) \,.\, [\beta_1(1-\alpha)] + Q_{\mathcal{T}}(\tilde{X}_t) \,.\, [\beta_2(1-\alpha)]}. \tag{4.38}$$

**Deep Abnormality at Continuous Level (DCLA)**

This abnormality indicator is based on the similarity between state prediction ($\tilde{X}_t^*$) of the winning particle (the one with the highest weight) and observation ($\tilde{Z}_t$). It consists of a set of abnormalities corresponding to each sub-carrier incorporated in the Generalized States ($\tilde{\mathbf{X}}$), such as:

$$DCLA = \{DCLA_{f_1}, DCLA_{f_2}, \ldots, DCLA_{f_n}\}, \tag{4.39}$$

where,

$$DCLA_{f_n} = d(\mathcal{A}\tilde{Z}_t, \mathcal{B}\tilde{X}_t^*) = \sqrt{(I_{f_n} - I_{f_n}^*)^2 + (Q_{f_n} - Q_{f_n}^*)^2}, \tag{4.40}$$

where $n \in \{1,2,\ldots,d\}$ represents the sub-carriers index and $d$ is the total number of the sensed sub-carriers. In (4.40), $\mathcal{A}$ and $\mathcal{B}$ are $2d \times 2d$ matrices and their elements are zeros except those that matches the associated sub-carrier and are switched to **1** in order to pick up the corresponding $I$ and $Q$ components. For example, regarding the indicator of sub-carrier $f_n$, elements $\mathcal{A}_{n,n}$, $\mathcal{A}_{d+n,d+n}$ and $\mathcal{B}_{n,n}$,$\mathcal{B}_{d+n,d+n}$ will be set to **1**.

### 4.2.4 Abnormality Characterization

In the previous functionality, message-passing in the HDBN from slice to slice is exploited to calculate the abnormality measurements, while here the message-passing in the same slice (intra-slice) will be used to calculate the Generalized Errors (as shown in Fig. 4.2-b) from which the jammer can be characterized at multiple levels by means of machine learning.

#### Jammer Characterization at Discrete Level

At this level, the radio can characterize the abnormal situation by analysing the superstates' evolution based on the predictive messages ($\pi(\tilde{S}_t^m)$) and that based on the diagnostic messages ($\lambda(\tilde{S}_t^m)$). In this way two sets of superstates ($\mathbb{S}^\pi$ and $\mathbb{S}^\lambda$) are created, the first one contains the predicted superstates while the second contains the observed superstates. At each time instant $t_j$ (the time when the jammer is detected) these superstates can be obtained as follows:

$$\begin{cases} \tilde{S}_{t_j}^\pi = \underset{\tilde{S}_{t_j}^m \in \mathbb{S}}{\operatorname{argmax}}\, y(\tilde{S}_{t_j}^m), \\ \tilde{S}_{t_j}^\lambda = \underset{\tilde{S}_{t_j}^m \in S}{\operatorname{argmax}}\, \lambda(\tilde{S}_{t_j}^m), \end{cases} \tag{4.41}$$

where $y(\tilde{S}_t^m)$ is defined in (4.15), $\tilde{S}_{t_j}^\pi \in \mathbb{S}^\pi$ and $\tilde{S}_{t_j}^\lambda \in \mathbb{S}^\lambda$. Comparing between $\tilde{S}_{t_j}^\pi$ and $\tilde{S}_{t_j}^\lambda$ can help to understand how the jammer is affecting the superstates evolution. If $\tilde{S}_{t_j}^\lambda$ is not equal to $\tilde{S}_{t_j}^\pi$, this means that the jammer shifts the signal from $\tilde{S}_{t_j}^\pi$ to $\tilde{S}_{t_j}^\lambda$, otherwise the signal is manipulated by the jammer but kept in the expected superstate. In other words, the radio expects (predicts) that the signal's sample (i.e. OFDM symbol) will fall in a certain superstate based on the dynamic rules learned in previous experience (related to model $m$). However, during attacks, the jammer shifts the signal's sample to another superstate or even manipulate the sample but keeps it inside the predicted superstate (the case when $\tilde{S}_{t_j}^\pi$=$\tilde{S}_{t_j}^\lambda$). This cross-correlation between the predictive support $\pi(\tilde{S}_{t_j}^m)$ and the diagnostic support $\lambda(\tilde{S}_{t_j}^m)$ allows the radio to understand the jammer's effect on the superstates (at discrete level) of the learned dynamic model and to discover the jammer's strategy or the dynamic rules it is following to attack the signal.

### Jammer Characterization at Continuous Level

Characterizing the attack at the continuous level helps the radio to understand the jammer's force in terms of $I$ and $Q$ values and how much the jammer shifted the signal from one superstate to the other or from a specific superstates' centroid. This depends on the characterization done before at the discrete level. The characteristics at the discrete level can be forwarded towards the continuous level to calculate the generalized errors as follows:

$$\mathbb{D}_{t_j} = \begin{cases} \overbrace{\mu\big(\operatorname*{argmax}_{\tilde{S}_{t_j}^m \in \mathbb{S}} \lambda(\tilde{S}_{t_j}^m)\big) - \tilde{X}_{t_j}^{\lambda}}^{\textbf{Generalized Error 1 } (\varepsilon^1_{\tilde{X}_{t_j}})} & \text{if } \tilde{S}_{t_j}^{\pi} = \tilde{S}_{t_j}^{\lambda}, \\ \underbrace{\mu\big(\operatorname*{argmax}_{\tilde{S}_{t_j}^m \in S} \lambda(\tilde{S}_{t_j}^m)\big) - \mu\big(\operatorname*{argmax}_{\tilde{S}_{t_j}^m \in S} \pi(\tilde{S}_{t_j}^m)\big)}_{\textbf{Generalized Error 2 } (\varepsilon^2_{\tilde{X}_{t_j}})} & \text{if } \tilde{S}_{t_j}^{\pi} \neq \tilde{S}_{t_j}^{\lambda}, \end{cases} \tag{4.42}$$

where $\mathbb{D}_{t_j}$ is the generalized error containing the $I$ and $Q$ values and the corresponding derivatives at multiple sub-carriers. In (4.42), if the jammer manipulates the signal but keep it in the same superstate (the expected one), the generalized error is equal to the mean value of $(\tilde{S}_{t_j}^{\lambda})$ subtracted from the generalized state associated with the most probable superstate in $\pi(\tilde{S}_{t_j})$. Otherwise, if the jammer shifts the signal from one superstate to another one, the generalized error is equal to the mean value of the current superstate $(\tilde{S}_{t_j}^{\lambda})$ subtracted from the mean value of the predicted superstate $(\tilde{S}_{t_j}^{\pi})$. The I-Q voting theory is employed to vote for the most probable IQ values encoded in $\mathbb{D}_{t_j}$ and obtained from (4.42). The radio will vote to similar $\mathbb{D}_{t_j}$ values, where the votes along with $\mathbb{D}_{t_j}$ and the corresponding superstates (the predicted ones) will be stored in a cell to be used later on. To understand how much the jammer shifted the signal with respect to the center of the expected superstate, the radio picks the $\mathbb{D}_{t_j}$ value which has the maximum number of votes from the cell stored during the real-time process and extract consequently the corresponding derivatives that realize the jammer's force $(U_{jammer})$.

#### Jammer Characterization at Observation Level (Observational Characterization)

the characteristics obtained at the discrete level are forwarded towards the observation level to calculate the generalized error at this level ($\varepsilon_{\tilde{Z}_t}$) and explain such error as well. From the higher level, the radio can know which superstates of the model are affected by the jammer. Calculating the distance from the superstates' centroid allows to extract the source of the cause (jammer) that affected the shift noticed at higher levels. So, $\varepsilon_{\tilde{Z}_t}$ can be calculated in the following way:

$$\tilde{Z}_t^J = \overbrace{\tilde{Z}_t - H\mu(\underset{\tilde{S}_t \in S}{\operatorname{argmax}}\, \lambda(\tilde{S}_t))}^{\textbf{Generalized Error 3 } (\varepsilon_{\tilde{Z}_t})}, \tag{4.43}$$

which represent the jammer's Generalized State ($\tilde{Z}_t^J$) from which the radio can extract the jamming signal.

### 4.2.5 Incremental Learning of new models

The information obtained in the previous steps will be used here to incrementally learn the new model at two hierarchical levels, at the discrete level by updating the transition matrix of model ($m = 1$) and at the continuous level by updating the linear model associated to the reference model ($m = 1$).

#### Update Transition Matrix

The difference between variables whose belief is given by $\lambda(\tilde{S}_t^m)$ and $\pi(\tilde{S}_t^m)$ denotes the discrete Generalized Error ($\varepsilon(\tilde{S}_t^m)$) which represents the *innovation* provided at the discrete level by PF. As mentioned before $\lambda(\tilde{S}_t^m)$ is a vector containing $L$ elements and it is the same for all the particles propagated by the PF at time instant $t$. $\pi(\tilde{S}_t^m)$ is a 1x$L$ vector picked from the transition matrix and $\varepsilon(\tilde{S}_t^m)$ is a $K$x$L$ matrix where $K$ is the total number of the elements in set $\mathbb{S}$ which is defined in (4.16). Therefore $\varepsilon(\tilde{S}_t^m)$ can be defined as:

$$\varepsilon(\tilde{S}_t^m) = \begin{bmatrix} \varepsilon_1 \\ \vdots \\ \varepsilon_K \end{bmatrix} = \begin{bmatrix} \lambda_1 - \pi_{\mathbb{S}_k 1}, & \lambda_2 - \pi_{\mathbb{S}_k 2}, & \dots & \lambda_L - \pi_{\mathbb{S}_k L} \\ \vdots & \vdots & \vdots & \vdots \\ \lambda_1 - \pi_{\mathbb{S}_K 1}, & \lambda_2 - \pi_{\mathbb{S}_K 2}, & \dots & \lambda_L - \pi_{\mathbb{S}_K L} \end{bmatrix}, \tag{4.44}$$

where $\mathbb{S}_k$ denotes the $k$-th element (superstate) in the set $\mathbb{S}$. $\varepsilon(\tilde{S}_t^m)$ is a zero-mean vector which contains positive and negative elements. At each time instant, $t_j$ (i.e., when the jamming attack takes place, which is obtained from the abnormality signal classification), the $i^{th}$ row vector related to the winning particle extracted from the transition matrix is updated

following:

$$\Pi_{t_j}^{'} = \Pi + \varepsilon(\tilde{S}_{t_j}^{m}) = \begin{bmatrix} \pi_{11}+\varepsilon_1(1) & \pi_{12}+\varepsilon_1(2) & \dots & \pi_{1L}+\varepsilon_1(L) \\ \vdots & \vdots & \vdots & \vdots \\ \pi_{K1}+\varepsilon_K(1) & \pi_{K2}+\varepsilon_K(2) & \dots & \pi_{KL}+\varepsilon_K(L) \end{bmatrix} = \begin{bmatrix} \pi_{11}^{'} & \dots & \pi_{1L}^{'} \\ \vdots & \ddots & \vdots \\ \pi_{K1}^{'} & \dots & \pi_{KL}^{'} \end{bmatrix}. \tag{4.45}$$

After considering the whole time where we detect jamming signal, the updated transition matrix will be:

$$\Pi^{''} = \mathbb{E}\left[\Pi_{t_j}^{'}, \dots, \Pi_{t_j+e}^{'}\right], \tag{4.46}$$

where $\Pi^{''}$ denotes the mean value of all the saved versions of the updated transition matrix $\Pi'$ and $t_j + e$ is the time instant when the jammer ends the attack.

### Update Dynamic Model

after characterizing the jammer at the continuous level to learn the rules it followed to attack the commands, we can update the dynamic rules of the original dynamic model by adding the jammer's force. Therefore, the updated dynamic model which represents the new situation (signal + jammer) can be written as follows:

$$\tilde{X}_t = A\tilde{X}_{t-1} + B\left(U_{\tilde{S}_{t-1}^{m}} + U_{jammer}\right) + w_t, \tag{4.47}$$

where $A$, $B$ and $w_t$ are the same as in (4.6) and $U_{jammer}$ is obtained from jammer characterization at continuous level in Sec.4.2.4. In this way by using the updated dynamic model ($m$+1), the radio will be able to predict the jammer's effect on the commands in terms of $I$ and $Q$ values at multiple sub-carriers.

### Relation between incremental learning and the detected abnormalities

Abnormalities are the surprising patterns in the observation that are previously not seen. Detecting abnormalities means that the radio is surprised by the new measurement. This surprise is due to the fact that the radio was expecting to receive signals that are generated according to the dynamic model encoded in its brain but in fact, improbable signals are received. The analytical meaning of incremental learning is related to the Free-energy principle [253]. The objective is to minimize the free-energy (i.e. the prediction error amount) by continuous correction of the dynamic model that represents the radio environment. This can be done by encoding a new probabilistic representation or by updating the current version of that representation. According to [253], the free-energy can be defined as a function of sensory data (observations) and dynamic model states (prediction):

$$\mathcal{F}(\tilde{Z},\mathcal{M}) = <\mathcal{L}(t)>_q - \mathcal{H}(t), \tag{4.48}$$

where the free-energy comprises the energy ($<\cdot>$) expected under a certain density $q$ (that consists the statistical parameters of the generative model) and its entropy $\mathcal{H}$. $\mathcal{L}$ is the dynamic model encoded in the radio's brain that generates expected data samples and their causes. In the proposed framework (the generative model as HDBN), the free-energy can be reduced after calculating the generalized errors between top-down and bottom-up messages passing among hierarchical levels and consequently learning an appropriate model ($\mathcal{M}$) that optimize the predictions about hidden states generating observations ($\tilde{Z}$) and thus minimizing the free-energy. Hence, the solution is given by:

$$\mathcal{M} = \underset{\mathcal{M}}{\operatorname{argmin}} \mathcal{F}(\tilde{Z},\mathcal{M}), \tag{4.49}$$

which leads to the following optimization problem expressed in terms of abnormality measurements defined before:

$$\begin{cases} \pi(\tilde{S}_t^m) = \underset{\pi(\tilde{S}_t^m)}{\operatorname{argmin}} \Big( D_{KL}\big(\pi(\tilde{S}_t^m)||\lambda(\tilde{S}_t^m)\big) \Big), \\ \pi(\tilde{X}_t) = \underset{\pi(\tilde{X}_t)}{\operatorname{argmin}} -ln \Big[ \mathcal{BC}\big(\pi(\tilde{X}_t(\tilde{S}_t^m)), \lambda(\tilde{X}_t(\tilde{S}_t^m))\big) \Big]. \end{cases} \tag{4.50}$$

The objective is to improve the radio's predictive ability at hierarchical levels. Prediction at the discrete level can be optimized if the measurement votes for the most probable particles (generated by the PF) all the time as discussed in Subsection 4.2.31. During abnormal situation, the most probable particles are not voted by the observation providing by that a high $D_{KL}$ as shown in (4.21). This means that the dynamic transition matrix ($\pi(\tilde{S}_t^m)$) must be changed in a way that it will be supported by the observation ($\lambda(\tilde{S_t}^m)$) which leads to minimize the $\mathcal{KLD}_\mathcal{N}$ defined in (4.20), and thus the overall $\mathcal{KLDA}$ will decrease. The rules of how the dynamic transition matrix must be updated is shown in Subsections 4.2.4 and 4.2.5. At the continuous level, the abnormality measurement can be optimized if the overlap area between the prediction $\pi(\tilde{X}_t)$ and the measurement $\lambda(\tilde{X}_t)$ in the normal region $\mathcal{R}_N$ is quasi-1 as discussed in Subsection 4.2.3. During abnormal situation the overlap is very small (especially when the jammer attacks with high power). Thus, the prediction must be shifted towards the measurement, maximizing by that the overlap in the normal regions ($\mathcal{R}_N$) and minimizing it in the abnormal region ($\mathcal{R}_A$). Such shift realize the jammers power which can be learned during the jammer characterization process discussed in Subsection 4.2.4 and consequently update the linear dynamic model as shown in Subsection 4.2.5.

### 4.2.6 Internal Action

The jammer characterization done in the previous step at different hierarchical levels allows the radio to understand the jammer's nature of how (power), when (time) and where (frequency) it is attacking the commands. This offers several benefits to the radio including self-decision and self-action that support the radio to enhance the physical layer security. After extracting the jammer's signal the radio can suppress the jammer from the current observation and then overcome the issue of false commands or high error probabilities. This self-correction of the jammed signal realizes an auto-defence technique against the attacker threat without the help of other entity in the network. This reduces the time to act against the jammer, rather than sending feedback and then waiting for a response which increases the time of action during the real-time process. The Generalized Error defined in (4.43), produces the generalized state vector of the jammer ($\tilde{Z}_t^J$), which by the way can be fed to an unsupervised technique to be clustered allowing the radio not only to study statistically the effect of the jammer on the received commands (interaction b/w jammer and user) but also to study how the jammer's dynamics are evolving with time using probabilistic reasoning. This can be very useful when the jammer changes the strategy of the attacks, e.g. changes the output power while attacking the radio (to be investigated in future work). Regrading the proposed scenario we supposed that the jammer's output power will not change during the attacks. Thus, we supposed that there is only one cluster (superstate) of the jammer's dynamic model which encodes all the components of the $\tilde{Z}_t^J$. As done before after obtaining the superstates of the normal signal (commands) we can calculate the mean and covariance of the stand-alone superstate of the jammer from which the jammer's control vector ($U_{jammer}$) can be obtained. In addition, the radio can extract (or estimate) the jammer's signal from the $\tilde{Z}_t^J$ at multiple sub-carriers. $\tilde{Z}_t^J$ consists of the signal and the noise and can be expressed as:

$$\tilde{Z}_t^J = \hat{J}_t + w_t, \tag{4.51}$$

thus, the jammer's signal can be estimated using:

$$\hat{J}_t = \tilde{Z}_t^J - \hat{w}_t, \tag{4.52}$$

where $\hat{J}_t$ is the estimated signal of the jammer extracted from $\tilde{Z}_t^J$, and $\hat{w}_t$ can be estimated using the reference model ($m = 1$). After extracting the jammer's signal, the radio can decide to suppress its effect (internal action) from observation before entering to the demodulator and decoder blocks. The observation that represents the jammed signal can be expressed as:

$$Z_t = Y_t + J_t + w_t, \tag{4.53}$$

where $Z_t$ is the jammed signal, $Y_t$ the normal signal, $J_t$ the jammer's signal and $w_t$ is the channel's noise. Therefore, the corrected signal $Z_t^{\dagger}$ will be decomposed as:

$$Z_t^{\dagger} = Z_t - \hat{J}_t = Z_t - \tilde{Z}_t^J + \hat{w}_t. \tag{4.54}$$

## 4.3 System Model

The system model depicted in Fig. 4.3 consists of a cellular-connected UAV with a 4G antenna and GPS receiver, acting as aerial user equipment and served by the ground Base Station (BS). A human operator commands and controls the UAV using the LTE cellular connectivity. Commands are sent to the UAV through BS via the Downlink (DL) channel. We assume that the ground-to-UAV link is always a Line-Of-Sight (LOS) under an Additive White Gaussian Noise (AWGN) channel condition. The 3GPP path loss model defined in [254] is adopted under the Rural-macro with aerial vehicles (RMa-AV) scenario. We consider the DL channel under the threat of a terrestrial jammer which aims to send false commands to alter the trajectory and take control of the UAV. The propagation model consisting of the LTE downlink transmitter, receiver and jammer is shown in Fig. 4.4. The BS continuously sends a Radio Frame (RF) of 10 ms duration to the active users (already synchronized with BS) in the cell. The BS allocates a specific number of sub-carriers to each user for a predetermined time which are referred to as Physical Resource Blocks (PRBs). We supposed that the GPS measures the 3D position every 50 ms and the UAV receives one PRB every 50 ms as well (assuming that the BS follows the third allocation scheme for UAV command & control (C2) data as mentioned in [255]) since the 3GPP specifies that efficient management of a UAV would require a maximum of 100 kb/s for C2 data, latency of 50 ms and inter-arrival time (defined also as Transmission Time Interval TTI) of 100 ms [256]. The commands (Pitch, Yaw, and Roll) are sent in the PRB over 9 consecutive sub-carriers in the frequency domain within 1 OFDM symbol in the time domain. We call these REs as a Resource Vector (RV) as shown in Figs. 4.5-4.6. The remaining sub-carriers and OFDM symbols of the PRB are related to other information sent to the UAV. For our analysis, only the RV is considered in which we are interested in studying the command signals. However, this can be simply extended to consider the whole PRB in future investigation. We assume that the jammer is smart and is aware of the transmission protocol and the resource allocation strategy performed by the BS. Hence, the jammer can locate and identify the PRBs allocated to the UAV inside the radio spectrum and attacks it consequently.

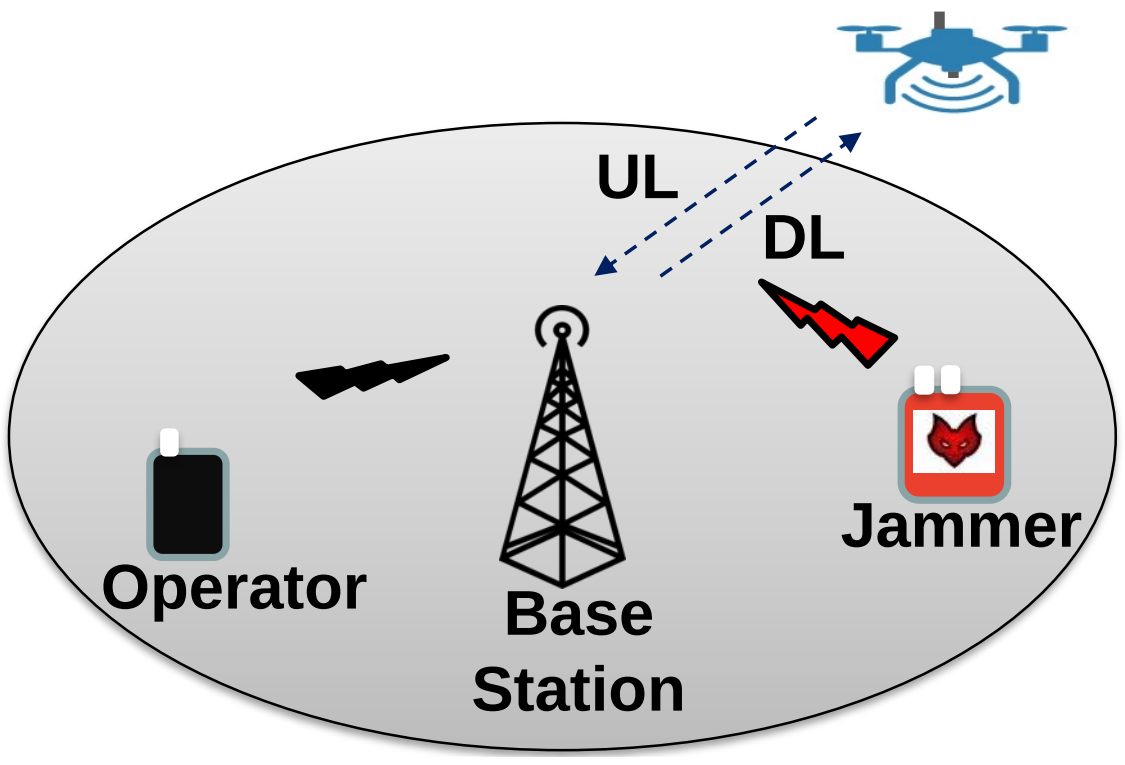

Fig. 4.3 Illustration of the system model.

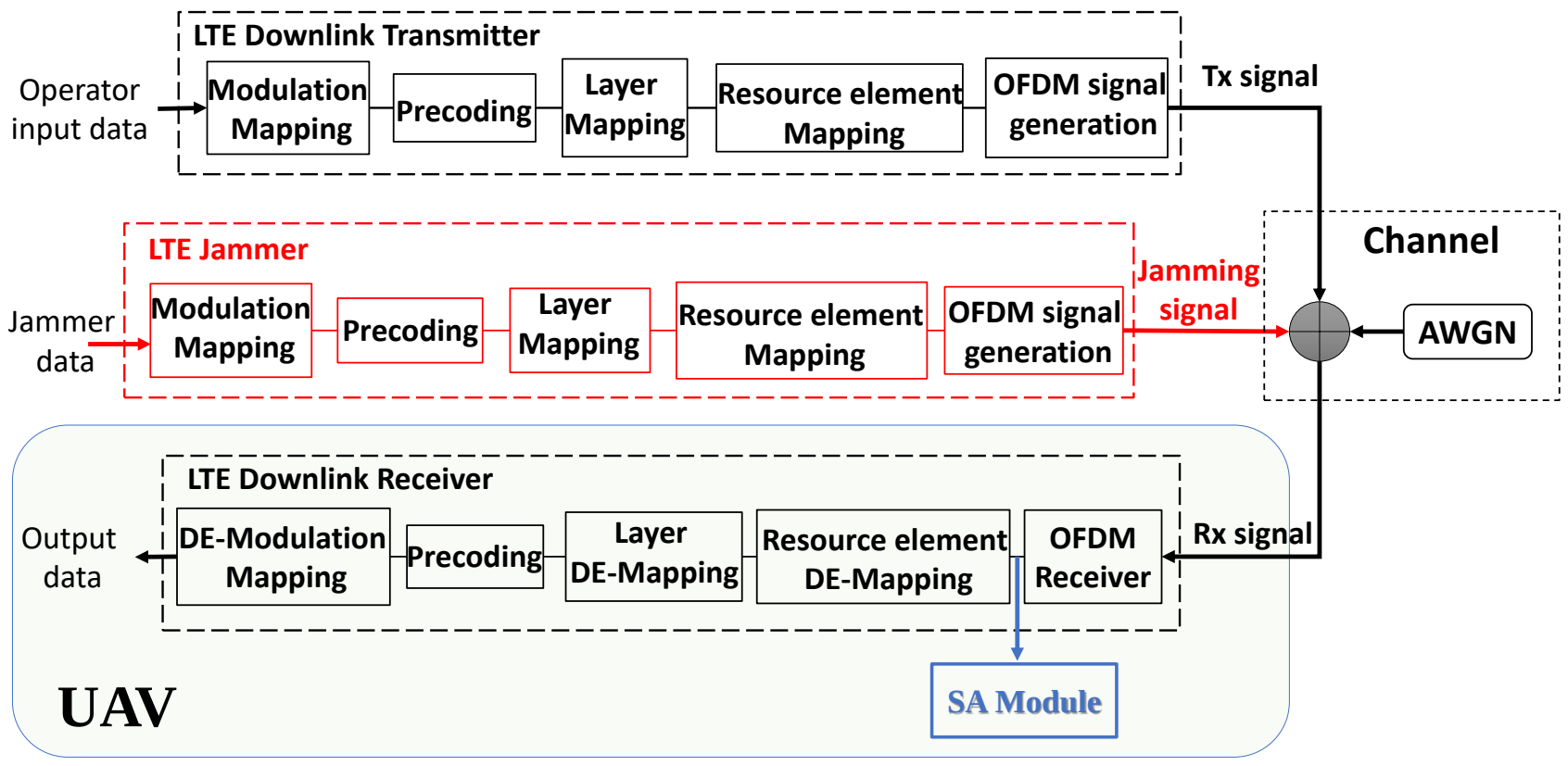

Fig. 4.4 Illustration of the Propagation Model including the Operator, Jammer and the UAV where the SA module is installed.

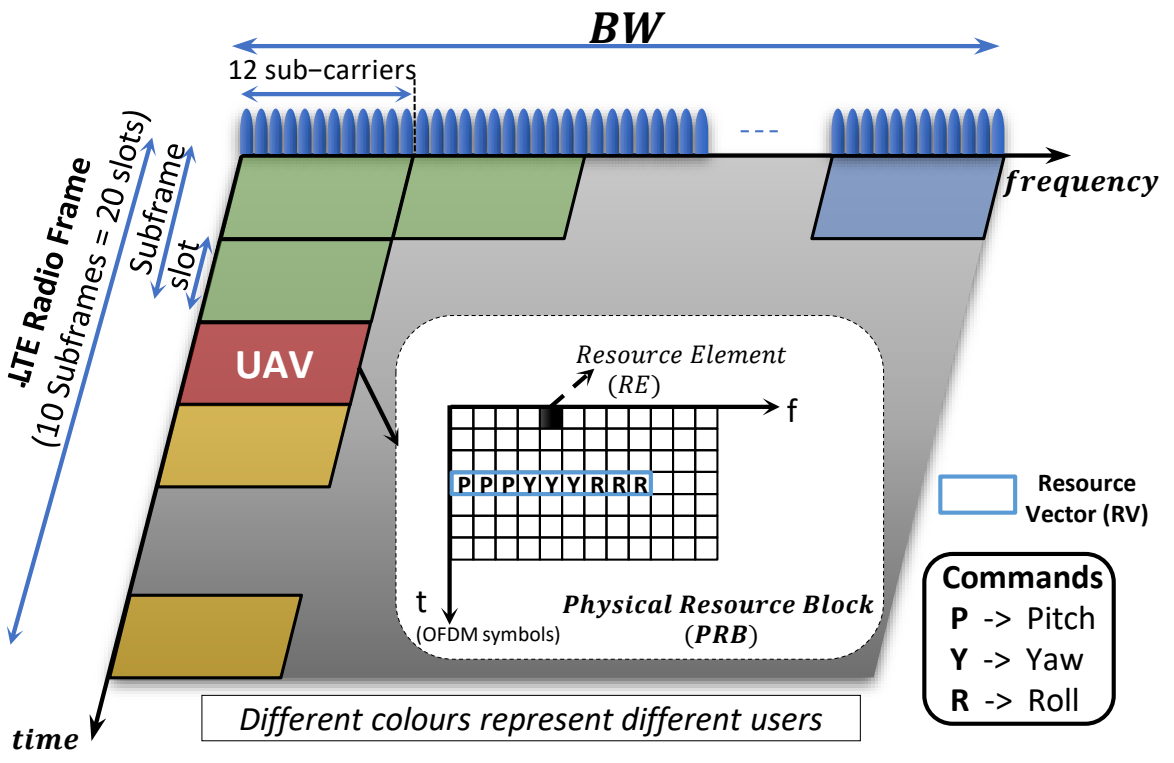

Fig. 4.5 LTE Physical resource allocation and RF structure.

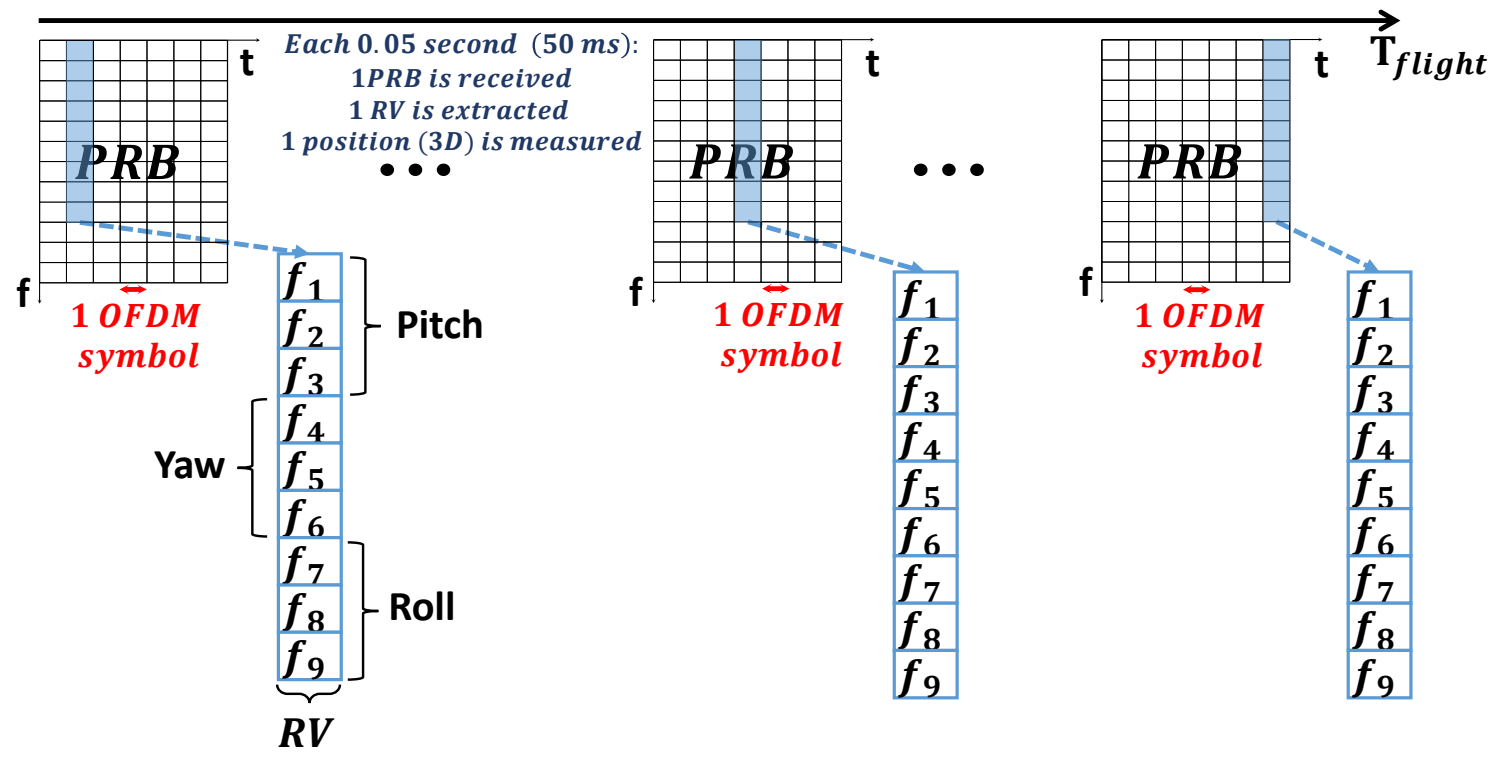

Fig. 4.6 Timing of the PRBs and RVs received by the UAV.

In our study, the data is extracted after the OFDM receiver block. More precisely, after the FFT, where the output of this block consists of all the Resource Elements (REs) which represent the time-frequency grid. At this level, the UAV can scan and sense the whole REs of the grid and capture the $IQ$ data without any extra hardware (by exploiting the FFT). The SA module is installed at this level too enabling the UAV to perform all its functionalities presented in the following section.

### 4.3.1 Computational Complexity Analysis

Fig. 4.7 illustrates the signalling process involved in the SA module installed on the UAV and Table. 4.1 describes the procedures of each functionality and the corresponding execution phase. Moreover, to further clarify the workload required by the UAV to execute the mentioned procedures, the computational complexity is concluded in Table. 4.2. In the considered framework, the workload of the UAV can be divided into two components. The first one is the workload during the flight (real-time or online) and the second is the workload after the flight (offline). At the very beginning of the learning process (the first iteration in Fig. 4.1 of the manuscript), the UAV starts with a simplified HDBN model of only two levels where the linear dynamic model assumes that generalized continuous variables of the radio environment are not changing. In this case the UAV will perform predictions only at the continuous level (predictions of Generalized state vectors of $4 \times d$ dimensions where $d$ is the number of the sub-carriers sensed by the UAV). Therefore at the first iteration the UAV performs $M$ predictions, where $M$ is the total number of OFDM symbols received by the UAV during the total flight time. Hence, the complexity of the first iteration is $O(M)$. In addition, since the UAV during the first iteration is detecting abnormalities all the time, it needs to memorize all the generalized errors. After that the UAV will perform clustering of these errors and learn the first generative dynamic model offline. Thus it will not affect the complexity and power consumption. In the successive iterations (refer to Fig. 4.1 in the manuscript) the UAV can switch between the dynamic models stored in its brain. So, by facing a new experience the UAV performs $m$ predictions ($m$ is the number of models stored in the radio's brain) at the discrete level and $N$ predictions at the continuous level each time instant $t$ (or each OFDM symbol). Thus the complexity of the successive iterations is $O(m \times N \times M)$. The complexity of detecting the jammer at two hierarchical levels concerns calculating: the Kullback-Leibler-Divergence ($KLDA$) at the discrete level and the Bhattacharyya distance ($CLA$) at continuous level. The complexity of $KLDA$ is $O(nL \times M)$ where $n$ is the total number of superstates in $\mathbb{S}$ (defined in $(19)$ in the manuscript) and $L$ is the total number of superstates in $S^m$ (defined in $(5)$ in the manuscript). The complexity of $CLA$ is $O(2d \times M)$ where $d$ is the number of the sensed sub-carriers. It is worth noting that the predictions and generation of abnormalities must be done before the time of arrival, i.e. if the UAV receives a set of commands each $50ms$ it must perform the predictions and determines whether or not a jammer is attacking within this time. So, the execution of this process and how fast it is depends on the processors on-board the UAV.

Table 4.1 Execution Phase of each functionality of the proposed SA module

| Functionality | Procedures | Execution Phase |
|---|---|---|
| Radio Environment Perception | predictions | during flight (online) |
| Jammer detection | binary classification | during flight (online) |
| Jammer Characterization | generalized errors calculation and storage | during flight (online) |
| Jammer Characterization | discovering the jammer's dynamic rules | after flight (offline) |
| Incremental Learning | update levels and storage | after flight (offline) |
| Internal Action | jammer suppression | during flight (online) |

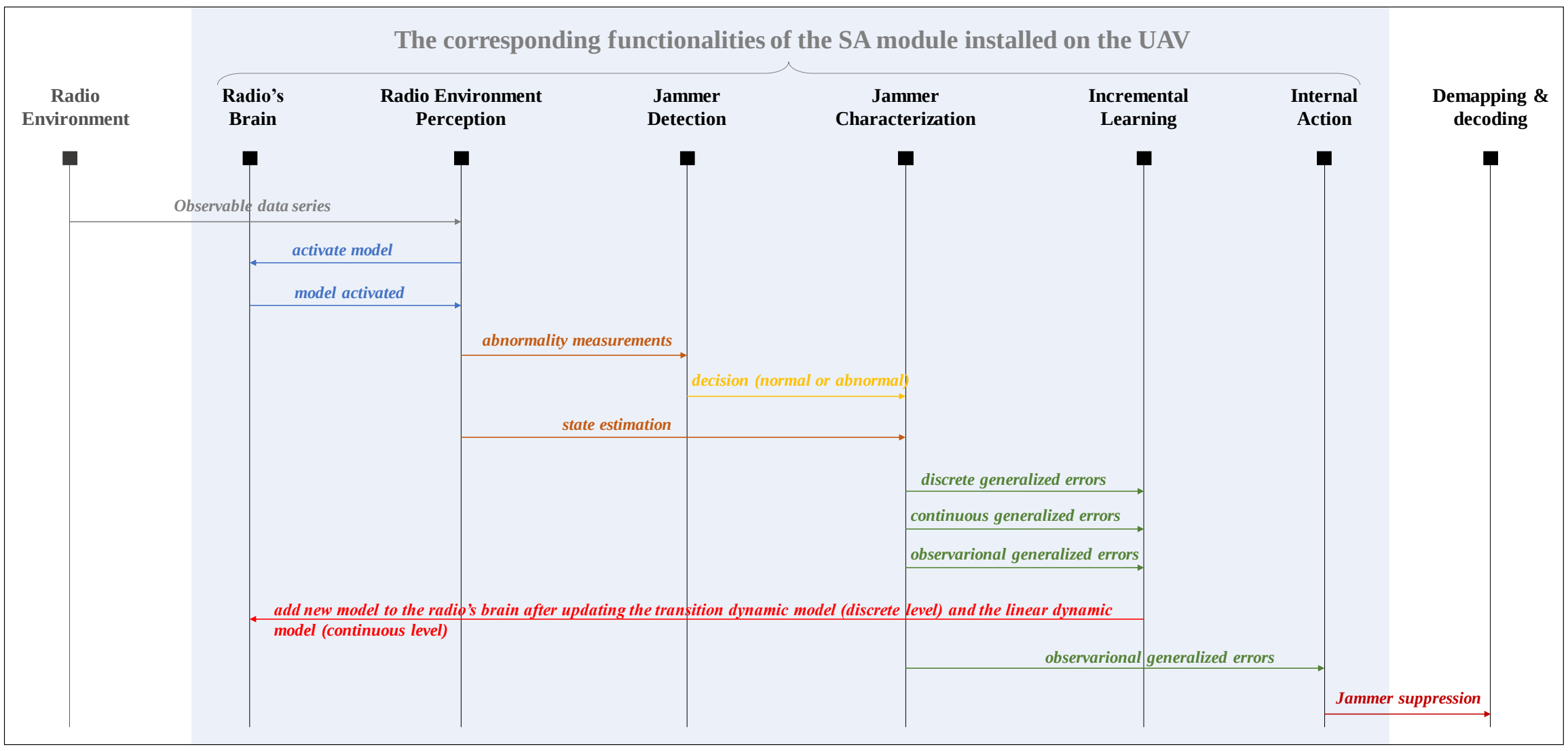

Fig. 4.7 Illustration of the signalling involved in the SA module.

The proposed SA module contains jammer characterization and incremental learning functionalities which allow the UAV to learn and update the knowledge *offline*. By doing so, the complexity can be reduced significantly. Moreover, in the proposed framework, the UAV is not transmitting or exchanging any signal with other entity in the network, which usually has a significant impact on the energy consumption and can impose an additional burden on the UAV.

Table 4.2 Complexity analysis of the proposed SA module

| **Functionality** | **Complexity Order** | **Workload Phase** |
|---|---|---|
| Radio Environment Perception (First iteration) | $O(M)$ | during flight (online) |
| Radio Environment Perception (Successive iterations) | $O(m \times N \times M)$ | during flight (online) |
| Jammer Detection ($D_B$) | $O(2d \times M)$ | during flight (online) |
| Jammer Detection ($KLD$) | $O(nL \times M)$ | during flight (online) |
| Jammer Characterization (prediction error calculation) | $O(M)$ | during flight (online) |
| Internal Action | $O(M)$ | during flight (online) |

## 4.4 Experimental Results

We conduct an extensive Monte Carlo simulation to evaluate the performance of the proposed framework using simulated data. Firstly, the trajectory of a quadcopter UAV is simulated based on [257]. A relationship between the commands and velocities of the UAV at different angles (Pitch, Yaw and Roll) is studied to generate the appropriate bits for simulating the LTE signal. Similarly, the altered trajectory is also extracted from the jammed signal. The LTE signal is generated according to the 3GPP specifications [258], and the important parameters are defined in Table 4.3. The flight time of the UAV is $T_{flight} = 30s$ consisting of 600 samples due to the fact that the position is measured by the GPS every 50 ms. In addition, the UAV receives a PRB every 50ms and extracts the RV that contains a set of commands sent over 9 consecutive sub-carriers in 1 OFDM symbol. Thus, during the $T_{flight}$ the UAV will receive 600 sets of commands, corresponding to 600 OFDM symbols in time domain (Fig. 4.6). Each received set of commands indicate how the UAV will move in the 3D space. The normal signal and the jammer are QPSK modulated. The output of the QPSK modulator for both is normalized based on the average power. The normal signal has average power $P_S = 1W$. While the average power of the jammer is $P_J$.
Four different situations are considered, the first one is related to the **Reference Situation**

Table 4.3 LTE Simulation Parameters

| Parameter | Value |
|---|---|
| BW | 1.4 MHz |
| Duplex mode | FDD |
| $\Delta f$ | 15 kHz |
| Number of PRBs per BW | 6 |
| Sampling frequency | 1.92 Mhz |
| $N_{FFT}$ | 128 |
| OFDM symbols per slot | 7 |
| CP length | normal |
| SNR | 15 dB |
| Modulation | QPSK |
| Channel | AWGN |
| Total Radio Frames | 600 |

representing a normal behaviour of the signal that carries original commands sent by the operator as shown in Fig. 4.8-a. The corresponding UAV trajectory is depicted in Fig. 4.10-a. The remaining situations are concerning the smart jammer behaviours in attacking the commands who has an average power $P_J = 1W$, and they are listed as follows: **Situation 1**, the jammer is attacking consecutively starting from time (in terms of OFDM symbols) $t = 200$ till $t = 400$ as shown in Fig. 4.8-b, where the altered UAV trajectory during the jamming attacks is shown in Fig. 4.10-b. In **Situation 2**, the jammer behaves in a dynamic fashion by attacking from $t = 1$ till $t = 200$, and from $t = 400$ till $t = 600$ as shown in Fig. 4.9-a, while, Fig. 4.9-b (**Situation 3**) illustrates a faster dynamic behaviours of the jammer. UAV trajectories during Situation 2 and Situation 3 are depicted in Fig. 4.10-d and Fig. 4.10-e.

Initially, the radio (UAV) does not have any knowledge (null memory) about the surrounding environment. Thus, at the beginning stage which is the *first iteration* exhibited in Fig. 4.1, the UAV predicts the future states of the spectrum supposing that the signals' states are static and do not change with time by employing UKF. Such an assumption leads to high abnormalities all the flight time since the UAV fails to predict the real states of the signal, as shown in Fig. 4.11-a. Accordingly, based on these predictions and by using the innovations (derivatives) produced by the UKF, the UAV will form and store the generalized errors. Then it will perform an unsupervised clustering method (the GNG algorithm) in an offline manner to learn and memorize the first generative model which represent the dynamics of the received commands during the normal situation. After that, the UAV is capable to predict the future states of the commands at multiple sub-carriers. This can be verified by calculating the abnormality signal during the normal situation. If the abnormality is quasi-zero, i.e. the

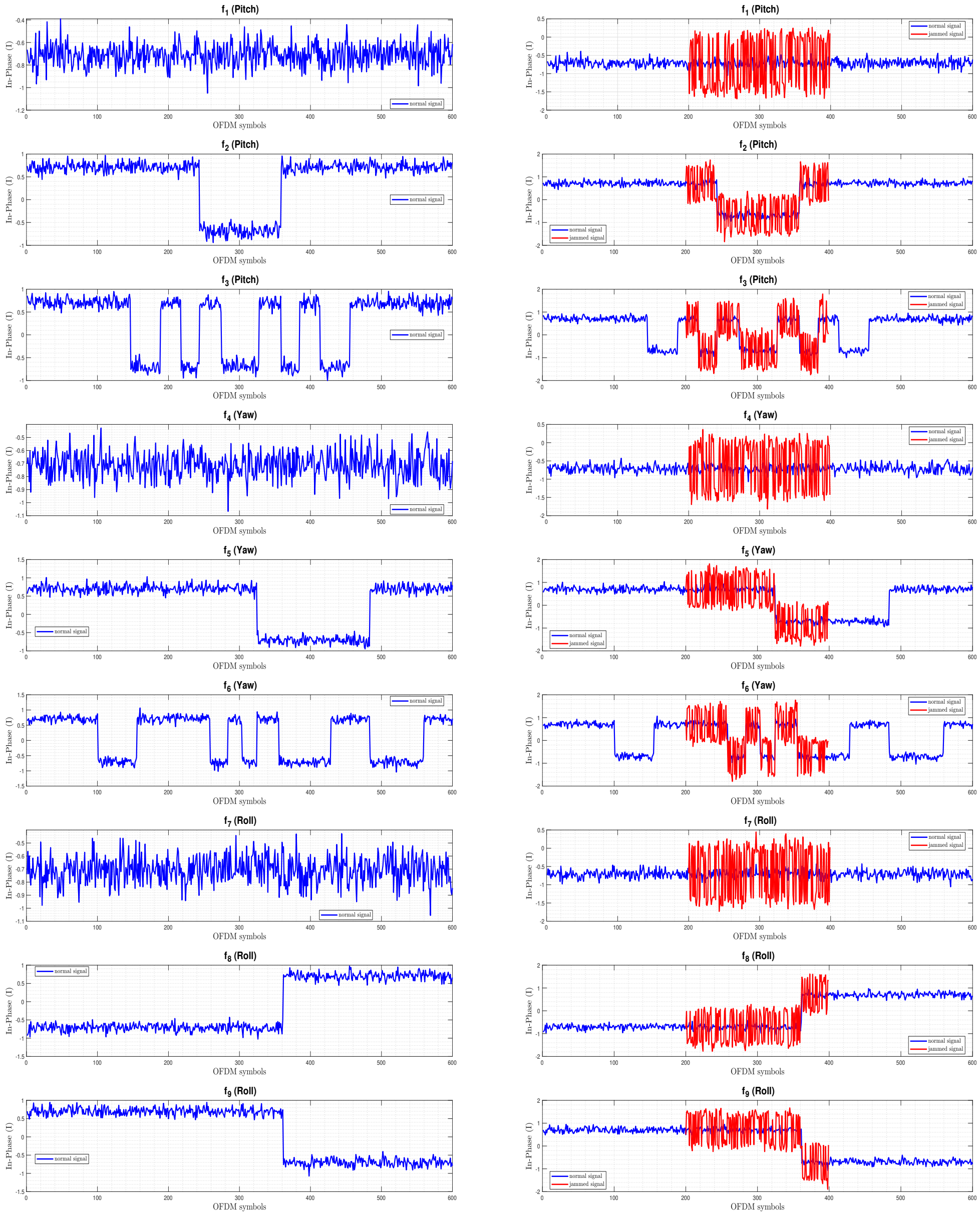

(a) Reference Situation (b) Situation 1

Fig. 4.8 Received Commands at multiple sub-carriers at different situations.

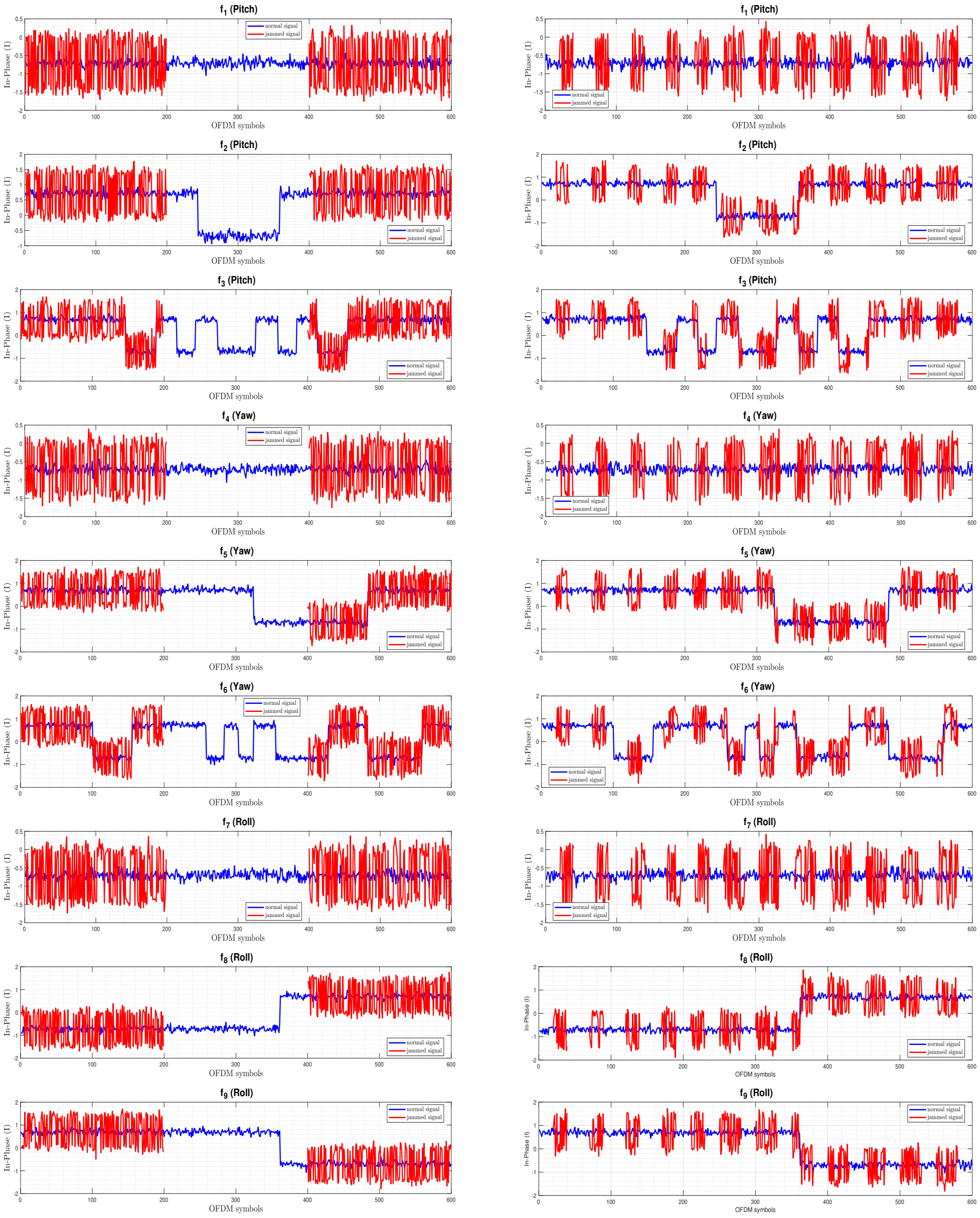

(a) Situation 2 (b) Situation 3

Fig. 4.9 Received Commands at multiple sub-carriers at different situations.

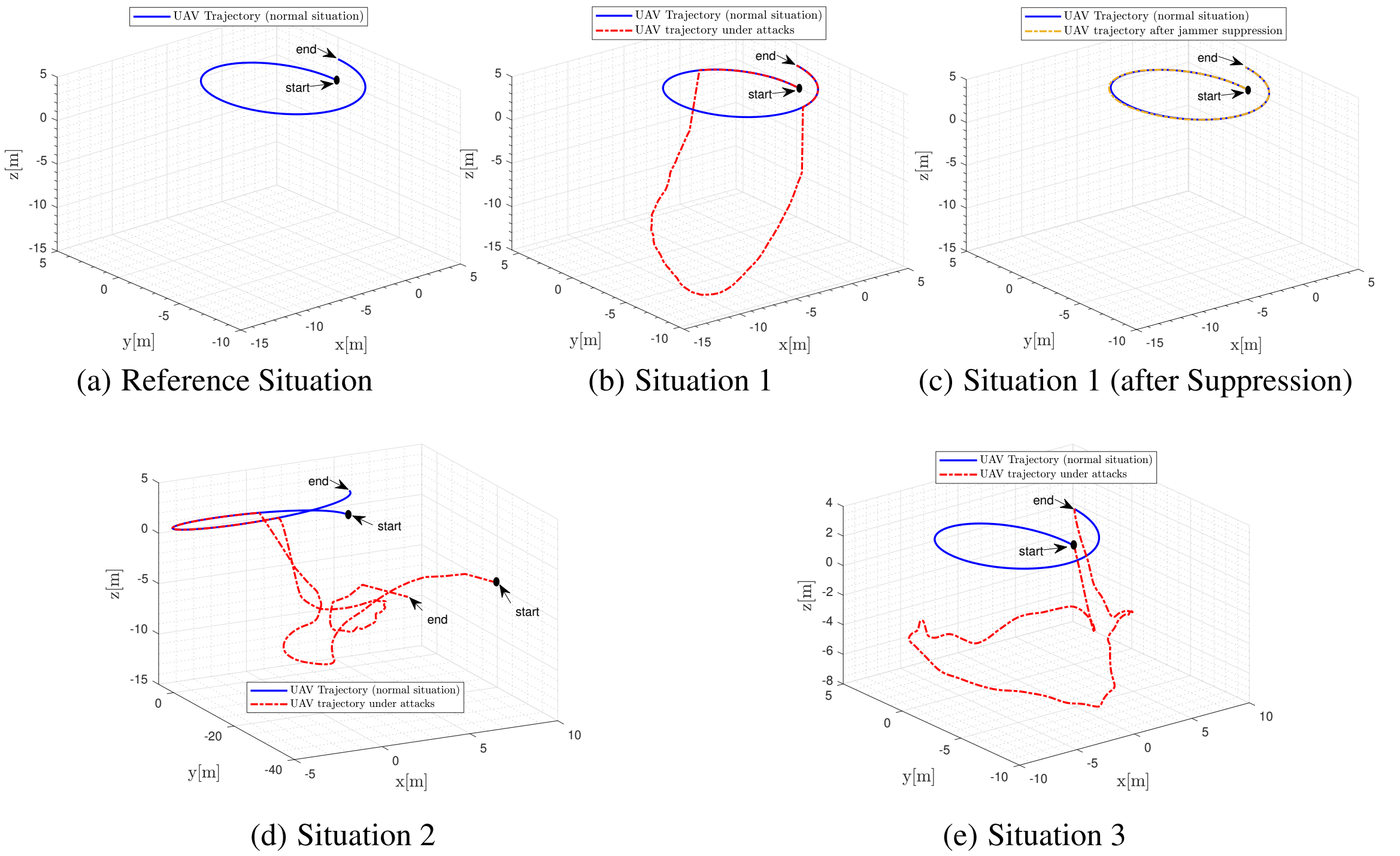

(a) Reference Situation (b) Situation 1 (c) Situation 1 (after Suppression)

(d) Situation 2 (e) Situation 3

Fig. 4.10 UAV Trajectories.

learned model succeeded to capture the dynamic rules of the signal and allowed the UAV to perform correct predictions as shown in Fig. 4.11-b where the abnormality at the continuous level defined in (4.25) is showed. Testing new observations $Z_t$ and predicting eventually, could follow the same rules with which the dynamic model has been learned from previous experience (reference situation) when the jammer was absent or could deviate due to the new rules caused by the jammer. Thus, in Situation 1, the UAV can detect any attack while predicting the future states of the commands and receiving the observations as shown in Fig. 4.12-a-b. As well as extracting the jammer's signal (Fig. 4.13-a) and act accordingly by mitigating its effect on the received commands (Fig. 4.13-b) which leads to auto-correction of the altered trajectory in Fig. 4.10-c.

To evaluate the performance of suppression fairly among all the conducted experiments, we calculated the mean square error (MSE) between the estimated jamming signals and the original ones (depicted in Fig. 4.14(a)) and between the suppressed command signals and the original commands (depicted in Fig. 4.14(b)) over different Jamming-to-Signal-Ratios (JSRs). We can observe that the proposed approach can estimate the jammer and mitigate its effect efficiently, which is verified by the low MSE values shown in Fig. 4.14. In addition, from the figure, we can see that the MSE decreases as the JSR increases, meaning that jammer attacking with higher powers can be better estimated than jammers attacking with

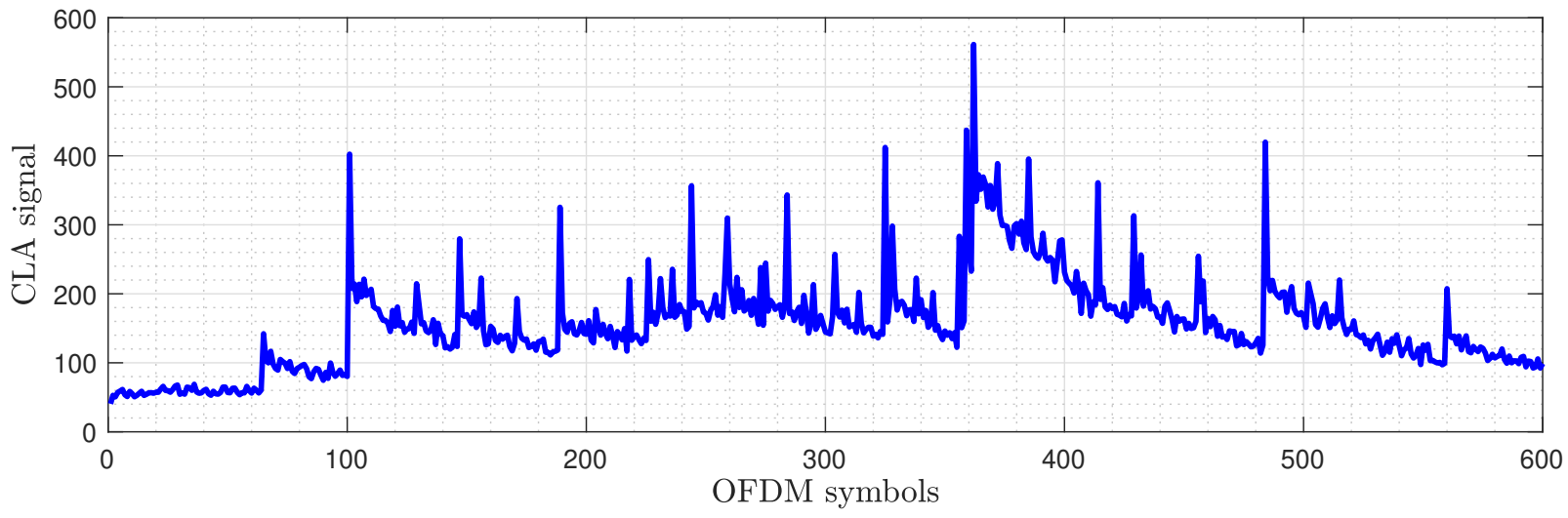

(a) UAV with empty brain (null memory)

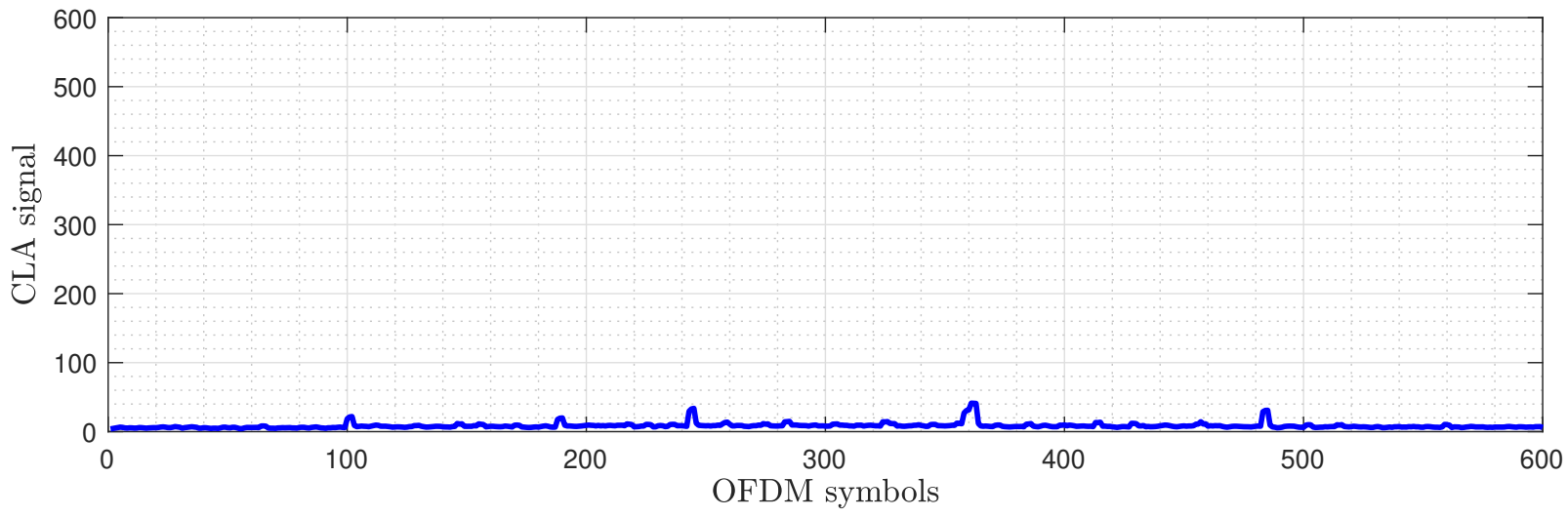

(b) After learning and stored the first dynamic model

Fig. 4.11 Abnormality Measurements at the continuous level.

lower powers. Then, accurate estimations of the jammer yield accurate suppressions of its effect on the received commands.

After this situation (i.e., Situation 1), the UAV can study the new behaviour (detected jammer) and learn incrementally a new dynamic model which represents the interaction of the jammer and the normal signal (commands). Facing a new situation (Situation 2), where the same jammer (detected before) is attacking the commands allow the radio to recognize it and predict its future activity inside the radio spectrum since it has already learned the rules of attacking. In this situation the UAV's memory contains two dynamic models: the UAV will *i)* switch between these models; *ii)* and select the best one that fits the observation. Fig. 4.12-c-d shows the abnormality measurements related to the reference model (learned during the reference situation) and that related to multiple models (reference model and the one learned incrementally that represents the situation where the jammer is ON). As expected the abnormality at both levels is decreased, which means that the UAV succeeded to learn the dynamic rules of the jammer in attacking the commands (as discussed in section 4.2.5) after characterizing the attacks at multiple level (discussed in section 4.2.4). Switching among multiple models means that once the UAV detects the jammer, it switches to the model encoding the interaction between the jammer and UAV, allowing to predict the expected behaviour of the jammer in attacking the commands. Therefore, those predictions explain the received stimuli and lead to a decrease in the abnormality. Additionally, another

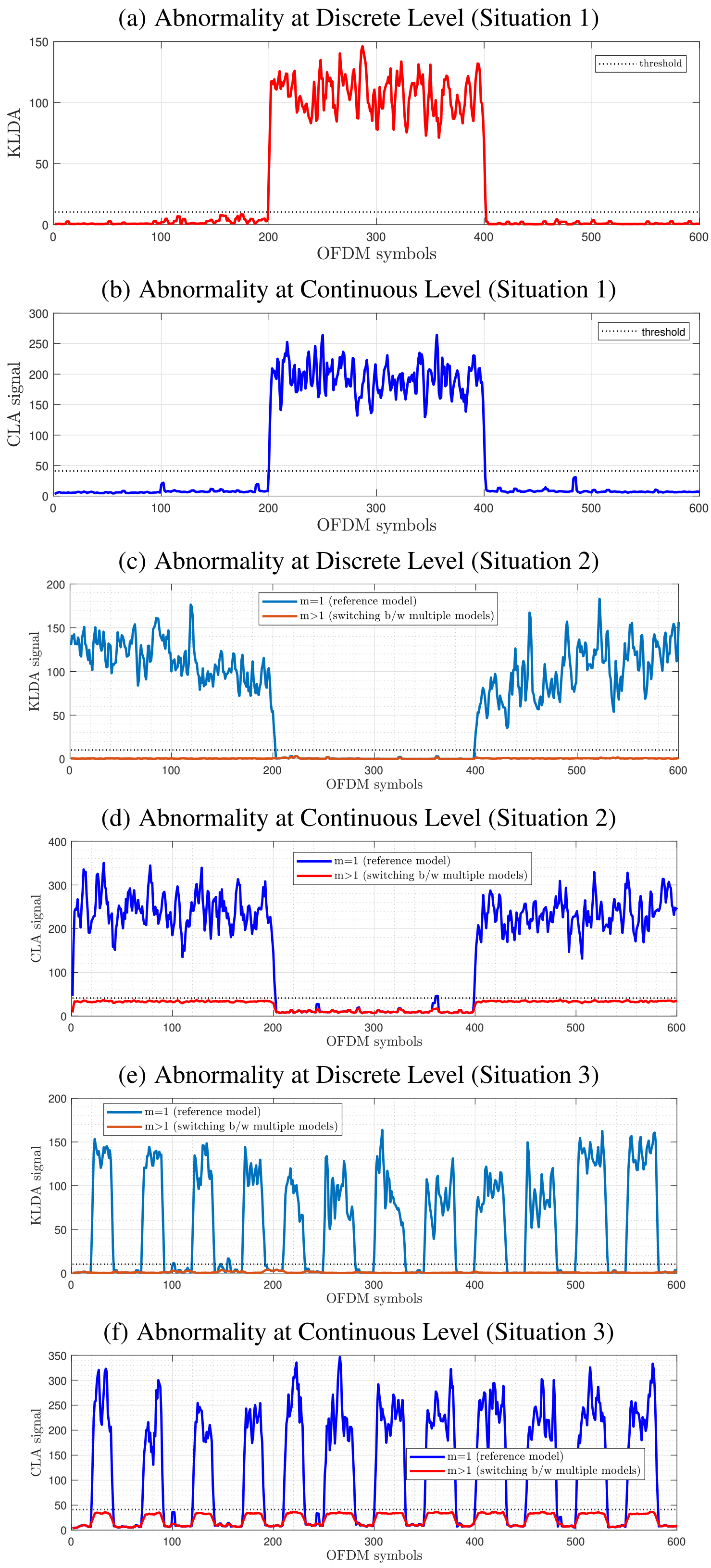

Fig. 4.12 Jammer Detection at hierarchical levels during different situations.

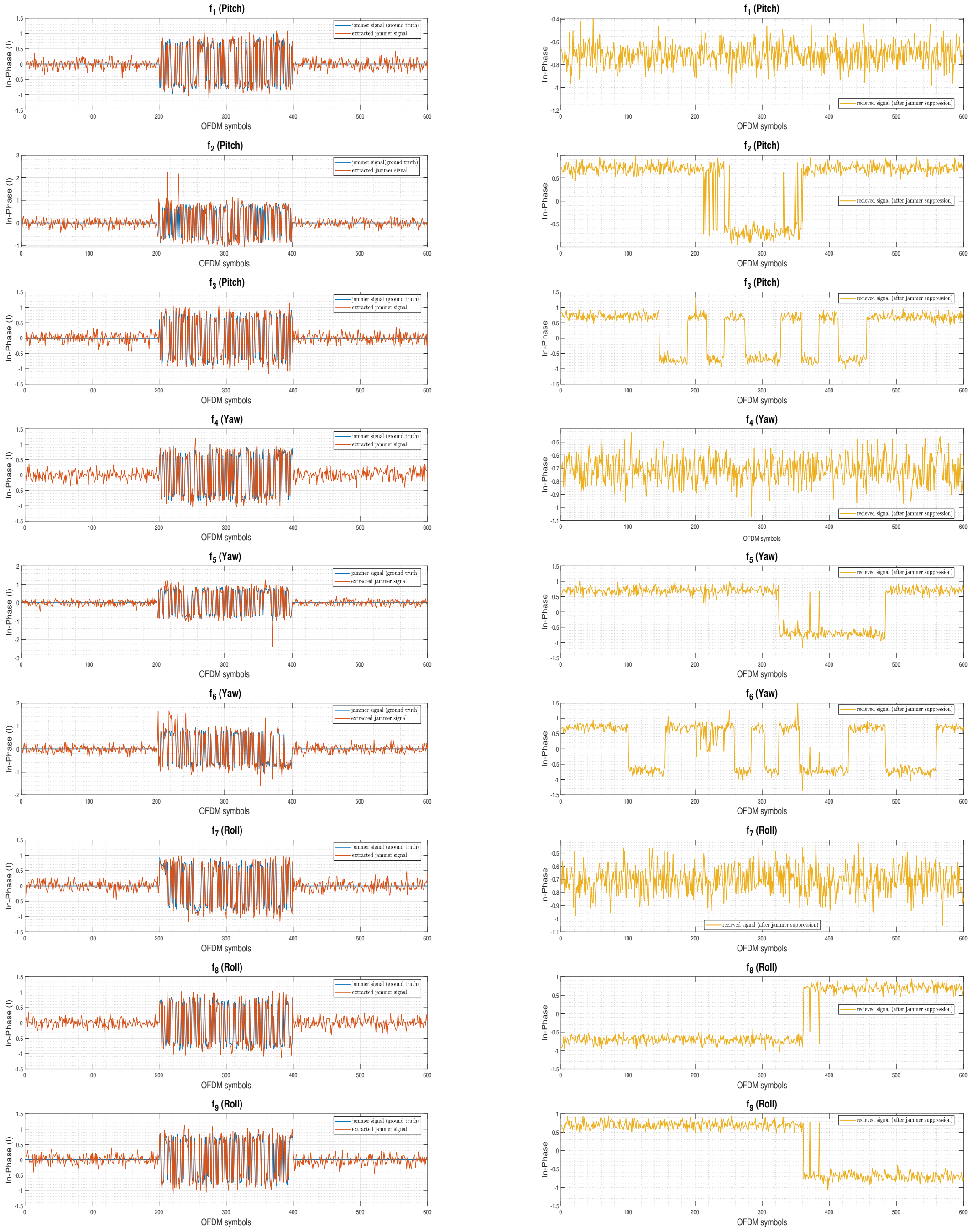

(a) Jammer's signal Extraction (b) Received signal after jammer suppression

Fig. 4.13 Jammer extraction and jammer suppresion after characterizing the jamming attacks.

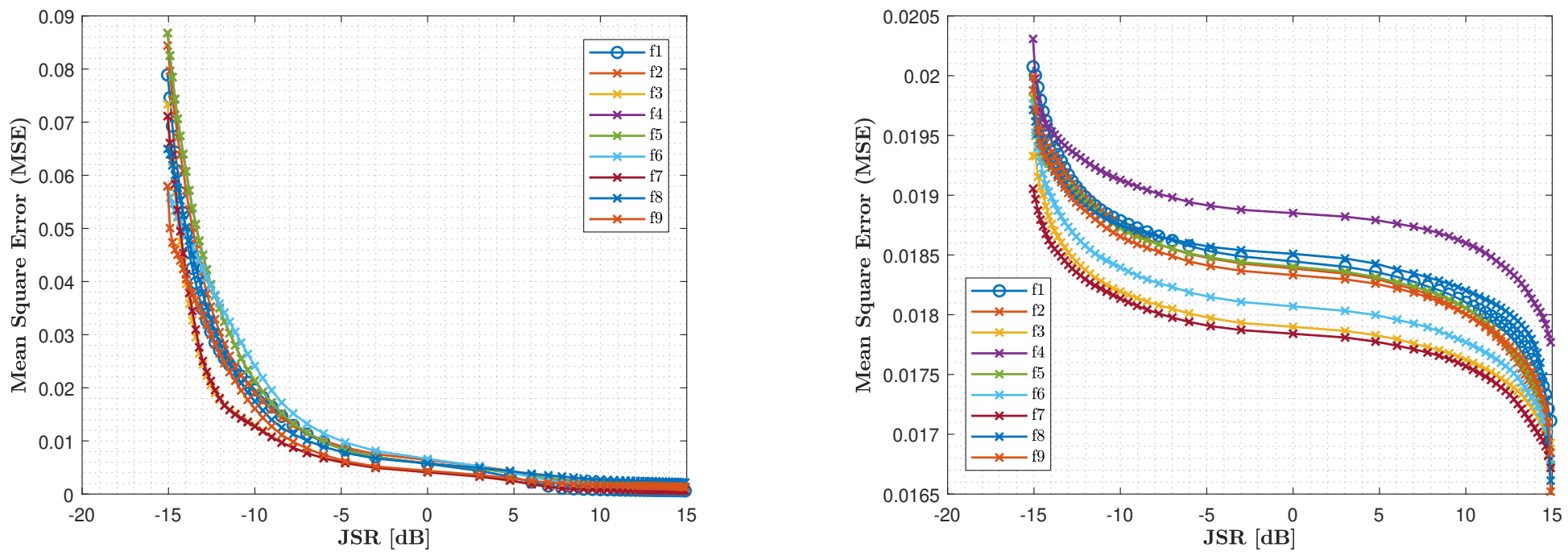

(a) (b)

Fig. 4.14 Mean Square Error (MSE) of: (a) the estimated jamming signals versus JSR and (b) the suppressed command signals versus JSR.

new experience (Situation 3) has been tested by considering fast dynamics of attacking the commands by the same jammer. Fig. 4.12-e-f, confirms that the UAV has been succeeded in characterizing the jammer in question and learn its model incrementally, and then to predict its effect in future situations. Accordingly, Fig. 4.12(c)(d)(e)(f) validate the discussion done in sections 4.2.4 and 4.2.5.

Further experiments are tested to validate the proposed approach, by varying the Jamming-to-Signal-Ratio (JSR) from $-15dB$ to $+15dB$. In all these experiments the jammer attacks dynamically all the 9 sub-carriers in frequency domain and certain OFDM symbols in time domain. Particularly, from $t = 1$ till $t = 150$, from $t = 200$ till $t = 350$, and from $t = 400$ till $t = 550$. In order to evaluate the performance of the proposed framework in detecting the jamming attacks referring to the reference model learned by the UAV, we used a range of confidence thresholds to build the corresponding ROC curves illustrated in Fig. 4.23-a along with the Area Under Curve (AUC) (Fig. 4.23-b) and Accuracy (ACC) (Fig. 4.23-c). The ROC curves show that the MJPF filtering on the reference model provides high detection probability ($P_d$) at both levels even when the jammer attacks with very low power. For example, with JSR = -5 dB, the probability of detection is almost 100% for both KLDA and CLA measurements. The high detection probabilities can be explained by the fact that the predictions performed at the higher level of the HDBN using PF were precise and accurate almost of the time considered in the analysis. The probability of detecting the jammer at the discrete level is calculated by using: $P_d = Pr\big\{KLDA > (1-\zeta)\alpha\big\}$. We estimated the acceptance ratio ($\zeta$) as $\zeta = 0.8$ after observing the predicted superstates and the observed ones. This also implies that the learned model is accurate in predicting the discrete variables as can be observed from Fig. 4.12-a where the KLDA signal is quasi-zero (i.e. observation matches the prediction) in the time instants when the jammer is OFF. Directly estimating $\alpha$ is difficult here since the UAV does not has any prior knowledge about the jammer. However, it can be estimated by supposing that half of the probability mass function of the observation ($\lambda(\tilde{S}_t)$) falls in the normality region and the other half in the abnormality region (the concept of $\alpha$ is discussed in Section 4.2.31). Thus, we set $\alpha = 0.5$. In this case, the threshold is adapted to different situations. On the other hand, the probability of detecting the jammer at the continuous level is calculated by using: $P_d = Pr\left\{ CLA > \sqrt{Q_{\mathfrak{T}}(\tilde{X}_t)\ .\ [\beta_1(1-\alpha)] + Q_{\mathfrak{T}}(\tilde{X}_t)\ .\ [\beta_2(1-\alpha)]} \right\}$. In the numerical results, we chose $Q_{\mathfrak{T}} = 0.2$ which is estimated after observing the $CLA$ signal during normal situation shown in Fig. 4.11-b where the abnormal signal is quasi-0 implying that the learned generative model is accurate to predict continuous variables. Also here $\alpha$ is the area of $\lambda(\tilde{X}_t)$ falling inside the normal regions as discussed in Section 4.2.32 and it is difficult to estimate it as claimed before so we set $\alpha = 0.5$ while $\beta_1 = 1$ and $\beta_2 = 0$. In addition, a comparison with the traditional Energy detector (ED) is provided as shown in Fig. 4.23 that shows how the proposed approach beats the Energy detector (ED) in detecting the jammer at different JSRs.

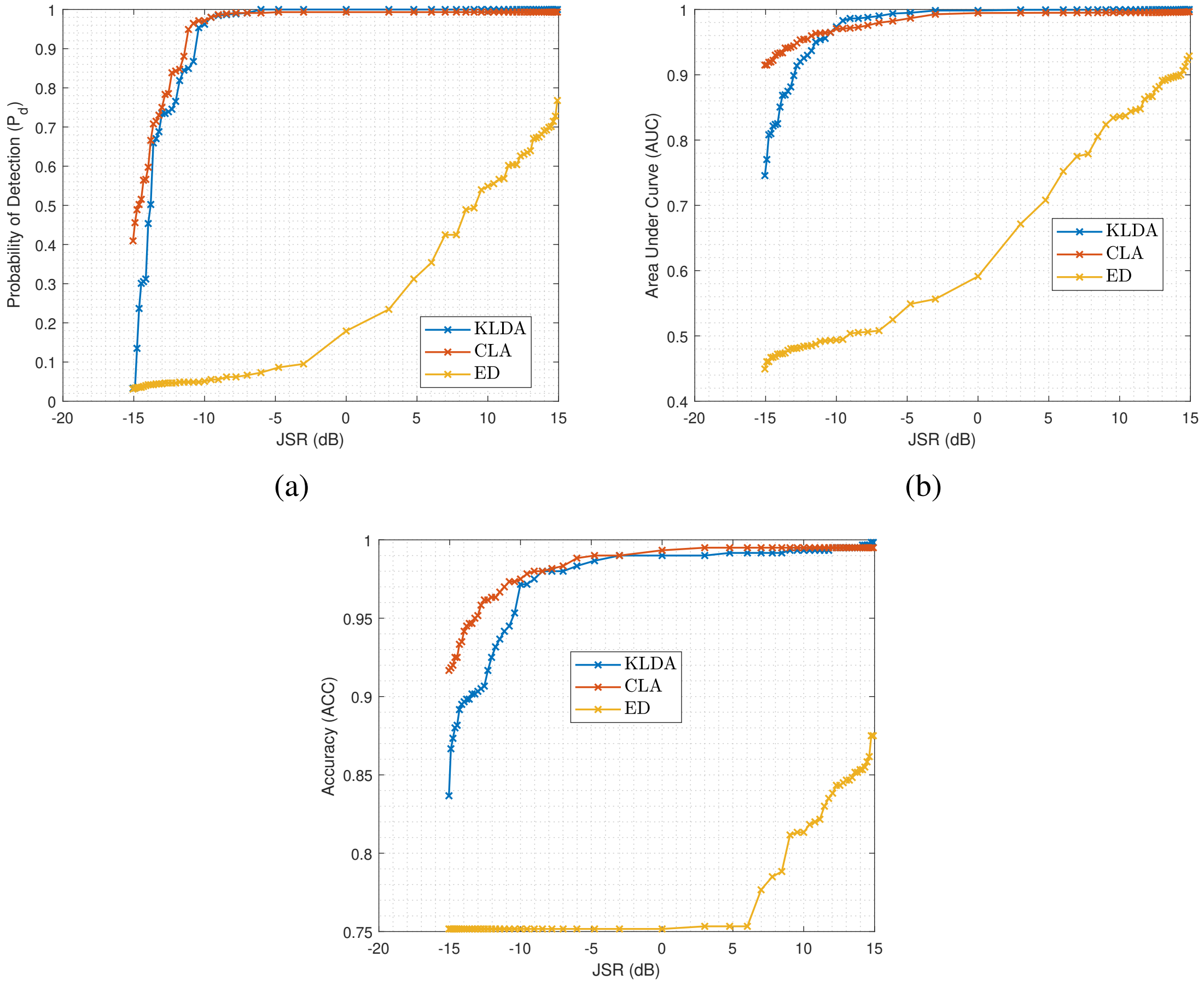

(c)

Fig. 4.15 Performance comparison between the SA module (KLDA, CLA) and Energy detector (ED): ROC curves (a) and the corresponding AUC (b) and ACC (c).

The characterization of the jammer is performed by the UAV during the first interval (from 1 to 150 OFDM symbols) and the successive periods are used to validate if the UAV succeeded to capture the dynamics of such attacks. This capability can be verified by calculating the Root Mean Square Error (RMSE) of the abnormality measurements (KLDA and CLA) and comparing these errors in two cases. The first case ($m = 1$) is when the UAV relay on the reference model (the first dynamic model) while the second case ($m > 1$) is based on switching between the reference model and the one learned incrementally which encodes the dynamic rules of the jammer in question. RMSE is the difference between prediction and evidence realizing the prediction accuracy, high RMSE means that the evidence does not match the prediction while low RMSE stands for the fact that prediction matches the evidence. Also, RMSE depends on the abnormality level which increases as the jammer's power increase as shown in Fig. 4.16 (blue curves). In the second case, by switching between two models the UAV can predict the jammer's effect in future OFDM symbols and cause a decrement of the RMSE respectively. This can be seen in Fig. 4.16 (red plots) where the RMSE of KLDA (Fig. 4.16-a) and RMSE of CLA (Fig. 4.16-b) converge and are somehow stable and decreased with respect to the RMSE related to the reference model. This implies that the abnormality level is stable too since the UAV is predicting correctly the jammer's activity and not detecting unexpected behaviours any more (not surprised any more).

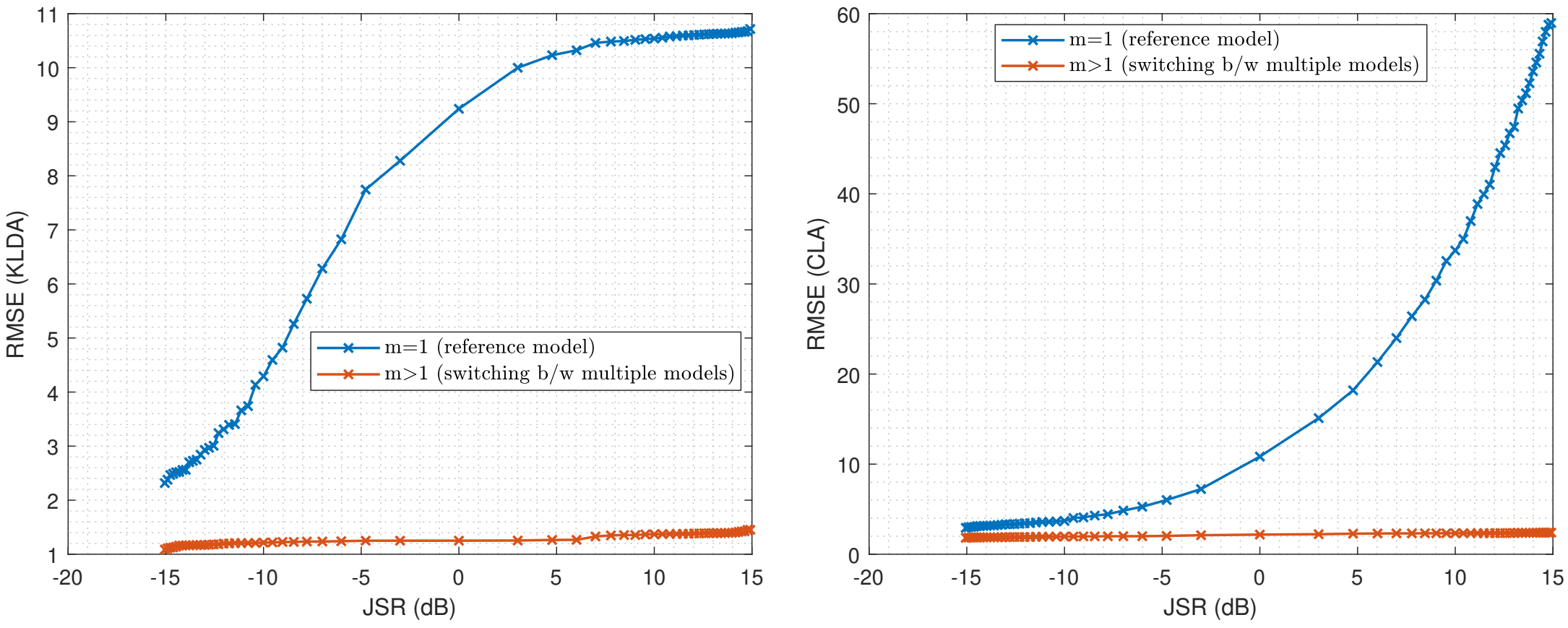

(a) (b)

Fig. 4.16 RMSE comparison versus JSR by using only the reference model (blue curves) and switching between two models (red curves). (a) RMSEs of KLDA. (b) RMSEs of CLA.

### 4.4.1 Test different jammers with different modulations

This section will answer the following question: Does the change of jammer's modulation influence the framework performance? In fact, the jammer's modulation scheme will not influence too much the performance of the proposed framework in detecting the attacks since the reference dynamic model is learned without any prior knowledge about the jammer. The jammer will change the signal's distribution in different ways depending on the strategy it is following in attacking the signal and the modulation it is using. For example, a QPSK jammer will shift a specific data sample in 4 directions while a 32-PSK jammer will shift it in 32 directions. But in any case, the UAV using the SA module can detect both of the jammers regardless their modulation since the presented jammer detection methods at different levels measure the difference between the observation and the prediction without taking into account the direction of the shift that the jammer caused. It is to note that the shift is studied at the discrete level to understand how the jammer affected the discrete level of the reference dynamic model and the energy of such shift at the continuous level to understand the jammer's power that caused such shift. However, analysing the direction of the shifts caused by the jammer at the continuous level is essential for the jammer classification (or modulation recognition) task which will be investigated in Chapter 5. Nevertheless, different modulation schemes used by the jammer (8-*PSK*, 16-*QAM*, 32-*PSK* and 64-*QAM*) are tested to analyse the performance of the proposed framework in detecting the attacks as shown in Fig. 4.17 and Fig. 4.18.

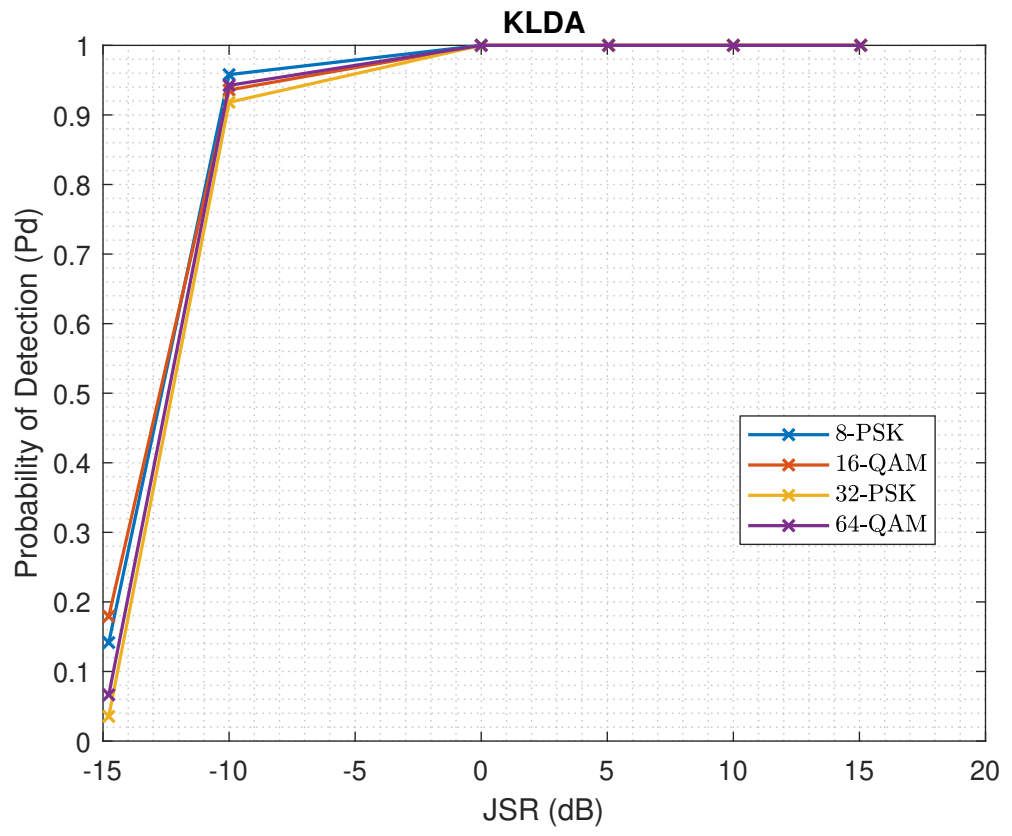

Fig. 4.17 Detection of multiple jammers using different modulation schemes at the discrete level using the KLDA abnormality measurement.

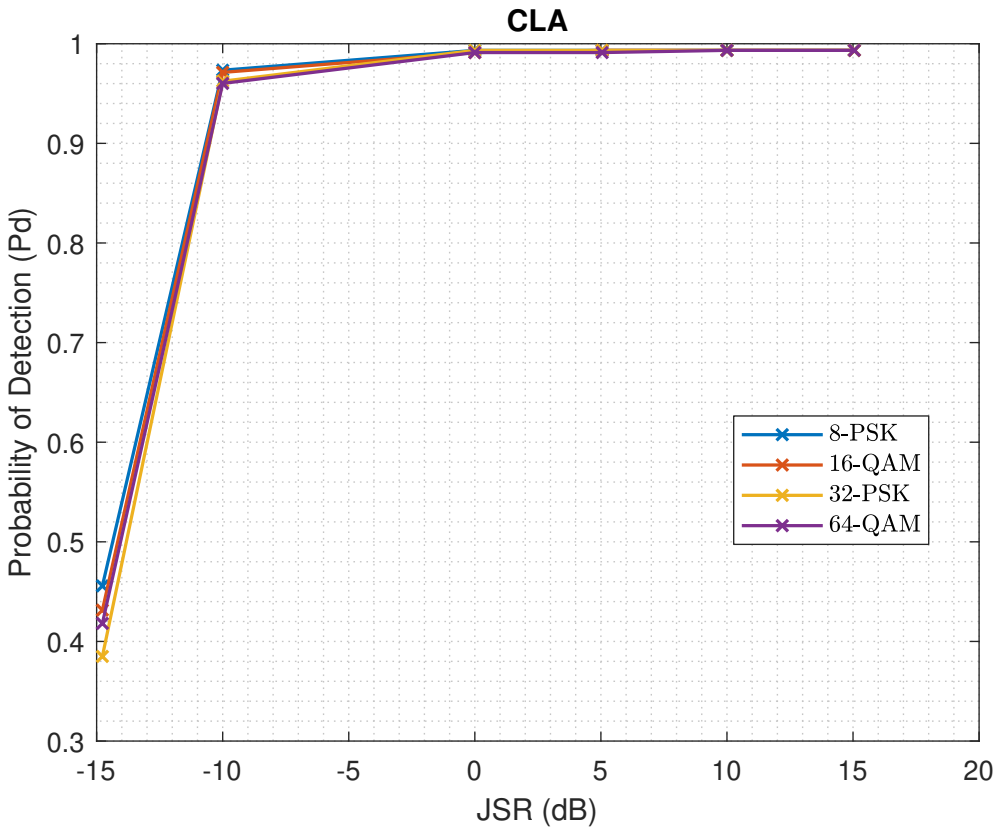

Fig. 4.18 Detection of multiple jammers using different modulation schemes at the continuous level using the CLA abnormality measurement.

### 4.4.2 Deep Jammer detection

In this section we evaluate the UAV's capability in detecting and locating the jammer deeply in both time and frequency domains. First we consider three different situations to show visually the obtained results then we tested additional experiments by varying the JSR.

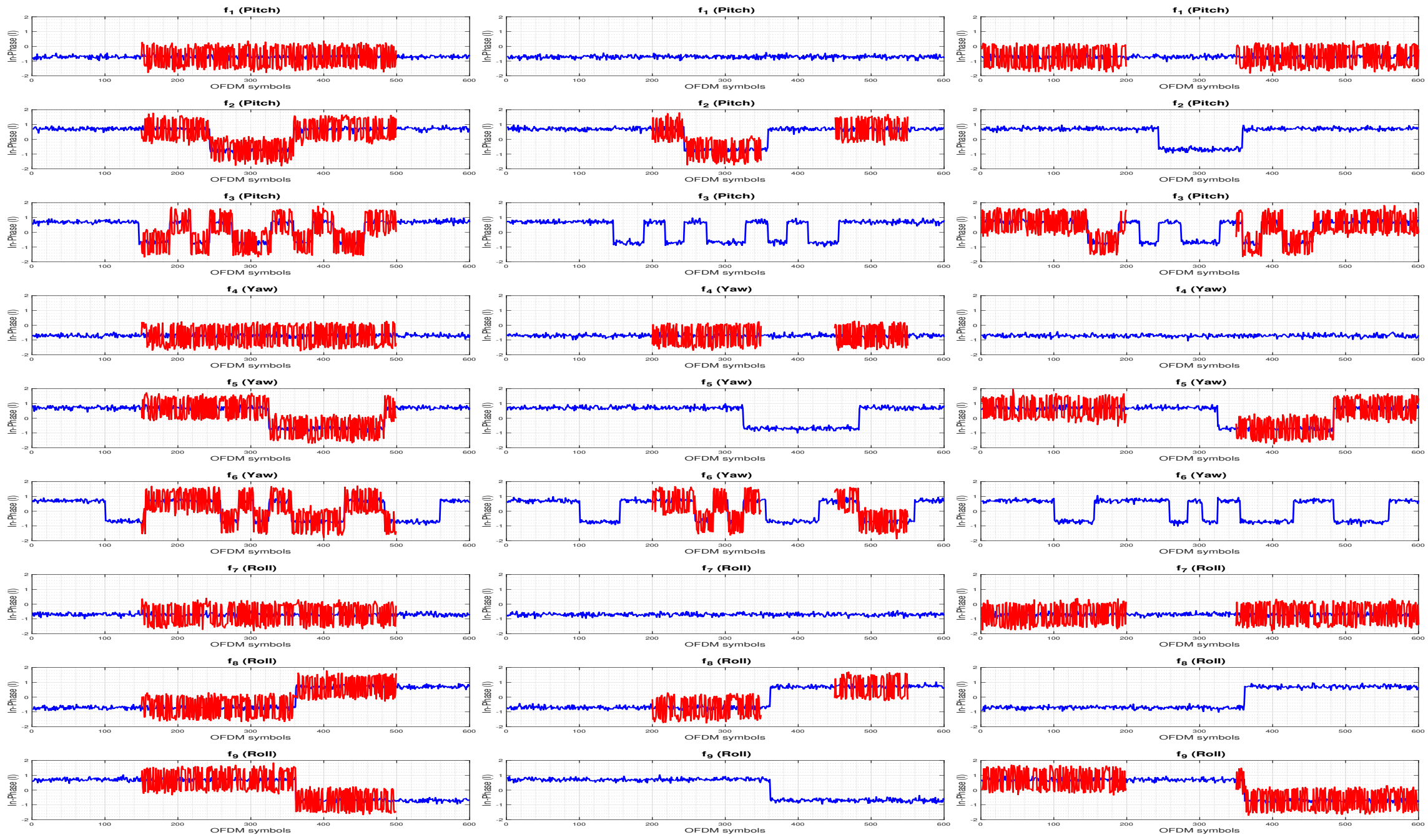

(a) Situation 4 (b) Situation 5 (c) Situation 6

Fig. 4.19 Received Commands during different situations. Blue and red colors represent the normal and jammed signal respectively.

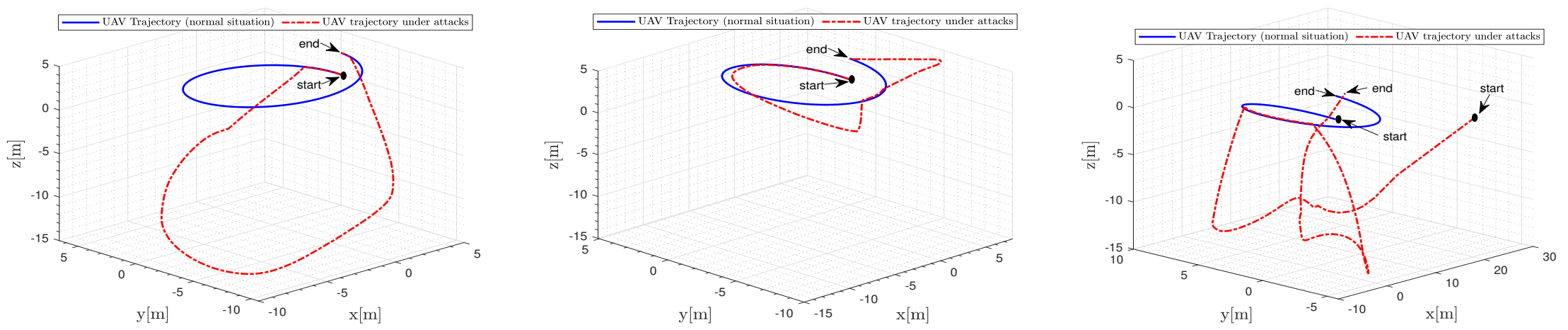

(a) Situation 4 (b) Situation 5 (c) Situation 6

Fig. 4.20 UAV trajectory in different situations. Blue and red colors represent the UAV trajectory without and with jammer attack respectively.

***(a)* Situation 4**: the jammer is attacking continuously all sub-carriers with average power $P_J$=1W starting from time (in terms of OFDM symbols) *t=150* till *t=500* (refer to Fig. 4.19-a). Where the altered trajectory is shown in Fig. 4.20-a.
***(b)* Situation 5**: the same jammer ($P_J$=1W) behaves in different way by attacking from *t=200* till *t=350* and from *t=450* till *t=550* sub-carriers $f_2$, $f_4$, $f_6$, $f_8$ (see Fig. 4.19-b) and the affected trajectory is shown in Fig. 4.20-b.
***(c)* Situation 6**: the jammer changes its strategy by attacking sub-carriers $f_1$, $f_3$, $f_5$, $f_7$, $f_9$ from *t=1* till *t=200* and from *t=350* till *t=600* with $P_J$=1W as shown in Fig. 4.19-c and alters the trajectory as in Fig. 4.20-c.

At each time instant, $t$, the UAV predicts the next set of commands that might receive at multiple sub-carriers based on the learned DBN model. Observing new signals in different radio situations could follow the same rules with which the dynamic model has been learned in previous experience (without jammer) or could deviate due to new rules caused by external force (jammer). UAV can identify any received command online affected by a jammer using the abnormality indicator DCLA provided by the MJPF at the continuous level. At the continuous level, if the probability of having the prediction near the measurement is high, then the DCLA signal is low (below threshold) and vice-versa. At this level, the UAV is able not only to detect the jammer in time domain but also to locate it in frequency domain and knowing exactly which sub-carriers are targeted by the jammer as shown in Fig. 4.21-4.22 (abnormality indicator for each sub-carrier, $DCLA_{f_n}$). Optimized thresholds for distinguishing normal and abnormal data samples are obtained by testing the reference situation (jammer off). The thresholds of DCLA are obtained by calculating the mean value plus standard deviation of DCLA (from Fig. 4.21-a). To validate the proposed approach we conducted and tested further experiments by varying the Jamming-to-Signal-Ratio (JSR) from $-16dB$ to $16dB$. The jammer is following the same strategy in all these experiments by attacking dynamically from $t$ = *1* till $t$ = *150*, from *t=200* till *t=350*, and from *t=400* till *t=550* all the sub-carriers. The proposed method as shown in Fig. 4.23 achieves high

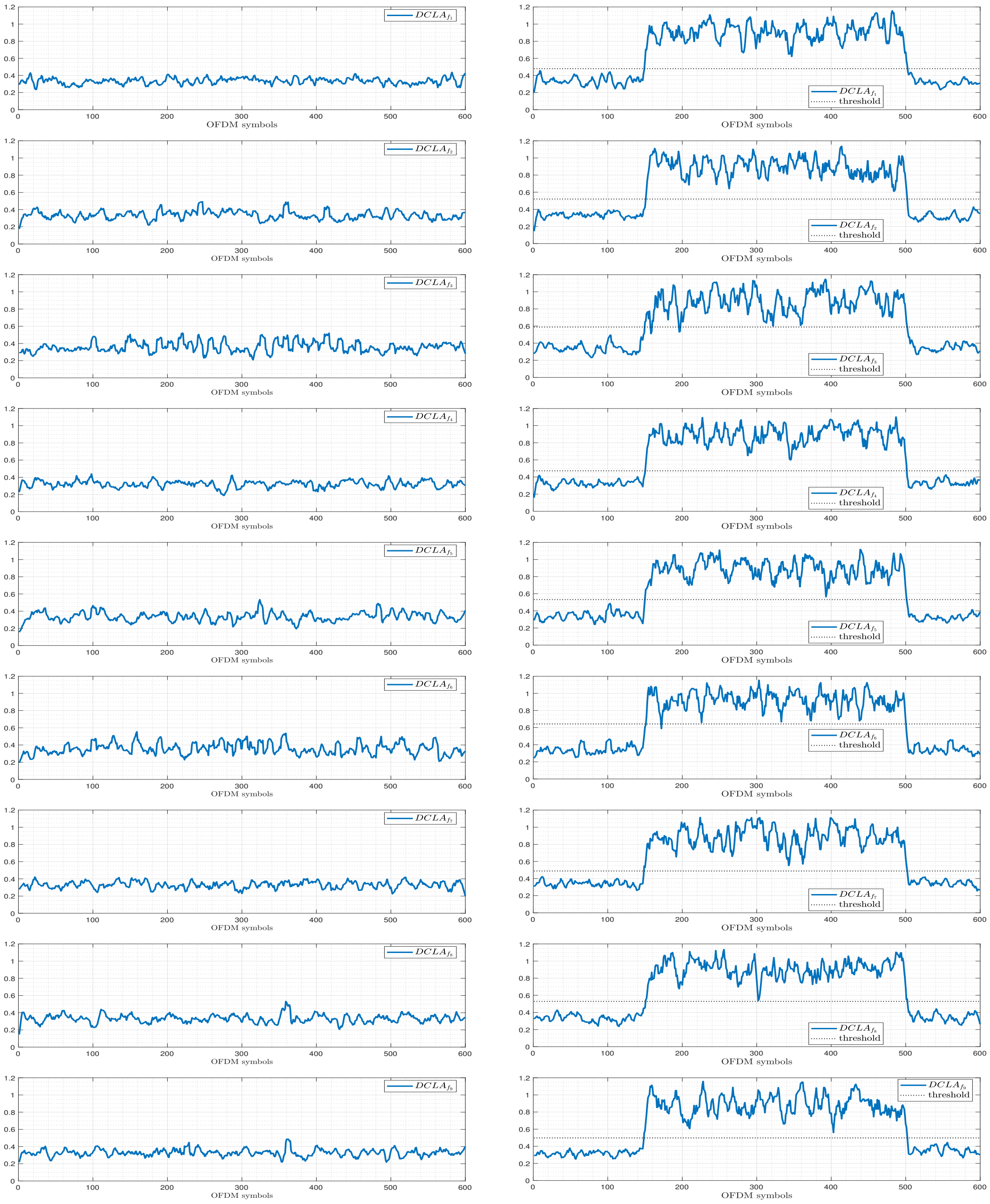

(a) Reference Situation (b) Situation 4

Fig. 4.21 Deep Abnormality at the continuous level during different situations.

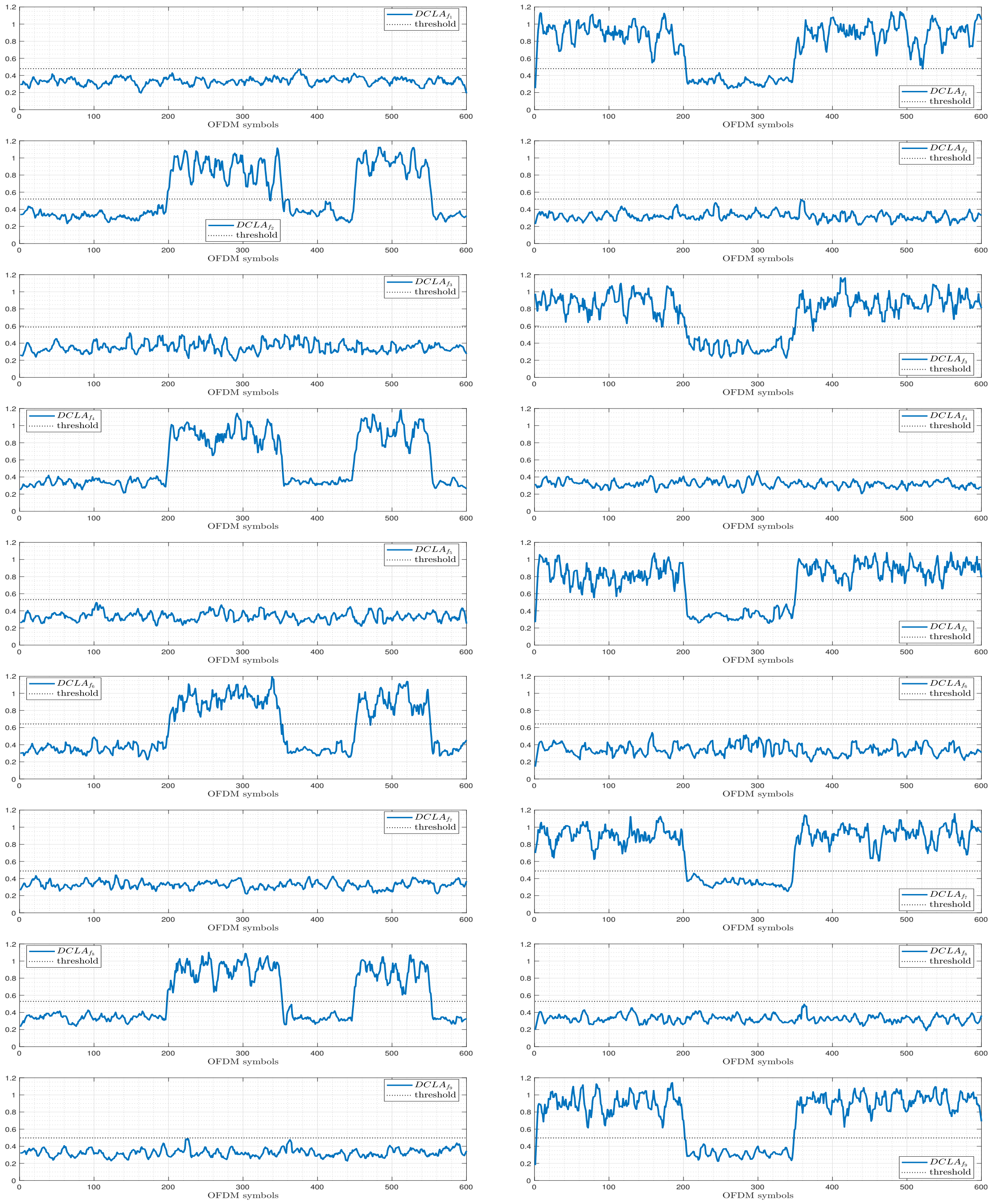

(a) Situation 5

(b) Situation 6

Fig. 4.22 Deep Abnormality at the continuous level during different situations.

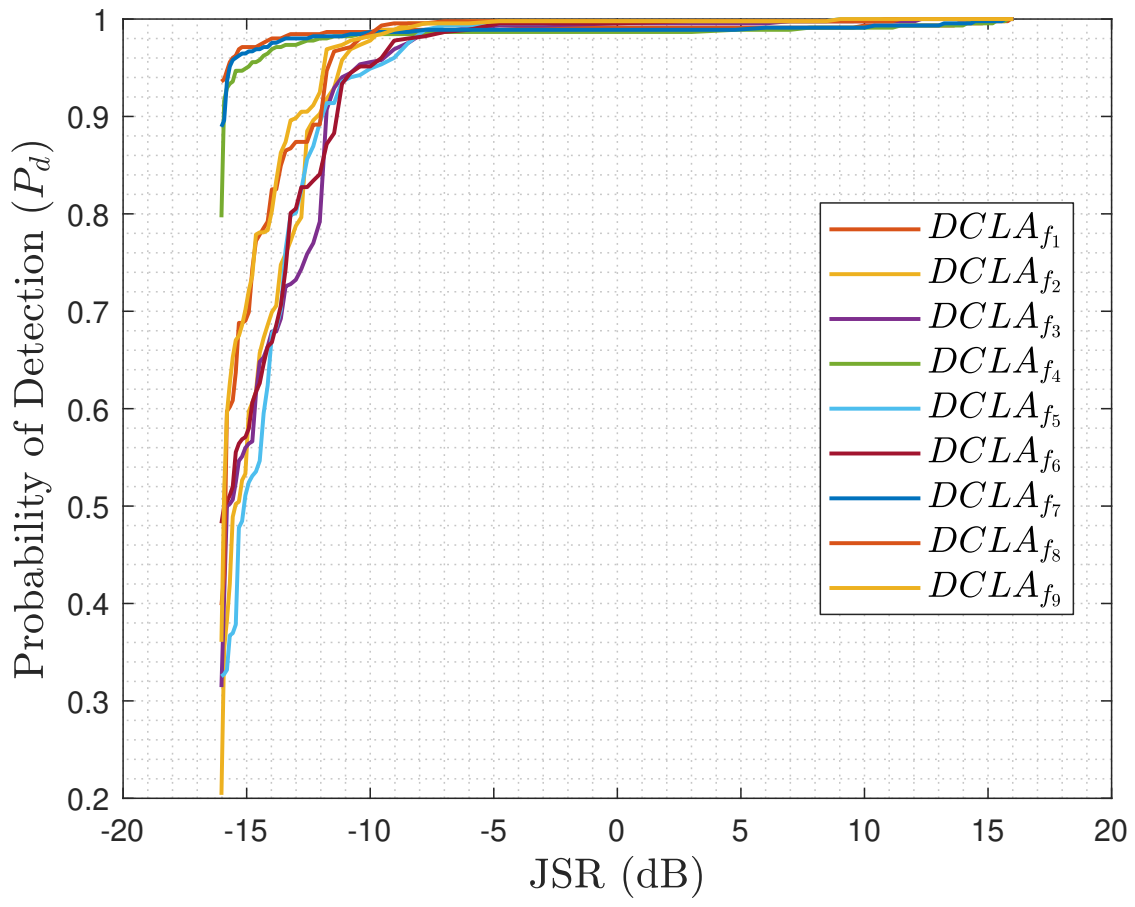

(a) Probability of Detection

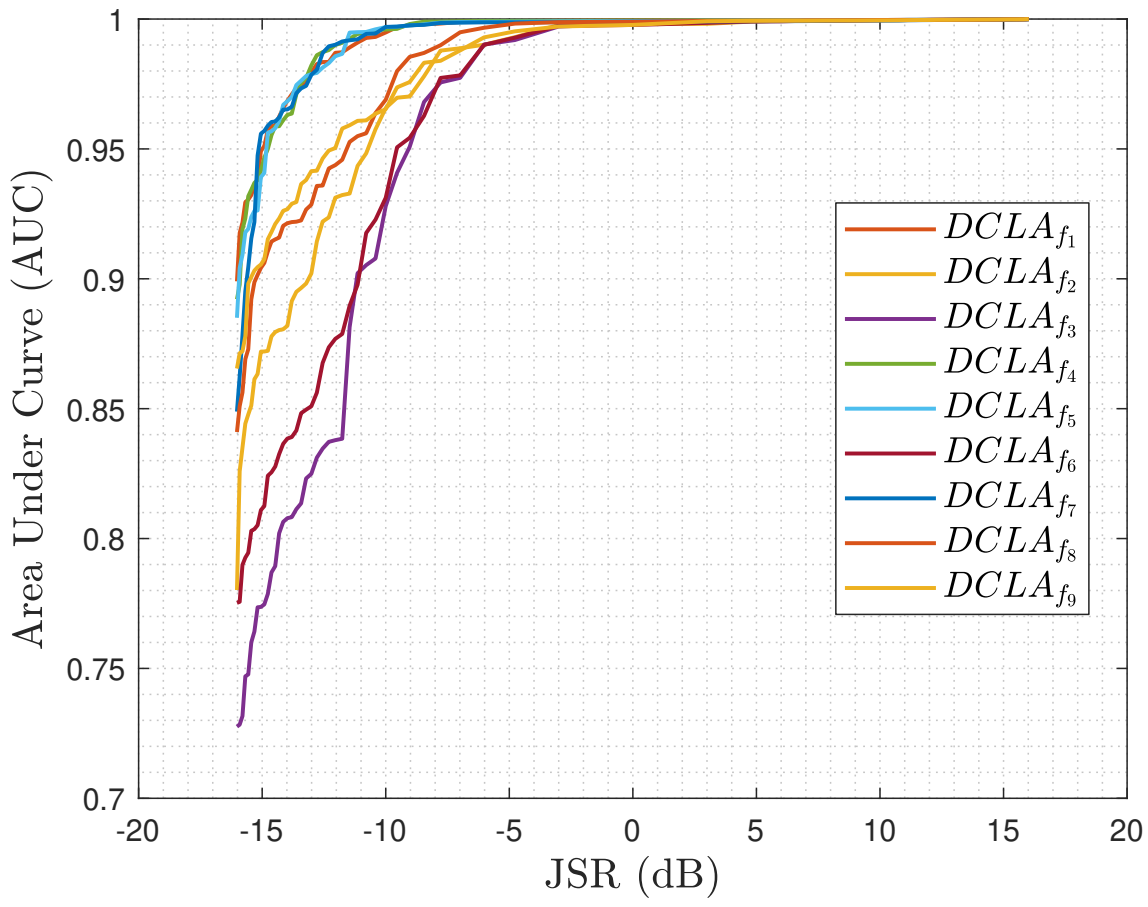

(b) Area Under Curve (AUC)

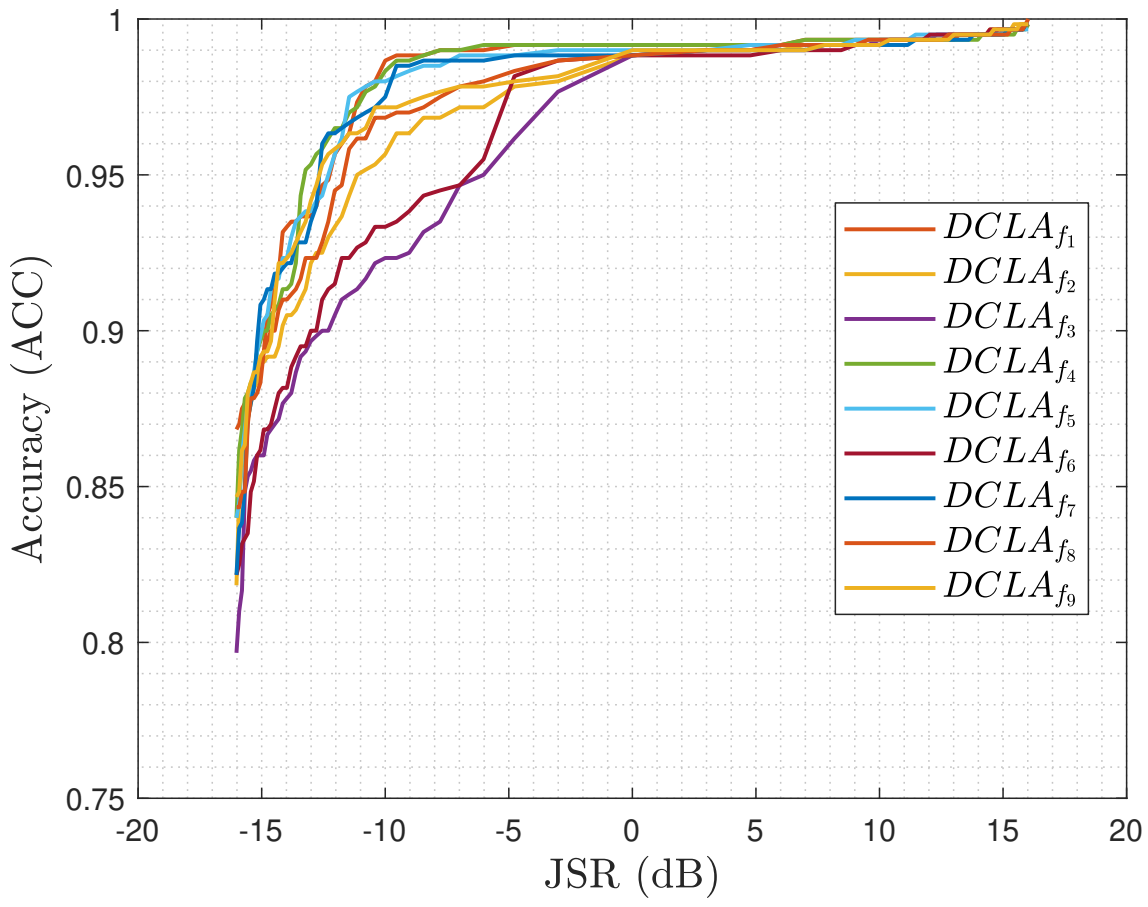

(c) Accuracy (ACC)

Fig. 4.23 ROC curves: (a) Probability of Detection and the corresponding (b) Area Under Curve (AUC) and (c) Accuracy (ACC) as varying JSR.

accuracy in detecting and locating malicious attacks in both time and frequency domains even when the jammer attacks with very low power. Moreover, using DCLA to detect the jammer deeply at multiple sub-carriers is fundamental to mitigate its effect on the received commands.

### 4.4.3 Test Fading Channel and different SNR values

The 3GPP defined different models for the LOS probability, pathloss, shadow-fading and fast-fading in the Release-15 to characterize the channels between an Aerial User Equipment (UE) and a terrestrial Base Station (BS). In this section, we evaluate the proposed approach under fast fading condition with respect to the specifications and modified parameters for RMa-AV scenario defined in [254] (Table B.1.2.-1) and different SNR values (from $-15dB$ to $+15dB$) with fixed $JSR$ (6dB). As shown in the figures (Fig. 4.24) below, the UAV augmented with the SA module is still able to learn a good representation of the radio environment and consequently detect jamming attacks. The performance at the discrete level degrades faster as SNR decreases with respect to that at the continuous level due to the fact that the transitions of the signal between the superstate (discrete level) becomes faster, moreover the clusters at low SNR are more concentrated around the origin which makes it difficult to capture the dynamic changes at the discrete level accurately.

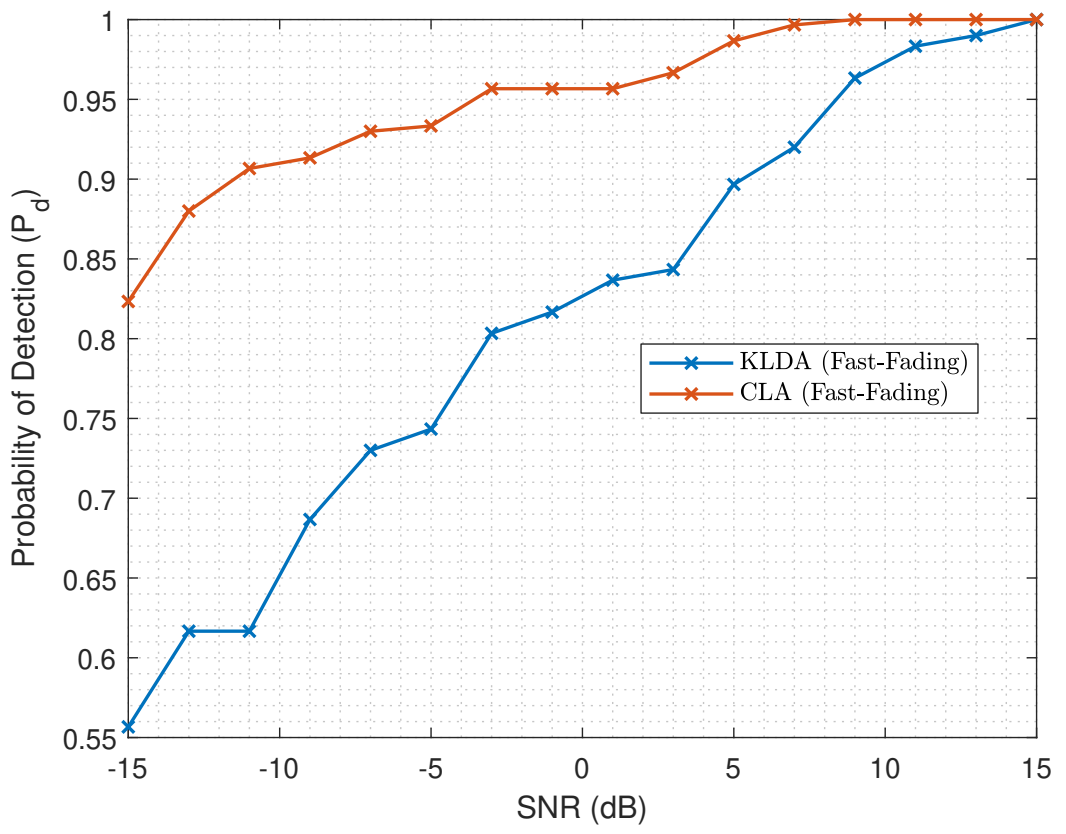

Fig. 4.24 Probability of detection ($P_d$) and Accuracy (ACC) as varying SNR under Fast-Fading condition.

## 4.5 Conclusion

In this chapter, we proposed to introduce an emergent data-driven Self-Awareness (SA) module to enhance the physical layer security in CR where the cognitive-UAV scenario is used as case study. This module allows the radio to build up its own memories incrementally by observing the stimulus received from the radio environment and learning with reasoning a hierarchical representation of such observation. The radio augmented with SA is capable to predict the radio environment and identify any abnormality within a received signal. The SA module is also capable of characterizing the abnormal situation caused by a jammer while studying the rules of how (Power), when (Time) and where (Frequency) the jammer is attacking. Additionally, an incremental learning process is proposed for the radio to learn a new model that represents the new situation, decide and act efficiently by suppressing the jamming signal. All these tasks are performed by the radio itself in an incremental approach without any external supervision, and the proposed framework is generalized enough to be employed in different radio applications. We show that the proposed HDBN framework accurately characterizes the jammer's behaviours in different situations while the probability of detection is significantly high even at low Jamming-to-Signal-Ratio. The results also show that after learning the jammer's behaviors, the UAV with the proposed framework can correctly predict the future activities of the jammer, which can eventually help in mitigating any future attacks. Next chapters will analyse the jammer classification in a more practical multiple jammers scenario as well as introducing interactive learning models between the jammer and the user to design an anti-jamming technique.

# Chapter 5

# Signal Identification and Automatic Modulation Classification

The integration of Cognitive Radio (CR) with Unmanned Aerial Vehicles (UAVs) is an effective step towards relieving the spectrum scarcity and empowering the UAV with a high degree of intelligence. The dynamic nature of CR and the dominant line-of-sight links of UAVs poses serious security challenges and make the CR-UAV prone to a variety of attacks as malicious jamming. Joint jammer detection and automatic jammer classification is a powerful approach against the physical layer threats by identifying multiple jammers attacking the network that realize a crucial stage towards efficient interference management. This chapter extends the SA module presented in Chapter 4 by proposing a novel method for joint detection and automatic classification of multiple jammers attacking with different modulation schemes. The method is based on learning a representation of the radio environment encoded in a Generalized Dynamic Bayesian Networks (GDBN) whilst multiple GDBN models represent various jamming signals under different modulation schemes. The CR-UAV performs multiple predictions online in parallel and evaluates multiple abnormality measurements based on a Modified Markov Jump Particle Filter (M-MJPF) to select the best-fit model that explains the detected jammer and recognize the modulation scheme accordingly. The simulated results demonstrate that the proposed GDBN-based method outperforms Long Short-Term Memory (LSTM), Convolutional Neural Network (CNN) and Stacked Autoencoder (SAE) in terms of classification accuracy and achieves a higher degree of explainability of its own decisions by interpreting causes and effects at hierarchical levels under the Bayesian learning and reasoning processes. Furthermore, this chapter introduces for the first time the automatic modulation conversion which is a prospective candidate technology in the future wireless communications. The proposed framework is based data-driven approach following the inherent intelligent capability of generalized filtering integrated with transport planning.

## 5.1 Introduction

The advent of the Unmanned Aerial Vehicles (UAVs) and its recent rapid growth in a myriad of applications have got plenty of interest to be leveraged in the fifth-generation (5G) technology [32]. Owing to the dynamic deployment flexibility, high mobility and strong Line-of-Sight (LoS) communication links of UAVs, they are regarded as an important complement to the terrestrial networks from the sky [111]. However, UAV-based communications will face several problems such as spectrum scarcity due to the explosively increasing number of connected UAVs [259], energy-efficiency due to the on-board limited battery lifetime [260, 261] and physical layer security (e.g. jamming attacks) due to the open nature of ground-to-air wireless channels and the dominant LoS propagation links [262]. Cognitive Radio (CR) is considered as one of the most promising solutions that can tackle the aforementioned problems due to its capability in pursuing its own goals autonomously, learning the radio environment, monitoring and predicting the environmental changes and infer the appropriate action that can be performed [263]. A series of recent studies have investigated the integration of CR and UAVs (i.e. Cognitive UAV Radios) for different aspects such as, communication capacity and Quality of Service (QoS) improvement [264, 265], collision avoidance stability [266], trajectory optimization [267], energy harvesting [268], spectrum scarcity [269], energy efficiency optimization [270, 271], interference coordination [272], data dissemination [273], joint sub-carrier and power allocation [274, 275], and for secure communications [276]. Moreover, studies on UAV-aided networks supported by machine learning has been investigated in [277, 278].

UAV communications are susceptible to jamming attacks by terrestrial malicious nodes distributed over a large area on the ground that can exploit the strong LoS channels to launch powerful attacks and interfere with the UAV resulting in communication failure [279]. In addition, smart jammers equipped with cognitive capabilities can pose more security threats [185]. They can sense the radio spectrum and discover the UAV's transmission policy to update their attack strategy and force the UAV to learn wrong behaviours and take misleading actions. Thus, enhancing the physical layer security is of great concern to ensure reliable communications and successfully deploy cognitive UAV Radios. This chapter focuses on the joint detection and classification of multiple jammers attacking the UAV's control and command link. Jammer detection is the first essential step to determine the radio situation, while jammer classification is an important stage towards an efficient interference management solution [280].

In the previous chapter, we introduced the concept of Self-Awareness (SA) in CR to empower the radio with a brain for high-level intelligence. The SA module allows the radio to reach the capability of learning a representation of the radio environment encoded in a generative dynamic model to be stored in the radio's brain. Studies from neuroscience have shown that the brain can efficiently use sensory information to resolve uncertainty about its computations and the surrounding world by representing sensory signals *probabilistically* in the form of probability distributions [281]. Inspired by such a brain functionality (known as the 'Bayesian Brain Hypothesis'), we propose to equip CR with a *probabilistic* Generative dynamic model, such as Dynamic Bayesian Networks (DBNs) using Generalized variables, i.e., Generalized DBN (GDBN), encoding knowledge about the radio itself and the structural regularities from its external milieu variations via sensory signals. GDBNs describe in a probabilistic manner how a given signal might have been generated by predicting new data samples and inferring the hidden states that caused the observed signal. Since CR operates in stochastic wireless environments under uncertainty, using a powerful statistical tool as Bayesian filtering is fundamental to dealing with uncertainty and performing inference and estimation of environmental states efficiently. The motivation of studying Bayesian filtering for environmental state estimation is that it is optimal in a conceptual sense [282]. Bayesian Filtering employed on GDBNs allows the radio to evaluate the situation through different abnormality measurements at multiple hierarchical levels and understand if the situation is normal or abnormal (e.g. detecting normal and jamming signals). If an abnormality is detected, the radio can characterize it to discover the new rules and encode them incrementally in a new dynamic model. However, an important question that needs to be addressed here is *when* the radio must learn a new model based on the current experience? Abnormality detection is not enough to answer this question. In contrast, abnormality classification is an indispensable functionality towards this understanding by comparing multiple abnormality signals generated by multiple models already learned in previous experiences and evaluating how much the current situation differs from them.

In this chapter, we extend the SA module by adding the Abnormality Classification functionality to jointly detect and classify multiple jammers according to their modulation scheme. Initially, the Cognitive-UAV begins with null memory without any a priori knowledge about the radio environment, supposing that no signals present and observations are due to a stationary noise process, i.e., a process evolving according to static rules. Then the Cognitive-UAV starts to build up knowledge about the environment by exploiting the Generalized Errors (i.e. prediction errors) at the state level to discover the real dynamic rules of how the signals (related to the commands) are behaving inside the radio spectrum. These errors can be clustered in an unsupervised manner to learn the corresponding reference GDBN model under a normal situation (i.e. where the jammer is absent). The cognitive-UAV can use the acquired reference GDBN model in future experiences to predict the commands that it is supposed to receive under normal circumstances by employing a Modified Markov Jump Particle Filter (M-MJPF). Consequently, it can detect any jamming attack using abnormality measurements at hierarchical levels as well as calculating the new Generalized Errors once an abnormality is detected. This computational scheme assumes that the cognitive-UAV generates probabilistic predictions continually on what commands come next based on the rules encoded in the reference model and compares those predictions at different hierarchical levels with the UAV's real communications stimuli that lead to the computation of hierarchical Generalized Error signals. These Generalized Error signals are of great importance to understand why the current dynamic model can not explain the current radio situation and how we can update the model to adapt to the abnormal situation. In addition, those errors can be informative enough to understand the cause behind the abnormality and provide a way to extract the jammer's signal. Exploiting the Generalized Errors allows extracting the jammer's signal and guides the cognitive-UAV to learn a separated GDBN model for each detected jammer. In this way, the cognitive-UAV's brain consists of a reference GDBN model representing the command signals that the UAV is expecting to receive in a typical radio situation and a set of multiple GDBN models representing the jammers' behaviours incrementally learned in previous experiences under different modulation schemes in abnormal radio situations. The link between the reference model and the other ones is described by the Generalized Errors provided by the reference model and used by the set of models as observations. In other words, the UAV uses the reference model to infer the hidden states of the radio environment, detect abnormalities in case of attack and calculate the Generalized Errors. Those errors can be used as observations by the other set of models (representing the jammers) while performing multiple predictions in parallel to evaluate the best GDBN model (inside the set) that better explains the current observation (i.e. Generalized Error provided by the reference model) and recognize the modulation scheme of the jammer consequently. The classification task is formulated in terms of objective function that maximizes the Bayesian model evidence (or marginal likelihood), which is the probability of observing signals conditioned to a model generating those signals or to minimize the surprise (i.e. abnormality). This means that we will test different hypotheses (i.e. models) and weighting them to select the model that has the greatest evidence and minimum surprise (i.e. abnormality).

Besides, 6G is expected to have a high-level hierarchy involving low bit-rate IoT devices to Gigabit rate connectivity (e.g., backhaul links). In such a cooperative communication scenario, intermediate or relay nodes are deployed to assist the source and the intended destination (i.e., a Macro Base Station MBS). In this case, a Self-Aware relay node (SAN) receiving signals with low-order modulation from IoT sensors can convert the modulation format of received signals into a higher-order format. The SAN then redirects the transported signals (with higher order modulation by performing modulation conversion) over the relay backhaul link between SAN and MBS, thus improving transmission capacity and spectral efficiency. Therefore, modulation conversions, which find its origin in optical communications [283], can be a prospective candidate technology in future generation wireless communications. In addition, automatic modulation classification is a possible application where modulation format conversion can play a significant role, providing an explainable approach to identifying the received signal and the type of information it carries. Wherefore, this chapter introduces for the first time the automatic modulation conversion in wireless communications which allows an AI-enabled node to predict signals' dynamics of different modulation schemes and explain how it can be transported (converted) with the minimal effort and forwarded with higher spectral efficiency. To achieve this goal, we propose to integrate the Generalized Filtering framework by Transport Planning to learn the way of converting low-order modulations to high-order modulations, which has also been validated by performing the automatic modulation classification. Simulation results demonstrate the effective performance of our novel framework on converting and classifying multiple modulation formats.

The main contributions of this chapter are as follows: *i)* we propose an efficient learning mechanism within the Growing Neural Gas (GNG) to capture the dynamic transitions of the radio signal modulated under certain modulation scheme; *ii)* we formulate the modulation classification problem in terms of an objective function that aims to minimize the surprise (i.e. abnormality) by testing different models learned by the radio and weighting them to select the model that causes the minimum surprise and thus that better explains the modulation scheme of the detected jamming signals; *iii)* extensive simulations verify that the proposed GDBN-based framework for automatic jamming signal classification performs with superiority classification accuracy than LSTM, CNN and SAE; *iv)* the GDBN models can achieve higher interpretability than Deep Learning-based models since they can explain the predictions explicitly at hierarchical levels and use the abnormality measurements and Generalized Errors as self-information to keep learning by understanding incrementally; ; *v)* we show how the radio is capable of predicting the dynamics of different modulation formats and of explaining how the dynamic rules evolve by transporting low-order into high-order modulations using the acquired transport maps.

## 5.2 Related Work

Radio Signal Classification is an important task in many communications systems [284]. It is mainly based on Automatic Modulation Classification (AMC) that servers as an intermediate step between signal detection and signal demodulation. AMC is widely used in both civil and military fields and finds applications in Cognitive Radio (CR) for efficient spectrum management, and secure communications [285, 286]. Traditional approaches for modulation classification include Likelihood-Based (LB) approach and Feature-Based (FB) approach [287–289]. The LB approach is based on comparing the likelihood ratio of the received signal with a threshold. The LB is optimal in the Bayesian sense by minimizing the probability of false classification. However, it is computationally complex and requires an estimation of parameters (e.g. channel parameters) to calculate the likelihood probability, which is not always possible in real radio scenarios as in CR. Also, the performance degrades in the presence of phase and frequency offset. The FB approach does not require an estimation of parameters, and it is based on some features as the variance of the centered normalized signal amplitude, phase and frequency. Thus it is less complicated compared to the LB approach and easy to use. Even though it is sub-optimal, however, a proper design allows achieving optimal performance.

Deep learning-based methods for AMC are extensively investigated in the literature. In [286], a Long Term Short Memory (LSTM) is used for this purpose where the data-augmentation methods (i.e. rotation, flip, and Gaussian noise) are studied to cope with small datasets by expanding the data and thus improving the robustness of deep learning models and classification accuracy as well. However, expanding the dataset might lead to several problems as increasing latency which is vital in some applications as vehicular communications. An improved Convolutional Neural Network (CNN)-based automatic modulation classification network (IC-AMCNet) is proposed in [290] where different types of layers as convolutional, dropout and Gaussian noise are applied for regularization and to overcome the overfitting issue. In addition, a small number of filters is used in each layer to reduce the processing time. Authors in [291] proposed a gated recurrent, residual network (GrrNet) for modulation classification consisting of a ResNet extractor module, fusion module and GRU-based classification module. However, both [290], and [291] used supervised training by feeding the networks with the signal features along with the labels that indicate the modulation scheme of the input. This may require a significant effort to label large amounts of training examples that can be expensive and time-consuming. Interesting research has been conducted in [292] to study the visualization methods for deep learning-based radio modulation classifiers (based on CNN and LSTM) and thus to understand the modulation classification mechanism for better interpretability. However, such visualization techniques do not exploit the extracted radio features in an unsupervised way, allowing the radio to encode the dynamic changes between different modulation schemes, which enhance the learning and perception processes of the radio.

In [293], a compressive convolutional neural network (CCNN) is proposed for AMC where different regular constellation images and contrast-enhanced grid constellation images are generated from received signals and used as network inputs. In addition, a compressive loss constraint is proposed to train the CCNN to capture high-dimensional features as well as utilizing the intra-class compactness and inter-class separability to enhance robustness performance for a different order of modulations. Simulation results showed the superior classification compared with RNN, DNN and CNN. Other works also converted the radio signal into images, e.g. Choi-Williams time-frequency distribution (CWD) image [294], Feature Point (FP) images [295], Contour Stellar Image (which gets more color feature compared to the Constellation Diagram) [296], amplitude spectrums of bispectrum (ASB) images [297], cyclic spectrum images [298]. The studies mentioned above have obtained promising results in modulation classification. However, they require high computational processing to convert signals to images that can be unfeasible in the UAV scenario, and they might lose important information and ignore crucial details by passing from time-frequency representation to image representation.

In this chapter, we propose a novel GDBN-based method for AMC that can overcome the drawbacks of other existing methods in the literature discussed previously since *i)* it does not require involving data augmentation (as in [286]) to create bigger datasets that might increase latency. *ii)* It follows a data-driven unsupervised approach by allowing the radio to build up knowledge about the radio spectrum from null memory (without using explicit labels of the input signals as done in [290, 291]). Hence, radio's autobiographical memories are grown up incrementally by observing real-time data and learning autonomously from the extracted generalized errors. *iii)* Deep learning approaches ([286, 290–292]) where the hidden variables are at a sub-symbolic level considered as a black box cannot provide a high level of explainability of their decisions, creating results that are hard to understand. The proposed approach has a higher degree of interpretability that can determine and associate causes and effects at hierarchical levels thanks to the underlying Bayesian learning and reasoning processes. In addition, it achieves a higher degree of explainability of its own decisions where hidden variables used in the generalized model make it possible to draw explicit causal dynamic probabilistic relationships among continuous signals and their symbolic higher-level counterparts to study how significance each parameter in contributing to the final decision. *iv)* It relies on raw In-Phase (I) and Quadrature (Q) components of radio signals that are easy to extract (unlike [293–297] that convert radio signals into images) and offers flexibility in implementing the proposed framework in different systems and environments.

## 5.3 System Model

The system model depicted in Fig. 5.1 extends that illustrated in Fig. 4.3 and consists of a cellular-connected UAV, Base Station (BS), a UAV operator, and multiple terrestrial jammers aiming to attack the Command and Control (C2) link by sending false commands to alter the trajectory and take control of the UAV. The jammers are smart; they can identify and locate the resources allocated to the UAV by the BS inside the radio spectrum and attack consequently using different modulation schemes. Signals of all jammers and the UAV operator are generated according to the propagation model that is shown in Fig. 4.4. The positions of BS, UAV and jammers are denoted by $q^g = [x^g, y^g, z^g]$, $q^{j_k} = [x^{j_k}, y^{j_k}, z^{j_k}]$ and $q^u_t = [x^u_t, y^u_t, z^u_t]$, respectively. The UAV's position $q^u_t$ varies with time while $q^g$ and $q^{j_k}$ are fixed positions where $j_k \in K$ represents the $k$-th jammer adopting the $k$-th modulation scheme and $K$ denotes the number of candidate modulation schemes. The line-of-sight (LOS) channels of the UAV communication links become more dominant than other channel impairments, such as small scale fading and shadowing, if the altitude of the UAV is much higher than that of the terrestrial users or BS [299]. We adopt the 3GPP Rural-macro with aerial vehicles (RMa-AV) scenario under LOS conditions [254]. In addition, the doppler frequency shift caused by the UAV mobility is assumed to be compensated at the receiver as in [299]. Thus, the path loss model ($PL^{bu}_{t|dB}$) for the LOS RMa-AV from the BS to the UAV at a given time instant is described as:

$$PL^{bu}_{t|dB} = 20\log_{10}\left(\frac{40\pi d_{bu,t} f_c}{3}\right) + \min(0.03\kappa, 10)\log_{10}(d_{bu,t}) + \min(0.044\kappa, 14.77) + 0.002\log_{10}(\kappa) d_{bu,t}, \quad (5.1)$$

where $\kappa$ represents the average building height gain and $d_{bu,t}$ is the 3D between BS and the UAV at time slot $t$ defined as:

$$d_{bu,t} = \sqrt{(x^g - x^u_t)^2 + (y^g - y^u_t)^2 + (z^g - z^u_t)^2}. \quad (5.2)$$

The path loss model defined in (5.1) can be transformed into the linear domain according to:

$$PL^{bu}_t = 10^{\frac{PL^{bu}_{t|dB}}{10}}. \quad (5.3)$$

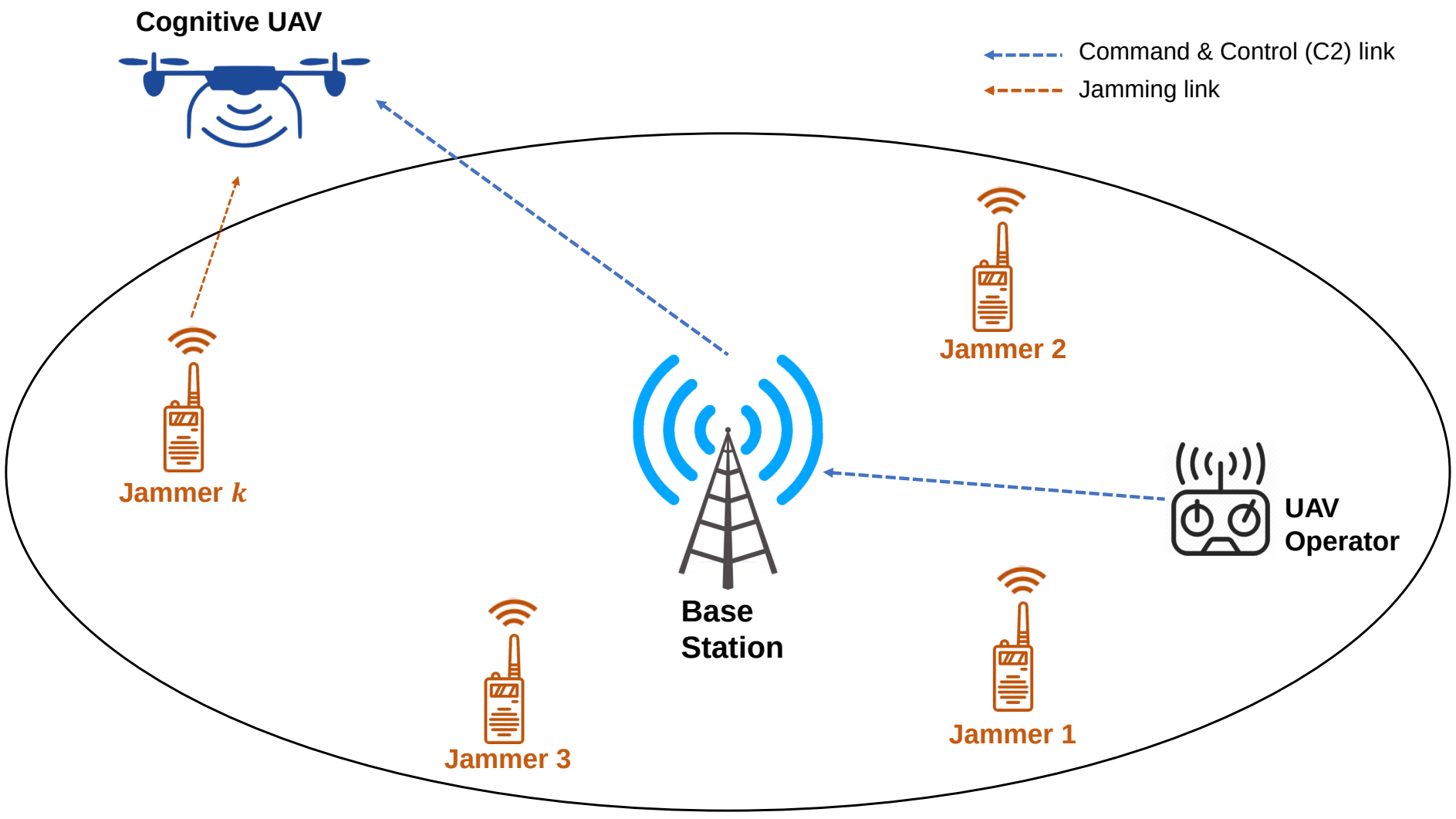

Fig. 5.1 Illustration of the system model.

Likewise, the path loss model ($PL^{j_k u}_{t|dB}$) from the $k$-th jammer and the UAV is expressed as:

$$PL^{j_k u}_{t|dB} = 20\log_{10}\left(\frac{40\pi d_{j_k u,t} f_c}{3}\right) + \min(0.03\kappa, 10)\log_{10}(d_{j_k u,t}) + \min(0.044\kappa, 14.77) + 0.002\log_{10}(\kappa) d_{j_k u,t}, \quad (5.4)$$

where $d_{j_k u,t}$ is the 3D between the $k$-th jammer and the UAV at time slot $t$ defined as:

$$d_{j_k u,t} = \sqrt{(x^{j_k} - x^u_t)^2 + (y^{j_k} - y^u_t)^2 + (z^{j_k} - z^u_t)^2}. \quad (5.5)$$

The path loss model defined in (5.4) can be transformed into the linear domain according to:

$$PL^{j_k u}_t = 10^{\frac{PL^{j_k u}_{t|dB}}{10}}. \quad (5.6)$$

Accordingly, the channel gain ($h^{bu}_t$) from the BS to the UAV at a given time instant is described as:

$$h^{bu}_t = \frac{1}{PL^{bu}_t}, \quad (5.7)$$

and the channel gain ($h^{j_k u}_t$) from the $k$-th jammer to the UAV is:

$$h^{j_k u}_t = \frac{1}{PL^{j_k u}_t}. \quad (5.8)$$

The UAV receives a PRB each 50ms where the Pitch, Yaw, and Roll commands are trans-mitted over 9 consecutive sub-carriers (frequency domain) within 1 OFDM symbol (time domain) and it is equipped with a GPS receiver and RF antenna which are supposed to be synchronized, i.e. the UAV receives one PRB and measures the 3D position (by the GPS receiver) every 50 ms as shown in Fig. 4.5 and Fig. 4.6. The UAV extracts the commands' features (i.e., $IQ$ components) after the OFDM receiver block by exploiting the FFT. At this level (i.e. the output of FFT), all the Resource Elements (REs) in the time-frequency grid can be scanned without any extra hardware or computation by reusing the hardware of FFT cores [300]. In our study, we considered the REs carrying commands (i.e., RV in Fig. 4.6) solely because we aim to analyze the command signals only. However, this can be simply extended in future investigations to consider the whole PRB.

## 5.4 Problem Formulation

At each instant $t$, the cognitive-UAV will receive a set of commands (i.e., C2 data) and move accordingly. Our objective is to give the cognitive-UAV the capability to self-evaluate the received commands by identifying if the commands are corrupted by the jammer (i.e., jammer detection) and, consequently, learn jammer dynamics to predict its behaviour in the future and recognize the modulation scheme of the detected jammer (i.e., jammer classification).

The jammer detection task can be formulated in terms of a binary hypothesis testing problem given as:

$$\begin{cases} \mathcal{H}_0 : \mathrm{z}_t = h_t^{bu}\mathrm{x}_t + \mathrm{n}_t, \\ \mathcal{H}_1 : \mathrm{z}_t = h_t^{bu}\mathrm{x}_t + h_t^{j_k u}\mathrm{x}_t^j + \mathrm{n}_t, \end{cases} \tag{5.9}$$

where hypothesis $\mathcal{H}_0$ and $\mathcal{H}_1$ indicate the absence and presence of the jammer, respectively. $\mathrm{z}_t$ is the received C2 signal, $\mathrm{x}_t$ is the desired signal, $\mathrm{x}_t^j$ is the jamming signal and $\mathrm{n}_t$ is an additive white Gaussian noise. The proposed approach is based on learning a reference dynamic model ($\mathcal{M}_0$) explaining the normal situation (without jamming attacks) under hypothesis $\mathcal{H}_0$. During testing, cognitive-UAV performs predictions $\mathrm{x}_t^*$ conditioned on the learned model $\mathcal{M}_0$ and characterized by the posterior $\mathrm{P}(\mathrm{x}_t^*, \mathcal{M}_0|\mathrm{z}_t)$. Jammer can be detected by comparing the similarity between predictions and likelihood ($\mathrm{P}(\mathrm{z}_t|\mathrm{x}_t^*)$) using a probabilistic distance $\mathcal{D}$ (i.e., abnormality) between $\mathrm{P}(\mathrm{x}_t^*, \mathcal{M}_0|\mathrm{z}_t)$ and $\mathrm{P}(\mathrm{z}_t|\mathrm{x}_t^*)$.

On the other hand, the jamming modulation classification task can be formulated as a classification problem with K modulation schemes. The aim of the classifier is to identify $\mathrm{x}_t^j$ of the received signal $\mathrm{z}_t$ from a learned set $\mathcal{M} = \{\mathcal{M}_1, \mathcal{M}_2, \dots, \mathcal{M}_K\}$ of $K$ dynamic models representing $K$ possible modulation schemes and give out $\mathrm{P}(\mathrm{x}_t^j \in \mathcal{M}_k|\mathrm{z}_t)$ where $\mathcal{M}_k \in \mathcal{M}$. Since, all modulation schemes are equally likely, then the optimal classifier is the maximum-

log-likelihood classifier finding the maximum among $K$ conditional probabilities $\mathrm{P}(\mathrm{z}_t|\mathrm{x}_t,\mathcal{M}_k)$, according to:

$$\hat{k} = \underset{\mathcal{M}_k}{\operatorname{argmax}} \log \mathrm{P}(\mathrm{z}_t|\mathrm{x}_t,\mathcal{M}_k), \tag{5.10}$$

where $k = 1,2,\ldots,K$. It is to note that (5.10) is equivalent to finding the minimum among $K$ probabilistic distances $\mathcal{D}\left[\mathrm{P}(\mathrm{z}_t|\mathrm{x}_t^{j*}),\mathrm{P}(\mathrm{x}_t^{j*},\mathcal{M}_k|\mathrm{z}_t)\right]$, according to:

$$\hat{k} = \underset{\mathcal{M}_k}{\operatorname{argmin}} \mathcal{D}\left[\mathrm{P}(\mathrm{z}_t|\mathrm{x}_t^{j*}),\mathrm{P}(\mathrm{x}_t^{j*},\mathcal{M}_k|\mathrm{z}_t)\right]. \tag{5.11}$$

where $\mathrm{x}_t^{j*}$ is the predicted jammer signal based on model $\mathcal{M}_k$. Thus, by using a similar approach described by Friston for Bayesian filtering in [301], it can be shown that each likelihood using a different model can be approximated by searching the minimum free energy configuration as an upper-bound estimate. Finding the maximum upper bound is equivalent to obtaining the highest value of (5.10). The free energy can be expressed as the negative of the distance in (5.11). So, obtaining the model that minimizes (5.11) is an approximation to (5.10).

## 5.5 Proposed Automatic Jamming Modulation Classification

### 5.5.1 Radio Environment Representation

In our approach, we use a generalized state-space model [1] to represent the radio environment. We assume, that the observed signal $\tilde{Z}_t$ is a linear combination of one latent generalized state $\tilde{X}_t$ that represents the direct cause of the observation and a multivariate generalized Gaussian noise $\tilde{v}_t$ ($\tilde{v}_t \sim \mathcal{N}(0, \Sigma_{\tilde{v}_t})$) and defined as follows:

$$\tilde{Z}_t = H\tilde{X}_t + \tilde{v}_t, \tag{5.12}$$

where $H \in \mathbb{R}^{d\times d}$ is the matrix that maps hidden states to observations. The generalized observation $\tilde{Z}_t \in \mathbb{R}^d$ comprises the signal's states in terms of $I$ and $Q$ components and the corresponding first-order temporal derivatives ($\dot{I}$, $\dot{Q}$), where $d$ is the space dimensionality and it is equal to the total number of sub-carriers where the commands are transmitted. The evolution of the hidden generalized states $\tilde{X}_t$ can be approximated as a linear combination of the previous state $\tilde{X}_{t-1}$ by the control vector (or force) $U_{\tilde{S}_t}$ and formulated as follows:

$$\tilde{X}_t = A\tilde{X}_{t-1} + BU_{\tilde{S}_t} + \tilde{w}_t, \tag{5.13}$$

where $A \in \mathbb{R}^{d\times d}$ and $B \in \mathbb{R}^{d\times d}$ are the dynamic model and control model matrices, respectively. $U_{\tilde{S}_t}$ is associated with the generalized superstate $\tilde{S}_t$ where $\tilde{X}_t$ is expected to be found and $\tilde{w}_t$ is the generalized process noise such that $\tilde{w}_t \sim \mathcal{N}(0, \Sigma_{\tilde{w}_t})$.

[1]Random variables involved in the Generalized state-space model are represented in terms of Generalized coordinates of motion. The latter consists of the variable per se and its n-th order temporal derivatives as proposed by Karl Friston [301]. In this work, we consider only generalized motion up to order 1 (i.e., the $1^{st}$-order derivatives), and we refer to random variables (discrete and continuous) in generalized coordinates as generalized superstates (discrete) and generalized states (continuous).

The generalized superstates ($\tilde{S}_t$) consisting of the superstate ($\tilde{S}_t$) and the corresponding event are the deep hidden causes generating observations ($\tilde{Z}_t$) and direct hidden causes affecting generalized states ($\tilde{X}_t$). Superstates are the neurons (or clusters) obtained after employing an unsupervised clustering method on the input signals (discussed in the following section). Those neurons represent discrete regions of the physical signal where each neuron contains a set of homogeneous IQ data samples (i.e., samples with very close characteristics). So, the radio environment evolves probabilistically, occupying a finite set of discrete states. Each generalized superstate ($\tilde{S}_t$) is assumed to follow a Gaussian distribution, and so it can be represented by its sufficient statistics, namely, generalized mean (i.e., mean on the IQ samples and mean on the IQ derivatives) and covariance. We know that continuous data samples (i.e., generalized states) inside each superstate evolve according to the same dynamic rule, i.e., the control vector $U_{\tilde{S}_t}$, that guides the evolution of generalized states at the lower level. This motivates the choice of using a linear approximation in (5.13) to model the dynamic evolution of the generalized states (describing the physical signals' dynamics).
At a high abstraction level (discrete level), the evolution of generalized superstates can be expressed in the following form:

$$\tilde{S}_t = f(\tilde{S}_{t-1}) + \tilde{w}_t, \tag{5.14}$$

where $f(.)$ is a non-linear function that describes the relationship between the previous superstate and the current superstate, realizing the dynamics of how the signal is transiting among the discrete regions and its evolution over time.

### 5.5.2 Learning Stage

GDBN encodes conditional dependencies explicitly among random variables at multiple levels, allowing the radio to understand the cause-effect relationships, which endows the radio to explain predictions, extract errors, and adapt to dynamic environmental changes. In addition, GDBN provides a probabilistic inference that is computationally efficient (i.e., they can specify dependencies only when necessary, leading to a significant reduction in the cost of inference). Moreover, due to its capability to provide discriminative property to assess received signal characteristics, GDBN can be used as a joint generative and discriminative model. Motivated by the above discussion, we propose to learn a GDBN as a representation of the radio environment. GDBN can model dynamic processes describing the signal's temporal evolution at hierarchical levels. GDBN provides a graphical structure representing hidden and observed states in terms of random state variables encoding the conditional dependencies among them and specifying a compact parameterization of the model. Two sets of parameters can represent it. The first includes the number of nodes in each time slice and the corresponding topology, while the second set consists of the conditional probability distributions (CPDs) described by edges of the network. Learning a GDBN consists of parameter learning and structure learning. The former is the process of learning the distributions of discrete or continuous hidden variables in the GDBN, while the latter uses data to learn the links among random variables in the GDBN. Both parameter and structure learning depends on the generalized state-space model in question. The proposed GDBN consists of three levels. The discrete level stands for the discrete variables describing the discrete regions of the signal. The medium level stands for the continuous states encoded inside each discrete region, and the bottom level stands for the observation.

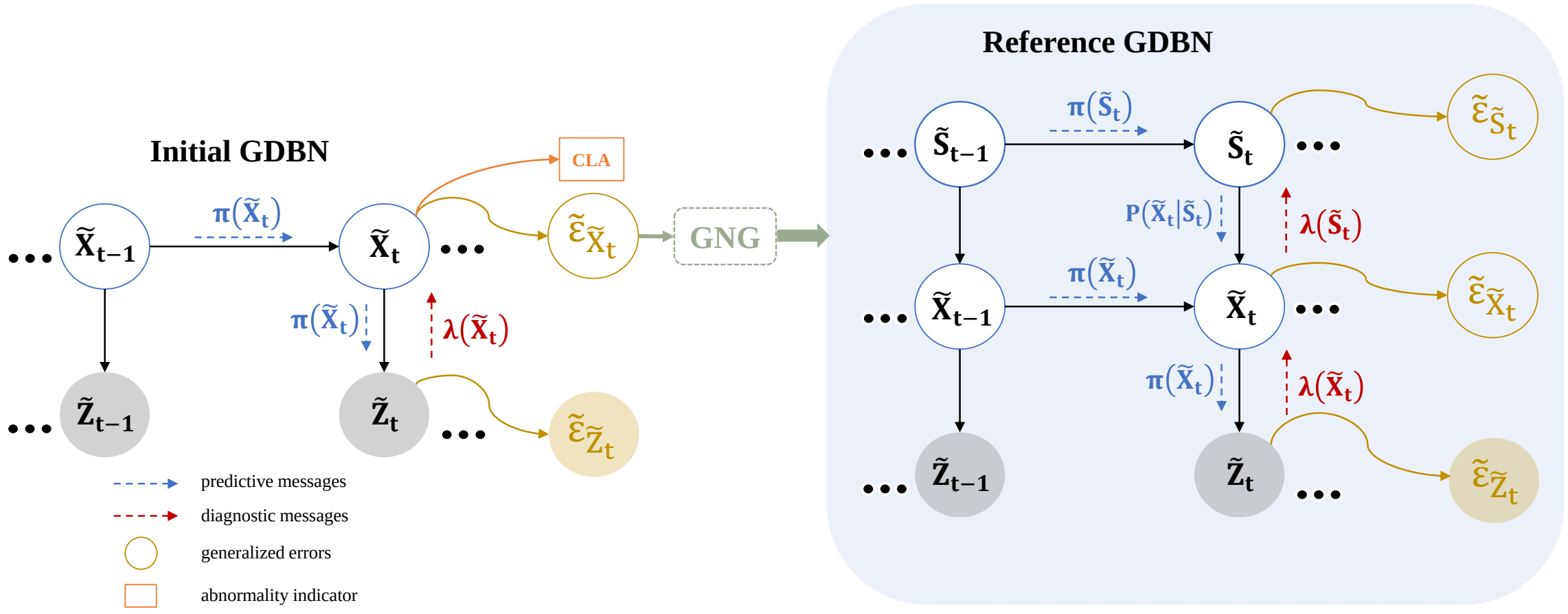

Fig. 5.2 Schematic that illustrates the process of learning incrementally by exploiting the generalized errors. The cognitive-UAV starts perceiving the environment using an initial GDBN model with a static assumption about the signal's evolution. While the cognitive-UAV is predicting and sensing the environment, it can calculate the generalized errors ($\tilde{\varepsilon}_{\tilde{X}_t}$) and stored them to perform the clustering after finishing the current experience.

The cognitive-UAV aims to learn and encode the radio environment representation in a GDBN under a normal radio situation. Initially, it starts with null memory without prior knowledge about the surrounding radio environment assuming that signals are evolving according to static rules. Thus, the cognitive-UAV starts perceiving the radio environment using an initial GDBN (consisting solely of the observation and state levels) on which an Unmotivated Kalman Filter (UKF) is employed (i.e., a null force filter with a static assumption about the environmental states) to predict the continuous signal using the following equation:

$$\tilde{X}_t = A\tilde{X}_{t-1} + \tilde{w}_t, \tag{5.15}$$

and then interpreting the received generalized observations $\tilde{Z}_t$ that comprises the variable and its generalized coordinates of motion coming from the receiver. In fact, since the signals inside the radio spectrum are following a certain dynamic behavior, the cognitive-UAV will detect abnormalities all the time and calculate the generalized errors ($\tilde{\varepsilon}^{[1]}_{\tilde{Z}_t}$) which are the differences between predictions and observations and it is expressed as:

$$\tilde{\varepsilon}^{[1]}_{\tilde{Z}_t} = \tilde{Z}_t - H\tilde{X}_t. \tag{5.16}$$

The UKF works by predicting the generalized states ($\tilde{X}_t$), projecting this into the measurement space and taking the difference between the current observed generalized measurement ($\tilde{Z}_t$) and the predicted one. This difference is known as innovation, which is computed in the measurement space. Thus, to project this difference back to the generalized state-space we must use the following formula:

$$\tilde{\varepsilon}^{[1]}_{\tilde{X}_t} = H^{-1}\tilde{\varepsilon}^{[1]}_{\tilde{Z}_t} = H^{-1}(\tilde{Z}_t - H\tilde{X}_t) = H^{-1}\tilde{Z}_t - \tilde{X}_t. \tag{5.17}$$

The generalized errors ($\tilde{\varepsilon}^{[1]}_{\tilde{X}_t}$) that capture the real dynamics of the signal are used as input to an unsupervised clustering technique, the Growing Neural Gas (GNG) (refer to Fig. 5.2). GNG encodes the generalized errors into discrete regions described by a set of neurons or superstates **S**, such that:

$$\mathbf{S} = \{S_1, S_2, \ldots, S_M\}, \tag{5.18}$$

where $M$ is the total number of neurons. After obtaining the neurons, we analyzed how the signal is transiting between them to learn the transition matrix $\Pi$ by estimating the transition probabilities $\pi_{ij} = P(S_t = i | S_{t-1} = j)$ over a period of time (i.e. the training time), where

$i, j \in \mathbf{S}$. The $M \times M$ transition matrix is defined as:

$$\Pi = \begin{bmatrix} \pi_{11} & \pi_{12} & \dots & \pi_{1M} \\ \pi_{21} & \pi_{22} & \dots & \pi_{2M} \\ \vdots & \vdots & \ddots & \vdots \\ \pi_{M1} & \pi_{M2} & \dots & \pi_{MM} \end{bmatrix}. \quad (5.19)$$

Thus, the generalized superstates ($\tilde{\mathbf{S}}$) can be expressed in terms of current discrete variable $S_t$ and the corresponding event $e_t^{ij}$ in the following way:

$$\tilde{S}_t = [S_t \; \dot{S}_t] = [S_t \; e_t^{ij}]. \quad (5.20)$$

An event can be described as a change at the discrete level (i.e., the transition from a certain superstate to a new one), such that:

$$e_t^{ij} = (S_{t-1} = i, S_t = j) \mid i \neq j. \quad (5.21)$$

The null event can be defined as $e_t^0$ when $i = j$. Furthermore, since the radio environment is dynamic and varies with time, estimating the temporal (i.e., time-varying) transition matrix $\Pi_\tau$ is of great interest. $\Pi_\tau$ encodes not only the possible transitions (transition probabilities) at the discrete level but also when those transitions or events will occur (i.e., the time required for a particular event to occur) and defined as:

$$\Pi_\tau = \begin{bmatrix} \pi_{11,\tau} & \pi_{12,\tau} & \dots & \pi_{1M,\tau} \\ \pi_{21,\tau} & \pi_{22,\tau} & \dots & \pi_{2M,\tau} \\ \vdots & \vdots & \ddots & \vdots \\ \pi_{M1,\tau} & \pi_{M2,\tau} & \dots & \pi_{MM,\tau} \end{bmatrix}, \quad (5.22)$$

where $\pi_{ij,\tau} = P(S_t = i | S_{t-1} = j, \tau)$ realizing a new condition in transiting to the new superstate $S_t = i$ after being in $S_{t-1} = j$ for a certain time (i.e., $\tau$). It is worth noting that as the dynamics of the signal become faster the time $\tau$ become smaller. So, the time-varying transition matrix encodes how transition probabilities vary with time; some probabilities increase and others decrease as time evolves, allowing to keep tracking the dynamic changes in the environment.

Each generalized superstate $\tilde{S}_m$ ($\tilde{S}_m \in \mathbf{S}$) is assumed to follow a multivariate Gaussian distribution with dimensionality $4d$. Thus, it can be represented by its sufficient statistics as generalized mean ($\tilde{\mu}_{\tilde{S}_m}$) and covariance ($\Sigma_{\tilde{S}_m}$). $\tilde{\mu}_{\tilde{S}_m}$ consists of the mean value $\mu_{\tilde{S}_m}$ describing the average of all the data samples encoded in this superstate $\tilde{S}_m$ in terms of $I$ and $Q$ as well as the mean $\dot{\mu}_{\tilde{S}_m}$ describing the average of the corresponding derivatives ($\dot{I}$, $\dot{Q}$) and it is defined as:

$$\tilde{\mu}_{\tilde{S}_m} = [\mu_{\tilde{S}_m}, \dot{\mu}_{\tilde{S}_m}], \tag{5.23}$$

such that,

$$\mu_{\tilde{S}_m} = \frac{1}{Y}\sum_{y=1}^{Y} s_y, \quad \dot{\mu}_{\tilde{S}_m} = \frac{1}{Y}\sum_{y=1}^{Y} \dot{s}_y, \tag{5.24}$$

where $Y$ is the total number of samples in $\tilde{S}_m$, $\mu_{\tilde{S}_m} \in \mathbb{R}^{2d}$, $\dot{\mu}_{\tilde{S}_m} \in \mathbb{R}^{2d}$, $s_y \in \tilde{S}_m$ representing the $I$, $Q$ samples of the distribution and $\dot{s}_y \in \tilde{S}_m$ representing the $\dot{I}$, $\dot{Q}$ samples. $\Sigma_{\tilde{S}_m}$ is a $4d \times 4d$ matrix defined as:

$$\Sigma_{\tilde{S}_m} = \begin{bmatrix} \sigma_1\sigma_1 & \dots & \sigma_1\sigma_{4d} \\ \vdots & \ddots & \vdots \\ \sigma_{4d}\sigma_1 & \dots & \sigma_{4d}\sigma_{4d} \end{bmatrix}, \tag{5.25}$$

where $\sigma_l$ is the variance of each component $l \in \{1,2,\dots,4d\}$ calculated as follows:

$$\sigma_l = \begin{cases} \mathbb{E}\big[(s^l - \mu_{\tilde{S}_m})(s^l - \mu_{\tilde{S}_m})^\intercal\big] & \text{if } 1 \le l \le d, \\ \mathbb{E}\big[(\dot{s}^l - \dot{\mu}_{\tilde{S}_m})(\dot{s}^l - \dot{\mu}_{\tilde{S}_m})^\intercal\big] & \text{if } d < l \le 2d. \end{cases} \tag{5.26}$$

### 5.5.3 Testing Stage

GDBN can decompose data with complex and non-linear dynamics into segments that are explainable by simpler dynamical units. The Modified Markov Jump Particle Filter (M-MJPF) (which is an evolved version of the MJPF introduced in [249]) is a specific class of switching dynamic systems employed on the learned GDBN model to discover the dynamical units and explain their switching behaviour and their dependency on both observations and discrete/continuous hidden states during the real-time process. As mentioned previously, M-MJPF like MJPF uses a combination of Particle Filter (PF) and a bank of Kalman Filters (KFs) to predict the generalized superstates (at discrete level) and generalized states (at the continuous level), respectively. MJPF has been modified (i.e., M-MJPF) to involve multiple generalized errors in the filtering process in order to explain the detected abnormalities at multiple levels (both continuous and discrete), characterize the cause of such abnormalities (i.e., the jammer) and consequently learn the new emergent rules occurred in the environment (to be discussed in Sec.5.5.4 and Sec.5.5.5). Also, the M-MJPF has been adapted to capture discriminatory features, allowing the use of multiple abnormalities for jammer classification (to be discussed in Sec.5.5.6).

The M-MJPF within the Bayesian Filtering framework provides two probabilistic inference modes: predictive or causal inference (top-down) and diagnostic inference (bottom-up). The predictive inference is based on passing predictive messages in a top-down manner, where predictions are performed based on the acquired knowledge in previous experience. The diagnostic inference is based on propagating likelihood messages after receiving the real measurement in a backward manner from bottom to up, where the likelihood messages evaluate how much the observation matches the predictions at hierarchical levels to update the belief in hidden variables accordingly. PF relies on a proposal density encoded in the learned transition matrix to sample a set of particles realizing the predicted superstates at the discrete level. Initially, PF propagates $N$ equally weighted particles (<.>) associated with a specific superstate, such that:

$$< \tilde{S}_t^n, W_t^n > \sim < \pi(\tilde{S}_t), 1/N >, \; n \in N. \tag{5.27}$$

It is worth noting that in our scenario, there is no need to use a large number of particles since the discrete level consists of a finite number of discrete regions. Thus, it is sufficient to use few particles to represent the posterior accurately (unlike the continuous space which may need a huge number of particles to represent the posterior correctly). After that, a KF is employed for each particle $(.^n)$ to predict $\tilde{X}_t$. The prediction at this level (continuous level) is guided by the prediction performed at the higher level as pointed out in (5.13) and can be expressed in terms of the conditional probability $P(\tilde{X}_t|\tilde{X}_{t-1}, \tilde{S}_t)$. In (5.13), the control vector $(U_{\tilde{S}_t})$ which realize the dynamic flow of the signal starting from the previous state is encoded in the generalized mean value defined in (5.23), hence $U_{\tilde{S}_t} = \dot{\mu}_{\tilde{S}_m}$ which by the way depends on the predicted generalized superstate $(\tilde{S}_t)$ at the discrete level. The posterior probability associated with the predicted generalized state is given by:

$$\pi(\tilde{X}_t) = P(\tilde{X}_t, \tilde{S}_t|\tilde{Z}_{t-1}) = \int P(\tilde{X}_t|\tilde{X}_{t-1}, \tilde{S}_t)\lambda(\tilde{X}_{t-1})d\tilde{X}_{t-1}, \tag{5.28}$$

where $\lambda(\tilde{X}_{t-1})$=$P(\tilde{Z}_{t-1}|\tilde{X}_{t-1})$. Accordingly, a message backward propagated from the bottom-level to the higher levels once a new evidence $\tilde{Z}_t$ is received can be exploited to adjust the expectations in hidden variables and estimate the posterior probability $P(\tilde{X}_t, \tilde{S}_t|\tilde{Z}_t)$ which is defined as:

$$P(\tilde{X}_t, \tilde{S}_t|\tilde{Z}_t) = \pi(\tilde{X}_t)\lambda(\tilde{X}_t). \tag{5.29}$$

Consequently, the likelihood message $\lambda(\tilde{S}_t)$ is propagated towards the top-level to update the belief in the hidden discrete variable by updating the weights according to:

$$W_t^n = W_t^n \lambda(\tilde{S}_t), \tag{5.30}$$

$\lambda(\tilde{S}_t)$ is a discrete probability distribution represented by:

$$\lambda(\tilde{S}_t) = \lambda(\tilde{X}_t)P(\tilde{X}_t|\tilde{S}_t) = P(\tilde{Z}_t|\tilde{X}_t)P(\tilde{X}_t|\tilde{S}_t), \tag{5.31}$$

where $P(\tilde{X}_t|\tilde{S}_t) \sim \mathcal{N}(\mu_{\tilde{S}_m}, \Sigma_{\tilde{S}_m})$ denotes a Gaussian distribution with mean $\mu_{\tilde{S}_m}$ and covariance $\Sigma_{\tilde{S}_m}$. While, $\lambda(\tilde{X}_t) \sim \mathcal{N}(\mu_{\tilde{Z}_t}, R)$ denotes a Gaussian distribution with mean $\mu_{\tilde{Z}_t}$ and covariance $R$. The multiplication between $\lambda(\tilde{X}_t)$ and $P(\tilde{X}_t|\tilde{S}_t)$ can be estimated by calculating the Battacharyya distance ($D_B$) as follows:

$$D_B\big(\lambda(\tilde{X}_t), P(\tilde{X}_t|\tilde{S}_t = \tilde{S}_k)\big) = -ln \int \sqrt{\lambda(\tilde{X}_t)P(\tilde{X}_t|\tilde{S}_t = \tilde{S}_k)} d\tilde{X}_t, \tag{5.32}$$

where $\tilde{S}_k \in \tilde{S}$. The vector $D_\lambda$ containing all the $D_B$ values between $\lambda(\tilde{X}_t)$ and all the superstates in the set $\tilde{S}$ is here estimated as:

$$D_\lambda = \Big[D_B\big(\lambda(\tilde{X}_t), P(\tilde{X}_t|\tilde{S}_t = \tilde{S}_1)\big), \dots, D_B\big(\lambda(\tilde{X}_t), P(\tilde{X}_t|\tilde{S}_t = \tilde{S}_L)\big)\Big]. \tag{5.33}$$

Therefore, the vector $\lambda(\tilde{S}_t)$ in terms of probability can be computed as:

$$\lambda(\tilde{S}_t) = \left[\frac{1/D_\lambda(1)}{1/\sum_{l=1}^{L} D_\lambda(l)}, \dots, \frac{1/D_\lambda(L)}{1/\sum_{l=1}^{L} D_\lambda(l)}\right] \tag{5.34}$$

After updating the weights, particles with very low weights are abandoned while particles with high weights are kept and multiplied so that all particles have equal weight; this process is known as sequential importance resampling (SIR). The logic of the M-MJPF is reported in **Algorithm 1** (see Appendix A.1).

### 5.5.4 Hierarchical Abnormality measurements and Generalized errors

We have seen that predictive and diagnostic reasoning can be used to estimate a joint posterior at different hierarchical levels. An additional process can be done here to evaluate the differences between two messages arriving at a given node and:

- estimate the surprise (i.e. the abnormality) using a proper probabilistic distance (e.g. Bhattacharyya distance, Kullback–Leibler divergence).

- calculate the generalized errors by subtracting the stochastic variables related to predictions and observations.

### Discrete Level

This level describes the signal's evolution at a high level of abstraction. In order to evaluate to what extent the current signal's evolution matches the predicted one based on the learned and encoded dynamics in the reference GDBN, we used the Symmetric Kullback-Leibler Divergence ($D_{KL}$) to calculate the similarity between the two messages (that represent discrete probability distributions) entering to node $\tilde{S}_t$, namely, $\pi(\tilde{S}_t)$ and $\lambda(\tilde{S}_t)$ which is formulated as:

$$KLDA = \sum_{i \in \mathcal{S}} Pr(\tilde{S}_t = i) D_{KL}\big(\pi(\tilde{S}_t)||\lambda(\tilde{S}_t)\big) + \sum_{i \in \mathcal{S}} Pr(\tilde{S}_t = i) D_{KL}\big(\lambda(\tilde{S}_t)||\pi(\tilde{S}_t)\big), \quad (5.35)$$

where $Pr(\tilde{S}_t)$ is the probability of occurrence of each superstate picked from the histogram at time instant $t$ and calculated as follows:

$$Pr(\tilde{S}_t) = \frac{fr(\tilde{S}_t = i)}{N}, \quad (5.36)$$

where $fr(.)$ is the frequency of occurrence of a specific superstate $i$ and $N$ is the total number of particles propagated by PF and $\mathcal{S}$ is the set consisting of all the winning particles, such that:

$$\mathcal{S} = \big\{ i | Pr(\tilde{S}_t) > 0 \big\},\ i \in \mathbf{S}. \quad (5.37)$$

The jammer detection decision is made by comparing $KLDA$ to a threshold ($\psi$). Thus, the hypothesis testing problem defined in (5.9) can be rewritten as:

$$\mathcal{H}_0 : KLDA < \psi, \quad \mathcal{H}_1 : KLDA > \psi. \quad (5.38)$$

In addition, the generalized errors ($\tilde{\varepsilon}_{\tilde{S}_t}$) associated with the abnormality indicator (5.35) allows to understand how the jammer affected the discrete level of the reference model. Thus, after detecting the jammer at the discrete level using (5.35), it is possible to explain why we noticed a high abnormality by calculating the difference between the diagnostic message $\lambda(\tilde{S}_t)$ and the predictive message $\pi(\tilde{S}_t)$, such that:

$$\tilde{\varepsilon}_{\tilde{S}_t} = \lambda(\tilde{S}_t) - \pi(\tilde{S}_t), \quad (5.39)$$

**Continuous Level**

This level describes the continuous evolution of the signal guided by the evolution at the discrete level. Measuring the distance between the predictive message $\pi(\tilde{X}_t)$ and $P(\tilde{X}_t|\tilde{S}_t)$ using $D_B$ defined as:

$$CLB = -\ln\Bigg(\mathcal{BC}\big(\pi(\tilde{X}_t), P(\tilde{X}_t|\tilde{S}_t)\big)\Bigg), \tag{5.40}$$

where

$$\mathcal{BC} = \int \sqrt{\pi(\tilde{X}_t)P(\tilde{X}_t|\tilde{S}_t)}d\tilde{X}_t, \tag{5.41}$$

is the Bhattacharyya Coefficient. *CLB* allows evaluating if the predictions at the continuous level match the predictions at the discrete level and thus explains if the signal's dynamics at both the discrete and continuous level evolve according to the rules learned before in a way that it can explain the received signal.
Moreover, it is possible to understand how much the observation supports the predictions using the second abnormality detector at this level defined as:

$$CLA = -\ln\Bigg(\mathcal{BC}\big(\pi(\tilde{X}_t), \lambda(\tilde{X}_t)\big)\Bigg), \tag{5.42}$$

where

$$\mathcal{BC} = \int \sqrt{\pi(\tilde{X}_t)\lambda(\tilde{X}_t)}d\tilde{X}_t. \tag{5.43}$$

Thus, jammer detection at the continuous level is made by comparing *CLA* to a threshold ($\eta$) which is stated as:

$$\mathcal{H}_0 : CLA < \eta, \quad \mathcal{H}_1 : CLA > \eta. \tag{5.44}$$

The abnormality indicators mentioned above can be used to evaluate the radio situation and discover if something wrong occurred in the radio environment that violates the dynamic rules learned in previous experience. However, computing the generalized errors at the continuous level allows discovering the new force (related to the detected jammer) present in the surrounding environment and understanding how much it changed the evolution at the continuous level. The generalized errors at this level are based on the difference between the lateral predictive message $\pi(\tilde{X}_t)$ and the hierarchical messages coming from the bottom level that are projected on the discrete space and on the continuous space. As mentioned before (in Section 5.5.2), the generalized error ($\tilde{\varepsilon}^{[1]}_{\tilde{X}_t}$) projected on the continuous space and associated with (5.42) is defined in (5.17). On the other hand, it would be possible to calculate the Generalized Errors $\tilde{\varepsilon}^{[2]}_{\tilde{X}_t}$ (associated with (5.40)) between the continuous and the observation level by subtracting the posterior from the real measurement that is projected on the discrete level and formulated in the following way:

$$\tilde{\varepsilon}_{\tilde{X}_t}^{[2]} = \begin{cases} \tilde{\mu}\left(\underset{\tilde{S}_t \in S}{\operatorname{argmax}} \lambda(\tilde{S}_t)\right) - \tilde{X}_t \quad if \;\; \tilde{S}_t^{\pi} = \tilde{S}_t^{\lambda}, \\ \tilde{\mu}\left(\underset{\tilde{S}_t \in S}{\operatorname{argmax}} \lambda(\tilde{S}_t)\right) - \tilde{\mu}\left(\underset{\tilde{S}_t \in S}{\operatorname{argmax}} \pi(\tilde{S}_t)\right) \quad if \;\; \tilde{S}_t^{\pi} \neq \tilde{S}_t^{\lambda}, \end{cases} \tag{5.45}$$

where $\tilde{S}_t^{\pi} = \underset{\tilde{S}_t \in S}{\operatorname{argmax}} \pi(\tilde{S}_t)$ is the expected superstate and $\tilde{S}_t^{\lambda} = \underset{\tilde{S}_t \in S}{\operatorname{argmax}} \lambda(\tilde{S}_t)$ is the observed superstate. The distinction between these errors at the continuous level is that the first ($\tilde{\varepsilon}_{\tilde{X}_t}^{[1]}$) is used by KF to correct the predictions and adapt to the new situation during the testing phase, while the second ($\tilde{\varepsilon}_{\tilde{X}_t}^{[2]}$) is used off-line after finishing the experience to discover the dynamic behaviour of the detected jammer that can be encoded in a new dynamic model.

#### Observation Level

At this level we can calculate two generalized errors as well. The first one is related to the difference between actual measurement and prediction projected on the measurement space as defined in (5.16).

On the other hand, since we know which superstates of the model are affected by the jammer (from the discrete level), calculating the distance from the superstates' centroid allows to extract the source of the cause (jammer) that affected the shift noticed at higher levels. So, $\tilde{\varepsilon}_{\tilde{Z}}^{[2]}$ can be calculated in the following way:

$$\tilde{\varepsilon}_{\tilde{Z}_t}^{[2]} = \tilde{Z}_t - H\tilde{\mu}(\underset{\tilde{S}_t \in S}{\operatorname{argmax}} \lambda(\tilde{S}_t)), \tag{5.46}$$

which represent the jammer's signal. This can be explained by the fact that the received signal $\tilde{Z}_t$ in an abnormal situation consists of both the normal signal that the UAV is supposed to receive and the jamming signal. So, subtracting the received jammed signal from its estimated superstate (at the top level) gives the new force signal (i.e. jammer). It is important to recall that estimating the new emergent force is possible since we represented the random hidden variables in generalized coordinates of motion (including the state per se and the corresponding temporal derivative).

### 5.5.5 Jammer extraction and learning dynamic models

The generalized errors at the continuous level and the observation level can be used to extract the jammer's dynamic rules as well as the jammer's signal, which can be used to learn the corresponding dynamic model by clustering those errors following the same approach seen before (to learn the reference GDBN model) for each jamming signal under the $k$-th modulation scheme (see Fig. 5.3). The generalized errors representing the jamming signal under the $k$-th modulation scheme are clustered using GNG, which provides a set $\mathbf{S}^{(\mathbf{k})}$ of discrete regions as mentioned before. Following the same mechanism we used to learn the reference model, i.e., estimating the transition matrix $\Pi^{(k)}$, time-varying transition matrix $\Pi_{\tau}^{(k)}$ and the statistical properties of each super-state in $\mathbf{S}^{(\mathbf{k})}$, we obtain a set $\mathcal{S}_{\mathcal{M}}$ of jamming dynamic models describing the jammers dynamic behaviours under different modulations, such that:

$$\mathcal{S}_{\mathcal{M}} = \{\mathcal{M}_1, \mathcal{M}_2, \dots, \mathcal{M}_K\}. \tag{5.47}$$

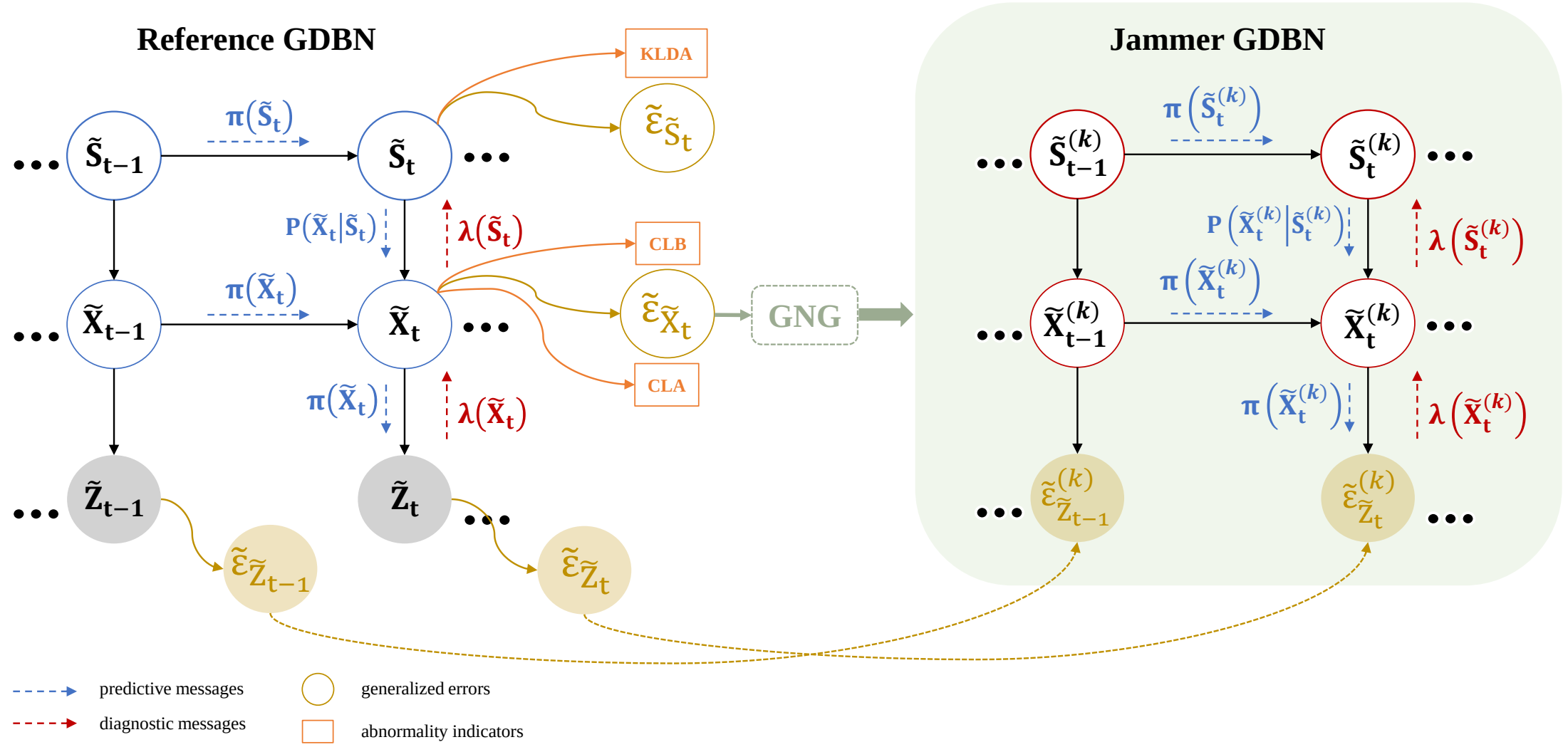

Fig. 5.3 Schematic that illustrate the process of learning a separated model of the detected jammer under the $k$-th modulation mode by exploiting the Generalized Errors. Also, it illustrates the relationship between the reference GDBN and the jammer's GDBN under the $k$-th modulation scheme. The reference GDBN model acts as the generative model that infers the direct cause and deep cause of its own observations and as the process of generating observations for the jammer's model.

However, here we propose to learn additional statistical properties for each $\tilde{S}_m^{(k)} \in \mathcal{M}_k$ (where $\mathcal{M}_k \subset \mathcal{S}_{\mathcal{M}}$), namely, a set $\tilde{\mu}_{\tilde{S}_m^{(k)}}$ of *conditional generalized mean values* defined as:

$$\tilde{\mu}_{\tilde{S}_m^{(k)}} = \left[\tilde{\mu}_{\tilde{S}_m^{(k)}|\tilde{S}_1^{(k)}}, \tilde{\mu}_{\tilde{S}_m^{(k)}|\tilde{S}_2^{(k)}}, \dots, \tilde{\mu}_{\tilde{S}_m^{(k)}|\tilde{S}_M^{(k)}}\right], \tag{5.48}$$

where the *conditional control vectors* ($U_{\tilde{S}_m^{(k)}}$) are encoded such that:

$$U_{\tilde{S}_m^{(k)}} = \left[ U_{\tilde{S}_m^{(k)}|\tilde{S}_1^{(k)}}, U_{\tilde{S}_m^{(k)}|\tilde{S}_2^{(k)}}, \dots, U_{\tilde{S}_m^{(k)}|\tilde{S}_M^{(k)}} \right], \tag{5.49}$$

and a set $\Sigma_{\tilde{S}_m^{(k)}}$ of *conditional covariance matrices* defined as:

$$\Sigma_{\tilde{S}_m^{(k)}} = \left[ \Sigma_{\tilde{S}_m^{(k)}|\tilde{S}_1^{(k)}}, \Sigma_{\tilde{S}_m^{(k)}|\tilde{S}_2^{(k)}}, \dots, \Sigma_{\tilde{S}_m^{(k)}|\tilde{S}_M^{(k)}} \right]. \tag{5.50}$$

This additional information allows understanding not only the dynamic random changes at the discrete level (through the transition probabilities encoded in the transition matrix) but also to discover and represent the force that generated those changes and the rules by which the signal is shifting among them. This realizes the key to predict the dynamic changes of different modulation modes efficiently.

### 5.5.6 Online Automatic Jamming modulation Classification (AJC)

In order to recognize the correct modulation scheme of the detected jammer (i.e. current observation), the UAV will perform multiple predictions in parallel using the learned and stored models in $\mathcal{S}_{\mathcal{M}}$ during the training process and the corresponding statistical properties (defined in (5.48), (5.49) and (5.50)). Thus, at each time instant $t$, we have multiple predictions related to multiple GDBN models, where each model $\mathcal{M}_k$ explains the dynamics of the jammer modulated under the *k-th* modulation scheme (refer to Fig. 5.4). The UAV can evaluate which of these predictions explain the current radio situation by using the abnormality measurement defined in (5.42) applied to the jammer model and defined as:

$$Abn_k = -\ln\left( \mathcal{BC}\big(\pi(\tilde{X}_t^{(k)}), \lambda(\tilde{X}_t^{(k)})\big) \right), \tag{5.51}$$

where

$$\mathcal{BC} = \int \sqrt{\pi(\tilde{X}_t^{(k)})\lambda(\tilde{X}_t^{(k)})} d\tilde{X}_t^{(k)}. \tag{5.52}$$

A set of abnormalities $\mathcal{S}_{Abn}$ is available at each time instant $t$, such that:

$$\mathcal{S}_{Abn}(t) = \{Abn_1, Abn_2, \dots, Abn_K\}. \tag{5.53}$$

The classifier at the UAV is supposed to recognize correctly the modulation scheme of the received signal from a set ($\mathcal{S}_{mod}$) of candidate modulations denoted by integer values, such

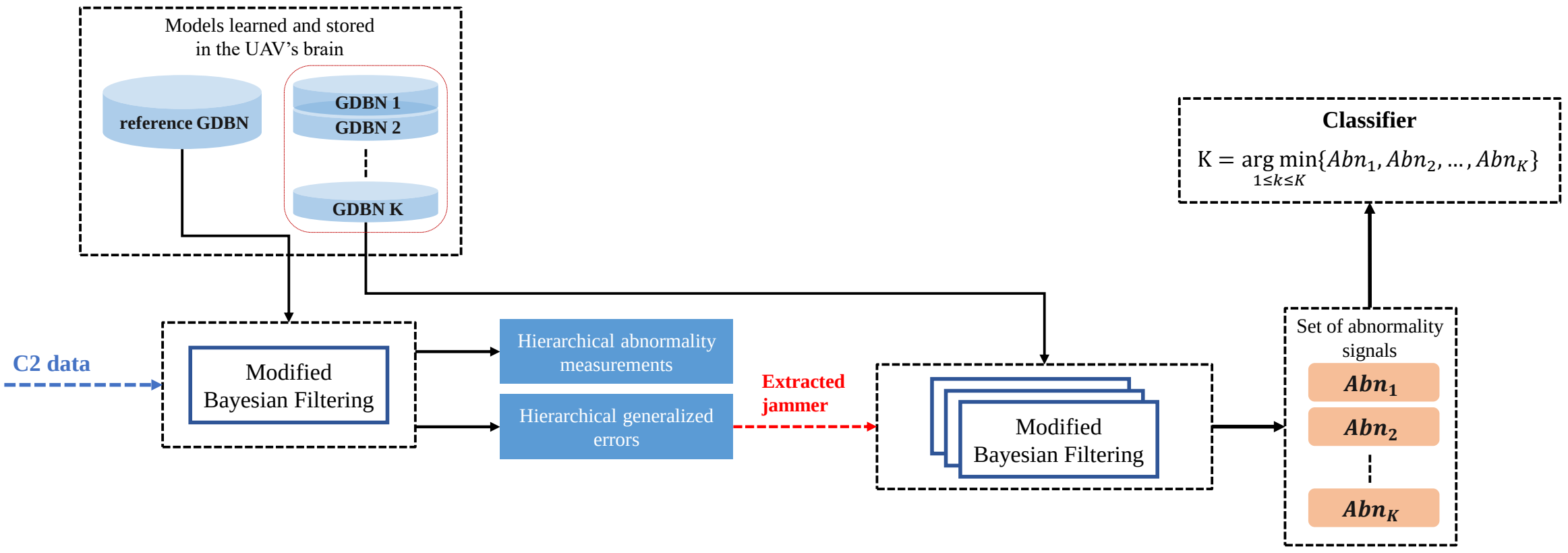

Fig. 5.4 GDBN-based Jamming Modulation Classification Framework.

that:

$$\mathcal{S}_{mod} = \{1, \ldots, K\}. \tag{5.54}$$

Then, the modulation classification can be made by comparing between all the abnormality values and selecting the index of the minimum abnormality in the set $\mathcal{S}_{Abn}(t)$ to recognize the modulation scheme, which is given by:

$$\hat{k}(t) = \underset{1 \leq k \leq K}{\text{argmin}} \{\mathcal{S}_{Abn}\}, \text{where } \hat{k}(t) \in \mathcal{S}_{mod}, \tag{5.55}$$

where $\hat{k}(t)$ is the index of the jamming model $\mathcal{M}_{\hat{k}(t)} \in \mathcal{S}_{\mathcal{M}}$ providing the minimum abnormality (i.e., $Abn_{\hat{k}(t)}$). The probability of correct classification $P_{cc}$ can be used as a performance metric to evaluate the classification task, and it is expressed as follows:

$$P_{cc} = \frac{1}{T} \sum_{t=1}^{T} P(\hat{k}(t) = k(t)|k(t)), \tag{5.56}$$

where $T$ is the total testing time and $P(\hat{k}(t) = k(t)|k(t))$ is the probability that the modulation scheme is correctly predicted as $k(t)$ at time $(t)$. The AJC method is summarized in **Algorithm 2** (see Appendix A.2).

### 5.5.7 Complexity Analysis

In this section, we analyse the complexity of Algorithm 1 and Algorithm 2. The main operations in Algorithm 1 are the joint prediction at multiple levels and the computation of abnormality measurements. At each time instant $t$, the cognitive-UAV performs $N$ predictions at the discrete level and $N$ predictions at the continuous level (i.e., predictions of Generalized state vectors of $4 \times d$ dimensions where $d$ is the number of the sub-carriers sensed by the UAV). Therefore, considering the whole time $T$ (i.e., the total number of OFDM symbols received by the UAV during the total flight time), the time computational complexity is given by $\mathcal{O}(N \times T) + \mathcal{O}(N \times T)$. The complexity of detecting the jammer at two hierarchical levels concerns calculating the Kullback-Leibler-Divergence (*KLDA*) at the discrete level and the Bhattacharyya distance (*CLA* and *CLB*) at the continuous level. The complexity of *KLDA* is $O(I \times M \times T)$ where $I$ is the total number of winning superstates in $\mathcal{S}$ (defined in (5.37)) and $M$ is the total number of superstates in $\mathbf{S}$ (defined in (5.18)). The complexity of *CLA* is $O(2d \times T)$ where $d$ is the number of the sensed sub-carriers. It is worth noting that the predictions and generation of abnormalities must be made before the time of arrival, i.e. if the UAV receives a set of commands each $50ms$ it must perform the predictions and determine whether or not a jammer is attacking within this time. So, the execution of this process and how fast it is, depends on the processors onboard the UAV. The complexity of the main operations involved in Algorithm 1 are summarized in Table 5.1.

Table 5.1 Complexity Analysis of **Algorithm 1**

| **Procedure** | **Complexity Order** |
|---|---|
| Prediction at Discrete Level | $\mathcal{O}(N \times T)$ |
| Prediction at Continuous Level | $\mathcal{O}(N \times T)$ |
| Abnormality Measurement (KLDA) | $\mathcal{O}(I \times M \times T)$ |
| Abnormality Measurement (CLA) | $\mathcal{O}(2d \times T)$ |
| Abnormality Measurement (CLB) | $\mathcal{O}(2d \times T)$ |

In Algorithm 2, at each time instant $t$, predictions and jammer detection (through abnormality measurements) are performed by recalling Algorithm 1 (i.e., M-MJPF). Once an abnormality is detected (i.e., jammer is present), the algorithm uses all the available jamming models to provide multiple predictions at multiple levels. The complexity of the prediction operation at discrete level is given by $\mathcal{O}(K \times N \times T)$ where $K$ is the total number of jamming models learned so far, such that $K \in \mathcal{S}_{\mathcal{M}}$ and $N$ is the total number of particles propagated by PF. Likewise, the complexity of the prediction operation at the continuous level is $\mathcal{O}(K \times N \times T)$. The computational complexity of the abnormality measurement (*Abn*), which is based on the Bhattacharyya distance, is the same as for CLA and CLB and given by $\mathcal{O}(2d \times T)$ (refer to Table 5.2).

Moreover, in the proposed framework, the cognitive-UAV is not transmitting or exchanging any signal with other entities in the network, which usually has a significant impact on the computational complexity and can impose an additional burden on the UAV.

Table 5.2 Complexity Analysis of **Algorithm 2**

| **Procedure** | **Complexity Order** |
|---|---|
| Prediction at Discrete Level | $\mathcal{O}(K \times N \times T)$ |
| Prediction at Continuous Level | $\mathcal{O}(K \times N \times T)$ |
| Abnormality Measurement ($Abn$) | $\mathcal{O}(2d \times T)$ |

# 5.6 Simulation Results and Discussion

## 5.6.1 Simulation setup

The proposed framework for joint detection and classification of multiple jammers is evaluated using simulated data. The UAV trajectory is simulated based on [257]. We study the relationship between the commands and the velocities of the UAV to generate the appropriate bits and consequently generate the LTE signal according to the 3GPP specifications [258] and the important parameters defined in Table 5.3. Similarly, the altered trajectory is extracted from the jammed LTE signal.

Table 5.3 Simulation Parameters

| Parameter | Value |
|---|---|
| BW | 1.4 MHz |
| Duplex mode | FDD |
| $\Delta f$ | 15 kHz |
| Number of PRBs per BW | 6 |
| Sampling frequency | 1.92 MHz |
| $N_{FFT}$ | 128 |
| OFDM symbols per slot | 7 |
| CP length | normal |
| SNR | [-20 dB, ..., +20 dB] |
| C2 Modulation | QPSK |
| Jammer Modulation | $\mathcal{S}_{mod}$ = {BPSK, QPSK, 8-PSK, 16-QAM, 32-PSK, 64-QAM, 256-QAM} |
| Jamming to Signal Power Ratio (JSR) | 6 dB |
| Channel | AWGN |
| Total Radio Frames | 600 |

The UAV flight time is $T_{flight}$=*30s* consisting of 600 samples (aka, 600 sets of commands corresponding to 600 OFDM symbols in time domain (Fig. 5.5-a)). In addition, the UAV extracts the RV from the received PRB every 50 ms, where the RV contains a set of commands transmitted over 9 consecutive sub-carriers in 1 OFDM symbol. Each set of commands will indicate the movement of the UAV in the 3D space.

The output of the digital modulators for both the normal signal and the jammers is normalized based on the average power. The considered situations are:

***(i)*** **Reference Situation**: representing the normal behaviour (without attacks) of the signal related to the original commands sent by the operator (see Fig. 5.5-a) which is used to learn the reference GDBN model. The UAV trajectory during this situation is depicted in Fig. 5.6-a. ***(ii)*** **Abnormal Situation**: during this situation the jammer uses 2 configurations. The first one (used in Section 5.6.2) is related to the jammer who is attacking continuously all the sub-carriers starting from time (in terms of OFDM symbols) $t = 300$ till $t = 600$ in different radio experiences adopting one modulation scheme from $\mathcal{S}_{mod}$ in each experience. While the second (used in Section 5.6.3) is related to the jammer who is attacking from $t = 1$ till $t = 300$ to evaluate the classification performance after learning the jamming models.

### 5.6.2 Learning Reference Model and jamming Models

Initially, the UAV starts perceiving the radio environment and predicting the environmental states using an initial GDBN model, supposing that the signals' dynamics are static. Such an assumption leads to high abnormalities all the time since the UAV fails to predict the actual states of the signals. Exploiting the Generalized Errors calculated during the abnormal situation (using (5.17)) allows the UAV to discover the real dynamics by clustering those errors in an unsupervised manner and store them in the reference GDBN model. After that, the UAV equipped with the reference GDBN can accurately predict the future states of the commands at multiple sub-carriers without being surprised anymore by the observations.

Fig. 5.7 verifies this where we can observe a high abnormality signal all the time by using the initial GDBN (due to the lack of knowledge about the environmental dynamics) and a quasi-zero abnormality signal by using the reference GDBN that encodes the dynamic rules of the signals allowing by that the UAV to perform correct predictions and so avoid surprising states.

After learning the reference GDBN model when the jammers are absent and by facing a new radio experience, the cognitive-UAV can predict the future commands that it is expecting to receive at multiple sub-carriers and consequently detect any jamming attacks at different hierarchical levels using the abnormality measurements (KLDA and CLA) defined in (5.35) and (6.24). We evaluate the detection performance of the proposed approach for multiple jammers with different modulation schemes in different radio conditions by varying the SNR from $-20$ dB to $+20$dB as shown in Fig. 5.8. It can be observed that the cognitive-UAV is capable of detecting the jammer efficiently at the continuous level (through the CLA) with high probability and high accuracy even at very low SNR values regardless of the modulation scheme adopted by the jammer. From the figure, we can also observe that the performance of detecting the jammer at the discrete level (through KLDA) degrades as the SNR decrease, this is due to the fact that the signal dynamics at low SNR become faster and thus the transitions among the discrete variables are speedy which make it difficult to capture the dynamic transitional rules efficiently. Moreover, rapid changes in the accuracy of KLDA in Fig. 5.8(b) from $-4$dB $\leq$ SNR $\leq +4$dB is due to the increased probability of false alarm, which is affected by the rapid transitions of the received signal among the discrete regions. However, the advantage of detecting the abnormality at multiple hierarchical levels is that when the performance degrades at the discrete level, we can rely more on the continuous level for better performance.

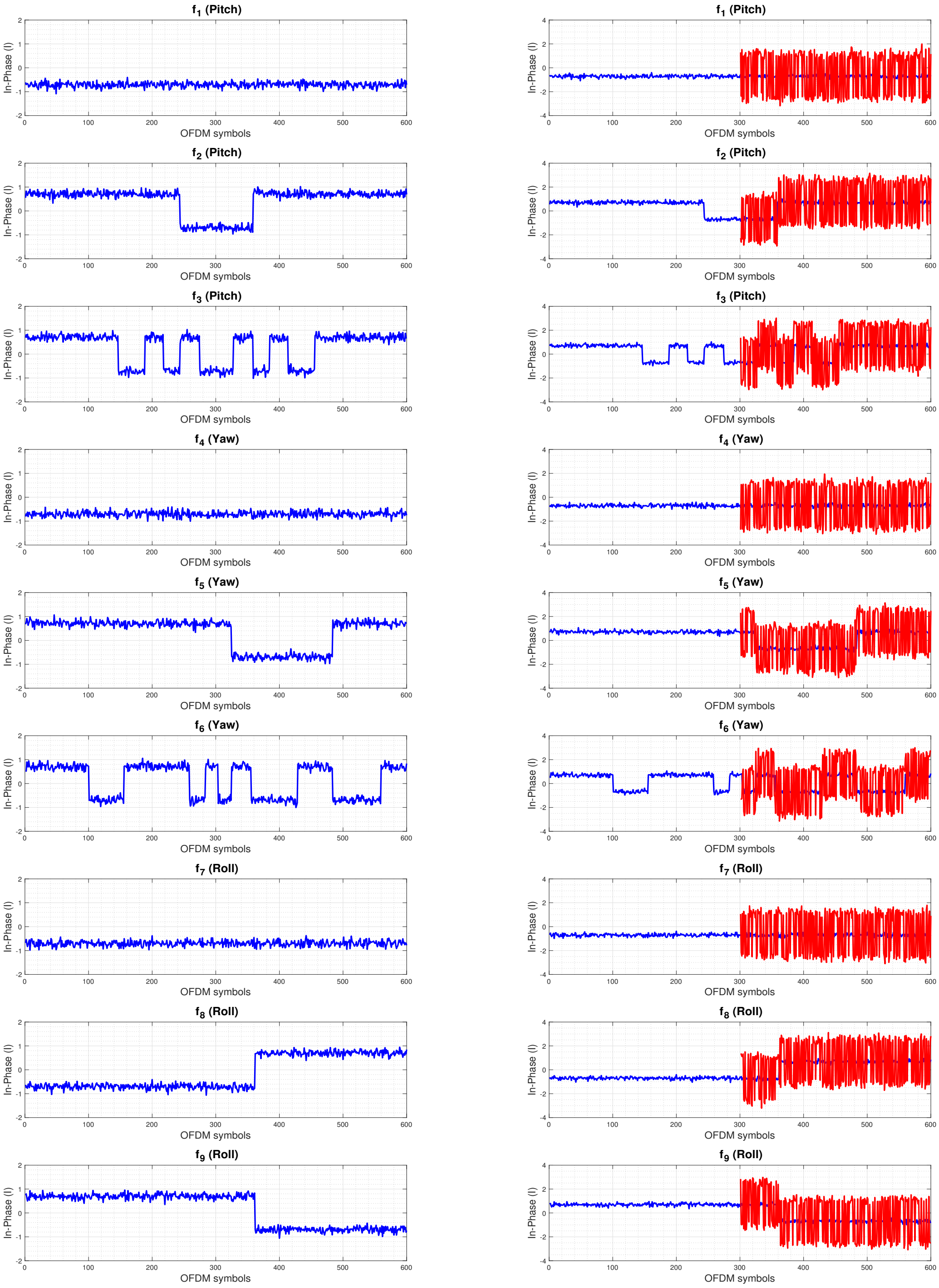

(a) Reference Situation (b) Abnormal Situation

Fig. 5.5 The received commands during different situations. (a) Received commands at multiple sub-carriers during normal situations. (b) An example of the received commands at multiple sub-carriers under jamming attacks (BPSK jammer SNR=14dB).

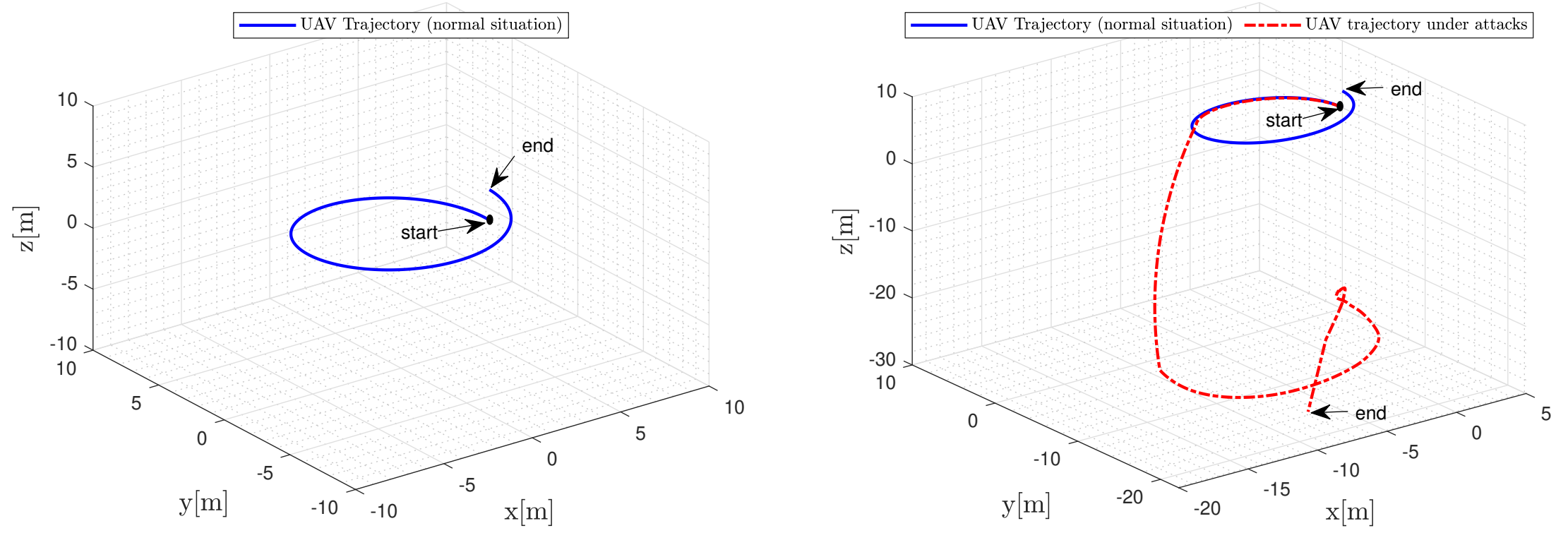

(a) Reference Situation (b) Abnormal Situation

Fig. 5.6 UAV trajectories during different situations. (a) UAV trajectory in a normal situation. (b) An example of UAV trajectory under jamming attacks. Blue and red colors represent the trajectory without and with jammer attacks, respectively.

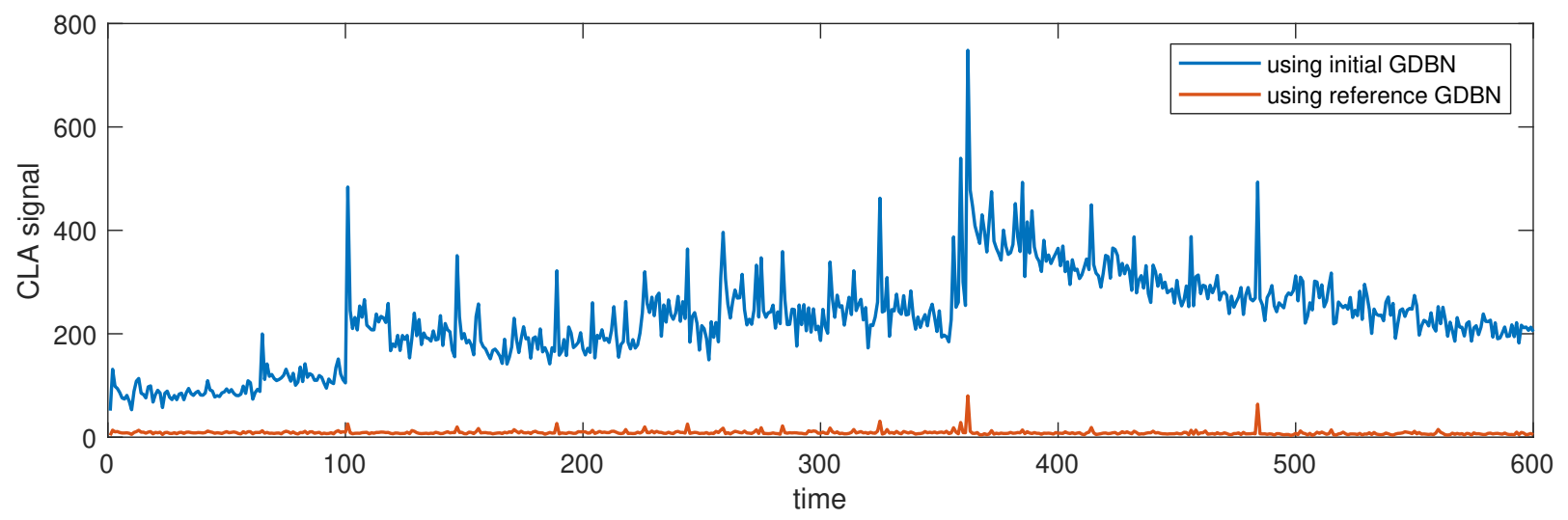

Fig. 5.7 Abnormality indicator at the continuous levels (defined in (6.24)) using the initial GDBN model and the reference GDBN model.

We showed that it is possible to extract and estimate the jamming signal after detecting its malicious activities on the ongoing communication by exploiting the generalized errors defined in (5.46). Fig. 5.9 shows some examples of the I/Q time domain plot of the extracted jammers under different modulation schemes at sub-carrier $f_1$ and 10dB SNR. Fig. 5.10 shows the scatter plots of the extracted jammer and the corresponding ground truth.

The estimated jamming signals in different radio experiences are used to learn separated GDBN models encoding the jamming behaviours under different modulation schemes. After employing the unsupervised method (GNG) to cluster the extracted jammers, we obtain a set of GDBN models forming the set $\mathcal{S}_{\mathcal{M}}$ as defined in (5.47). In this way, the UAV's brain consists of the reference model that describes what commands the UAV is expecting to receive under normal circumstances and another set of models ($\mathcal{S}_{\mathcal{M}}$) representing the dynamic behaviour of multiple jammers using different modulation schemes. In this way, the UAV predicts the future commands using the reference model, calculates the abnormality measurements and the generalized errors. If an abnormality is occured, the UAV will perform multiple predictions in parallel and calculates the corresponding abnormality measurements. The UAV compares among the multiple abnormality measurements and pick the index of the minimum one which is associated with the index of the jamming models in the $\mathcal{S}_{\mathcal{M}}$ to recognize the modulation scheme of the detected jammer.

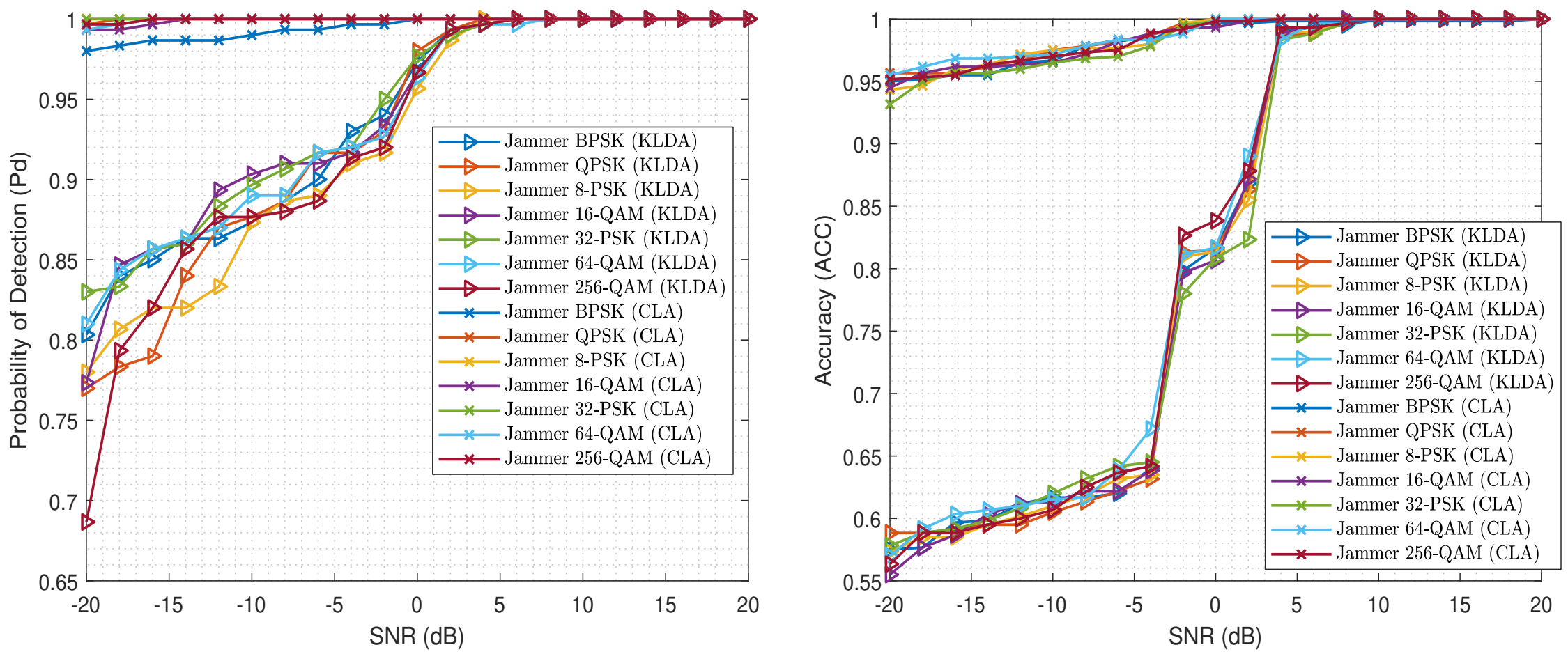

(a) (b)

Fig. 5.8 Performance of detecting multiple jammers modulated under different modulation scheme as varying the SNR. (a) Probability of Detection at hierarchical levels. (b) Detection Accuracy at hierarchical levels.

### 5.6.3 Online Classification Process

In Fig. 5.11, we showed the classification accuracy of the proposed GDBN for each modulation scheme in the candidate set ($\mathcal{S}_{mod}$). We can observe that GDBN achieves high classification accuracy for most of the modulation schemes, especially at $SNR > 5dB$. The low accuracy at low SNRs ($SNR < 0dB$) for the majority of the modulation schemes in $\mathcal{S}_{\mathcal{M}}$ can be explained by the fact that at low SNR the data samples of each modulation are concentrated around the origin (in the complex IQ plane), and the dynamics at low SNR become very fast which makes it difficult to discover and capture these dynamic rules that are

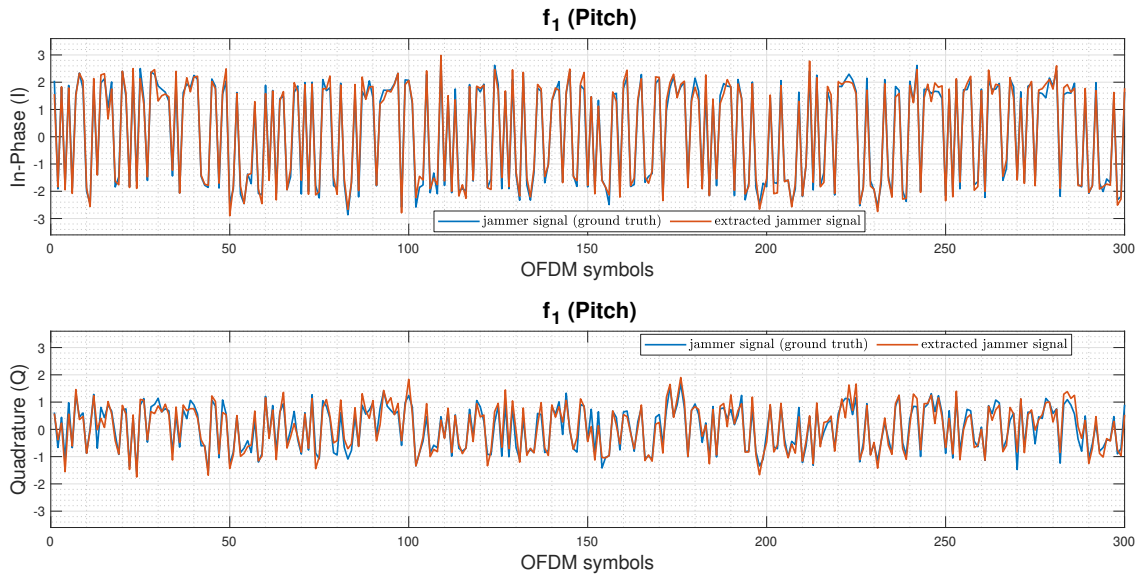

(a) Jammer BPSK

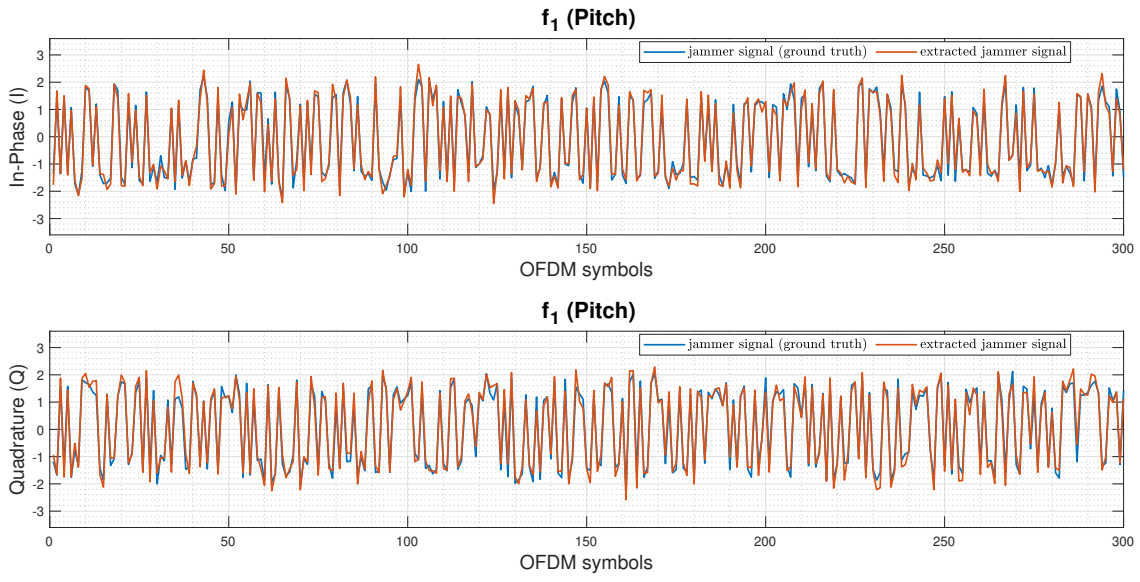

(b) Jammer QPSK

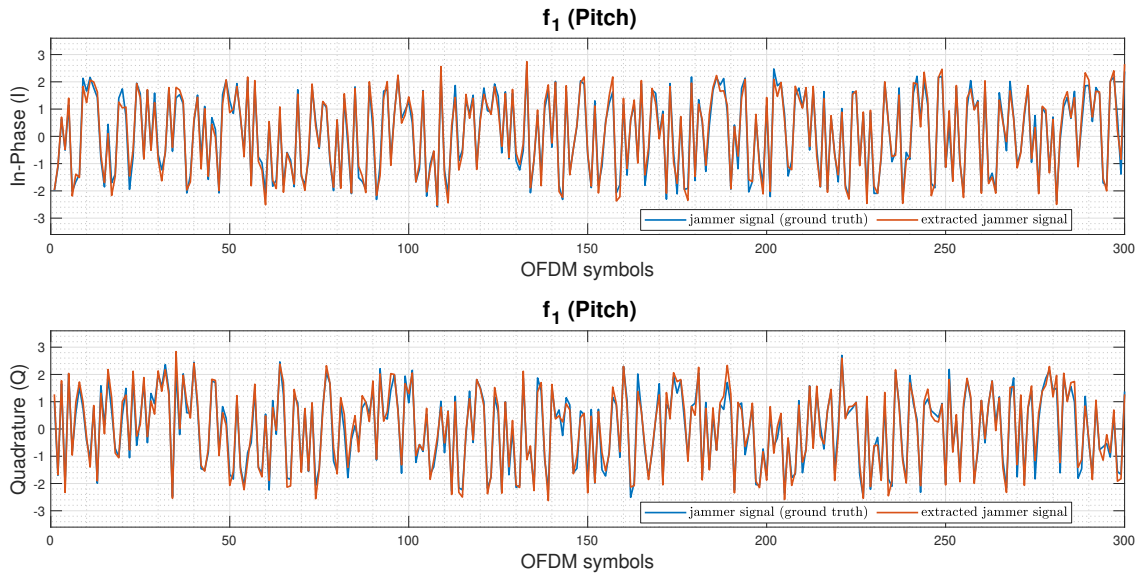

(c) Jammer 8-PSK

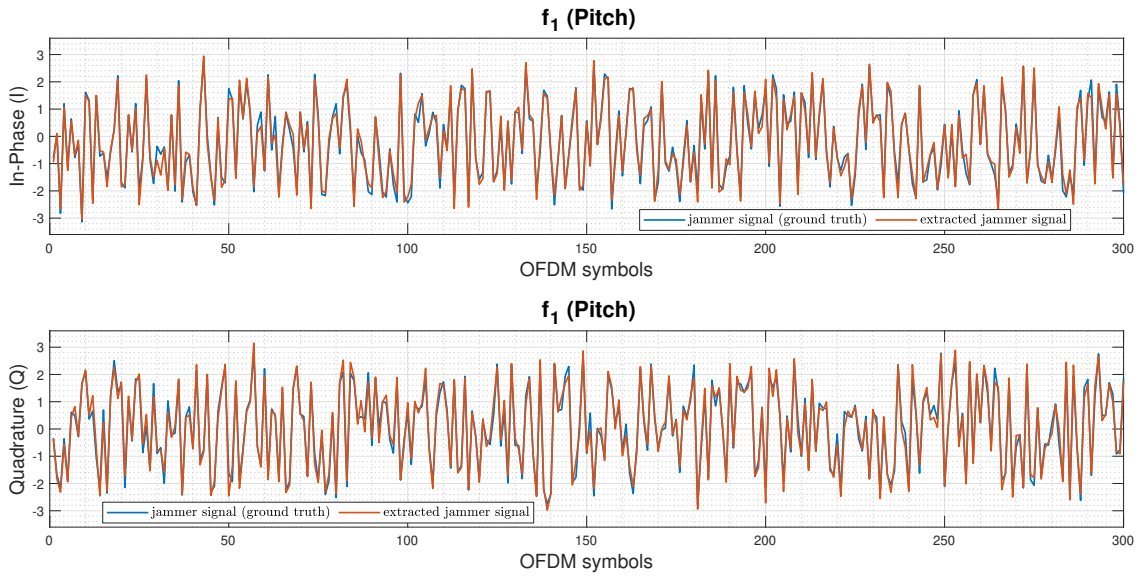

(d) Jammer 16-QAM

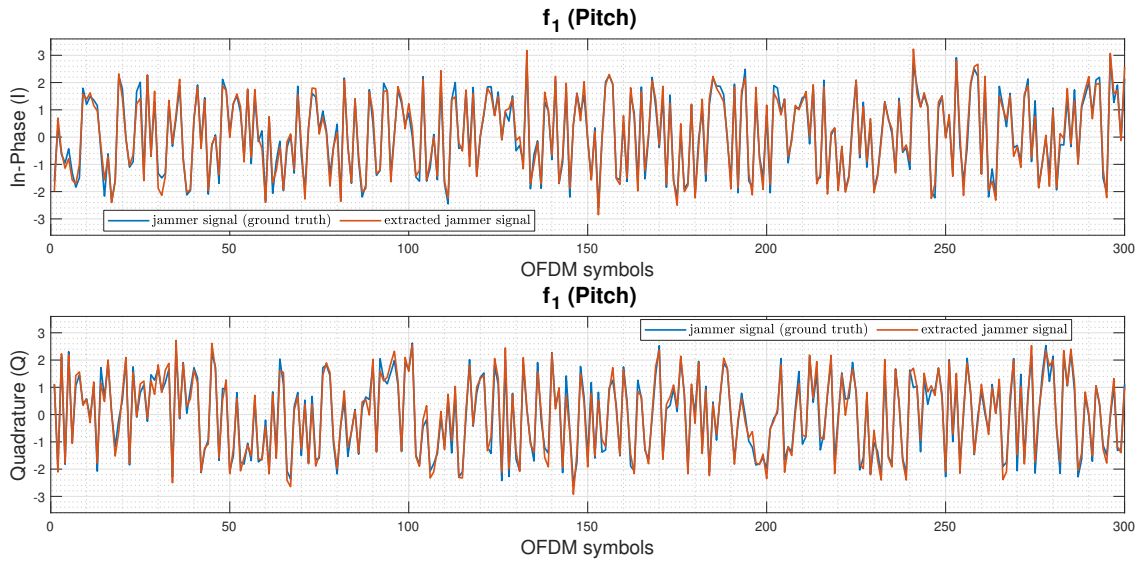

(e) Jammer 32-PSK

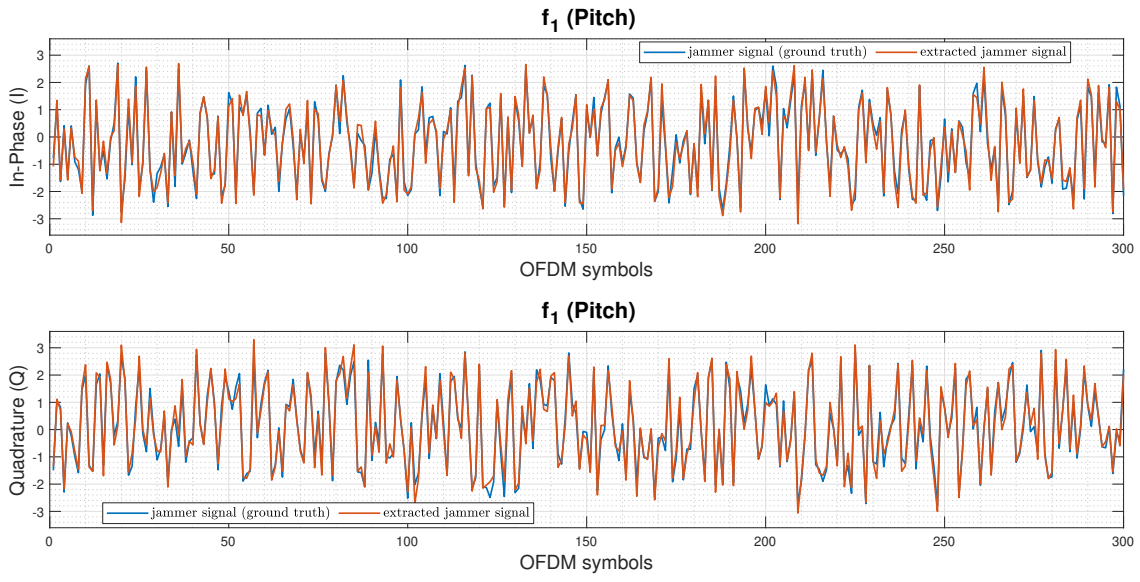

(f) Jammer 64-QAM

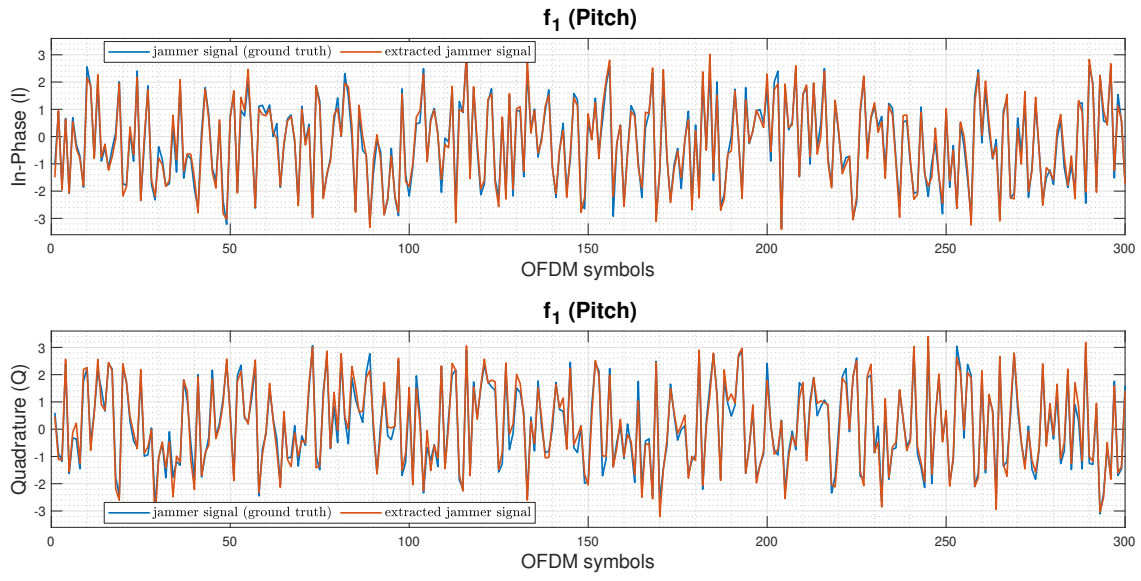

(g) Jammer 256-QAM

Fig. 5.9 I/Q time domain plot of the extracted jamming signals (based on $\tilde{\varepsilon}^{[2]}_{\tilde{Z}_t}$ defined in (5.46)) under different modulation schemes and SNR=10 dB at sub-carrier $f_1$ and of the ground truth jamming signals.

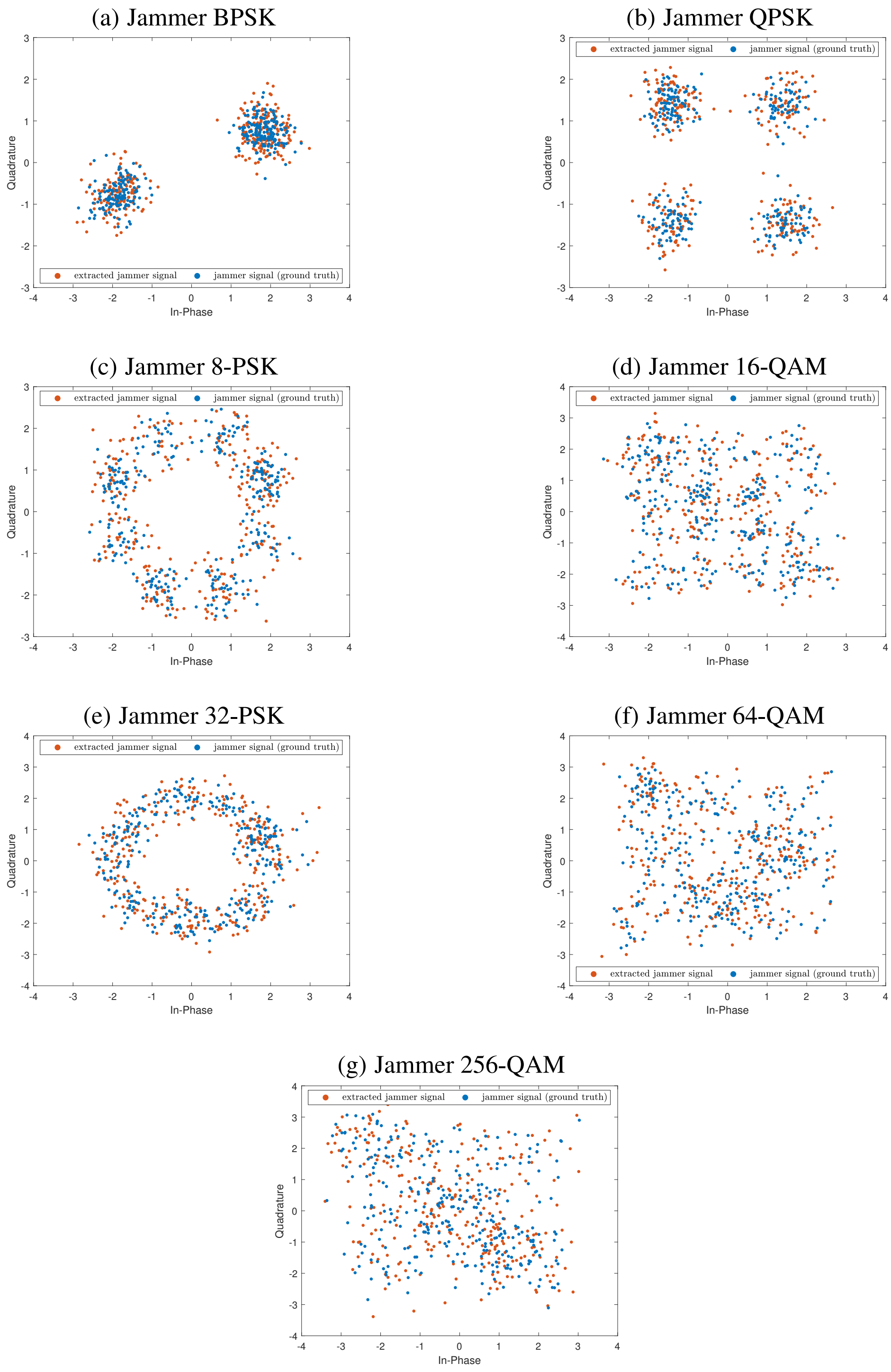

Fig. 5.10 Scatter plot of the extracted jamming signals (based on $\tilde{\varepsilon}^{[2]}_{\tilde{Z}_t}$ defined in (5.46)) and the corresponding ground truth at sub-carrier $f_1$ and 10 dB SNR.

encoded in the GDBN model in an efficient way. Some examples of the resultant confusion matrices at various SNR ratios are exhibited in Fig. 5.12.

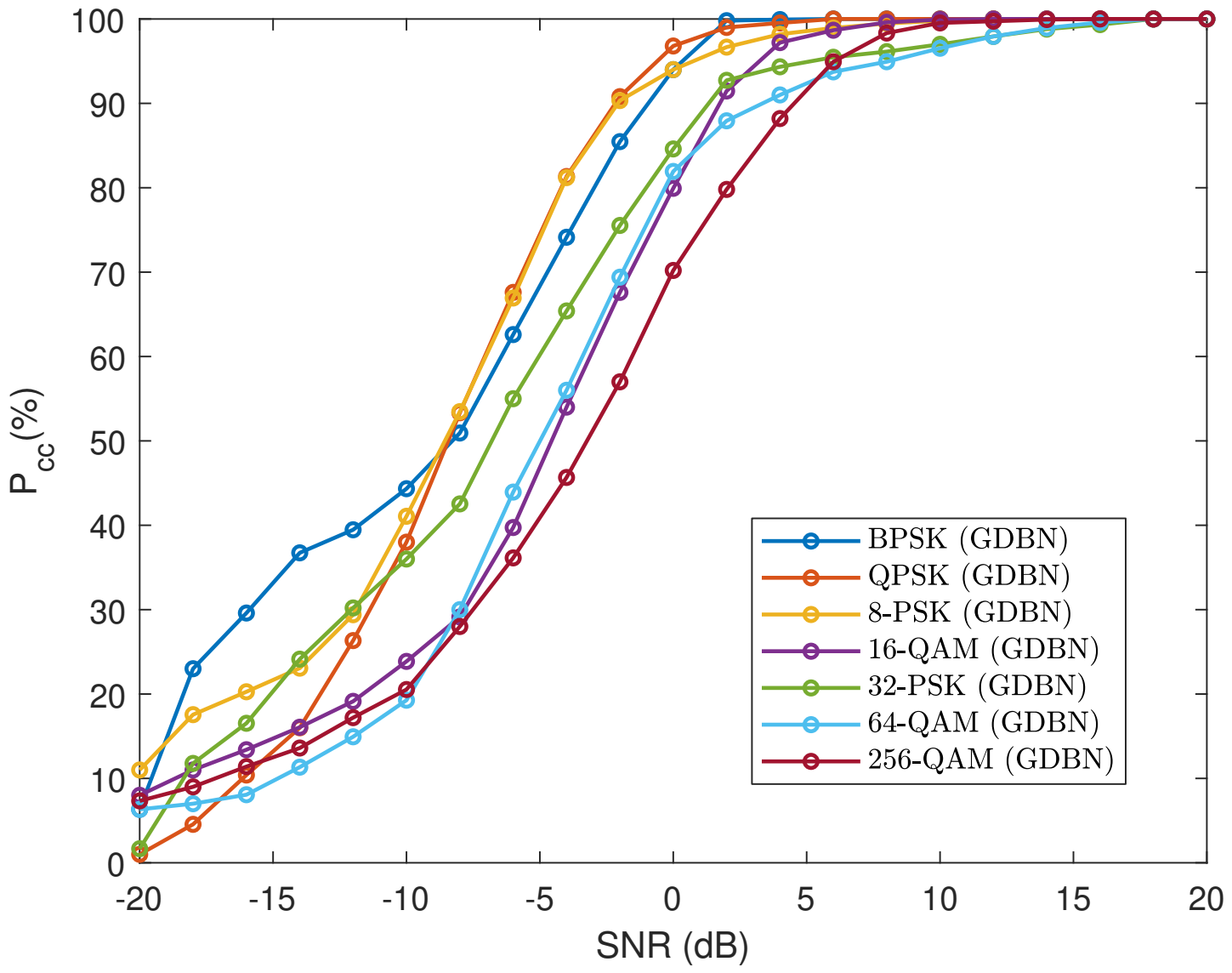

Fig. 5.11 Performance evaluation of the proposed GDBN-based framework: Probability of correct classification for each modulation scheme versus SNR.

In addition, we compare the performance of the proposed GDBN with the Convolutional Neural Network (CNN), the Long Short Term Memory (LSTM) and the Stacked AutoEncoder (SAE). We followed the same approach used to learn the GDBN (thus using the same state vector used as input to the GNG to learn the GDBN) for all three CNN, LSTM and SAE for a fair comparison. For CNN, we used the same configuration (i.e. same number of layers) employed in [302], but with different input, here we used a state vector consisting of IQ components and the corresponding derivatives. While the LSTM used here has 3 layers, one LSTM layer, one fully connected layer, and finally, a dense softmax layer that maps the classified features to one of the available modulations schemes in $\mathcal{S}_{mod}$. The SAE consists of two autoencoders stacked on top of one another (trained in an unsupervised manner) and a softmax layer for classification (trained in a supervised fashion using labels for the training signals). Fig. 5.13, shows the performance comparison between the proposed GDBN, LSTM, CNN and SAE. It can be seen that the GDBN outperforms the other techniques in the majority of the available modulation schemes. This can be understood better by plotting the overall comparison performance, i.e., the average probability of correct classifications among all the $Pcc$ related to each modulation. The overall comparison is depicted in Fig. 5.14, and it shows that the proposed GDBN beats LSTM, CNN and SAE especially at $SNR > 5dB$. This means that the proposed approach succeeded to learn the dynamic proprieties (at hierarchical levels) of the signal under a certain modulation scheme, which allows predicting the future behaviour of the signal based on the rules encoded in that model. In addition, LSTM, CNN and SAE perform the supervised learning by using the input vector along with the labels of each modulation scheme during the learning process, while in the case of GDBN, we followed an unsupervised approach to learn the model. Also, we have seen that GDBN allows to learn the relationships among the random variables (at hidden layers) in the network explicitly and evaluate the situation using abnormality measurements which can be used as self-information by the radio itself to extract new features and learn emergent rules representing new radio situations incrementally. This is difficult in LSTM, CNN and SAE where the dependencies between the hidden variables at multiple layers are viewed as a black box, so results can not be explained. This limitation impacts the capability of learning by understanding which is crucial in CR to learn continually while observing the environment.

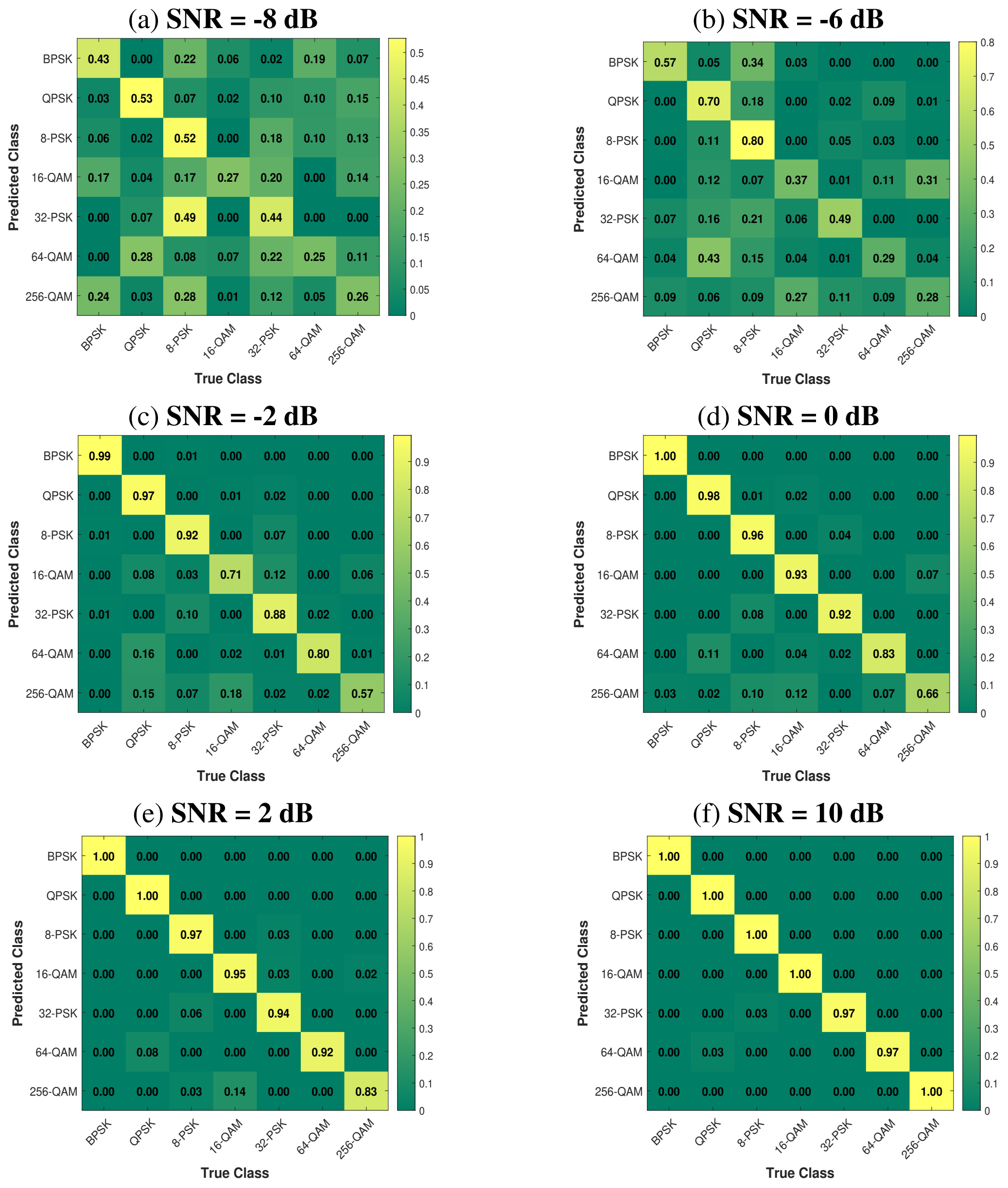

Fig. 5.12 Confusion matrices for the proposed GDBN method at various SNR values.

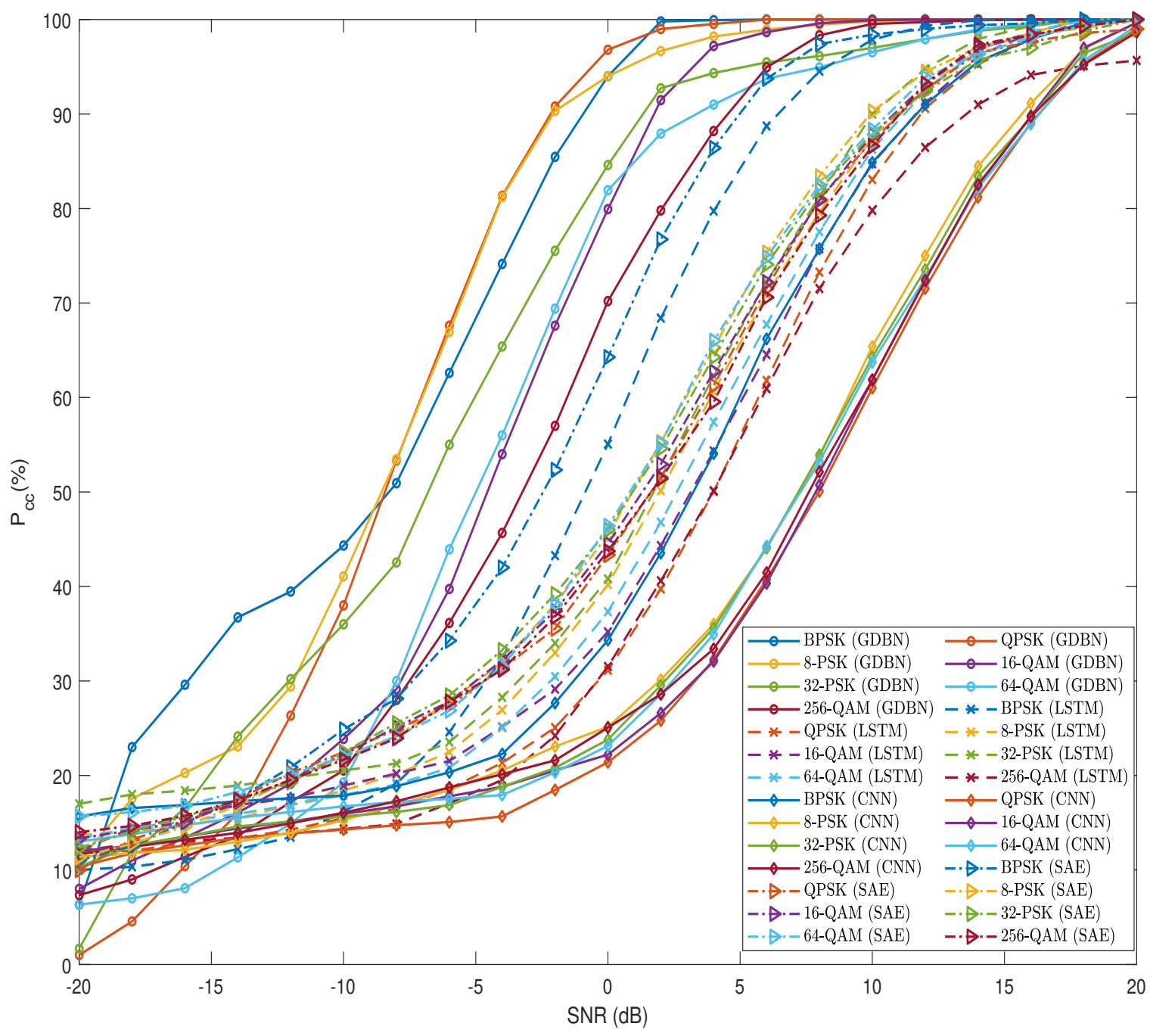

Fig. 5.13 Performance comparison between the GDBN, LSTM, CNN and SAE: Probability of correct classification for each modulation scheme versus SNR.

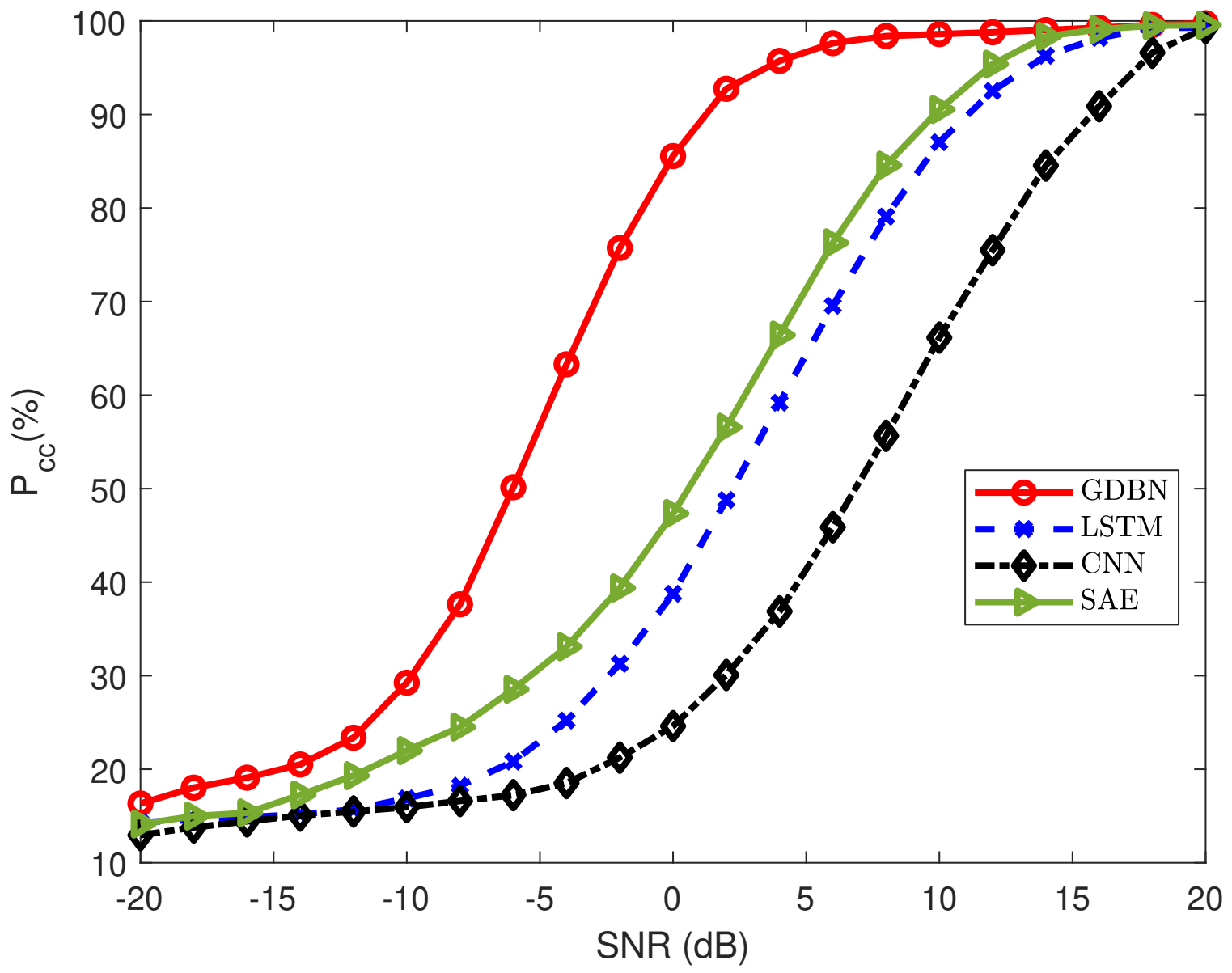

Fig. 5.14 Performance comparison between the GDBN, LSTM, CNN and SAE: Overall probability of correct classification versus SNR.

Furthermore, we analyzed the performance of the proposed framework to automatically classify the detected jammers by changing the number of neurons (i.e. the number of super-states representing the discrete level of the model) used to learn the jamming models. It is to note that in the previous results, we used a fixed number of neurons ($L = 4$) also when we compare with other methods. Considering the influence of the number of neurons on the classification process in addition to the influence of the SNR ratio is of great importance. We applied Bayesian optimization to improve the classification performance by using different $L$ values (related to $L$ models) and finding the model that returns the best classification accuracy ($P_{cc}$). The performance comparison of the jammer's GDBN models with a different number of neurons is shown in Fig. 5.15. We can observe from Fig. 5.15, that increasing the number of neurons ($L$) improves the classification accuracy for high order modulations (i.e., 32-PSK, 64-QAM and 256-QAM). This can be explained by the fact that having a high number of constellations can not be represented efficiently by few neurons since it deteriorates the capture of the dynamic transitions of the data samples under the high order modulations. At low SNR ratios, the confusion between different schemes is high due to the high interference caused by the channel, leading to low classification accuracy. The impact of the number of neurons on the classification accuracy can be better understood by evaluating the overall performance of the proposed approach in classifying various modulation schemes. Fig. 5.16 shows the overall probability of correct classification (over all modulation schemes) and gives a clear idea of how the performance changes as changing the $L$ parameter. It is clear that increasing the number of neurons ($L$) improves the overall classification accuracy.

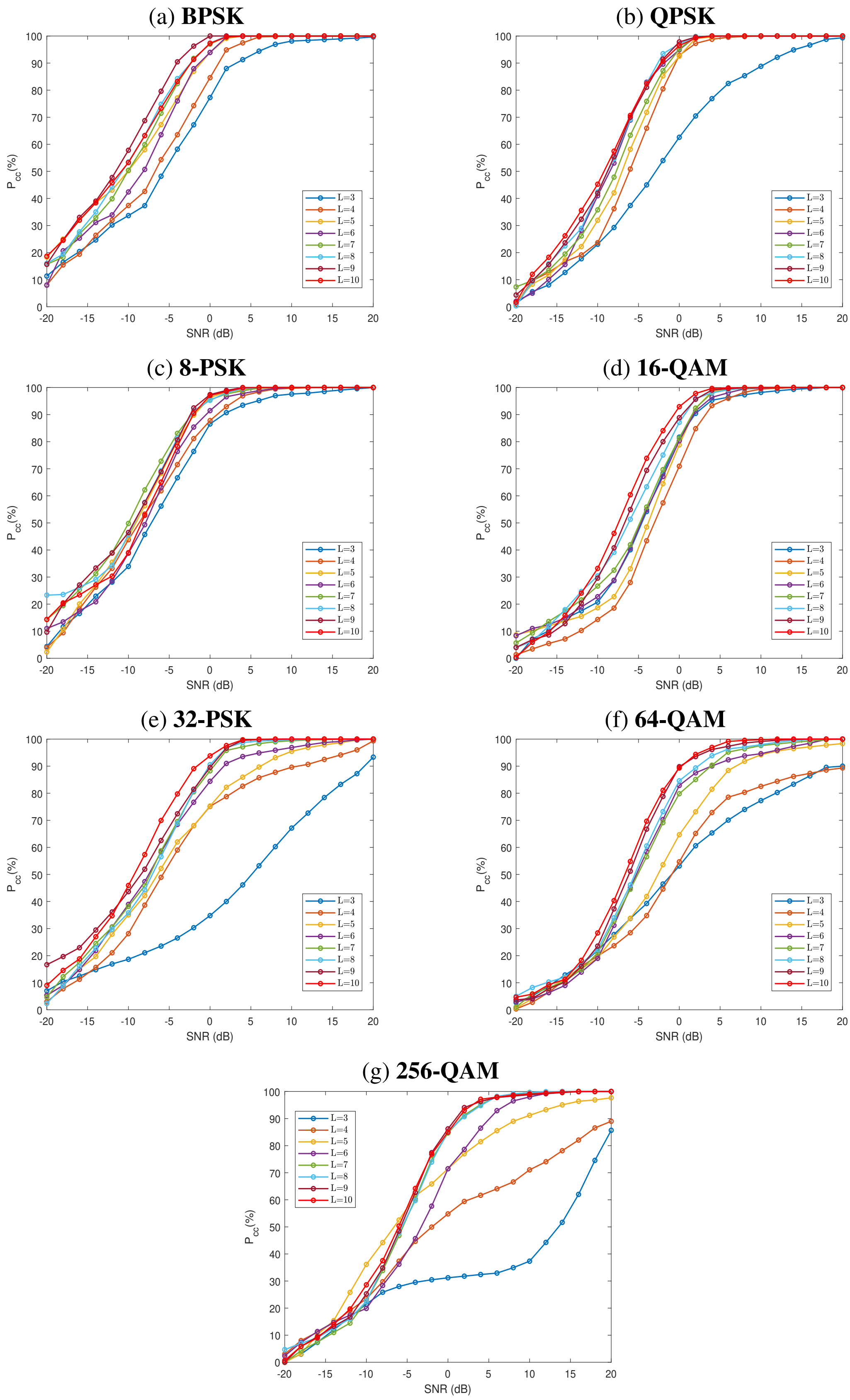

Fig. 5.15 Performance evaluation in terms of classification accuracy of the proposed GDBN method using different number of neurons ($L$) to learn the jamming models.

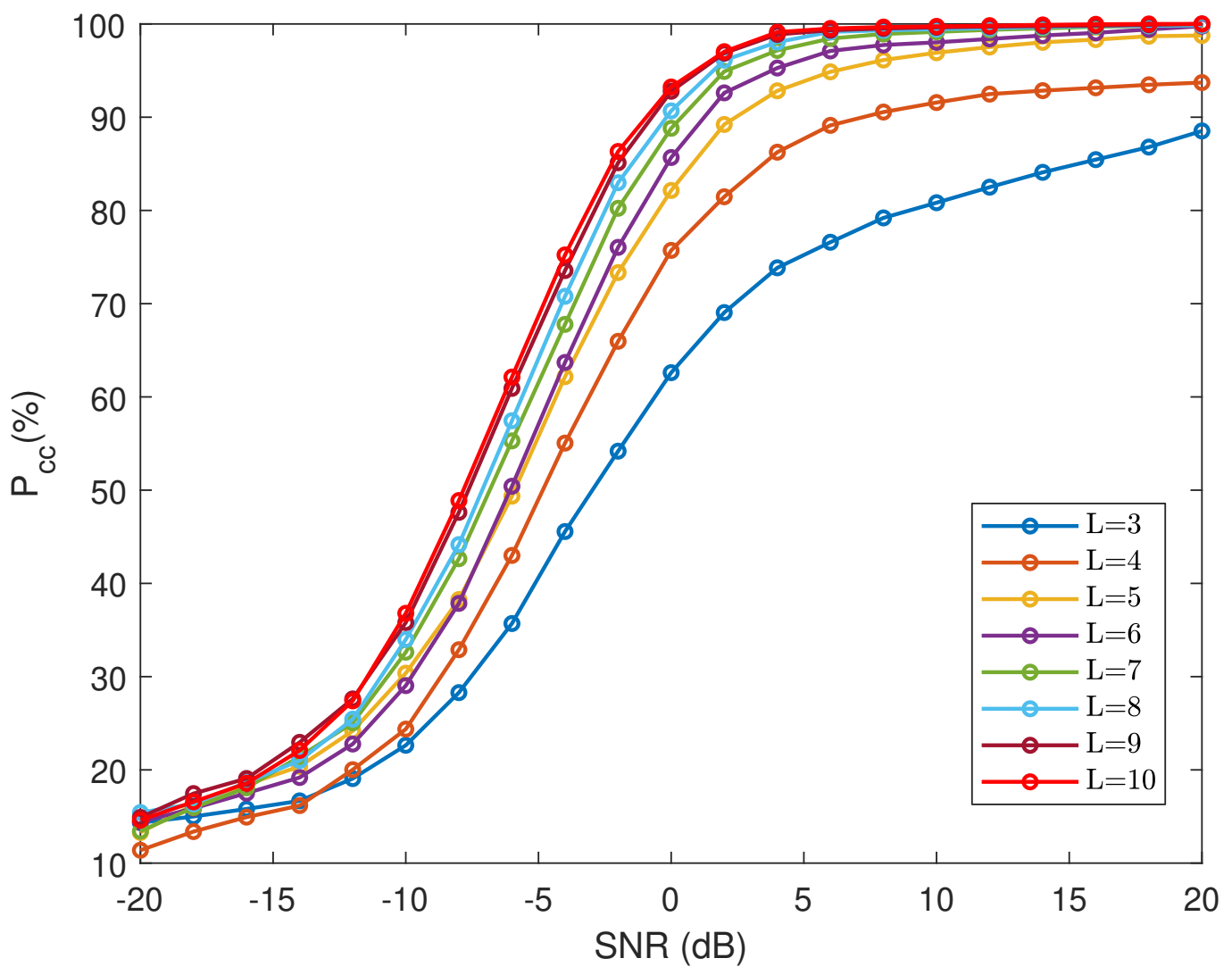

Fig. 5.16 Performance evaluation in terms of overall classification accuracy of the proposed GDBN method using a different number of neurons ($L$) to learn the jamming models.

## 5.7 RadioML Dataset

In this section, we validated the proposed approach for automatic modulation classification using a real dataset.

### 5.7.1 Real Dataset

We employed a real dataset named RadioML (version 2018[2]) [302] to assess the performance of the proposed GDBN-based AMC after running extensive simulations. The candidate set of the modulation schemes picked from the dataset is $\mathcal{S}_{mod}$={OOK, QPSK, 32-PSK, 16-QAM, 32-QAM, 64-QAM, 256-QAM}. The dataset was built using GNU radio block that includes different effects as center frequency offset, sample rate offset, selective fading and AWGN to simulate real-world radio conditions. The dataset consists of about 2 million examples (which we call events) under different SNR values. The SNR ranges from -20dB to +30dB with a step size of 2dB. In our study, each event is divided into two subsets 50% for training and 50% for testing, and the classification task is performed at each event to classify between single complex symbols. The challenge of this approach is the ability to perform accurate classification without requiring many symbols, which improve the latency and make it possible to recognize the modulation scheme in a real-time manner just by processing one symbol, which is crucial in the IoT networks.

[2]Dataset available on https://www.deepsig.ai/datasets

### 5.7.2 Results and discussion

During the training process, the radio learns a GDBN model for each modulation scheme. After this process, the radio possesses $K$ GDBN models stored in its brain, where each model encodes the dynamic behaviour of the corresponding modulation. During the testing process, the radio performs multiple predictions in parallel and calculate the abnormality indicator as defined in (5.51) where the classifier pick the minimum abnormality signal (among the $K$ abnormality signals) as defined in (5.55) to recognize the modulation scheme.

In Fig. 5.17, we showed the classification accuracy of the proposed GDBN for each modulation scheme in $\mathcal{S}_{mod}$. We can observe that GDBN achieves high classification accuracy for most of the modulation schemes, especially at SNR>5dB. The low accuracy at low SNRs (<0dB) for the majority of the modulation schemes in $\mathcal{S}_{mod}$ can be explained by the fact that at low SNR, the data samples of each modulation are concentrated around the origin (in the complex IQ plane) and thus the dynamics become very fast which make it difficult to discover and capture these dynamic rules that are encoded in the GDBN model in an efficient way.

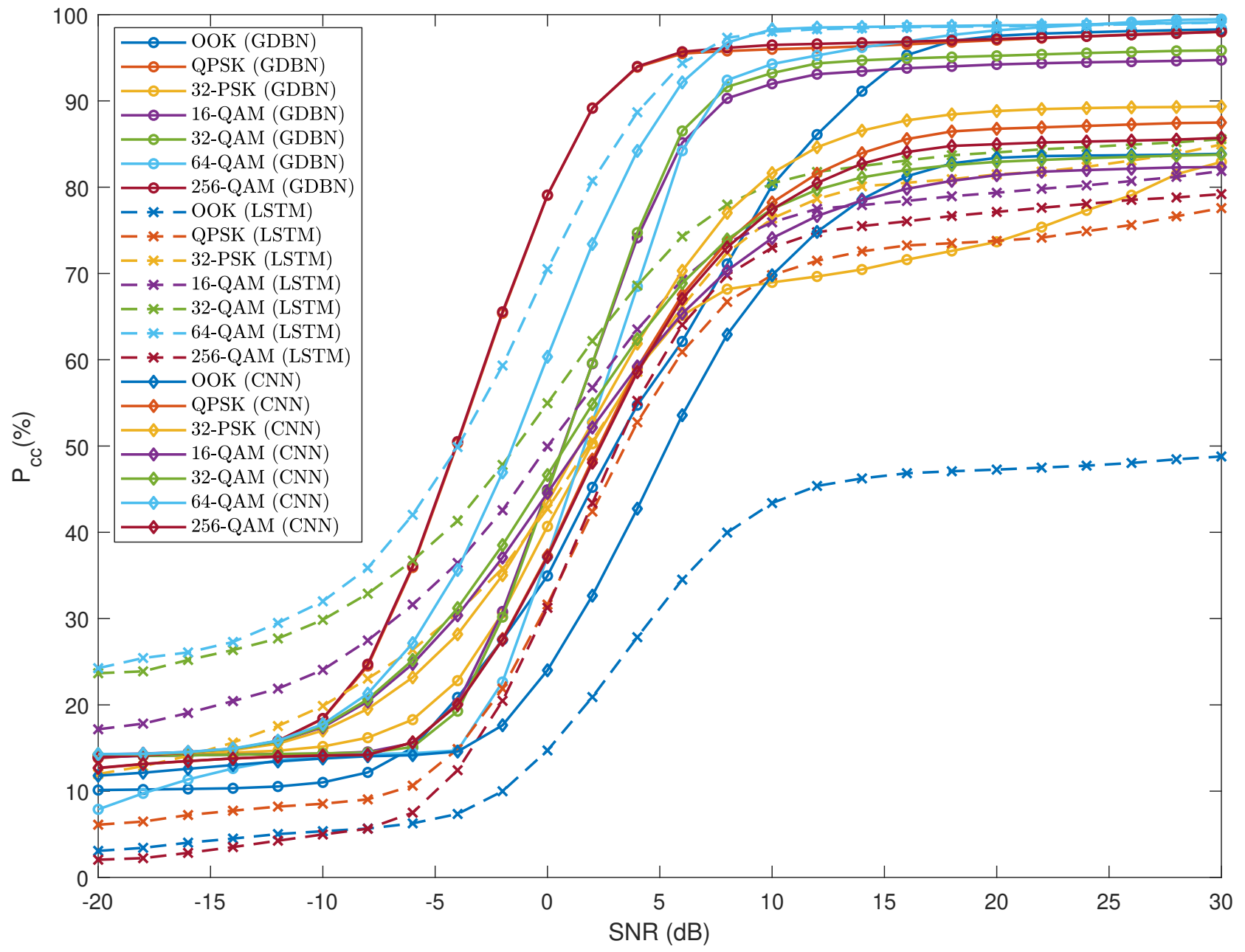

Fig. 5.17 The performance comparison between GDBN, LSTM and CNN.

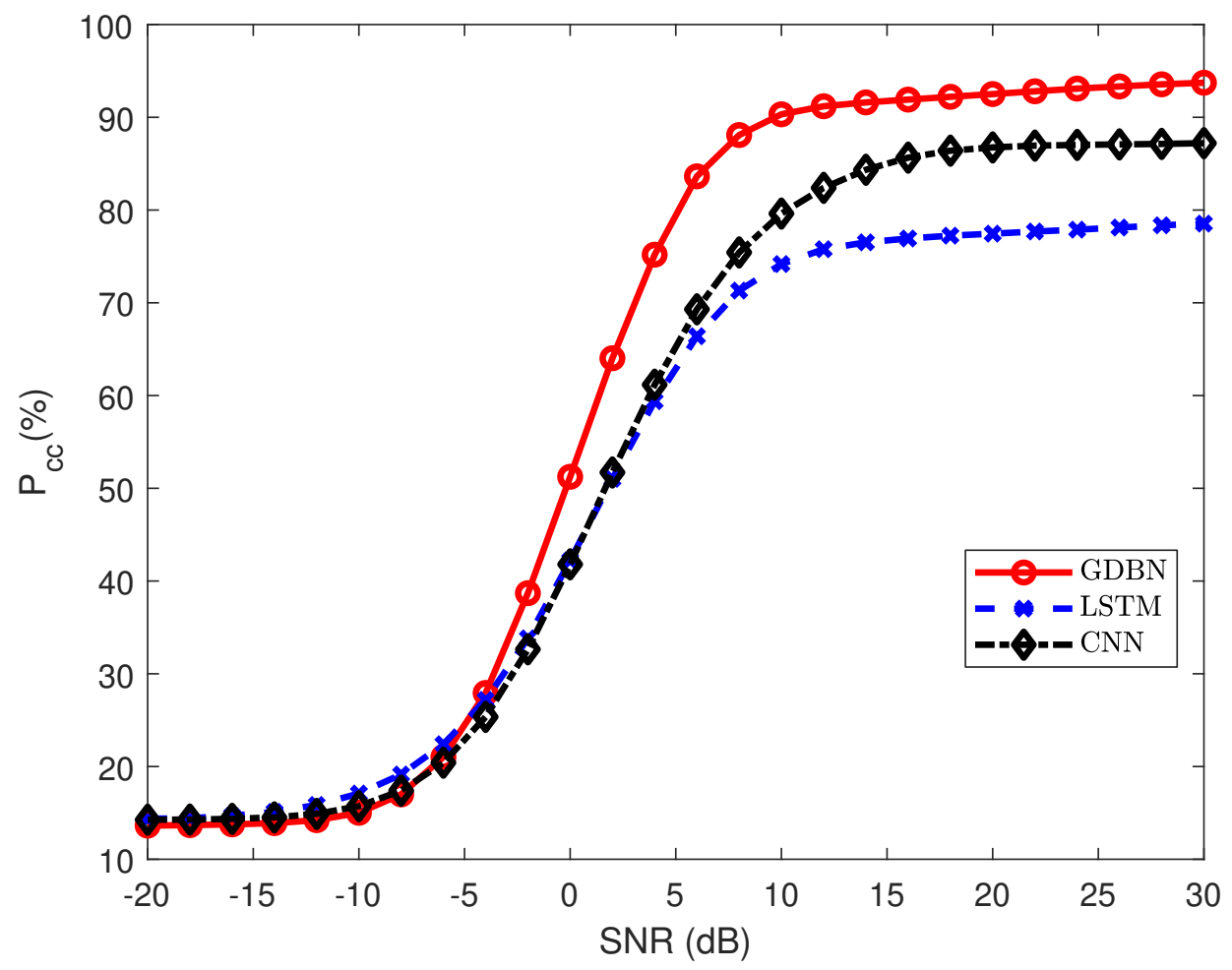

Fig. 5.18 The overall performance comparison among GDBN, LSTM and CNN.

In addition, we compared the performance of the proposed GDBN with the CNN and the LSTM. We followed the same approach used to learn the GDBN (thus using the same state vector used as input to the GNG to learn the GDBN) for both CNN and LSTM for a fair comparison. Moreover, for CNN, we used the same configuration (i.e. same number of layers) employed in [302], but with different input, here we used a state vector consisting of $IQ$ components and the corresponding derivatives. While the LSTM used here has three layers, one LSTM layer, one fully connected layer, and finally a dense softmax layer that maps the classified features to one of the available modulation schemes in $\mathcal{S}_{mod}$. Again, Fig. 5.17 shows the performance comparison between the proposed GDBN, LSTM and CNN. It can be seen that the GDBN outperforms the other techniques in the majority of the available modulation schemes. This can be understood better by plotting the overall comparison performance, i.e., the average probability of correct classifications among all the $P_{cc}$ related to each modulation. The overall comparison is depicted in Fig. 5.18, and it shows that the proposed GDBN beats both LSTM and CNN in the considered scenario, especially at SNR$>$5dB. This means that the proposed approach succeeded in learning the dynamic proprieties (at hierarchical levels) of the signal under a certain modulation scheme, which allows predicting the future behaviour of the signal based on the rules encoded in that model. In addition, LSTM and CNN perform the supervised learning by using the input vector along with the labels of each modulation scheme during the learning process, while in the case of GDBN, we followed an unsupervised approach to learn the model. Also, we have seen that GDBN allows to learn the relationships among the random variables (at hidden layers) in the network explicitly and evaluate the situation using abnormality measurements which can be used as self-information by the radio itself to extract new features and learn emergent rules representing new radio situations incrementally. This is difficult in the case of LSTM and CNN, where the dependencies between the hidden variables at multiple layers are viewed as a black-box, and thus results can not be explained. This limitation impacts the capability of learning by understanding which is crucial in CR to learn continually while observing the environment.

## 5.8 Proposed Framework for Joint Automatic Modulation Format Conversion and Classification (AMCC)

This section proposes a joint automatic modulation conversion and classification (AMCC) framework, which allows an AI-enabled wireless node to predict signals' dynamics of different modulation schemes and explain how it can be transported (converted) with minimal effort and forwarded with higher spectral efficiency. To achieve this goal, we propose a Generalized Filtering framework integrated by Transport Planning to learn the way of converting low-order modulations to high-order modulations, which has also been validated by performing the automatic modulation classification. Simulation results demonstrate the effective performance of our novel framework on converting and classifying multiple modulation formats. This section describes: 1) the representation of the wireless environment; 2) the process of learning the dynamic model where the top-level of the hierarchy can be represented in graphs from which the transport plan can be learned; 3) how the Double Generalized Dynamic Bayesian Network (DGDBN) integrated with the M-MJPF allows to convert the actual modulation format and recognize the modulation scheme of the sensory signals.

### 5.8.1 Signal Model and Problem Formulation

The modulation conversion task can be expressed as finding an optimal transport plan to transfer signals from source distributions to target ones. In this work we focus on converting low-order to high-order modulation formats. Lets take as an example the case of converting a source format (QPSK) to target format (16QAM), first it is important to find the similarities by mapping the $M_1 = 4$ constellations of QPSK to the $M_2 = 16$ constellations of 16QAM. Mapping is based on measuring the corresponding distances associated with the minimum force needed to move data samples from the source constellation to proper target ones. Second, it is important to synchronize the conversion process by performing re-timing or applying temporal delay since converting 1 symbol of QPSK needs 2 symbols of 16QAM as QPSK encodes 2 bits per symbol and 16QAM encodes 4 bits per symbol so to recover the 4 bits correctly in 16QAM we need to have 4 bits transmitted over 2 symbols in QPSK. The main idea is to convert the signal from QPSK to 16QAM and ensure the correct recovery of the original data sequence (bits) which can be validated through the calculation of the Bit-error-rate (BER).

The modulation classification task can be expressed as a classification problem with K modulation formats. The received/sensed signal by the AI-enabled radio can be stated as:

$$r_t = he^{j(2\pi ft+\theta)}\mathrm{s}_t^k + v_t, \tag{5.57}$$

where $h$ is the channel coefficient, $f$ is the frequency offset and $\theta$ is the phase offset and $v_t$ is the Additive Gaussian Noise (AWGN) which is drawn from a zero-mean normal distribution with variance ($\sigma_v^2$). $\mathrm{s}_t^{(k)}$ is the complex symbol under the $k$-th modulation format which can be represented as:

$$\mathrm{s}_t^k = \big[A_m \sum_n a_n g(t-nT_s)\big] cos\big(2\pi(f_c+f_m)t+\phi_0+\phi_m\big), \tag{5.58}$$

if transmitted signal is M-PSK. In (5.58), $A_m$, $a_n$, $T_s$, $f_c$, $f_m$, $\phi_0$ and $\phi_m$ are the modulation amplitude, symbol sequence, symbol period, carrier frequency, modulation frequency, initial phase, and modulation phase, respectively. The function $g(t)$ equals to 1 if $1 \leq t \leq T_s$ and to 0 otherwise. In case of M-QAM $\mathrm{s}_t^k$ can be represented as:

$$\mathrm{s}_t^k = \big[A_m \sum_n a_n g(t-nT_s)\big] cos(2\pi f_c t+\phi_0) + \big[A_m \sum_n b_n g(t-nT_s)\big] sin(2\pi f_c t+\phi_0), \tag{5.59}$$

where $a_n$, $b_n \in [(2m-1-\sqrt{M})], m = \{1,2,\ldots,\sqrt{M}\}$, and the two carriers are modulated by $a_n$ and $b_n$. The aim of the classifier is to identify $s_t$ from the observation $r_t$ and give out $\mathrm{P}(\mathrm{s}_t \in k|r_t)$ where $k \in K$.

The following section explains in detail the proposed framework for the joint modulation conversion and classification following a data-driven approach which can be applicable in a pure software manner using e.g., software defined radios.

## 5.8.2 RF representation

We assume that the hidden dynamics of the physical signals present in the radio environment and generated by various entity nodes (e.g., IoT sensors) can be cast at hierarchical levels and treated as conditionally dependent variables. We represented those variables in generalized coordinates of motion and employed Generalized Filtering [303], i.e., Bayesian filtering in generalized coordinates. The hierarchical causal models representing the signals' dynamics can be structured in a Generalized Dynamic Bayesian Network (GDBN) and formulated in terms of stochastic processes defined as:

$$\tilde{S}_t^{(e)} = f(\tilde{S}_{t-1}^{(e)}) + \tilde{w}_t, \tag{5.60}$$

$$\tilde{X}_t^{(e)} = g(\tilde{X}_{t-1}^{(e)}, \tilde{S}_t^{(e)}) + \tilde{w}_t = A\tilde{X}_{t-1}^{(e)} + BU_{\tilde{S}_t^{(e)}} + \tilde{w}_t, \tag{5.61}$$

$$\tilde{Z}_t^{(e)} = h(\tilde{X}_t^{(e)}) + \tilde{v}_t = H\tilde{X}_t^{(e)} + \tilde{v}_t. \tag{5.62}$$

The dynamics at the top level of hierarchy evolve according to (5.60) where $f(.)$ is a non-linear function determining the causal transitions at that level and depends on time-varying transition probabilities which is subject to random noise $\tilde{w}_t$ that is assumed to be drawn from a zero multivariate normal distribution with covariance $\Sigma_{\tilde{w}_t}$ such that $\tilde{w}_t \sim \mathcal{N}(0, \Sigma_{\tilde{w}_t})$. $(.)^{(e)}$ symbolises an entity node adopting a particular digital modulation format. The top level holds a belief about the level below and guides the prediction at that level after indicating the control vector ($U_{\tilde{S}_t^{(e)}}$) to be used as pointed out in (5.61). Thus, the Generalized States' (GS) evolution depends on the previous GS ($\tilde{X}_{t-1}^{(e)}$) and the force enocoded in $U_{\tilde{S}_t^{(e)}}$. In (5.61), $A \in \mathbb{R}^d$ and $B \in \mathbb{R}^d$ are the dynamic and control model matrices that parametrize the linear dynamics at the medium level of hierarchy where $d$ stands for the signal's features (I,Q components) dimensionality. Hence, (5.60) and (5.61) allow making inferences about hidden states causing sensory signals that can be interpreted by prior beliefs about the underlying model generating those signals. The bottom level of the hierarchy stands for the observed sensory signals modeled in (5.62) where H is the observation matrix that parametrize the observation model and maps hidden GSs ($\tilde{X}_{t-1}^{(e)}$) to Generalized Observations ($\tilde{Z}_{t-1}^{(e)}$); $\tilde{v}_t \sim \mathcal{N}(0, \Sigma_{\tilde{v}_t})$ is the measurement noise.

### 5.8.3 Learning GDBN

We assume that the radio starts perceiving the surrounding environment supposing that no signals are present in the spectrum and observations are only subject to random noise. At this stage ($1^{st}$ iteration), the radio adopts an initial GDBN model consisting of two levels, the GS level and observation level and employs an Unmotivated Kalman Filter (UKF) under this static assumption by using the following simplified dynamic model:

$$\tilde{X}_t^{(e)} = A\tilde{X}_{t-1}^{(e)} + \tilde{w}_t. \tag{5.63}$$

Predicting future states according to (5.63) leads to notice deviations between what the radio is expecting and what it is actually measuring all the time and allows to calculate innovations at the observation level following:

$$\tilde{y}_t^{(e)} = \tilde{Z}_t^{(e)} - H\tilde{X}_t^{(e)}. \tag{5.64}$$

Then, the innovations projected on the GS level can define the Generalized Errors (GEs) in the following form:

$$\tilde{\mathcal{E}}_{\tilde{X}_t^{(e)}} = [\tilde{X}_{t-1}^{(e)}, P(\dot{\mathcal{E}}_{\tilde{X}_t^{(e)}})] = [\tilde{X}_t^{(e)}, H^{-1}\tilde{y}_t^{(e)}]. \tag{5.65}$$

The GEs are treated by the radio as self-information to discover the emergent dynamic rules present in the environment (acquire knowledge about the surroundings) and build up its own long term memory. In order to learn the GDBN model, the radio clusters in an unsupervised manner the GEs ($\tilde{\mathcal{E}}_{\tilde{X}_t^{(e)}}$) calculated during its first operation in the field as mentioned previously and used the GNG for that purpose. GNG outputs a set of Generalized Superstates (GSS) $\mathcal{V}_e = \{\tilde{S}_1^{(e)}, \tilde{S}_2^{(e)}, \ldots, \tilde{S}_N^{(e)}\}$ Since each GSS in $\mathcal{V}_e$ is assumed to follow a multivariate Gaussian distribution, it can be represented by its sufficient statistics, i.e., the generalized mean $\tilde{\mu}_{\tilde{S}_n^{(e)}} = [\mu_{\tilde{S}_n^{(e)}}, \dot{\mu}_{\tilde{S}_n^{(e)}}]$ and covariance matrix $\Sigma_{\tilde{S}_n^{(e)}}$ where $n \in \{1, 2, \ldots, N_e\}$. Analysing the signal's dynamic transitions among the learned clusters allows estimating the transition probabilities $\pi_{ij} = P(\tilde{S}_t^{(e)} = i | \tilde{S}_{t-1}^{(e)} = j)$ to learn the transition matrix $\Pi \in \mathbb{R}^{N_e \times N_e}$

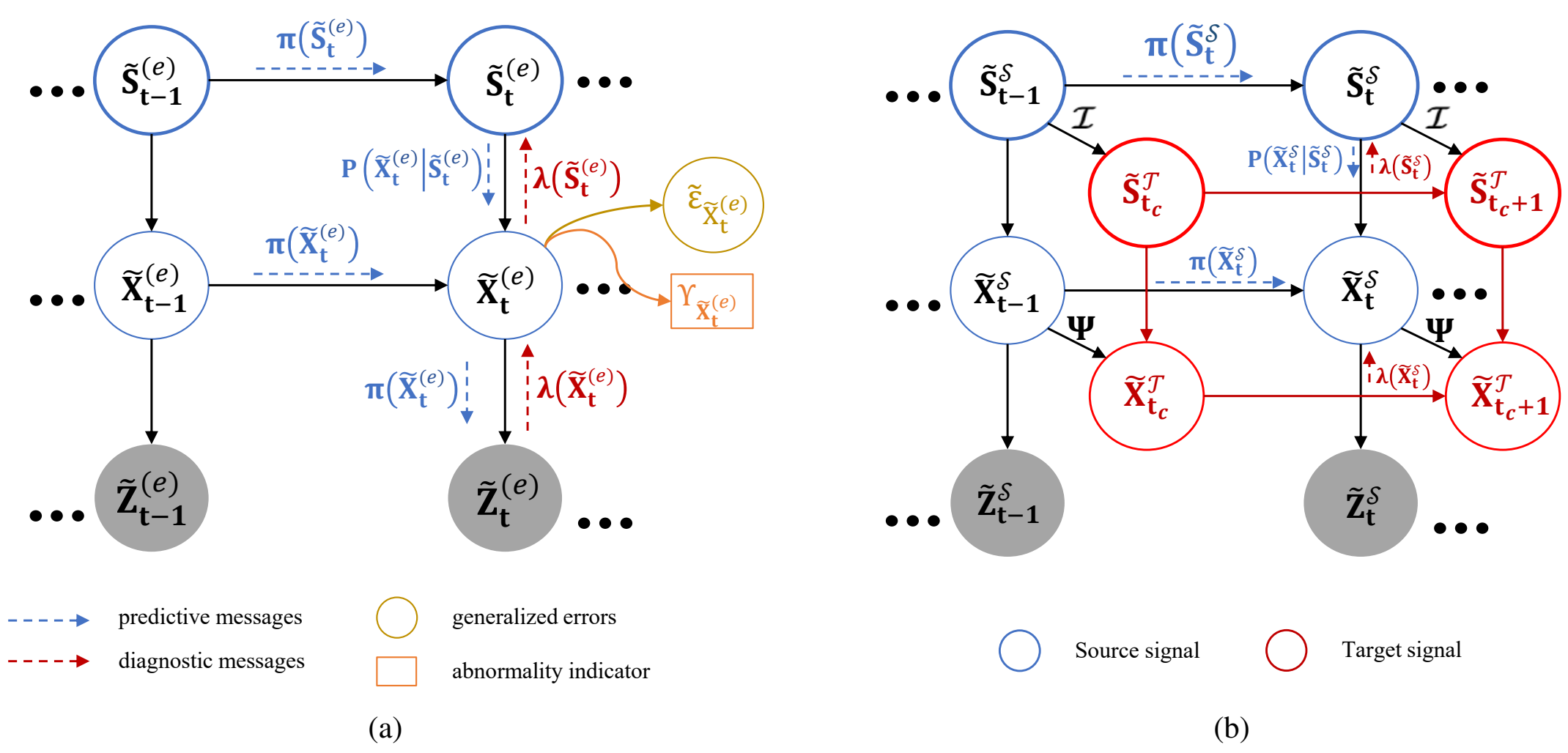

Fig. 5.19 The proposed graphical representations for joint prediction and conversion. (a) GDBN, (b) DGDBN.

defined as:

$$\Pi = \begin{bmatrix} \pi_{11} & \pi_{12} & \cdots & \pi_{1N} \\ \pi_{21} & \pi_{22} & \cdots & \pi_{2N} \\ \vdots & \vdots & \ddots & \vdots \\ \pi_{N1} & \pi_{N2} & \cdots & \pi_{NN} \end{bmatrix}, \tag{5.66}$$

where $\sum_{j}^{\mathrm{N}_e} \pi_{ij} = 1$, such that, $i, j \in \mathrm{N}_e$. Furthermore, estimating the time-varying transition matrix denoted by $\Pi_\tau$ is of great interest due to the dynamic nature of the radio environment that varies with time, such that:

$$\Pi_\tau = \begin{bmatrix} \pi_{11,\tau} & \pi_{12,\tau} & \cdots & \pi_{1N,\tau} \\ \pi_{21,\tau} & \pi_{22,\tau} & \cdots & \pi_{2N,\tau} \\ \vdots & \vdots & \ddots & \vdots \\ \pi_{N1,\tau} & \pi_{N2,\tau} & \cdots & \pi_{NN,\tau} \end{bmatrix}, \tag{5.67}$$

where $\pi_{ij,\tau} = \mathrm{P}(\tilde{S}_t^{(e)} = i | \tilde{S}_{t-1}^{(e)} = j, \tau)$ characterize a new condition on the transition from $i$ to $j$ which depends on the time elapsed ($\tau$) in state $i$. The learned neurons (or GSS), along with their sufficient statistics and transition matrices (forming the so-called Vocabulary) can be represented in a Graph where vertices express the GSSs and edges express the transitions among vertices, as we will discuss in the following section.

The learning procedure shown in this section is repeated during different experiences involving multiple entities (e) transmitting the same information but adopting different modulation formats to learn their correspondent vocabularies represented as different graphs. Since the transmitted information is the same, then it can become helpful to provide the AI-enabled radio with the capability of translating the languages of the transmitted signals, e.g., translating a QPSK signal carrying an image into a 16QAM signal carrying the same image. In this sense, the information carried by the two signals is the same but represented in different languages (modulation schemes).

### 5.8.4 Graph Matching and Transport Plan Learning

Assume that we have two entities denoted as source ($\mathcal{S}$) and target ($\mathcal{T}$). These entities can be represented as two graphs, denoted as $\mathcal{G}_\mathcal{S} = (\mathcal{V}_\mathcal{S}, \mathrm{E}_\mathcal{S})$ and $\mathcal{G}_\mathcal{T} = (\mathcal{V}_\mathcal{T}, \mathrm{E}_\mathcal{T})$. $\mathcal{V}_\mathcal{S}$ and $\mathrm{V}_\mathcal{T}$ are a set of $\mathrm{N}_\mathcal{S}$ and $\mathrm{N}_\mathcal{T}$ vertices, each with a set $\mathrm{E}_\mathcal{S}$ and $\mathrm{E}_\mathcal{T}$ of edges. We assume that each graph is connected, directed and edge weighted. The graphs are obtained from different sources (signals with different modulation formats) and have different structures. Thus, we need to define a notion of similarity (distance) to compare them. The aim of this comparison is to form a matching map (one-to-one or one-to-many) allowing to capture all the possible correspondences among the two graphs (i.e., among $\mathcal{V}_{\mathcal{T}}$ and $\mathcal{V}_{\mathcal{S}}$) and consequently to compute an optimal transport plan allowing to convert $\mathcal{G}_{\mathcal{S}}$ into $\mathcal{G}_{\mathcal{T}}$ (i.e., to transfer signals from $\mathcal{G}_{\mathcal{S}}$ to $\mathcal{G}_{\mathcal{T}}$). Following are the essential steps for learning the transport plan (refer to Fig. 5.20).

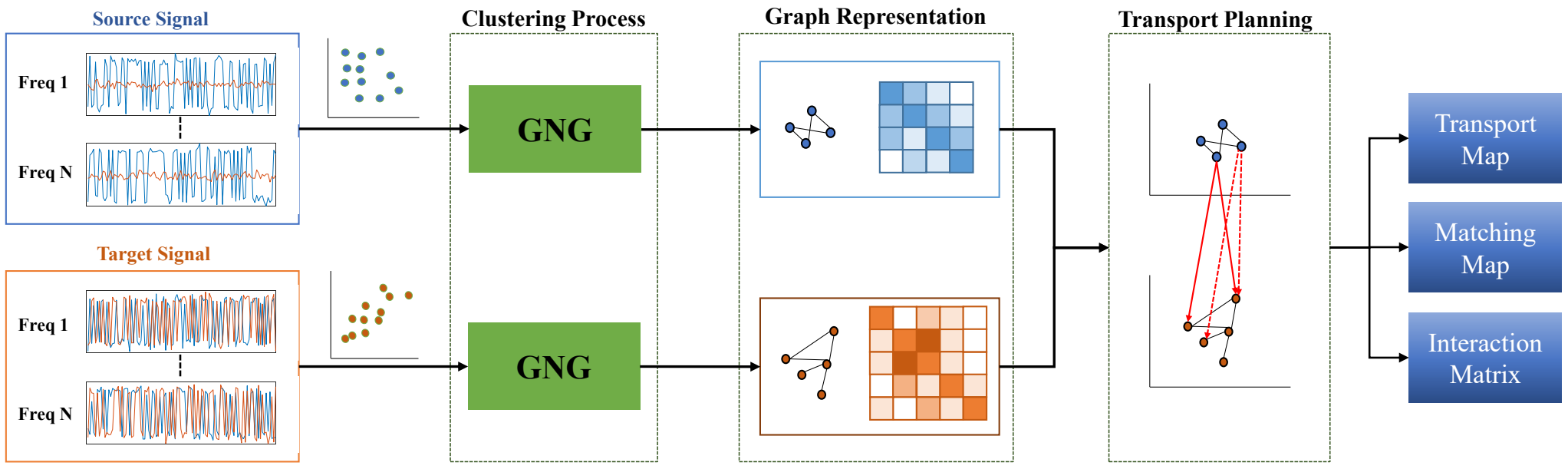

Fig. 5.20 Schematic illustrating the necessary steps to learn the transport plan.

**Graph Matching**. A possible approach to compare $\mathcal{G}_{\mathcal{S}}$ and $\mathcal{G}_{\mathcal{T}}$ is by calculating the divergence between the corresponding vertices of the two graphs encoded in sets $\mathcal{V}_{\mathcal{S}}$ and $\mathcal{V}_{\mathcal{T}}$. To this purpose we use the Kullback-Leibler divergence ($\mathcal{D}_{\mathcal{KL}}$) to measure how distant two distributions $\tilde{S}_i^{\mathcal{T}} \in \mathcal{V}_{\mathcal{S}}$ and $\tilde{S}_j^{\mathcal{T}} \in \mathcal{V}_{\mathcal{T}}$ are from each other according to:

$$\mathcal{D}_{\mathcal{KL}}(\tilde{S}_i^{\mathcal{S}} || \tilde{S}_j^{\mathcal{T}}) = \int \tilde{S}_i^{\mathcal{S}} \log\left(\frac{\tilde{S}_i^{\mathcal{S}}}{\tilde{S}_j^{\mathcal{T}}}\right) dS. \tag{5.68}$$

However, $\mathcal{D}_{\mathcal{KL}}$ is not a real distance metric since it is not symmetric and does not satisfy the triangle inequality. Thus, we used a symmetric version of the $\mathcal{D}_{\mathcal{KL}}$ defined as follows:

$$\mathcal{D}_{\mathcal{SKL}}(\tilde{S}_{\mathrm{i}}^{\mathcal{S}}, \tilde{S}_{\mathrm{j}}^{\mathcal{T}}) = \mathcal{D}_{\mathcal{KL}}(\tilde{S}_i^{\mathcal{S}} || \tilde{S}_j^{\mathcal{T}}) + \mathcal{D}_{\mathcal{KL}}(\tilde{S}_j^{\mathcal{T}} || \tilde{S}_i^{\mathcal{S}}). \tag{5.69}$$

Since vertices of the two graphs represent a multivariate Gaussian distributions, we use the approximated $\mathcal{D}_{\mathcal{KL}}$ [252] defined as:

$$\mathcal{D}_{\mathcal{KL}}(\tilde{S}_i^{\mathcal{S}} || \tilde{S}_j^{\mathcal{T}}) = \frac{1}{2}\left(\ln\left(\frac{|\Sigma_{\tilde{S}_j^{\mathcal{T}}}|}{|\Sigma_{\tilde{S}_i^{\mathcal{S}}}|}\right) - d + \mathrm{tr}(\Sigma_{\tilde{S}_j^{\mathcal{T}}}^{-1} \Sigma_{\tilde{S}_i^{\mathcal{S}}}) + (\mu_{\tilde{S}_i^{\mathcal{S}}} - \mu_{\tilde{S}_j^{\mathcal{T}}})^{\mathsf{T}} \Sigma_{\tilde{S}_j^{\mathcal{T}}}^{-1} (\mu_{\tilde{S}_i^{\mathcal{S}}} - \mu_{\tilde{S}_j^{\mathcal{T}}})\right). \tag{5.70}$$

Accordingly, we build a matching matrix $\mathcal{M} \in \mathbb{R}^{\mathrm{N}_{\mathcal{S}} \times \mathrm{N}_{\mathcal{T}}}$ encoding all the $\mathrm{N}_{\mathcal{S}} \times \mathrm{N}_{\mathcal{T}}$ normalized distances $\bar{\mathcal{D}}(\tilde{\mathrm{S}}_{\mathrm{i}}^{\mathcal{S}}, \tilde{\mathrm{S}}_{\mathrm{j}}^{\mathcal{T}})$ and expressed as:

$$\mathcal{M} = \begin{bmatrix} \bar{\mathcal{D}}(\tilde{\mathrm{S}}_1^{\mathcal{S}}, \tilde{\mathrm{S}}_1^{\mathcal{T}}) & \bar{\mathcal{D}}(\tilde{\mathrm{S}}_1^{\mathcal{S}}, \tilde{\mathrm{S}}_2^{\mathcal{T}}) & \dots & \bar{\mathcal{D}}(\tilde{\mathrm{S}}_1^{\mathcal{S}}, \tilde{\mathrm{S}}_{\mathrm{N}_{\mathcal{T}}}^{\mathcal{T}}) \\ \bar{\mathcal{D}}(\tilde{\mathrm{S}}_2^{\mathcal{S}}, \tilde{\mathrm{S}}_1^{\mathcal{T}}) & \bar{\mathcal{D}}(\tilde{\mathrm{S}}_2^{\mathcal{S}}, \tilde{\mathrm{S}}_2^{\mathcal{T}}) & \dots & \bar{\mathcal{D}}(\tilde{\mathrm{S}}_2^{\mathcal{S}}, \tilde{\mathrm{S}}_{\mathrm{N}_{\mathcal{T}}}^{\mathcal{T}}) \\ \vdots & \vdots & \ddots & \vdots \\ \bar{\mathcal{D}}(\tilde{\mathrm{S}}_{\mathrm{N}_{\mathcal{S}}}^{\mathcal{S}}, \tilde{\mathrm{S}}_1^{\mathcal{T}}) & \bar{\mathcal{D}}(\tilde{\mathrm{S}}_{\mathrm{N}_{\mathcal{S}}}^{\mathcal{S}}, \tilde{\mathrm{S}}_2^{\mathcal{T}}) & \dots & \bar{\mathcal{D}}(\tilde{\mathrm{S}}_{\mathrm{N}_{\mathcal{S}}}^{\mathcal{S}}, \tilde{\mathrm{S}}_{\mathrm{N}_{\mathcal{T}}}^{\mathcal{T}}) \end{bmatrix}, \tag{5.71}$$

where

$$\bar{\mathcal{D}}(\tilde{\mathrm{S}}_{\mathrm{i}}^{\mathcal{S}}, \tilde{\mathrm{S}}_{\mathrm{j}}^{\mathcal{T}}) = \frac{\mathcal{D}_{\mathcal{SKL}}(\tilde{\mathrm{S}}_{\mathrm{i}}^{\mathcal{S}}, \tilde{\mathrm{S}}_{\mathrm{j}}^{\mathcal{T}})}{\sum_{j}^{\mathrm{N}_{\mathcal{T}}} \mathcal{D}_{\mathcal{SKL}}(\tilde{\mathrm{S}}_{\mathrm{i}}^{\mathcal{S}}, \tilde{\mathrm{S}}_{\mathrm{j}}^{\mathcal{T}})}, \tag{5.72}$$

such that, $i = \{1, 2, \dots, \mathrm{N}_{\mathcal{S}}\}$ and $j = \{1, 2, \dots, \mathrm{N}_{\mathcal{T}}\}$.

**Graph Interaction**. While mapping a certain vertex in $\mathcal{G}_{\mathcal{S}}$ to the set of vertices in $\mathcal{G}_{\mathcal{T}}$ it might appear a 1-to-many mapping based on the similarity distance calculated in (5.70), i.e., the distance from 1-to-many is similar. Thus, if a certain source vertex in $\mathcal{G}_{\mathcal{S}}$ fires, it is more probable that one of the many vertices in $\mathcal{G}_{\mathcal{T}}$ will fire too. In order to capture the joint firing pattern we further integrate the time-varying interaction Matrix $\mathcal{I}$ within the optimal map to track the vertex firing between all the possible pair of vertices among the two graphs and defined as:

$$\mathcal{I} = \begin{bmatrix} \mathrm{P}(\tilde{\mathrm{S}}_1^{\mathcal{S}}, \tilde{\mathrm{S}}_1^{\mathcal{T}}) & \mathrm{P}(\tilde{\mathrm{S}}_1^{\mathcal{S}}, \tilde{\mathrm{S}}_2^{\mathcal{T}}) & \dots & \mathrm{P}(\tilde{\mathrm{S}}_1^{\mathcal{S}}, \tilde{\mathrm{S}}_{\mathrm{N}_{\mathcal{T}}}^{\mathcal{T}}) \\ \mathrm{P}(\tilde{\mathrm{S}}_2^{\mathcal{S}}, \tilde{\mathrm{S}}_1^{\mathcal{T}}) & \mathrm{P}(\tilde{\mathrm{S}}_2^{\mathcal{S}}, \tilde{\mathrm{S}}_2^{\mathcal{T}}) & \dots & \mathrm{P}(\tilde{\mathrm{S}}_2^{\mathcal{S}}, \tilde{\mathrm{S}}_{\mathrm{N}_{\mathcal{T}}}^{\mathcal{T}}) \\ \vdots & \vdots & \ddots & \vdots \\ \mathrm{P}(\tilde{\mathrm{S}}_{\mathrm{N}_{\mathcal{S}}}^{\mathcal{S}}, \tilde{\mathrm{S}}_1^{\mathcal{T}}) & \mathrm{P}(\tilde{\mathrm{S}}_{\mathrm{N}_{\mathcal{S}}}^{\mathcal{S}}, \tilde{\mathrm{S}}_2^{\mathcal{T}}) & \dots & \mathrm{P}(\tilde{\mathrm{S}}_{\mathrm{N}_{\mathcal{S}}}^{\mathcal{S}}, \tilde{\mathrm{S}}_{\mathrm{N}_{\mathcal{T}}}^{\mathcal{T}}) \end{bmatrix}, \tag{5.73}$$

where $\mathcal{I} \in \mathbb{R}^{\mathrm{N}_{\mathcal{S}} \times \mathrm{N}_{\mathcal{T}}}$, $\mathrm{P}(\tilde{\mathrm{S}}_k^{\mathcal{S}}, \tilde{\mathrm{S}}_l^{\mathcal{T}})$ is the joint firing probability of vertices $\tilde{S}_k^{\mathcal{S}}$ and $\tilde{S}_l^{\mathcal{T}}$ and $\sum_l^{\mathrm{N}_{\mathcal{T}}} \mathrm{P}(\tilde{\mathrm{S}}_k^{\mathcal{S}}, \tilde{\mathrm{S}}_l^{\mathcal{T}}) = 1$ such that $k \in \mathrm{N}_{\mathcal{S}}$ and $l \in \mathrm{N}_{\mathcal{T}}$.

**Learning the Optimal Transport Plan**. We aim to find a map $\Psi : \mathcal{G}_{\mathcal{S}} \rightarrow \mathcal{G}_{\mathcal{T}}$ which transports the mass from $\mathcal{V}_{\mathcal{S}}$ to $\mathcal{V}_{\mathcal{T}}$ while minimizing the mass transportation cost represented by the force needed to convert source data samples into target ones. Such required force is proportional to how much the vertices in the two graphs are distant from each other. It increases as the distance increase and viceversa. In this sense the optimal way of transporting the mass from $\mathcal{G}_{\mathcal{S}}$ to $\mathcal{G}_{\mathcal{T}}$ is to integrate the matching map encoded in (5.71) and the interaction matrix $\mathcal{I}$ defined in (5.73) and thus transforming the one-to-many mapping into one-to-one mapping. Consequently, since we know the optimal mapping between the vertices of the two graphs we can apply the proper transport map encoded in $\Psi$ which is defined as follows:

$$\Psi = \begin{bmatrix} \mathcal{U}_{\tilde{S}_1^{\mathcal{S}}, \tilde{S}_1^{\mathcal{T}}} & \mathcal{U}_{\tilde{S}_1^{\mathcal{S}}, \tilde{S}_2^{\mathcal{T}}} & \dots & \mathcal{U}_{\tilde{S}_1^{\mathcal{S}}, \tilde{S}_{N_{\mathcal{T}}}^{\mathcal{T}}} \\ \mathcal{U}_{\tilde{S}_2^{\mathcal{S}}, \tilde{S}_1^{\mathcal{T}}} & \mathcal{U}_{\tilde{S}_2^{\mathcal{S}}, \tilde{S}_2^{\mathcal{T}}} & \dots & \mathcal{U}_{\tilde{S}_2^{\mathcal{S}}, \tilde{S}_{N_{\mathcal{T}}}^{\mathcal{T}}} \\ \vdots & \vdots & \ddots & \vdots \\ \mathcal{U}_{\tilde{S}_{N_{\mathcal{S}}}^{\mathcal{S}}, \tilde{S}_1^{\mathcal{T}}} & \mathcal{U}_{\tilde{S}_{N_{\mathcal{S}}}^{\mathcal{S}}, \tilde{S}_2^{\mathcal{T}}} & \dots & \mathcal{U}_{\tilde{S}_{N_{\mathcal{S}}}^{\mathcal{S}}, \tilde{S}_{N_{\mathcal{T}}}^{\mathcal{T}}} \end{bmatrix}, \tag{5.74}$$

where $\mathcal{U}_{\tilde{S}_k^{\mathcal{S}}|\tilde{S}_l^{\mathcal{T}}}$ is the transport map encoding the transport force $\dot{\mu}_{\tilde{S}_k^{\mathcal{S}}|\tilde{S}_l^{\mathcal{T}}}$ and the transport uncertainty described by the covariance matrix $\mathcal{C}_{\tilde{S}_k^{\mathcal{S}}|\tilde{S}_l^{\mathcal{T}}}$, calculated in the following form:

$$\dot{\mu}_{\tilde{S}_k^{\mathcal{S}}|\tilde{S}_l^{\mathcal{T}}} = \mu_{\tilde{S}_l^{\mathcal{T}}} - \mu_{\tilde{S}_k^{\mathcal{S}}}, \tag{5.75}$$

$$\mathcal{C}_{\tilde{S}_k^{\mathcal{S}}|\tilde{S}_l^{\mathcal{T}}} = \mu_{\tilde{S}_l^{\mathcal{T}}\tilde{S}_k^{\mathcal{S}}} - \mu_{\tilde{S}_l^{\mathcal{T}}}\mu_{\tilde{S}_k^{\mathcal{S}}}. \tag{5.76}$$

**Synchronizing $\mathcal{G}_{\mathcal{S}}$ and $\mathcal{G}_{\mathcal{T}}$ by applying temporal delay**. As claimed in section 5.8.1, since source and target distributions (graphs) represent different modulation formats and the conversion follows low- to high-order manner, it is important to take into account of different symbol rates, so introducing some temporal delay in the transport process. The information transmitted by low-order modulation occupy more symbols with respect to higher order modulations. Thus, the time needed (i.e., the delay $t_c$) to perform the conversion can be calculated as a factor between source clusters ($N_{\mathcal{S}}$) and target clusters ($N_{\mathcal{T}}$) expressed as:

$$\gamma = \frac{\log_2(N_{\mathcal{T}})}{\log_2(N_{\mathcal{S}})}. \tag{5.77}$$

In this way, the radio understand that each $t_c = \gamma t$ it can convert the source distribution to the targeted one.

### 5.8.5 Joint RF Perception, Prediction and Conversion

The GDBN representation decomposes data with complex and non-linear dynamics into fragments that are explainable by simpler dynamical units. Switching dynamic systems as the Markov Jump Particle Filter (MJPF) [249] applied on the learned GDBN are capable of discovering the dynamical units and explain their switching behaviour. A Modified-MJPF (M-MPJF) is implemented here to perform joint predictions of GSs and GSSs by blending Particle Filter (PF) and Kalman Filter (KF) and providing various probabilistic inference modes (predictive and diagnostic) within Generalized Filtering. In the predictive inference mode each level of the proposed hierarchy holds predictions (beliefs) about the states of the level below. Those beliefs are signaled via predictive messages ($\pi(\tilde{\mathrm{S}}_t^{\mathcal{S}})$,$\pi(\tilde{\mathrm{X}}_t^{\mathcal{S}})$) in a top-down manner where they are compared against the sensory responses, resulting in multi-level abnormality indicators and Generalized Errors (GEs). These GEs are then fed back via diagnostic messages ($\lambda(\tilde{\mathrm{S}}_t^{\mathcal{S}})$,$\lambda(\tilde{\mathrm{X}}_t^{\mathcal{S}})$) from bottom-to-up the hierarchy to update the beliefs and thus improve future predictions and minimize future GEs.

In M-MJPF (employed on Double-GDBN in Fig. 5.19(b)), conversion allows to use a model learned from a signal modulated using a specific scheme to predict and update signals carrying the same information but modulated in a different way.

First, PF draws $Y$ equally weighted particles from the proposal distribution encoded in $\Pi$ to perdict the GSSs of $\mathcal{G}_{\mathcal{S}}$. Each propagated particle $\tilde{S}_{t,y}^{\mathcal{S}}$ can be used to predict $\tilde{S}_{t,y}^{\mathcal{T}}$ (i.e., converting $\tilde{S}_{t,y}^{\mathcal{S}}$ to $\tilde{\mathrm{S}}_{t,y}^{\mathcal{T}}$) after reaching the necessary time $t_c$ by using (5.73). Then, for each particle ($\tilde{S}_{t,y}^{\mathcal{S}}$) a KF is employed to predict the GS ($\tilde{\mathrm{X}}_{t,y}^{\mathcal{S}}$) of $\mathcal{G}_{\mathcal{S}}$ which depends on the predictions done at the level above as pointed out in (5.61). At this level converting $\tilde{\mathrm{X}}_{t,y}^{\mathcal{S}}$ to $\tilde{\mathrm{X}}_{t,y}^{\mathcal{T}}$ can be performed after reaching the required time $t_c$ by using the transport map encoded in (5.74) and extracting the proper transport force and transport uncertainty to update the dynamic model defined in (5.61) according to:

$$\tilde{\mathrm{X}}_{t_c,y}^{\mathcal{T}} = \mathbb{E}\Big[\big(\mathrm{A}\tilde{\mathrm{X}}_{t_c,y}^{\mathcal{S}} + \mathrm{B}\big(\mathrm{U}_{\tilde{\mathrm{S}}_{t_c,y}^{\mathcal{T}}} + \dot{\mu}_{\tilde{\mathrm{S}}_{t_c,y}^{\mathcal{S}}|\tilde{\mathrm{S}}_{t_c,y}^{\mathcal{T}}}\big)\big), \big(\mathrm{A}\tilde{\mathrm{X}}_{t_c-1,y}^{\mathcal{S}} + \mathrm{B}\big(\mathrm{U}_{\tilde{\mathrm{S}}_{t_c-1,y}^{\mathcal{T}}} + \dot{\mu}_{\tilde{\mathrm{S}}_{t_c-1,y}^{\mathcal{S}}|\tilde{\mathrm{S}}_{t_c-1,y}^{\mathcal{T}}}\big)\big), \ldots, \big(\mathrm{A}\tilde{\mathrm{X}}_{t_c-\gamma,y}^{\mathcal{S}} + \mathrm{B}\big(\mathrm{U}_{\tilde{\mathrm{S}}_{t_c-\gamma,y}^{\mathcal{T}}} + \dot{\mu}_{\tilde{\mathrm{S}}_{t_c-\gamma,y}^{\mathcal{S}}|\tilde{\mathrm{S}}_{t_c-\gamma,y}^{\mathcal{T}}}\big)\big)\Big], \tag{5.78}$$

where $\mathbb{E}[.]$ depicts the mean value. The idea of conversion is that predicting the source signal's hidden states allows to predict and infer those of the target signal so that we can predict (generate) one signal from the other using the transport plan.
The posterior probability associated with the predicted GS ($\tilde{\mathrm{X}}_{t,y}^{\mathcal{S}}$) and GSS and propagated towards the bottom level is given by:

$$\pi(\tilde{\mathrm{X}}_{t,y}^{\mathcal{S}}) = \mathrm{P}(\tilde{\mathrm{X}}_{t,y}^{\mathcal{S}}, \tilde{\mathrm{S}}_{t,y}^{\mathcal{S}} | \tilde{\mathrm{Z}}_{t-1}^{\mathcal{S}}) = \int \mathrm{P}(\tilde{\mathrm{X}}_{t,y}^{\mathcal{S}} | \tilde{\mathrm{X}}_{t-1,y}^{\mathcal{S}}, \tilde{\mathrm{S}}_{t,y}^{\mathcal{S}}) \lambda(\tilde{\mathrm{X}}_{t-1,y}^{\mathcal{S}}) d\tilde{\mathrm{X}}_{t-1,y}^{\mathcal{S}}, \tag{5.79}$$

where $\lambda(\tilde{\mathrm{X}}_{t-1,y}^{\mathcal{S}}) = \mathrm{P}(\tilde{\mathrm{Z}}_{t-1}^{\mathcal{S}} | \tilde{\mathrm{X}}_{t-1,y}^{\mathcal{S}})$.

Accordingly, once a new sensory response $\tilde{\mathrm{Z}}_t^{\mathcal{S}}$ is observed, diagnostic messages are fed upwards to update the posterior in the following way:

$$\mathrm{P}(\tilde{\mathrm{X}}_{t,y}^{\mathcal{S}}, \tilde{\mathrm{S}}_{t,y}^{\mathcal{S}} | \tilde{\mathrm{Z}}_t^{\mathcal{S}}) = \pi(\tilde{\mathrm{X}}_{t,y}^{\mathcal{S}}) \lambda(\tilde{\mathrm{X}}_{t,y}^{\mathcal{S}}). \tag{5.80}$$

Next, update the weights of the particles according to:

$$W_{t,y} = W_{t,y}\lambda(\tilde{\mathrm{S}}^{\mathcal{S}}_{t,y}), \tag{5.81}$$

and normalize by using the Sequential Importance Resampling (RIS). $\lambda(\tilde{\mathrm{S}}^{\mathcal{S}}_{t,y})$ is a discrete probability distribution represented by:

$$\lambda(\tilde{\mathrm{S}}^{\mathcal{S}}_{t,y}) = \lambda(\tilde{\mathrm{X}}^{\mathcal{S}}_{t,y})\mathrm{P}(\tilde{\mathrm{X}}^{\mathcal{S}}_{t,y}|\tilde{\mathrm{S}}^{\mathcal{S}}_{t,y}) = \mathrm{P}(\tilde{\mathrm{Z}}^{\mathcal{S}}_{t,y}|\tilde{\mathrm{X}}^{\mathcal{S}}_{t,y})\mathrm{P}(\tilde{\mathrm{X}}^{\mathcal{S}}_{t,y}|\tilde{\mathrm{S}}^{\mathcal{S}}_{t,y}). \tag{5.82}$$

### 5.8.6 Automatic Modulation Classification (AMC)

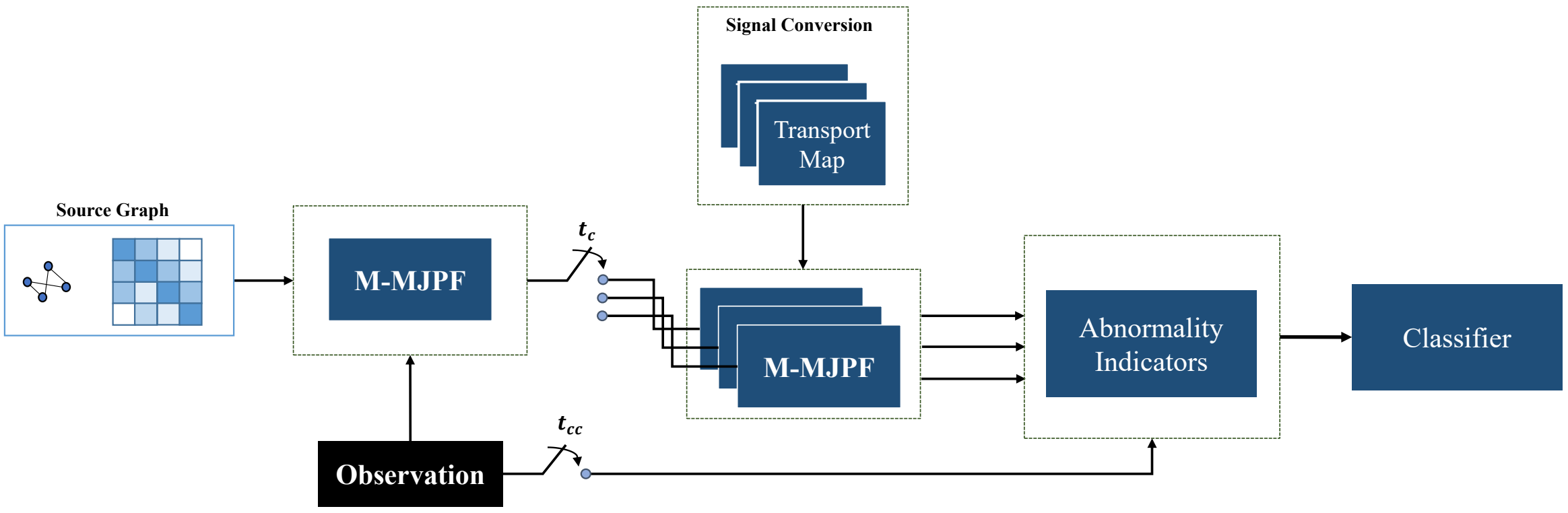

Fig. 5.21 Schematic illustrating the AMC process using Generalized Filtering and the learned transport plan.

In order to identify the correct modulation format (from the candidate set $\Theta$) of the received sensory signal (i.e., the current observation), the radio performs multiple predictions in parallel based on the rules learned from experience so far and evaluate which of the predictions better explain the current situation (refer to Fig. 5.21). The candidate set $\Theta = \{\mathcal{G}_{\mathcal{S}_1}, \mathcal{G}_{\mathcal{T}_2}, \dots, \mathcal{G}_{\mathcal{T}_K}\}$ contains one source modulation format and $K-1$ target modulation formats in ascending order where $K$ is the total number of formats. The radio starts predicting the dynamics of the source signal using the source graph ($\mathcal{G}_{\mathcal{S}_1}$) that encodes the rules of how the signal's dynamics evolve according to the source modulation format. Then, by using the optimal transport plan learned during training it can transfer (convert) the predicted source signal into target signals representing the dynamics of the other modulation formats (target ones). Since the radio does not has any prior knowledge about the sensory signals regarding the data rate and the modulation scheme adopted, it is also essential to apply temporal delay ($t_{cc}$) and synchronize the multiple predictions during online classification. So, the time required to perform the classification task is related to the time $t_{c_K}$ required converting the highest order modulation (i.e., $\mathcal{G}_{\mathcal{T}_K}$) in the set $\Theta$ such that, $t_{cc} = \alpha t_{c_K}$ where $\alpha > 1$. Thus, a set of multiple abnormality indicators associated with the multiple predictions (1 source prediction and $K-1$ target predictions) is available each $t_{cc}$, such that:

$$\Omega_{t_{cc}} = \{\Upsilon_{\tilde{X}_{t_{cc}}^{\mathcal{S}_1}}, \Upsilon_{\tilde{X}_{t_{cc}}^{\mathcal{T}_2}}, \dots, \Upsilon_{\tilde{X}_{t_{cc}}^{\mathcal{T}_K}}\}, \tag{5.83}$$

where,

$$\Upsilon_{\tilde{X}_{t_{cc}}^{(.)}} = -\ln\Big(\mathcal{BC}\big(\pi(\tilde{X}_{t_{cc}}^{(.)}), \lambda(\tilde{X}_{t_{cc}}^{\mathcal{S}_1})\big)\Big), \tag{5.84}$$

and $\mathcal{BC}(.)=\int \sqrt{\pi(\tilde{X}_{t_{cc}}^{(.)})\lambda(\tilde{X}_{t_{cc}}^{\mathcal{S}_1})} d\tilde{X}_{t_{cc}}^{\mathcal{S}_1}$ is the Bhattacharyya coefficient.
Then, the modulation classification can be made by selecting the index of the minimum abnormality in the set $\Omega_{t_{cc}}$, such that:

$$\hat{k}_{t_{cc}} = \underset{1\leq k\leq K}{\operatorname{argmin}}\, \Omega_{t_{cc}},\ \exists\, \hat{k}_{t_{cc}} \in \Theta. \tag{5.85}$$

### 5.8.7 Performance Metrics

Regarding the modulation format conversion task, the radio is committed to recover the raw data sequence (binary bits) correctly. Hence, the Bit Error Rate (BER) is computed to measure the number of bit errors between the original data sequence and the one recovered from the converted signals.

The probability of correct classification $P_{co}$ is used as performance metric to evaluate the AMC task, calculated as:

$$P_{co} = \frac{1}{T_{cc}} \sum_{t=1}^{T_{cc}} \Pr\big(\hat{k}_{t_{cc}} = k_{t_{cc}} | k_{t_{cc}}\big), \tag{5.86}$$

where $T_{cc}$ is the total time and $\Pr(.)$ is the probability that the modulation format is correctly predicted as $k_{t_{cc}}$ at time $t_{cc}$.

### 5.8.8 Numerical Simulations

In this section, we evaluate the performance of the proposed AMCC framework for random signal transmissions over AWGN channels. A data sequence of $24,000$ random bits is generated on which 4 different modulation formats are applied at the transmitter side namely, BPSK, QPSK, 16QAM and 64QAM where the output signals of all modulators are normalized based on the average power and have different data rates.

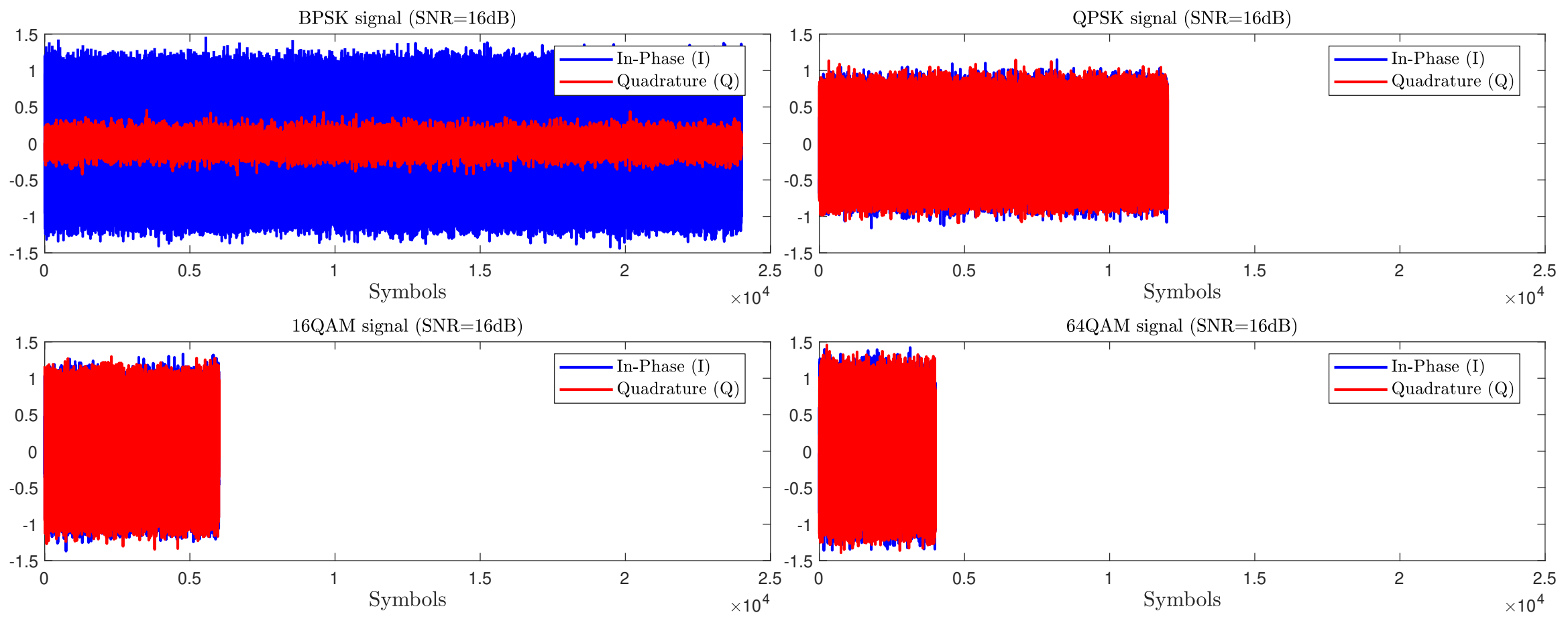

Fig. 5.22 Generated BPSK, QPSK, 16QAM and 64QAM signals in time domain.

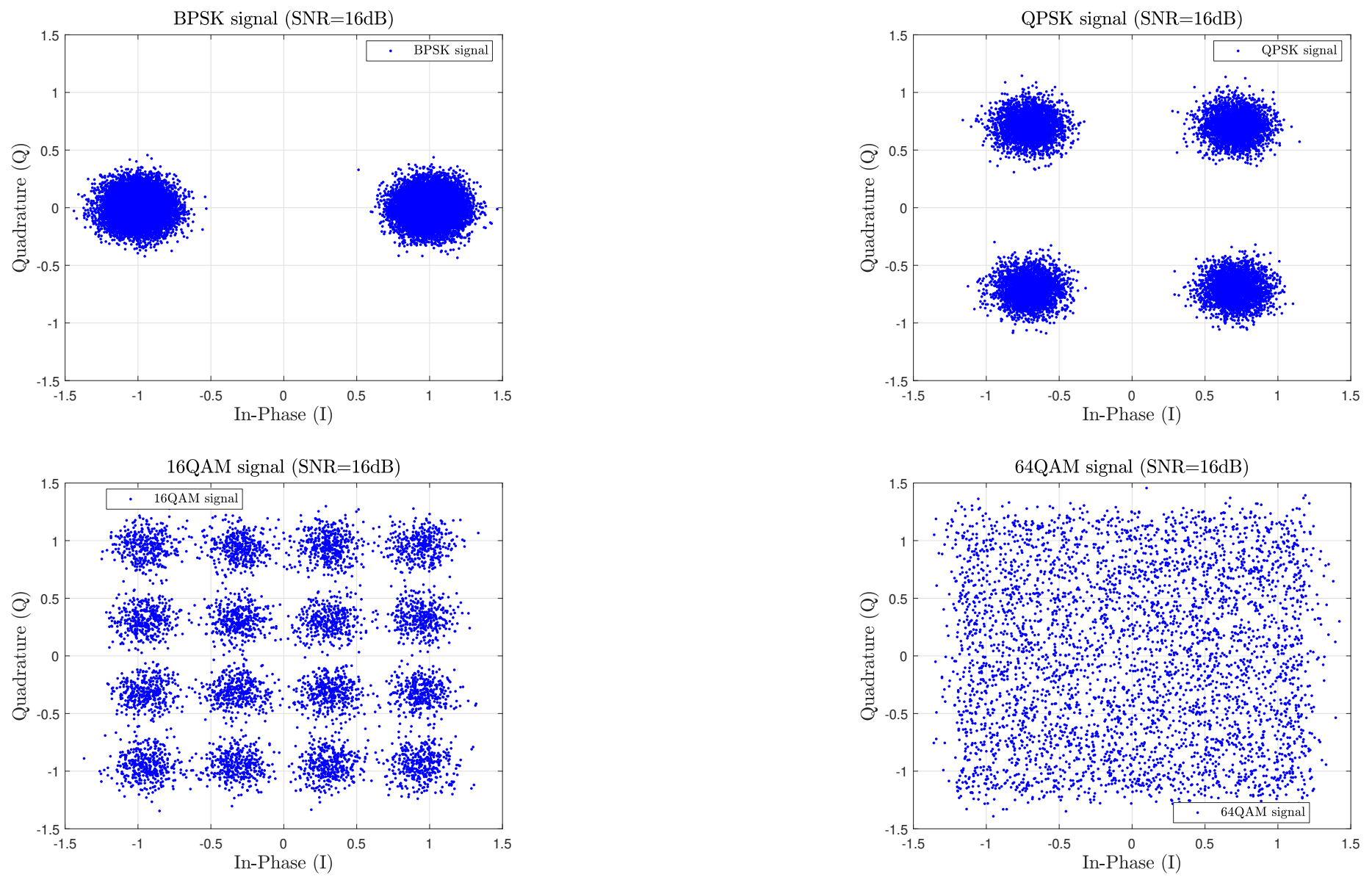

Fig. 5.23 Signal Constellations: Original signals at 16dB.

Fig. 5.22 depicts the generated signals in time domain from where we can observe the different data rates related to each modulation format. The signal constellations of the four modulation formats are shown in Fig. 5.23. Initially, the radio starts perceiving the environment supposing that no signals are present, by exploiting the generalized errors that encode the emergent dynamic rules it can cluster those errors to learn the corresponding dynamic models as illustrated in Fig. 5.24-Top. In this work we fixed the maximum number of clusters in GNG which is equal to the number of constellations of each modulation format.

The GNG output (clusters) representing the top level of abstraction is expressed into graph for each modulation format and visualized in Fig. 5.24-Bottom to perform the graph matching, graph interactions and to learn the optimal transport plan, respectively. Two configurations are tested: *i)* transferring data samples from a BPSK source graph to QPSK, 16QAM and 64QAM; *ii)* transferring data samples from a QPSK source graph to 16QAM and 64QAM. Fig. 5.25 illustrates some examples of the learned transport maps that can be used to convert a source distribution to a target one based on the transport forces defined in (5.75) and the corresponding uncertainty. Fig. 5.27 shows the converted constellations of QPSK, 16QAM and 64QAM starting from the predicted BPSK signal (source signal) Constellation diagrams of the predicted BPSK signal (source) and the converted QPSK, 16QAM, and 64QAM (targets) at SNR of 16dB and 8dB are illustrated in Fig. 5.27-5.29. It can be seen that the constellation points (data samples) of the converted signals spread in the correct way around the theoretical reference points ($+$ red symbols in the figure) in particular for QPSK and 16QAM. For 64QAM there are slight shifts from the reference points and this is logical due to the high number of constellations which also become more challenging by dealing with normalized signals (with the same power). It is obvious that decreasing the SNR value will introduce more distortions to the data samples and makes their spreading area (around the reference point) bigger. However, the converted data samples at smaller SNR (8dB) are always in line with the theoretical ones as it can be seen in Fig. 5.29. As claimed before, after converting a source modulation format to a target one we should be able to recover the original information (bits) transmitted over the source signal. So, if we transmit e.g., an image using BPSK, we must recover the same image after converting to QPSK and performing the demodulation and decoding process. To this purpose we measure the BER of the predicted and converted modulation formats and compare it with the corresponding analytical values. Fig. 5.30 shows the BER comparison where we can see that the BER rates of the converted signals are in line with those of the predicted and analytical signals which means that the converted signals are able to recover the original information without introducing additional noise (bit errors).

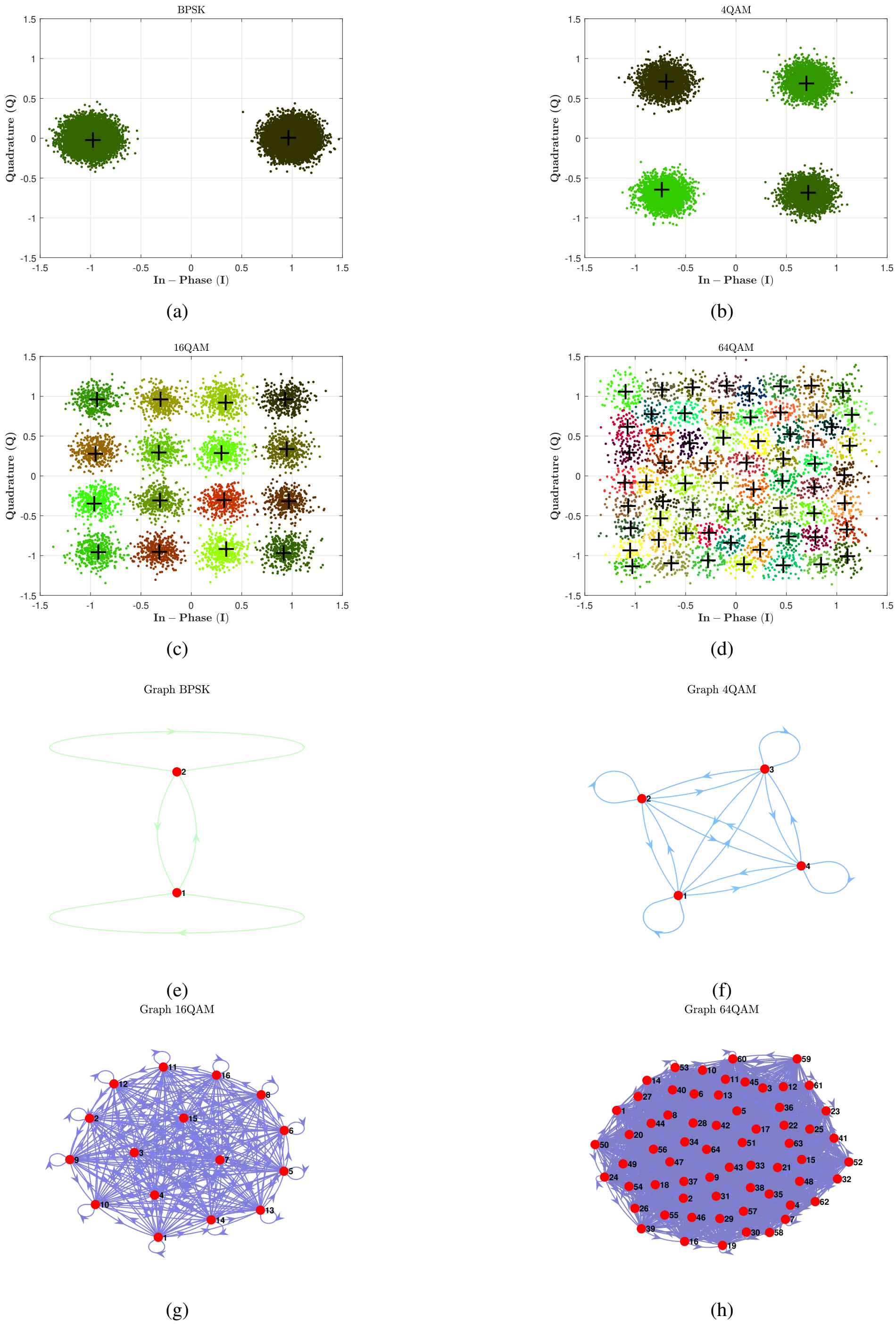

Fig. 5.24 Clustering Process: (Top) output of GNG using as input (a) BPSK signal, (b) QPSK signal, (c) 16QAM signal and (d) 64QAM signal. (Bottom) the corresponding graph representation of (e) BPSK clusters, (f) QPSK clusters, (g) 16QAM clusters and (h) 64QAM clusters.

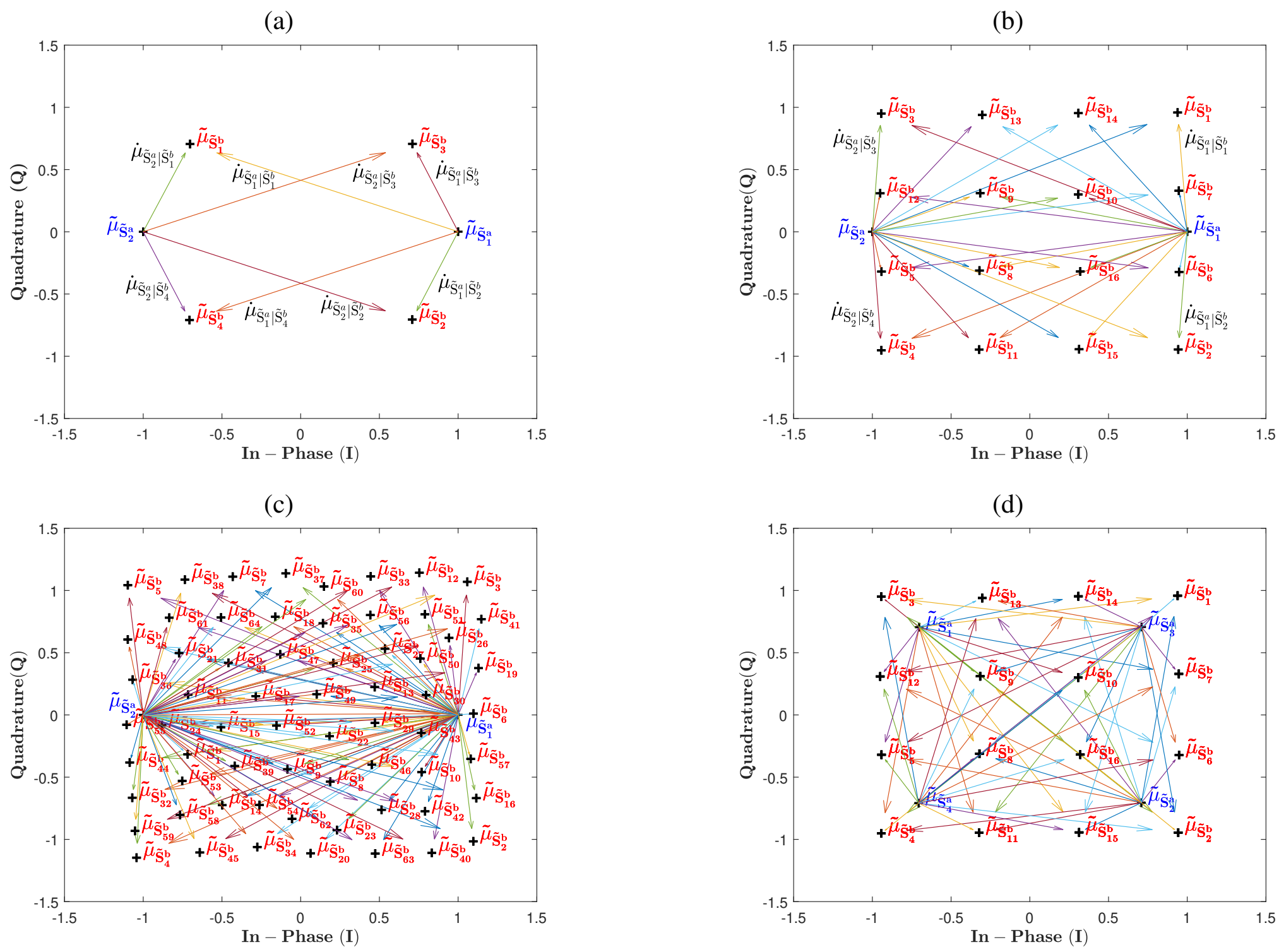

Fig. 5.25 Transport Planning: the learnt transport maps to convert (a) BPSK into QPSK, (b) BPSK into 16QAM, (c) BPSK into 64QAM and (d) QPSK into 16QAM.

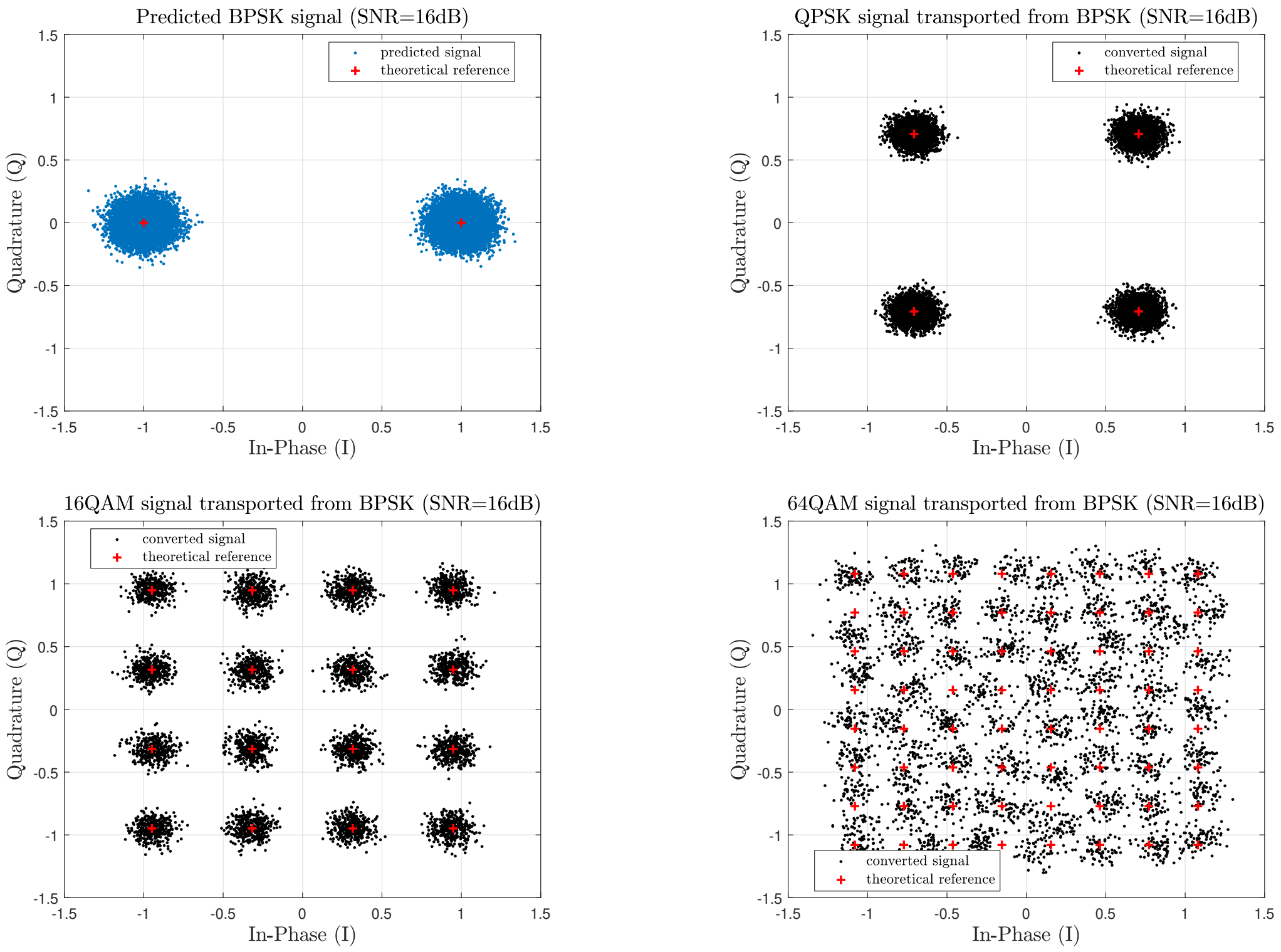

Fig. 5.26 Constellation diagrams of: (top to the left) the predicted BPSK signal and the converted signals QPSK (top to the right), 16QAM (bottom to the left), 64QAM (bottom to the right) signals at 16dB.

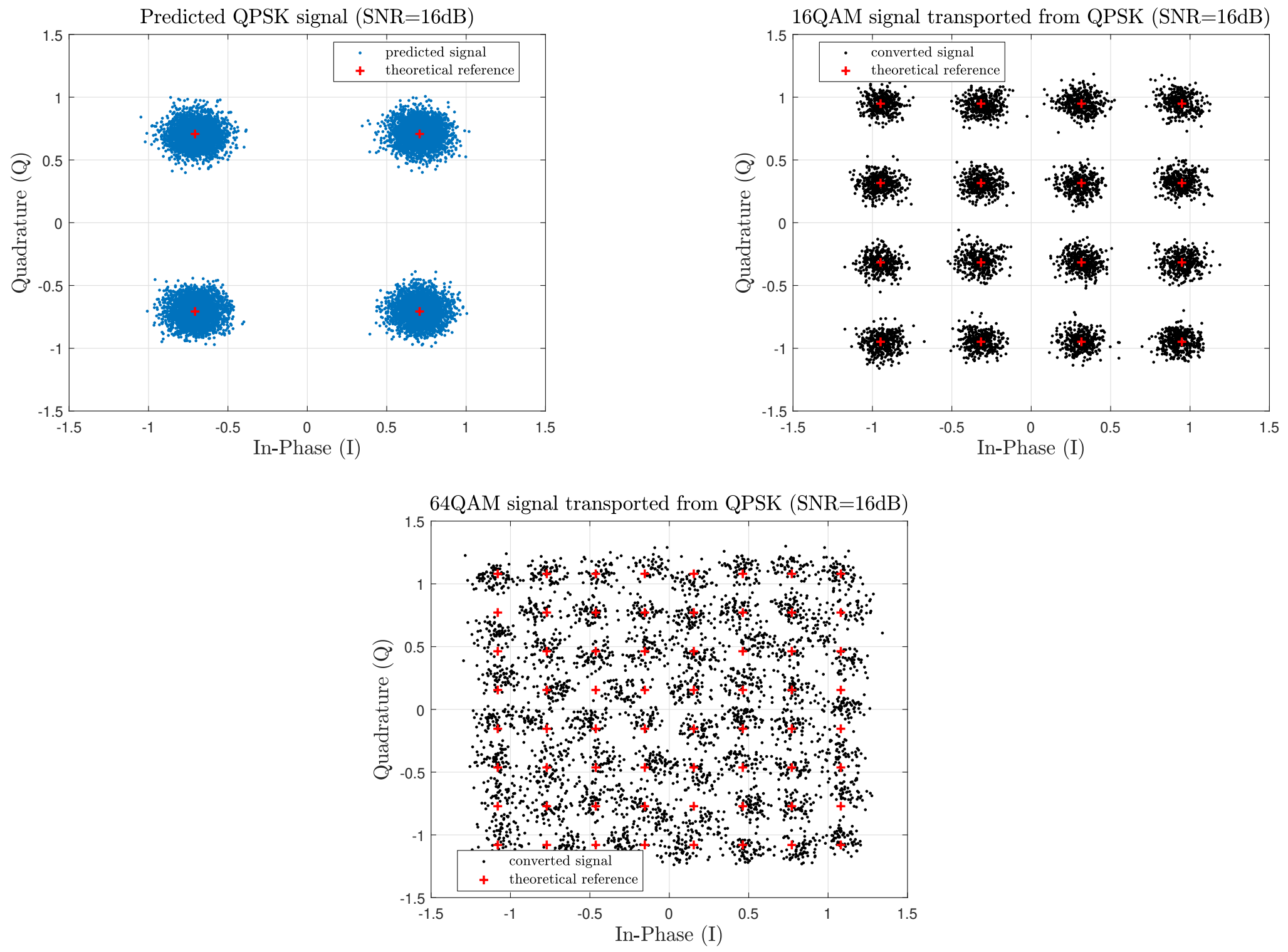

Fig. 5.27 Constellation diagrams of: (top to the left) the predicted QPSK signal and the converted signals 16QAM (top to the right), 64QAM (bottom) signals at 16dB.

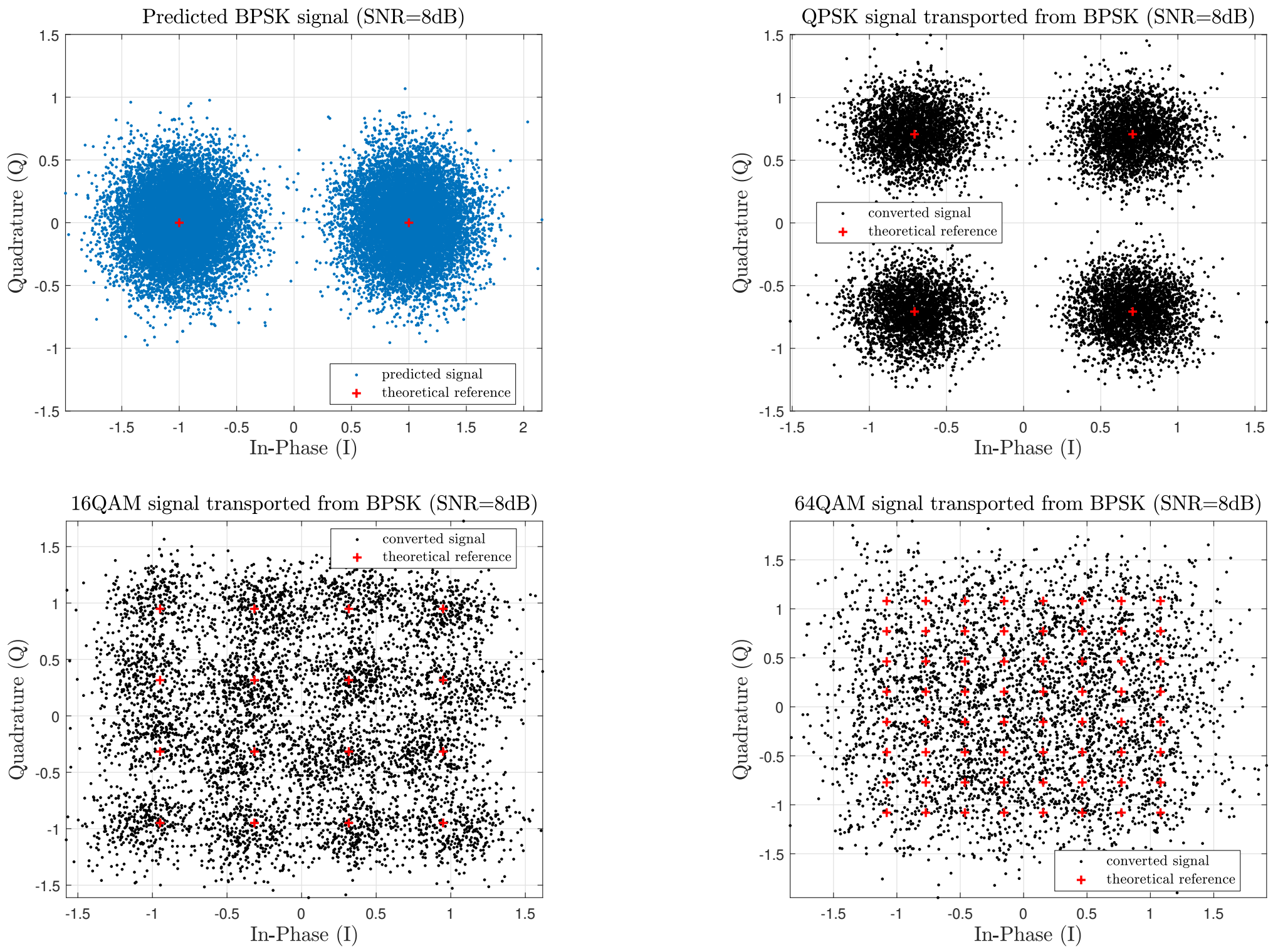

Fig. 5.28 Constellation diagrams of: (top to the left) the predicted BPSK signal and the converted signals QPSK (top to the right), 16QAM (bottom to the left), 64QAM (bottom to the right) signals at 8dB.

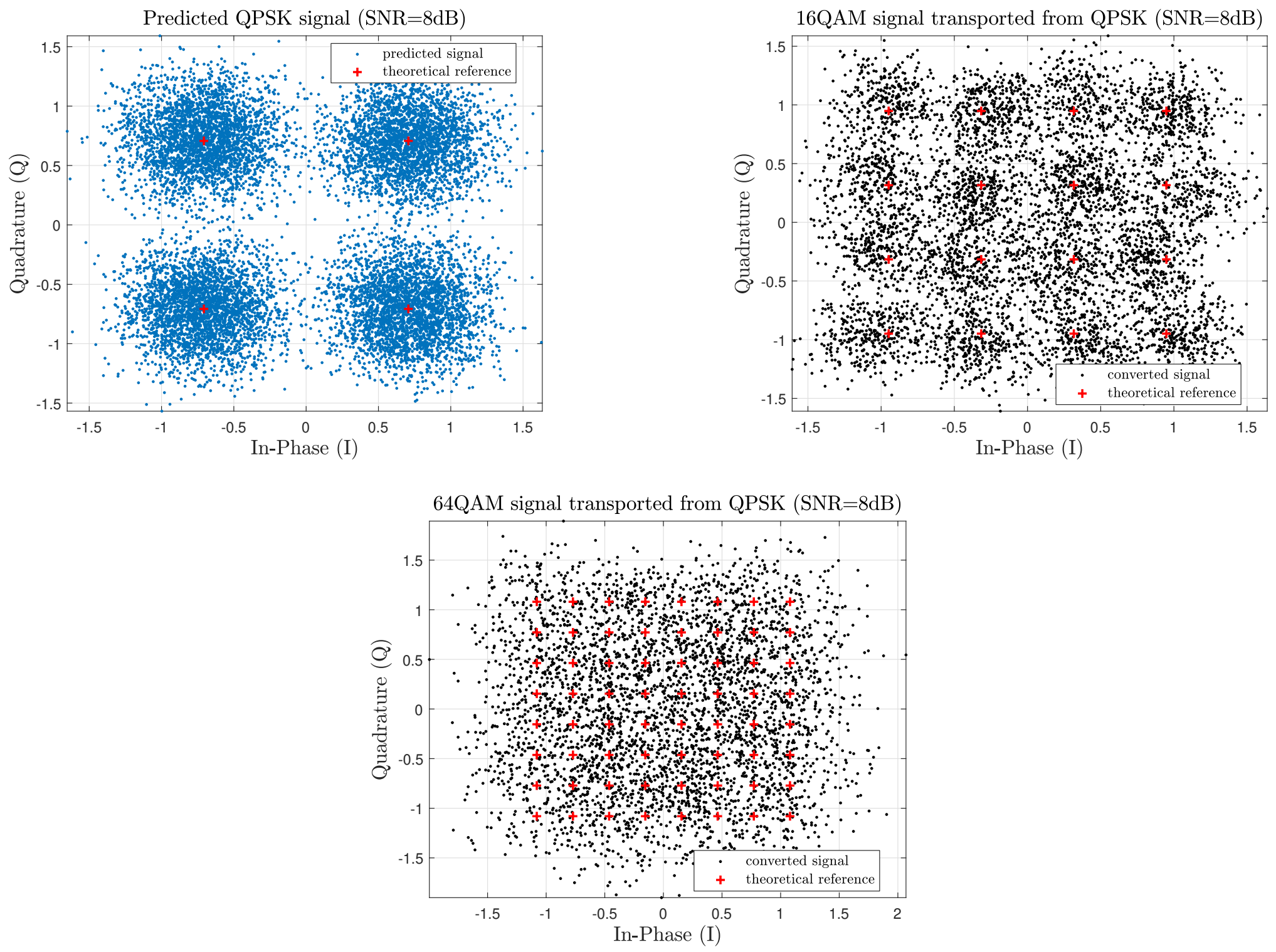

Fig. 5.29 Constellation diagrams of: (top to the left) the predicted QPSK signal and the converted signals 16QAM (top to the right), 64QAM (bottom) signals at 8dB.

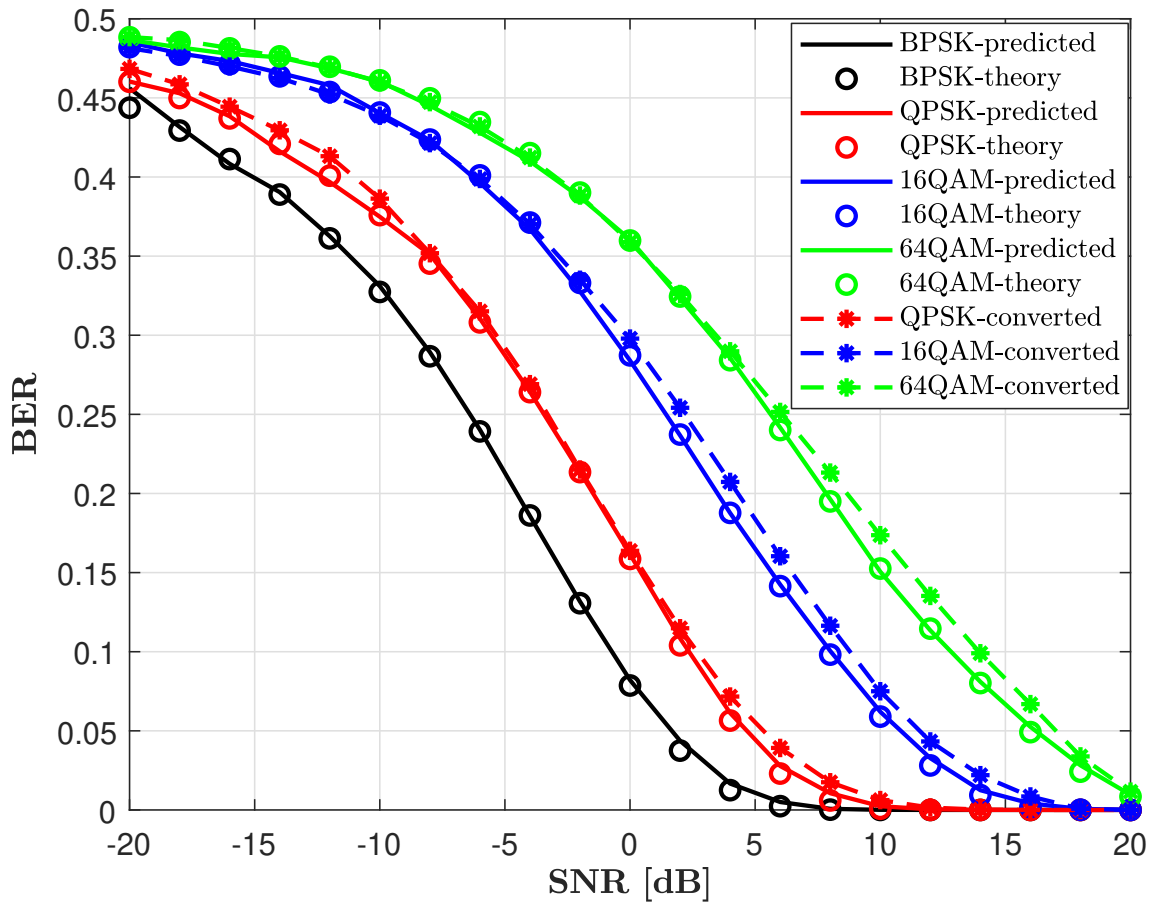

Fig. 5.30 BER comparison between predicted, converted and analytical signals as a function of SNR.

In addition, the performance of the modulation format conversion can be validated by performing the modulation classification task. Because predicting the dynamics of a source signal and converting them to explain how dynamic rules evolve among different modulation formats using only one dynamic model (source) accompanied with the transport maps must allow recognising the current observation. This joint approach in converting and classifying multiple modulation formats endows the AI-enabled radios with high explain-ability and interpret-ability.

The candidate set of the modulation formats considered in the classification task is $\Theta = \{\text{BPSK, QPSK, 16QAM, 64QAM}\}$. In the classification process, BPSK format represents the source signal and its dynamics are predicted using the corresponding dynamic model. From which the target formats can be predicted using the optimal transport plan. The time needed to convert source to target formats must be chosen carefully due to their different data rates. Converting BPSK to QPSK, 16QAM and 64QAM can be done each $t_{c_2}$=2$t$ (and so after predicting 2 BPSK symbols),$t_{c_3}$=4$t$ (and so after predicting 4 BPSK symbols) and $t_{c_4}$=6$t$ (and so after predicting 6 BPSK symbols), respectively. Likewise, the time ($t_{cc}$) required to perform the classification is proportional to the highest modulation order as mentioned in section 5.8.6 so $t_{cc}$=$\alpha t_{c_4}$=6$\alpha$ with $\alpha$=2. Fig. 5.31(a) shows the classification accuracy for each modulation scheme in the set $\Theta$ as a function of SNR. We can observe that the proposed approach achieves high classification accuracy at SNR>0 for each modulation scheme. At low SNR values the classification accuracy degrades especially for 16QAM and 64QAM due to the high interference caused by the channel and the fast dynamics that are difficult to be captured efficiently. Fig. 5.31(b) shows the overall accuracy, i.e., the average probability of correct classifications among all the $P_{cc}$ related to each modulation where we can see the effectiveness of the proposed approach in classifying different modulation formats from sensory signals.

## 5.9 Conclusion

In this chapter, we proposed a novel method for joint detection and classification of jamming attacks in a Cognitive-UAV-based radio application. The method is based on learning a dynamic model representing the radio environment under normal circumstances encoded in a GDBN model. The acquired knowledge encoded in the dynamic model can be used during the online phase to predict what the cognitive-UAV is supposed to receive, evaluate hierarchical abnormality measurements and generalised errors to explain the current situation by differentiating between normal and abnormal situations (i.e., jammer detection) and extract the jamming signal by exploiting the errors to encode it incrementally in a new GDBN model.

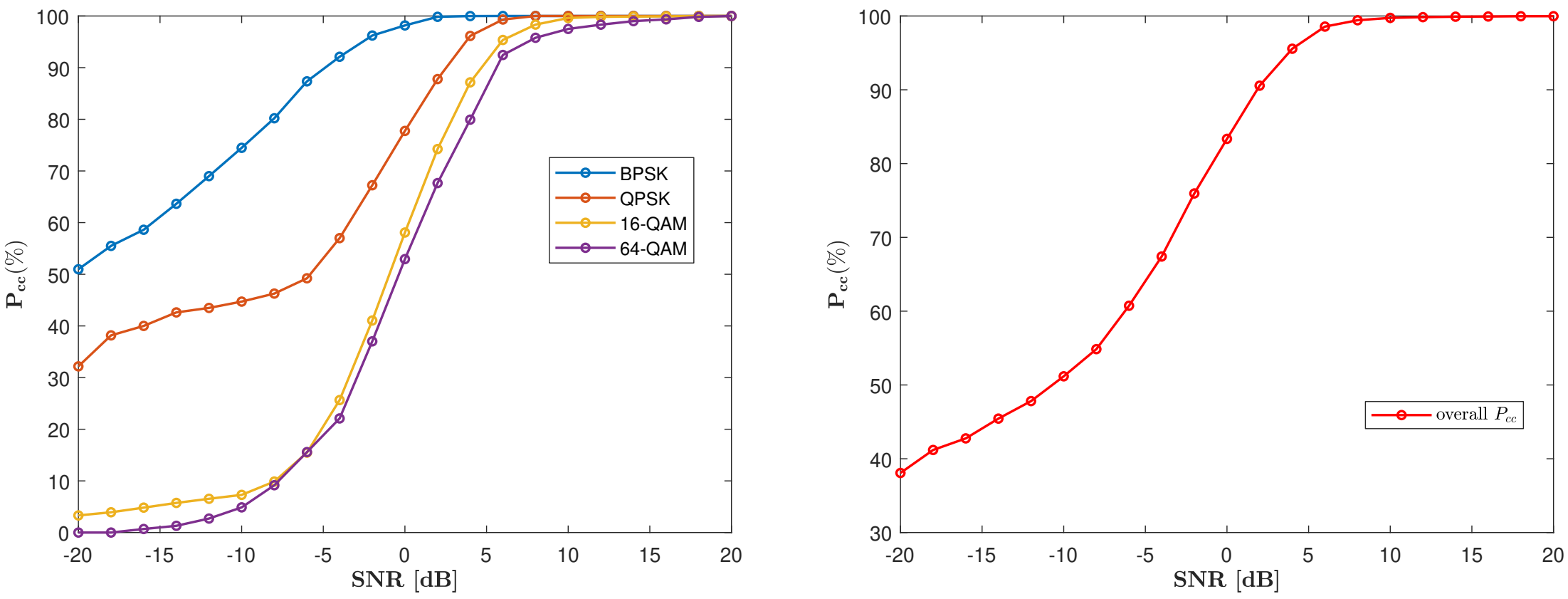

Fig. 5.31 Performance of modulation classification as a function of SNR: (a) classification accuracy for each modulation scheme and (b) overall classification accuracy.

Thus, in future experiences, if the cognitive-UAV detects a jamming attack, it can perform multiple predictions in parallel using the M-MJPF and selects the best model that explains the current situation to recognize the modulation scheme of the detected jammer. Simulation results showed that the proposed method achieves high probabilities in detecting jammer and high accuracy in classifying them and outperforms LSTM and CNN in classifying multiple jamming signals and different modulation schemes using a real dataset. In addition, the proposed approach provides interpretable results where multiple abnormality measurements and generalised errors can be used as self-information to keep learning incrementally.

Furthermore, in this chapter we propose applying transport planning concept for modulation conversion and classification, which is a prospective candidate technology in future wireless communications exhibiting true heterogeneity integrating cellular, space, UAV as well as satellite communications. The proposed automatic modulation conversion and classification (AMCC) framework is based data-driven approach following the inherent intelligent capability of Generalised filtering. Simulation results validate that the proposed approach allows transporting low-order to high-order modulation schemes without introducing distortions and performs modulation classification with high accuracy.

Next chapter will study the interaction between a cognitive user and jammer to design a novel resource allocation strategy for anti-jamming.

# Chapter 6

# A Novel Resource Allocation for Anti-jamming using Active Inference

This chapter proposes a novel resource allocation strategy for anti-jamming in a Cognitive-UAV based scenario using Active Inference (*AIn*). An Active Generalized Dynamic Bayesian Network (Active-GDBN) is proposed to represent the external environment that jointly encodes the physical signal dynamics and the dynamic interaction between UAV and jammer in the spectrum. We cast the action and planning as a Bayesian inference problem that can be solved by avoiding surprising states (minimizing abnormality) during online learning. Simulation results verify the effectiveness of the proposed *AIn* approach in minimizing abnormalities (maximizing rewards) and has a high convergence speed by comparing it with the conventional Frequency Hopping and Q-learning.

## 6.1 Introduction

Unmanned Aerial Vehicles (UAVs) are considered as potential candidate in future wireless technologies (beyond 5G and 6G) to complement the terrestrial wireless systems providing additional capacities to hotspot areas during temporary events and additional services to support the V2X and IoT applications [304–306]. Wireless communications are vulnerable to jamming attacks due to their inherent dynamic nature and openness of the wireless medium [307]. With the integration of UAVs, wireless communications are more prone to terrestrial malicious nodes (i.e., jammers) due to the high heterogeneity and dominant line-of-sight communication links of UAVs [279]. Jammers impact the physical layer by injecting disturbance signals over legitimate communication channels causing the damage of communication and degrading system's performance. Therefore, it is crucial to develop an anti-jamming strategy by performing dynamic channel allocation to reach robust connectivity and improve communication security. Cognitive Radio (CR) is a key technology to accomplish intelligent resource management in jamming scenarios due to its capability to monitor and learn the radio environment and adapt to dynamic environmental changes [308].

Detecting the existence of the jammer and understanding how it works realize the first important step towards avoiding jamming attacks. Jammers are categorised based on their jamming strategy as elementary or advanced [309]. Elementary jammers utilise a predefined strategy, such as random, constant and sweep. In contrast, the advanced jammers' strategy depends upon that adopted by its opponent. Smart jammers are a particular type of advanced jammers which adapt their jamming strategy based on the ongoing system's communication status [310].

Conventional anti-jamming solutions use fixed transmission patterns so they are unable to deal with dynamic jamming patterns in complicated radio environments with high uncertainty and unpredictable jamming behaviours [311]. Reinforcement Learning (RL) have attracted a lot of attention recently in the wireless communication field to design anti-jamming solutions in complex environments[312]. For instance, [313] used the Q-learning method in an autonomous CR which is supported by its spectrum awareness capability to detect and locate a sweep jammer to learn the optimal policy and avoid jamming attacks. In [314], a cooperative learning method for CR is proposed using on-policy synchronous Q-learning to deal with sweep and reactive jammers. However, Q-learning suffers from slow convergence if the state and action spaces are large, which leads to anti-jamming performance degradation. Deep Q-learning has been proposed to overcome such an issue as in [315] where a Deep Q-network is applied to determine whether to recommend that a secondary user in a CR network should leave a region under heavy jamming attacks or to remain and defeat a smart jammer by choosing a frequency hopping pattern. Moreover, A deep double Q-learning method is proposed in [316] to learn efficient communication policy for tackling different types of jammers.

RL methods are based on a reward signal coming from the environment as a feedback to evaluate the performed action. However, defining a proper reward function in complex and dynamic environments is a big challenge [317]. **Active Inference (*AIn*)** can overcome this challenging task by replacing reward functions with prior beliefs about desired sensory signals received from the environment. Thus, *AIn* agent can learn to describe how it expects itself to behave without getting a feedback from the environment. As a promising emerging theory from cognitive neuroscience, *AIn* provides a theoretical framework that supports the way biological agents perceive and act in the real world and offers an alternative to RL. *AIn* is a Bayesian framework towards explaining and understanding biological intelligence (in the brain) under the free-energy principle [318]. Minimizing the free-energy is a precondition for any self-organizing system to reach the equilibrium with its environment and resist tendency to disorder [40]. The maintenance of order can be achieved by avoiding surprises (or abnormalities) and ensuring that the states of the system remain within an evidence bound. The free energy minimization (and so surprise avoidance) involves two components: action and perception. Free energy depends on two probability densities: the one that generates the sensory signals (i.e, the generative model) and the other that represent in a probabilistic way its causes (called recognition density). Thus, free energy can be minimized by acting in the environment to change the sensory signals in the former case or by changing the internal states and consequently the recognition density in the latter case. According to the Bayesian Brain Hypothesis [281], the brain must represent sensory information and internal states in the form of probability distributions rather than on unitary estimates of parameter values to make effective judgements and guide actions efficiently.

Motivated by the above discussion, we propose in this chapter an *AIn* framework as a novel resource allocation strategy for anti-jamming in a Cognitive-UAV based scenario. Under the *AIn* framework, the Cognitive-UAV is endowed with a joint internal representation (generative model) of the external environment. The generative model encodes a joint representation of the physical signal realizing the commands that the cognitive-UAV is supposed to receive under normal circumstances and the available physical resources. This enables encoding the dynamic interaction between the UAV and the jammer in the spectrum. The objective is to learn the best set of actions performed by the UAV as interaction with a jammer that leads to the minimum surprise (positive reward). Such a representation goes over the necessity of mapping actions to signals' states directly (unlike the RL approach) and modelling them over a continuous state-space, which can be a complicated task in RL. There are four main rationals to use *AIn* approach over RL ([313], [314], [315], [316]): *i*) *AIn* operates in a pure belief-based setting allowing one to seek information about the environment and resolve uncertainty in a Bayesian-optimal fashion. *ii*) *AIn* enables speeding up the learning process by performing multiple updates simultaneously while adapting to the dynamic changes in the spectrum. *iii*) there is a dynamic balance between the exploration and exploitation due to the pure belief-based mode, while RL is driven by a value function that updates a single state action at each step. *iv*) In *AIn* the reliance on an explicit reward signal coming from the environment is not necessary; the reward is substituted by GEs that can be treated as self-information to avoid surprising states (i.e., states under attack) and reach the equilibrium. *To our best knowledge, this is the first work that adopts AIn for anti-jamming in Wireless Communications.*

## 6.2 System Model and Problem Formulation

Consider a cellular-connected UAV communicating with its respective Ground Base Station (GBS) to receive the tele-commands during a given mission of duration T over the Command and Control (C2) link which does not exceed a data rate of 100 Kbps [319], while a malicious terrestrial jammer transmits jamming signals with the intention of disturbing the legitimate UAV communications. The total mission duration is divided into equal time slots $t$, such that $t \in \{1, 2, \dots, \mathrm{T}\}$. The jammer may adopt constant, random or sweep jamming patterns during a certain experience. The UAV, GBS and jammer are denoted as $u$, $g$ and $j$, respectively. We consider a 3D Cartesian coordinate system where the coordinate of GBS and jammer are fixed at $o^g = [x^g, y^g, z^g]$ and $p^j = [x^j, y^j, z^j]$, respectively, while the time-varying coordinate of UAV at time instant $t$ is defined as $q^u_t = [x^u_t, y^u_t, z^u_t]$. In this chapter, we consider two different path loss models.

1. **3GPP RMa-AV path loss model (refer to 5.3):** the UAV communication link is assumed to be a LOS link under the RMa-AV scenario and the doppler frequency shift caused by the UAV mobility is assumed to be compensated at the receivers. Therefore, the path loss model ($\mathrm{PL_t}^{g,u}$) between the GBS and UAV at a given time instant follows the 3GPP path loss model which is defined in (5.3). Likewise, the path loss model ($\mathrm{PL_t}^{j,u}$) from jammer to the UAV is given in (5.6).

2. **Cellular to UAV angle–dependent (CtU-AD) path loss model [320]:** the path-loss model from the ground equipment (i.e., GBS or jammer) to UAV follows the cellular to UAV path-loss model dependent on the depression angles that extends the terrestrial path-loss model by adding an excess aerial path-loss, which can be expressed according to [321] as:
$$\mathrm{PL}_{\mathrm{t}}^{\mathrm{e,u}}(d_t, \theta_t) = \mathrm{PL}^{\mathrm{ter}}(d_t) + \eta(\theta_t) + \chi(\theta_t), \tag{6.1}$$
where $e \in \{g, j\}$, $\mathrm{PL}_{\mathrm{t}}^{\mathrm{ter}}(d_t) = 10\alpha \log(d_t)$ is the terrestrial path-loss of the point beneath the UAV, $\alpha$ is the terrestrial path-loss exponent that depends on the propagation environment and $d_t = \sqrt{(x^u_t - x^e)^2 + (y^u_t - y^e)^2}$ is the 2D distance between $e$ and $u$. In addition, $\eta(\theta_t) = C(\theta_t - \theta_0)\exp\big(-\frac{\theta_t - \theta_0}{D}\big) + \eta_0$ is the excess aerial path-loss and $\chi(\theta_t)$ is a zero-mean Gaussian variable with an angle-dependent standard deviation describing the shadowing effect such that $\chi(\theta_t) \sim \mathcal{N}(0, \sigma(\theta_t){=}a\theta_t + \sigma_0)$ where $C$ is the excess path-loss scaler, $D$ is the angle scaler, $\theta_0$ is the angle offset, $\eta_0$ is the excess path-loss offset, $a$ is the UAV shadowing slope, $\theta_t = \arctan\big(\frac{z^u_t - z^e_t}{d_t}\big)$ is the depression angle which approximately ranges from $(-2^\circ, 10^\circ)$ and $\sigma_0$ is the UAV shadowing offset.

The GBS assigns one Physical Resource Block (PRB) to the UAV each $t$ where C2 data are transmitted (including pitch, yaw and roll commands) [53]. The set of available links is denoted as $\mathcal{RB}$=$\{f_1,\ldots,f_n,\ldots,f_N\}$, $1 \leq \text{n} \leq N$, where $|\mathcal{RB}|$=$N$ is the total number of available PRBs that depends on the channel bandwidth $B$. To cope with the malicious jamming, the UAV aims to learn the best allocation strategy online by selecting the proper PRBs that are not targeted by the jammer while interacting with the environment and sending updated information to GBS to adapt to the environmental dynamic changes. Denote $\mathcal{H}_0$ and $\mathcal{H}_1$ as the hypotheses of the absence (i.e., UAV and jammer selected different PRBs) and presence (i.e., UAV and jammer selected the same PRB) of the jammer, respectively. The complex signal that is received at the UAV at time instant $t$ and over $f_n$ is given as:

$$r_{t,f_n} = \begin{cases} h^{g,u}_{t,f_n} x^{u}_{t,f_n} + v_t \quad under \ \mathcal{H}_0, \\ h^{g,u}_{t,f_n} x^{u}_{t,f_n} + h^{j,u}_{t,f_n} x^{j}_{t,f_n} + v_t \quad under \ \mathcal{H}_1, \end{cases} \tag{6.2}$$

where $x^{u}_{t,f_n}$ denotes the C2 signal, $h^{g,u}_{t,f_n} = \frac{1}{\text{PL}_t{}^{g,u}}$ is the channel gain from GBS to UAV, $x^{j}_{t,f_n}$ stands for the jammer's signal, $h^{j,u}_{t,f_n} = \frac{1}{\text{PL}_t{}^{j,u}}$ is the channel gain from jammer to UAV and $v_t$ is the random noise. The corresponding SINR at the UAV is given by:

$$\gamma_t = \frac{P^u_t h^{g,u}_{t,f_n}}{\alpha P^j_t h^{j,u}_{t,f_n} + \sigma^2}, \tag{6.3}$$

where $P^u_t$ is the transmitted power, $P^j_t$ is the jammer power, whose presence is denoted by $\alpha$ which is equal to 0 under $\mathcal{H}_0$ and equals to 1 under $\mathcal{H}_1$ and $\sigma^2$ is the noise power.

The anti-jamming defense problem can be formulated as a partially observable Markov decision process (POMDP) since the spectrum is only partially observable to the UAV (i.e., UAV cannot directly observe the environmental states but connects hidden states to observations in the form of probability distributions). A discrete-time POMDP that models the relationship between the UAV and its environment can be described as 7-element tuple $(S, X, \mathcal{A}, \mathcal{P}^u_\tau, \mathcal{P}^j_\tau, \Pi^{a^u}_\tau, \tilde{Z}_{t,f_n})$, where $S$ and $X$ are sets of the environmental hidden states, $\mathcal{A}$ is a set of actions where action is PRB selection ($a_t \in \mathcal{RB}$), $\mathcal{P}^u_\tau$ and $\mathcal{P}^j_\tau$ are the time-varying transition models for UAV and jammer, respectively. $\Pi^{a^u}_\tau$ is the AIn table that encodes the state-action couple and $\tilde{Z}_{t,f_n}$ are the observations received at each $t$ over $f_n$. During the offline training, UAV learns a dynamic model $\mathcal{M}$ encoding the dynamic rules that generate desired sensory signals (i.e., without jamming interference). During the active inference process (i.e., online learning), UAV predicts the environmental hidden states characterized by the posterior distributions $\text{P}(s^*_t{\in}S|z_t{\in}\tilde{Z}_{t,f_n}, \mathcal{M})$ and $\text{P}(x^*_t \in X|z_t \in \tilde{Z}_{t,f_n}, \mathcal{M})$ based on a prior belief (encoded in $\mathcal{M}$) and infers the actions most likely to generate preferred sensory signals (i.e., clean signals without jamming interference). Then, UAV can evaluate the situation after receiving the current observation $z_t$ and calculate the similarity between predictions and observations using a probabilistic distance $\mathcal{D}$ (i.e., abnormality indicator). If the similarity is high (i.e., $\mathcal{H}_0$), UAV can understand that the selected action has led to a desired states and to the reception of desired signals (without jamming attacks). If the similarity is low (i.e., $\mathcal{H}_1$), UAV can understand that the selected action is a bad action and updates $\Pi_\tau^{a^u}$ accordingly to avoid selecting actions that lead to surprising states (i.e., high abnormality). Therefore, while acting and sensing the spectrum, the UAV aims to minimise the cumulative abnormality:

$$\min_{a_t} \sum_{t=1}^{\mathrm{T}} \mathcal{D}\left( \mathrm{P}(s_t^*|z_t, \mathcal{M}), \mathrm{P}(z_t|s_t^*, \mathcal{M}) \right), \tag{6.4}$$

It is to note that (6.4) is equivalent to maximize the SINR.

## 6.3 Proposed Anti-jamming method

In this section, we propose an Active Inference method to solve the formulated anti-jamming problem. First, we define the radio environment and the UAV's model of that environment following generalized-state-space formulations. We use a generative model which jointly represents: the available resources and the spectrum occupancy of both cognitive-UAV and jammer; and the dynamics of the physical signals (i.e. desired observations learned offline). Then, we present the relationship between the internal model and its influence on the perception of the environmental states and on the decisions made by the UAV to select the best action that leads to minimum surprise (i.e., minimum abnormality).

### 6.3.1 Radio Environment Representation

We assume that the physical signals present in the radio environment can be cast at hierarchical levels using discrete and continuous Generalized-state-space models. The goal is to make inferences about hidden states and unknown parameters generating sensory signals, given only sensory observations. Generalized Filtering (i.e., Bayesian Filtering in Generalized coordinates) is used for that purpose since all random variables are represented in generalized coordinates of motion [303]. We assume that the spectrum is divided into $N$ resource blocks so containing $N$ sensory observations $\tilde{Z}_t$ which can be expressed as:

$$\tilde{Z}_t = \{\tilde{Z}_{t,f_1}, \tilde{Z}_{t,f_2}, \dots, \tilde{Z}_{t,f_N}\}. \tag{6.5}$$

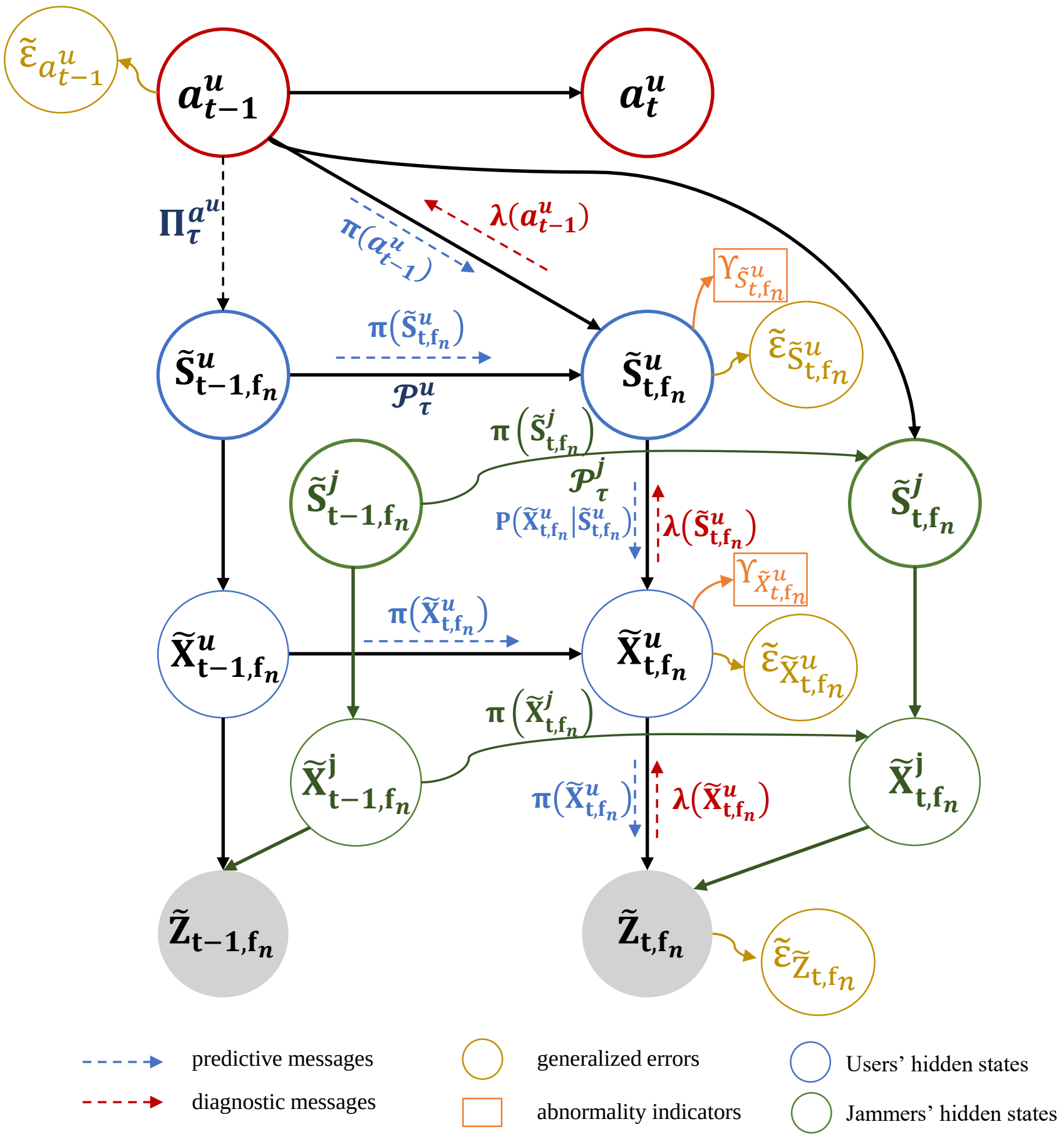

Fig. 6.1 Graphical representation of the proposed method. An Active-GDBN structure encoding: *i)* the dynamic rules that generate the preferred sensory signals used in off-line training phase for prediction and perception purposes. and *ii)* the possible actions that UAV can perform to achieve preferred sensory signals which is used during the online training for the active inference for joint perception-prediction purposes as well as for action.

The In-phase ($I$) and Quadrature ($Q$) components extracted at each $t$ from the received complex signal defined in (6.2) under $\mathcal{H}_0$ are grouped into a vector $\tilde{Z}_{t,f_n}$.

The multiple sensory signals at multiple resource blocks can be fused using a linearly conditionally dependent fusion mechanism according to:

$$\mathrm{P}(\tilde{Z}_t|\tilde{X}_{t,f_n}) = \alpha_{t,f_1}\mathrm{P}(\tilde{Z}_{t,f_1}|\tilde{X}_{t,f_1}) + \alpha_{t,f_2}\mathrm{P}(\tilde{Z}_{t,f_2}|\tilde{X}_{t,f_2}) + \cdots + \alpha_{t,f_N}\mathrm{P}(\tilde{Z}_{t,f_N}|\tilde{X}_{t,f_N}), \qquad (6.6)$$

where $\alpha_{t,f_n}$ denotes a discrete random variable that can take values either 1 or 0 which depends on the selected action ($a_{t-1}$) that indicates the sensing capability of the UAV and the range of frequencies it can scan, such that:

$$P(\alpha_{t,f_n}|a_{t-1} = f_i) = \delta(\alpha_{t,f_n} - f_i), \tag{6.7}$$

where, $\delta$ is the Dirac delta distribution. We assume that each sensory signal is a linear combination of one hidden Generalized State (GS) $\tilde{X}^u_{t,f_n}$ affected by additive random noise in a normal situation and by additional interference caused by the jammer in an abnormal situation and it is expressed as follows:

$$\tilde{Z}_{t,f_n} = H\tilde{X}^u_{t,f_n} + H\tilde{X}^j_{t,f_n} + \tilde{\upsilon}_{t,f_n}, \tag{6.8}$$

where $\tilde{Z}_{t,f_n} \in \mathbb{R}^d$ is the generalized observations including the signals' features in terms of $I$ and $Q$ components and the $1^{st}$-order temporal derivatives ($\dot{I}$, $\dot{Q}$). $\tilde{X}^u_{t,f_n}$ is the UAV's Generalized state (GS) and $\tilde{X}^j_{t,f_n}$ is the jammer's GS that is caused by $\tilde{S}^j_{t,f_n}$. $H \in \mathbb{R}^{d,d}$ maps hidden states to observations, $d$ is the space dimensionality, $f_n$ is the $n$-th resource block where $f_n \in \mathcal{RB}$ and $\tilde{\upsilon}_{t,f_n} \sim \mathcal{N}(0, \Sigma_{\tilde{\upsilon}_{t,f_n}})$. The dynamic evolution of $\tilde{X}^u_{t,f_n}$ is affected by the previous GS $\tilde{X}^u_{t-1,f_n}$ and guided by its direct cause $\tilde{S}^u_{t,f_n}$ which can be formulated as:

$$\tilde{X}^u_{t,f_n} = A\tilde{X}^u_{t-1,f_n} + BU_{\tilde{S}^u_{t,f_n}} + \tilde{w}_{t,f_n}, \tag{6.9}$$

where $A \in \mathbb{R}^{d,d}$, $B \in \mathbb{R}^{d,d}$ are the dynamic model and control model matrices, respectively. $U_{\tilde{S}^u_{t,f_n}}$ is the control vector and $\tilde{w}_{t,f_n}$ is the Generalized process noise such that, $\tilde{w}_{t,f_n} \sim \mathcal{N}(0, \Sigma_{\tilde{w}_{t,f_n}})$. The Generalized superstates are random discrete variables describing the discrete clusters of the UAVs' signals that evolve according to:

$$\tilde{S}^u_{t,f_n} = \mathrm{f}(\tilde{S}^u_{t-1,f_n}) + \tilde{w}_{t,f_n}, \tag{6.10}$$

where $\mathrm{f}(.)$ is a non-linear function describing the signals' dynamic transitions among the discrete variables and its evolution over time at a specific resource block $f_n$.

### 6.3.2 Offline Learning of desired observations

During training, we assume that the jammer is absent and the UAV aims to learn the dynamics of the desired observations (i.e., C2 signals without jamming interference) while sensing the spectrum. UAV starts perceiving the surroundings by sensing partially the spectrum (sense the allocated resources only) supposing that no signals are present and observations are subject to a stationary noise process that evolves according to static rules. UAV relays on (6.9) to predict the continuous signal's state where the force at sensing resource block ($f_n$) is $U_{\tilde{S}_{t,f_n}} = 0$, as no rules has been discovered yet. In case of active transmissions in $f_n$, UAV detects abnormalities all the time and calculates the Generalized Errors (GEs) projected on the GS space:

$$\tilde{\mathcal{E}}_{\tilde{X}^u_{t,f_n}} = \left[\tilde{X}^u_{t,f_n}, \mathrm{P}(\dot{\mathcal{E}}_{\tilde{X}^u_{t,f_n}})\right] = \left[\tilde{X}^u_{t,f_n}, H^{-1}\tilde{\mathcal{E}}_{\tilde{Z}_{t,f_n}}\right] = \left[\tilde{X}^u_{t,f_n}, H^{-1}\big(\tilde{Z}_{t,f_n} - H\tilde{X}^u_{t,f_n}\big)\right], \qquad (6.11)$$

where $\dot{\mathcal{E}}_{\tilde{X}^u_{t,f_n}}$ is the difference between predictions and observations that capture the dynamics of the signals present inside the spectrum and should be applied to $\tilde{X}^u_{t,f_n}$. GEs can be clustered in an unsupervised manner using the Growing Neural Gas (GNG) to learn the top level of abstraction allowing the UAV to predict signals' states at higher level which guide the prediction at lower levels. GNG produces a set of superstates (or clusters) encoding the GEs into discrete regions described by the set $S^u_{f_n}$, such that:

$$S^u_{f_n} = \{S^u_{1,f_n}, S^u_{2,f_n}, \dots, S^u_{M,f_n}\}, \qquad (6.12)$$

where $M$ is the total number of clusters associated with a specific resource block $f_n$.

Analysing the signal's dynamic transitions among the superstates allows to estimate the transition probabilities $\pi^u_{if_n|jf_n} = \mathrm{P}(S^u_{t,f_n} = i | S^u_{t-1,f_n} = j)$ where $i, j \in S^u_{f_n}$ and encoded in the transition matrix $\Pi^u_{f_n|f_n}$. Furthermore, since the signal states are dynamic and vary with time, estimating the time-varying transition matrix $\Pi^u_{f_n|f_n,\tau}$ is of great interest. $\Pi^u_{f_n,\tau}$ encodes the time-varying transition probabilities $\pi^u_{if_n|jf_n,\tau} = \mathrm{P}(S^u_{t,f_n} = i | S^u_{t-1,f_n} = j, \tau)$ realizing the condition of transiting to a new superstate $S^u_{t,f_n} = i$ after being in $S^u_{t-1,f_n} = j$ for a certain time $\tau$ which allows to keep tracking the dynamic changes in the environment. The generalized superstates then can be expressed in the following form: $\tilde{S}^u_{t,f_n} = [S^u_{t,f_n}\ \dot{S}^u_{t,f_n}] = [S^u_{t,f_n}\ e^{ij}_{t,f_n}]$, where $e^{ij}_{t,f_n}$ is the event occurred at time $t$. Moreover, each discrete variable $\tilde{S}^u_{m,f_n} \in \mathbf{S^u_{f_n}}$ is associated with statistical proprieties as generalized mean $\tilde{\mu}_{\tilde{S}^u_{m,f_n}}$ and covariance $\Sigma_{\tilde{S}^u_{m,f_n}}$. In this way, UAV has been trained to learn and encode the dynamic rules that generate desired sensory signals (i.e., without jamming attacks) using multiple sensory signals (over multiple resource blocks).

### 6.3.3 Active Inference stage (online learning)

The hierarchical dynamic causal models are formulated in terms of stochastic processes as defined in (6.8), (6.9), (6.10) and structured in an Active Generalized Dynamic Bayesian Networks (Active-GDBN) to model the dynamics of the corresponding nodes at multiple levels (refer to Fig. 6.1). Active-GDBN is a probabilistic graphical model with direct relationships and time-varying connectivity reflecting the cause-effect relationship among random variables and characterized by conditional dependencies in terms of probability distributions. The Active-GDBN allows to solve the POMDP to find the best set of actions by predicting the situation the UAV could encounter in the future, conditioned on the actions it executes. Thus, *AIn* provides a way, through planning as inference, to form beliefs about the future and describe the causal relationship among actions, hidden states and outcomes at multiple levels.

### Initialization

$\mathcal{P}^u_\tau$ and $\mathcal{P}^j_\tau$ are the $N \times N$ time-varying matrices encoding the possible transitions among the $N$ available resources performed by the UAV and encoding the UAV's belief about the possible actions that the jammer can perform, respectively. Since there is no a priori information concerning the jammer's behaviour inside the spectrum, the probability entries in both $\mathcal{P}^u_\tau$ and $\mathcal{P}^j_\tau$ are initially assigned equal values:

$$\mathcal{P}^u_\tau = \begin{bmatrix} \mathrm{P}(\Pi^u_{f_1|f_1,\tau}) & \cdots & \mathrm{P}(\Pi^u_{f_1|f_N,\tau}) \\ \vdots & \ddots & \vdots \\ \mathrm{P}(\Pi^u_{f_N|f_1,\tau}) & \cdots & \mathrm{P}(\Pi^u_{f_N|f_N,\tau}) \end{bmatrix}, \tag{6.13}$$

$$\mathcal{P}^j_\tau = \begin{bmatrix} \mathrm{P}(\Pi^j_{f_1|f_1,\tau}) & \cdots & \mathrm{P}(\Pi^j_{f_1|f_N,\tau}) \\ \vdots & \ddots & \vdots \\ \mathrm{P}(\Pi^j_{f_N|f_1,\tau}) & \cdots & \mathrm{P}(\Pi^j_{f_N|f_N,\tau}) \end{bmatrix}, \tag{6.14}$$

where $\mathrm{P}(\Pi^u_{f_r|f_q,\tau}) = \frac{1}{N}$, $\mathrm{P}(\Pi^j_{f_r|f_q,\tau}) = \frac{1}{N}$ $\forall r,q \in \mathcal{RB}$. $\Pi^{a^u}_\tau \in \mathbb{R}^{N,N}$ is a time-varying matrix encoding the probabilistic dependencies between states and actions representing the link $a^u_{t-1} \rightarrow \tilde{S}^u_{t-1,f_n}$ in the Active-GDBN that describes $\mathrm{P}(a^u_{t-1} = f_i|\tilde{S}^u_{t-1,f_k})$ and defined as:

$$\Pi^{a^u}_\tau = \begin{bmatrix} \mathrm{P}(a_1 = f_1|\tilde{S}^u_{t-1,f_1}) & \cdots & \mathrm{P}(a_N = f_N|\tilde{S}^u_{t-1,f_1}) \\ \vdots & \ddots & \vdots \\ \mathrm{P}(a_1 = f_1|\tilde{S}^u_{t-1,f_N}) & \cdots & \mathrm{P}(a_N = f_N|\tilde{S}^u_{t-1,f_N}) \end{bmatrix}, \tag{6.15}$$

where

$$\mathrm{P}(a^u_{t-1} = f_i|\tilde{S}^u_{t-1,f_k}) = \frac{1}{N} \quad \forall i,k \in \mathcal{RB}. \tag{6.16}$$

UAV's action depends on the state-action couple encoded in $\Pi^{a^u}_\tau$ and on its belief about the presence of the jammer in the radio spectrum encoded in $\mathcal{P}^j_\tau$.

### Action selection process

Initially, UAV performs random sampling to select the actions during the $1^{st}$ iteration as every possible action has the same probability ($\frac{1}{N}$) of being chosen. The selected action $a^u_{t-1}$ indicates what will be the next hidden state $\tilde{S}^u_{t,f_n}$ according to $\mathrm{P}(\tilde{S}^u_{t,f_n}|\tilde{S}^u_{t-1,f_n},a^u_{t-1})$. $\tilde{S}^u_{t,f_n}$ encodes the predicted cluster of the model and the activated PRB ($f_n$).

In the successive iterations, first, UAV predicts the future activity of the jammer implicitly according to $\mathcal{P}^u_\tau$. Then, it can adjust the action selection step by skipping the risky resources (i.e., resources expected with high probability to be targeted by the jammer in the near future). The action selection procedure depends on a certain policy adopted by the UAV according to:

$$a^{u*}_{t-1} = \mathrm{argmax}_{\tilde{S}^u_{t-1,f_k},\mathcal{P}^u_\tau(\tilde{S}^u_{t-1,f_k})}\, \pi(a^u_{t-1}), \tag{6.17}$$

where $\pi(a^u_{t-1})$=$\mathrm{P}(a^u_{t-1}|\tilde{S}^u_{t-1,f_k})$ is a specific row in $\Pi^{a^u}_\tau$ and $\mathcal{P}^u_\tau(\tilde{S}^u_{t-1,f_k})$ is a specific row selected from ($\mathcal{P}^u_\tau$) representing the dynamic model associated with ($\tilde{S}^u_{t-1,f_k}$) where the jammer's transitions are implicitly encoded. The model has prior belief about how a certain state ($\tilde{S}^u_{t-1,f_k}$) will evolve into another ($\tilde{S}^{u*}_{t,f_k}$) depending on the chosen action ($a^{u*}_{t-1}$) according to $\mathrm{P}(\tilde{S}^{u*}_{t,f_k}|a^{u*}_{t-1},\tilde{S}^u_{t-1,f_k})$, where $\tilde{S}^{u*}_{t,f_k}$ is the expected state associated with the selected action.

### Perception and joint state-prediction

After selecting the action that indicate the chosen PRB, UAV can relay on the corresponding transition matrix ($\Pi^u_{f_r|f_q,\tau}$) to perform the predictions by employing the M-MJPF that uses a combination of Particle Filter (PF) and a bank of Kalman Filters (KFs) to predict the generalized superstates and states, respectively. The M-MJPF within the Generalized Filtering provides three probabilistic inference modes; *top-down* causal inference, *lateral* temporal inference and *bottom-up* diagnostic inference. In the top-down mode, predictive messages are passed from hierarchically higher levels to lower levels where predictions are based on the rules learned in previous experiences and are supported by the lateral predictive messages between two consecutive time slices to explain the time-varying changes. In contrast, diagnostic messages travelling in the opposite direction from lower to higher levels encode the GEs signalling. PF starts by propagating $L$ particles ($<.>$) equally weighted based on the proposal density encoded in $\Pi^u_{f_r|f_q,\tau}$, such that: $<\tilde{S}^{u,l}_{t,f_n},W^l_t> \sim <\pi^u_{if_n|jf_n,\tau},\frac{1}{L}>$. For each particle $\tilde{S}^{u,l}_{t,f_n}$ a KF is employed to predict $\tilde{X}^u_{t,f_n}$. The prediction at this level is driven by the higher level as pointed out in (6.9) (where $U_{\tilde{S}^u_{t,f_n}} = \tilde{\mu}_{\tilde{S}^{u,l}_{t,f_n}}$) which can be expressed as $\mathrm{P}(\tilde{X}^u_{t,f_n}|\tilde{X}^u_{t-1,f_n},\tilde{S}^u_{t,f_n})$. The posterior probability associated with $\tilde{X}^u_{t,f_n}$ is given by:

$$\pi(\tilde{X}^u_{t,f_n}) = \mathrm{P}(\tilde{X}^u_{t,f_n},\tilde{S}^u_{t,f_n}|\tilde{Z}_{t-1,f_n}) = \int \mathrm{P}(\tilde{X}^u_{t,f_n}|\tilde{X}^u_{t-1,f_n},\tilde{S}^u_{t,f_n})\lambda(\tilde{X}^u_{t-1,f_n})d\tilde{X}^u_{t-1,f_n}, \tag{6.18}$$

where $\lambda(\tilde{X}^u_{t-1,f_n}) = \mathrm{P}(\tilde{Z}_{t-1,f_n}|\tilde{X}^u_{t-1,f_n})$.

Once a new sensory signal is received, diagnostic messages propagate in a bottom-up manner to adjust the expectations and update belief in hidden variables. Thus, posterior can be updated using:

$$\mathrm{P}(\tilde{X}^u_{t,f_n},\tilde{S}^u_{t,f_n}|\tilde{Z}_{t,f_n}) = \pi(\tilde{X}^u_{t,f_n})\lambda(\tilde{X}^u_{t,f_n}). \tag{6.19}$$

In addition, the likelihood message $\lambda(\tilde{S}^u_{t,f_n})$ can be used to update the particles' weights according to: $W^l_t = W^l_t\lambda(\tilde{S}^u_{t,f_n})$, where:

$$\lambda(\tilde{S}^u_{t,f_n}) = \lambda(\tilde{X}^u_{t,f_n})\mathrm{P}(\tilde{X}^u_{t,f_n}|\tilde{S}^u_{t,f_n}) = \mathrm{P}(\tilde{Z}^u_{t,f_n}|\tilde{X}^u_{t,f_n})\mathrm{P}(\tilde{X}^u_{t,f_n}|\tilde{S}^u_{t,f_n}), \tag{6.20}$$

and $\mathrm{P}(\tilde{X}^u_{t,f_n}|\tilde{S}^u_{t,f_n}) \sim \mathcal{N}(\mu_{\tilde{S}^u_m}, \Sigma_{\tilde{S}^u_m})$ denotes a multivariate Gaussian distribution. Also, the GE ($\tilde{\mathcal{E}}_{\tilde{S}^u_{t,f_n}}$) at the superstate level conditioned on transiting from $\tilde{S}^u_{t-1,f_n}$ can be expressed as:

$$\tilde{\mathcal{E}}_{\tilde{S}^u_{t,f_n}} = \left[\tilde{S}^u_{t-1,f_k}, \dot{\mathcal{E}}_{\tilde{S}^u_{t,f_n}}\right], \tag{6.21}$$

where $\dot{\mathcal{E}}_{\tilde{S}^u_{t,f_n}}$ is an aleatory variable whose probability density function is given by:

$$\mathrm{P}(\dot{\mathcal{E}}_{\tilde{S}^u_{t,f_n}}) = \lambda(\tilde{S}^u_{t,f_n}) - \pi(\tilde{S}^u_{t,f_n}), \tag{6.22}$$

representing the new force that can be used to update the original transition matrix ($\mathcal{P}^u_\tau$) and thus improve future predictions.

### Abnormality measurements

The abnormality indicators can be used to evaluate the radio situation and discover if something wrong occurred in the radio environment that violates the dynamic rules learned under normal circumstances. In order to evaluate to what extent the current signal's evolution at the discrete level matches the predicted one based on the learned and encoded dynamics in the model, we used an abnormality indicator ($\Upsilon_{\tilde{S}^u_{t,f_n}}$) based on the Symmetric Kullback-Leibler Divergence ($D_{KL}$) [53]. $\Upsilon_{\tilde{S}^u_{t,f_n}}$ calculates the similarity between the two messages that represent discrete probability distributions entering to node $\tilde{S}^u_{t,f_n}$, namely, $\pi(\tilde{S}^u_{t,f_n})$ and

$\lambda(\tilde{S}^u_{t,f_n})$, it is associated with (6.21) and can be formulated as:

$$\Upsilon_{\tilde{S}^u_{t,fn}} = \sum_{i \in \mathcal{S}} \mathrm{P}_r(\tilde{S}^u_{t,f_n} = i) D_{KL}\big(\pi(\tilde{S}^u_{t,f_n})||\lambda(\tilde{S}^u_{t,f_n})\big) + \sum_{i \in \mathcal{S}} \mathrm{P}_r(\tilde{S}^u_{t,f_n} = i) D_{KL}\big(\lambda(\tilde{S}^u_{t,f_n})||\pi(\tilde{S}^u_{t,f_n})\big), \tag{6.23}$$

where $\mathrm{P}_r(\tilde{S}^u_{t,f_n})$ is the probability of occurrence of each superstate picked from the histogram at time instant $t$ and calculated as follows: $\mathrm{P}_r(\tilde{S}^u_{t,f_n}) = \frac{fr(\tilde{S}^u_{t,fn}=i)}{N}$, where $fr(.)$ is the frequency of occurrence of a specific superstate $i$ and $N$ is the total number of particles propagated by PF and $\mathcal{S}$ is the set consisting of all the winning particles, such that: $\mathcal{S} = \{i|\mathrm{P}_r(\tilde{S}^u_{t,f_n}) > 0\}$, $i \in S^u_{f_n}$.

Likewise, it is possible to understand how much the observation supports the predictions at the GS level using:

$$\Upsilon_{\tilde{X}^u_{t,fn}} = -\ln\Big(\mathcal{BC}\big(\pi(\tilde{X}^u_{t,f_n}), \lambda(\tilde{X}^u_{t,f_n})\big)\Big) = \int \sqrt{\pi(\tilde{X}^u_{t,f_n})\lambda(\tilde{X}^u_{t,f_n})} d\tilde{X}^u_{t,f_n}, \tag{6.24}$$

where $\Upsilon_{\tilde{X}^u_{t,fn}}$ is associated with (6.11) and $\mathcal{BC}$ is the Bhattacharyya coefficient.

### Updating of action selection process

UAV perceives the surrounding radio environment through sensory observations and changes the environment through actions. Then it can infer the effects of the performed actions through observations due to the separation between the UAV and the environment through the Markov blanket [322].

After acting in the environment, UAV can save the consequence of the chosen action (i.e., the transition from $\tilde{S}^u_{t-1,f_k}$ to $\tilde{S}^{u*}_{t,f_k}$) in $\mathcal{P}^u_\tau$ and evaluate how much the sensory outcomes support predictions and thus evaluate if the performed action was good or bad by using the abnormality measurements defined in (6.23) and (6.24). In addition, it is possible to calculate the GE ($\tilde{\mathcal{E}}_{a^u_{t-1}}$) during abnormal situation to adapt the UAV's strategy in selecting actions and understand how it should behave in the future to avoid the jammer. $\tilde{\mathcal{E}}_{a^u_{t-1}}$ is the difference between observation and expectation which can be expressed as:

$$\tilde{\mathcal{E}}_{a^u_{t-1}} = \big[a^{u*}_{t-1}, \dot{\mathcal{E}}_{a^u_{t-1}}\big], \tag{6.25}$$

where $\dot{\mathcal{E}}_{a^u_{t-1}}$ depicts an aleatory variable representing the new force that should be applied to update $\pi(a^u_{t-1})$ and its probability density function is given by:

$$\mathrm{P}(\dot{\mathcal{E}}_{a^u_{t-1}}) = \lambda(a^u_{t-1}) - \pi(a^u_{t-1}), \tag{6.26}$$

that can be used as a metric alternative to the reward in RL. $\lambda(a^u_{t-1})$ is the diagnostic message travelling from $\tilde{S}^u_{t,f_n}$ towards $a^u_{t-1}$ and defined as:

$$\lambda(a^u_{t-1}) = \lambda(\tilde{S}^u_{t,f_n})\mathrm{P}(\tilde{S}^u_{t,f_n}|a^u_{t-1}). \tag{6.27}$$

$\lambda(a^u_{t-1})$ is a discrete probability distribution that holds information about the observed sensory signal and encoding the probabilities about how the states $\tilde{S}^u_{t,f_n}$ belonging to the available frequencies change based on the evidence, it is given by:

$$\lambda(a^u_{t-1}) = \begin{cases} \mathcal{P}_{\tau-1}(\tilde{S}^u_{t-1,f_n}) - \gamma^*, \ \mathit{if}\, a^u_{t-1} = a^{u*}_{t-1}, \\ \mathcal{P}_{\tau-1}(\tilde{S}^u_{t-1,f_n}) + \frac{\gamma^*}{N-1}, \ \mathit{if}\, a^u_{t-1} \neq a^{u*}_{t-1}, \end{cases} \tag{6.28}$$

where $\gamma$ depends on the the GE defined in (6.21), that is:

$$\gamma = \begin{cases} \gamma^* \ \mathit{if}\, \tilde{\mathcal{E}}_{\tilde{S}_{t,f_k}} \geq th, \\ 0 \ \ \mathit{if}\, \tilde{\mathcal{E}}_{\tilde{S}_{t,f_k}} < th, \end{cases} \tag{6.29}$$

where *th* is the threshold indicating whether the radio situation is normal or abnormal and the value of $\gamma^*$ depends on the abnormality indicators defined in (6.23) and (6.24). Hence, GE defined in (6.25) is proportional to (6.21) due to the messages propagated from lower level towards the higher levels, such that $\tilde{\mathcal{E}}_{a^u_{t-1}} = f(\tilde{\mathcal{E}}_{\tilde{S}^u_{t,fn}})$. When the user get surprised by the sensory outcomes after performing a certain action, it can use the prediction error signal to update the belief about jammer's transition model to improve future actions. The core idea is that the user occupying a piece of the spectrum should minimize the abnormality (surprise) associated with finding itself in unlikely states (states under attack). Jammer's dynamic model ($\mathcal{P}^j_\tau$) can be updated following:

$$\mathcal{P}^j_\tau(.\,,\tilde{S}^j_{t,f_n}) = \mathcal{P}^j_{\tau-1}(.\,,\tilde{S}^j_{t,f_n}) - \mathrm{P}(\dot{\mathcal{E}}^u_{a_{t-1}}), \tag{6.30}$$

In an abnormal situation, user and jammer are sharing the same resource block which means that they performed the same action. Thus, user should update $\Pi^{a^u}_\tau$ by decreasing the probability of selecting that action as follows:

$$\pi^*(a^u_{t-1}) = \pi(a^u_{t-1}) + \mathrm{P}(\dot{\mathcal{E}}^u_{a_{t-1}}), \tag{6.31}$$

and update $\mathcal{P}^u_\tau$ by decreasing the probability of transiting to $\tilde{S}^u_{t,f_k}$ from $\tilde{S}^u_{t-1,f_k}$ after choosing action $a^{u*}_{t-1}$ using the GE defined in (6.21) following:

$$\mathcal{P}^u_\tau(\tilde{S}^u_{t-1,f_k},\tilde{S}^u_{t,f_n}) = \mathcal{P}^u_{\tau-1}(\tilde{S}^u_{t-1,f_k},\tilde{S}^u_{t,f_n}) + \mathrm{P}(\dot{\mathcal{E}}_{\tilde{S}^u_{t,f_n}}). \tag{6.32}$$

Operating in a pure belief-based setting under *AIn* allows the user to carry out epistemic exploration (seeking information and resolving uncertainty) in a Bayesian-optimal fashion. Also, the reliance on an explicit reward signal is not necessary in *AIn*, where reward can be substituted by the generalized errors calculated by the Active-GDBN model that can be treated as self-information to reach the equilibrium.

## 6.4 Results and Discussion

In this section, simulation results are provided to evaluate the performance of the proposed approach for anti-jamming. In our analysis we consider the following three types of jammers: *1)* Constant jammer that acts on statistically pre-configured channels; *2)* Sweep jammer that attacks by sweeping among the available PRBs at each time slot; and *3)* Random jammer that selects uniformly random actions to attack the available resources at each time slot.
Following the guideline about the simulation setup for cellular-connected UAVs in [258, 254], we set the parameters for our simulator as: bandwidth (BW) varies from 1.4 MHz to 20 MHz such that BW={1.4 MHz, 3 MHz, 5 MHz, 10 MHz, 15 MHz, 20 MHz}; duplex mode is FDD; sub-carrier spacing equals 15 KHz; the number of PRBs per BW is {6, 15, 25, 50, 75, 100}; sampling frequency is 1.92 MHz; $N_{FFT}$ equals 128; OFDM symbols per slot is 7; CP lenght is normal; SNR equals $15dB$; C2 modulation is QPSK; jamming signal is QPSK modulated; jamming to signal power ratio (JSR) equals 6dB; and the total number of radio frames is 200. In addition, generation of C2 data, jamming signals and UAV trajectory are given in [53].

Two channel models are considered as discussed in Section 6.2 and simulation results are divided into two parts. In part I, we assume that the ground-to-UAV link is always a line-of-sight under an AWGN channel condition and we adopted the 3GPP path loss model under the RMa-AV scenario defined in [254]. Also, we consider different jamming hit rate (JHR) which is the percentage of jammed PRBs where JHR={20%, 40%, 60%, 80%}.
In part II, the cellular to UAV (CtU) angle-dependent (CtU-AD) path loss model is used in simulation. Under the CtU-AD model, we assume that the propagation environment is a typical suburban, the mean aerial speed is 4.8m/s, the BS height is 30m, the UAV height is 60m and the channel model parameters [321] are $\alpha = 3.04$, $\sigma_0 = 8.52$, $C = -23.29$, $\eta_0 = 20.70$, $\theta_0 = -3.61$, $D = 4.14$, $a = -0.41$, $\sigma_0 = 5.86$. Also, we consider a jamming hit rate (JHR), which is the percentage of jammed PRBs, of JHR=40%.

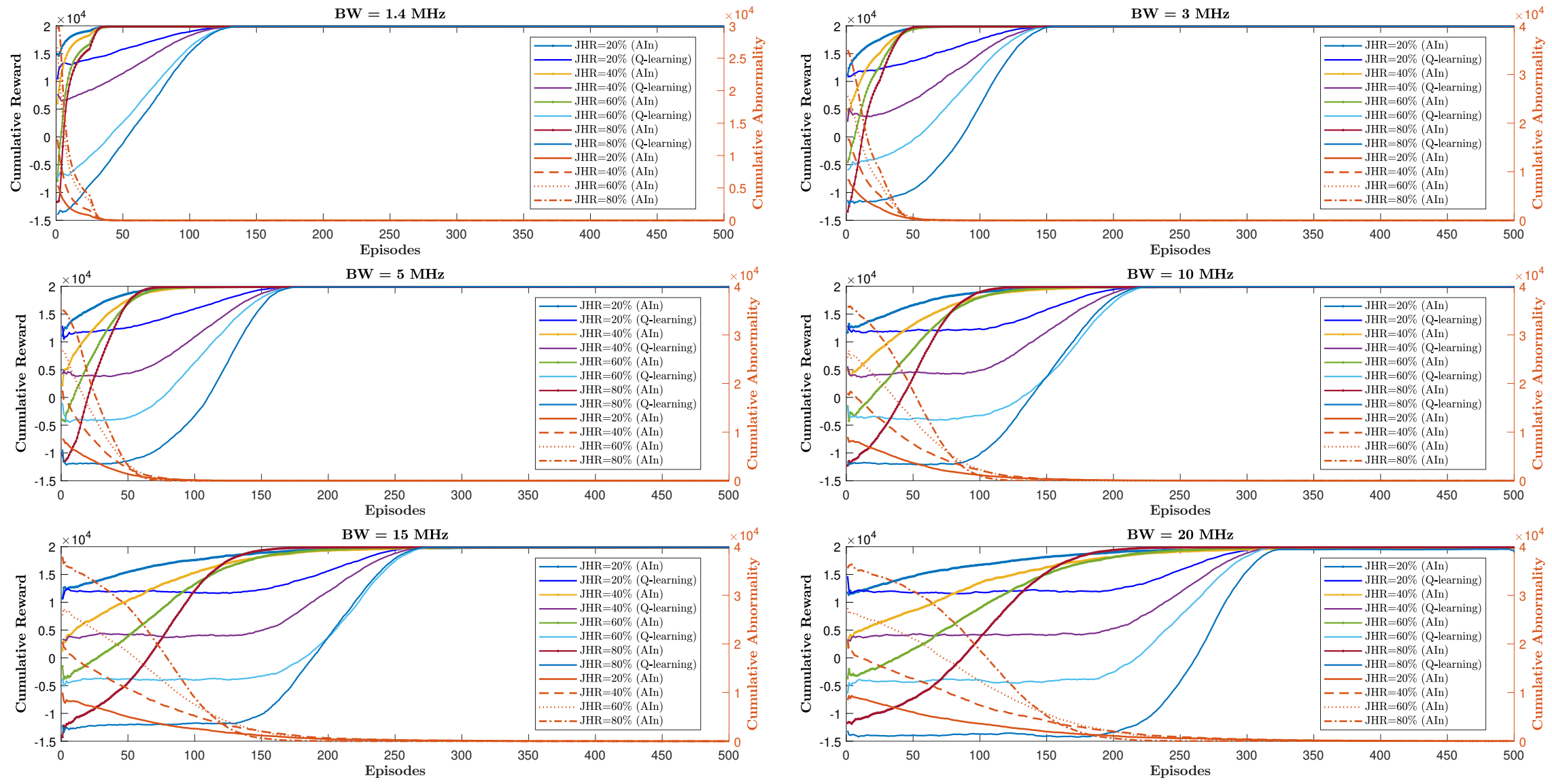

Fig. 6.2 Performance comparison of cumulative reward with the proposed *AIn* and the QL under the 3GPP RMa-AV model, *Constant* jamming strategy and different Bandwidth (BW).

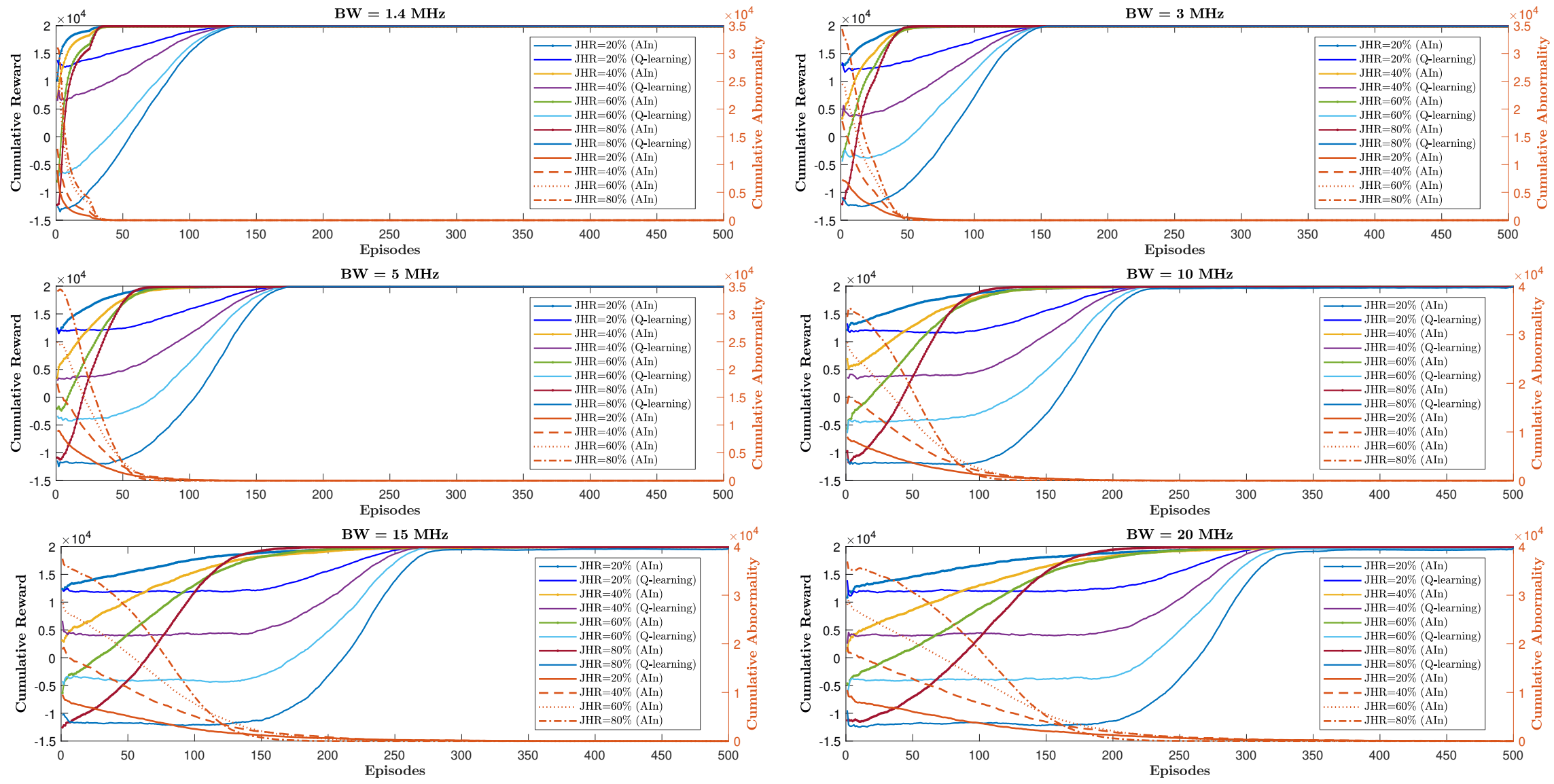

Fig. 6.3 Performance comparison of cumulative reward with the proposed *AIn* and the QL under the 3GPP RMa-AV model, *Random* jamming strategy and different Bandwidth (BW).

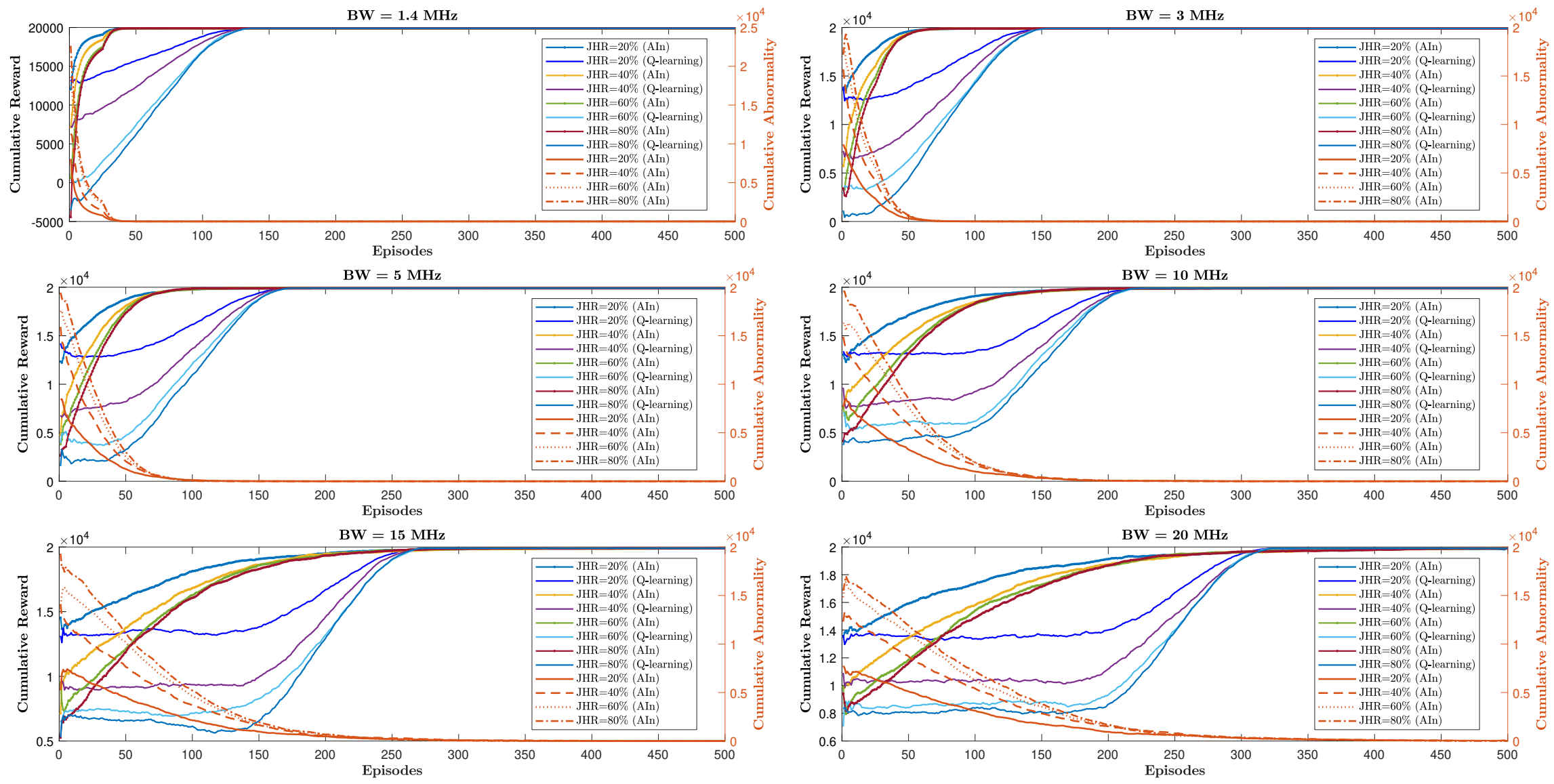

Fig. 6.4 Performance comparison of cumulative reward with the proposed *AIn* and the QL under the 3GPP RMa-AV model, *Sweep* jamming strategy and different Bandwidth (BW).

## 6.4.1 3GPP RMa-AV path-loss model (Part I)

Let us compare the performance of *AIn* in terms of cumulative abnormality (defined in (6.23)) and cumulative reward with that of random Frequency Hopping (FH-random) and Q-Learning (QL), as illustrated in Figs. 6.2-6.3-6.4. Here, the objective of *AIn* is to minimize abnormality while that of QL is to maximize reward. Thus, the reward is considered in *AIn* approach just for the sake of comparison with QL. We consider a binary reward which is equal to $-1$ under $\mathcal{H}_1$ and $+1$ under $\mathcal{H}_0$. Nevertheless, the relationship of these metrics is opposites to one another , i.e., when cumulative reward increases, cumulative abnormality decreases, as can be seen in Figs. 6.2-6.3-6.4. For a fair comparison with QL, we use time-varying q-tables to deal with the dynamic environmental changes. The exploration process in QL follows the $\varepsilon$-greedy policy with $\varepsilon = 1$ decaying to 0. It can be seen from the figures that *AIn* outperforms QL and FH-random under different BW values, JHRs and jamming strategies. *AIn* converges faster than the QL due to its capability in performing multiple updates in a pure belief-based mode. Another important outcome is that *AIn* approach reach balanced levels of exploration-exploitation that can maintain their orientation even if the environmental conditions change due to the fulfilled probabilistic updates.

We can observe that both *AIn* and QL reach the convergence for all the considered JHR values somehow at the same time but with different cumulative reward and cumulative abnormality values. The cumulative reward and abnormality values related to small JHR are greater than those related to high JHR and vice-versa before convergence. This can be explained by the fact that facing a jammer who is attacking with low JHR leads to less errors (wrong actions) as the agent has more possibilities to escape from the jammer compared to a situation where the jammer is attacking with high JHR and thus the number of safe resources becomes smaller which increase the probability of taking wrong actions that maximize the reward (minimize abnormality). Instead, making more errors allows to discover the jammer's strategy early which explains why different JHR values converge at the same time.

As shown in Figs. 6.2-6.3-6.4, when BW increases, the time needed to reach the convergence (for both *AIn* and QL) increases as the active agent needs more time to discover the environment and resolve uncertainty while observing partially the spectrum. The proposed approach can overpass this issue by activating multiple sensory signals which faster the process of seeking information about the environment and this will be studied in future work.

### 6.4.2 CtU-AD path-loss model (Part II)

In this part, we change the channel model by adopting the cellular to UAV angle-dependent path loss model as discussed in Section 6.2 to evaluate the performance of the proposed method and validate its flexibility in dealing with different channel conditions. Figs. 6.5-6.6-6.7 shows the performance comparison of cumulative reward and abnormality with the proposed *AIn*, the conventional Frequency Hopping (FH-random) and QL under different jamming patterns. We can observe that the proposed approach can explore the environment even when the channel condition varies, characterize the jammer and discover its strategy in attacking the PRBs. Figs. 6.8-6.9-6.10 depicts the cumulative SINR under different BW and jamming patterns achieved by the proposed *AIn* and compared it with FH-random and QL. By observing Figs. 6.5-6.6-6.7 and Figs. 6.8-6.9-6.10, we can notice that minimizing the abnormality (or maximizing the reward) leads to maximizing the SINR where the time needed to reach the convergence is equivalent to that in Figs. 6.5-6.6-6.7 and *AIn* beats both the FH-random and QL. This means that avoiding surprising states minimizes the abnormality and maximises reward and SINR.

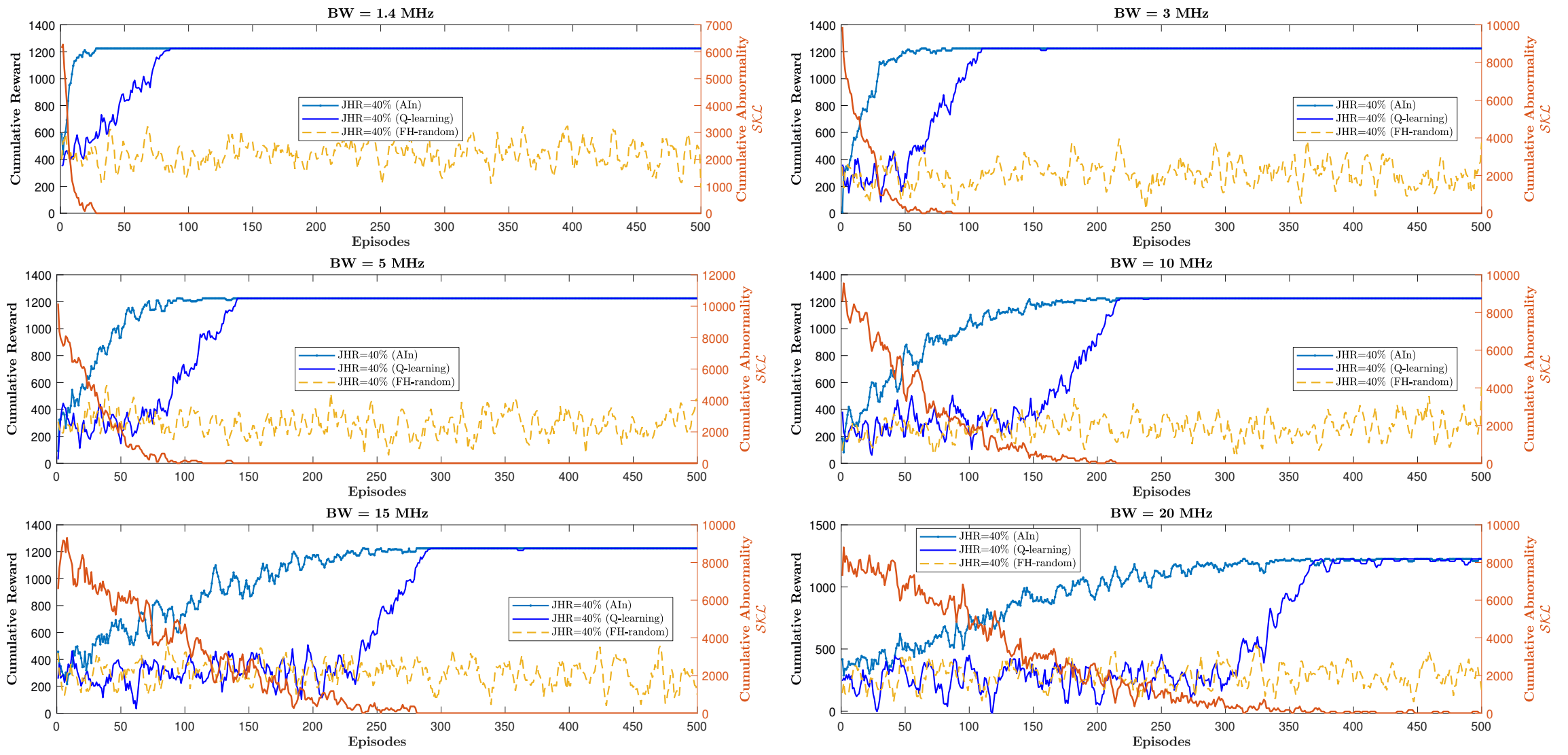

Fig. 6.5 Performance comparison of cumulative reward and abnormality with the proposed *AIn*, FH-random and QL under the CtU-AD mode, *Constant* jamming strategy and different Bandwidth (BW).

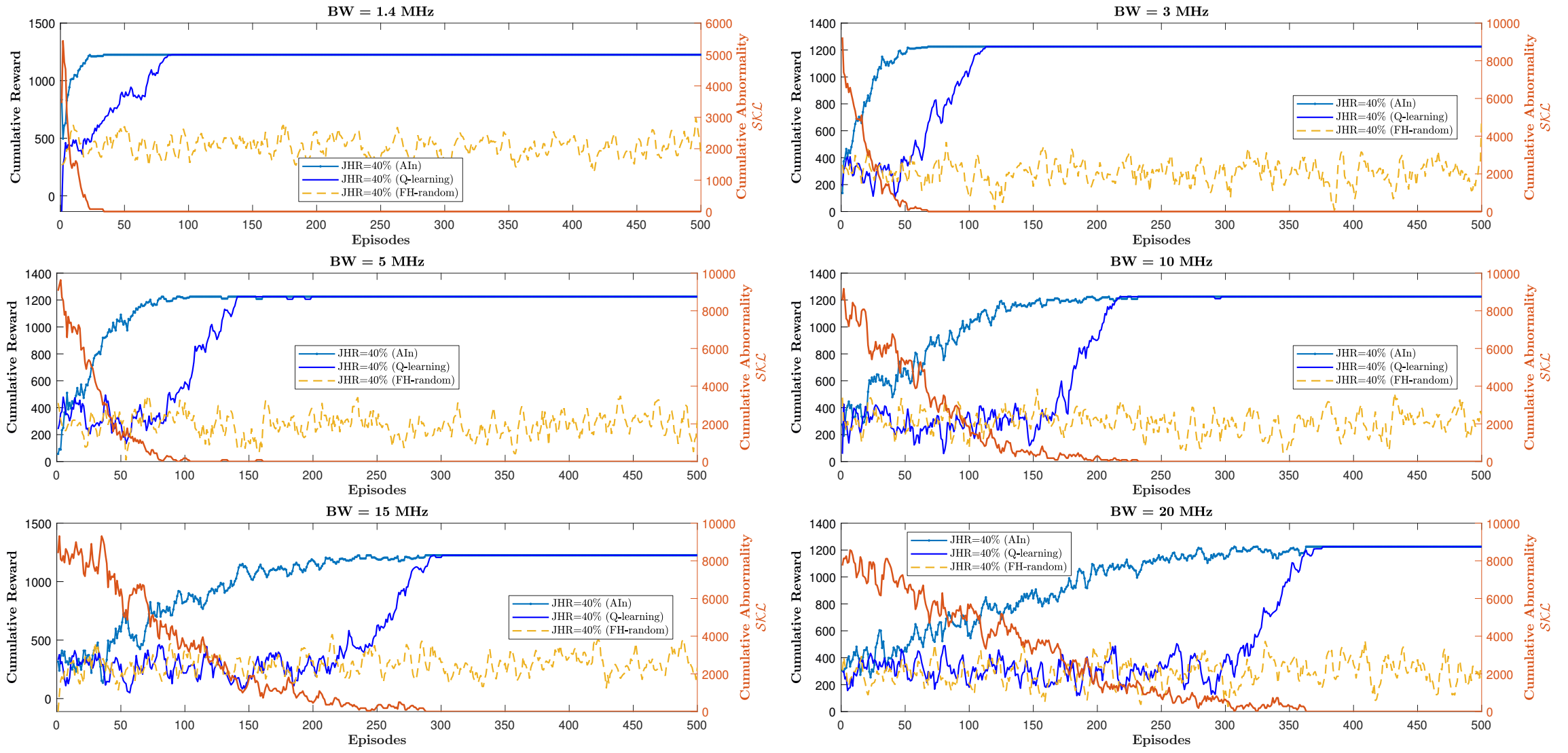

Fig. 6.6 Performance comparison of cumulative reward and abnormality with the proposed *AIn*, FH-random and QL under the CtU-AD mode, *Random* jamming strategy and different Bandwidth (BW).

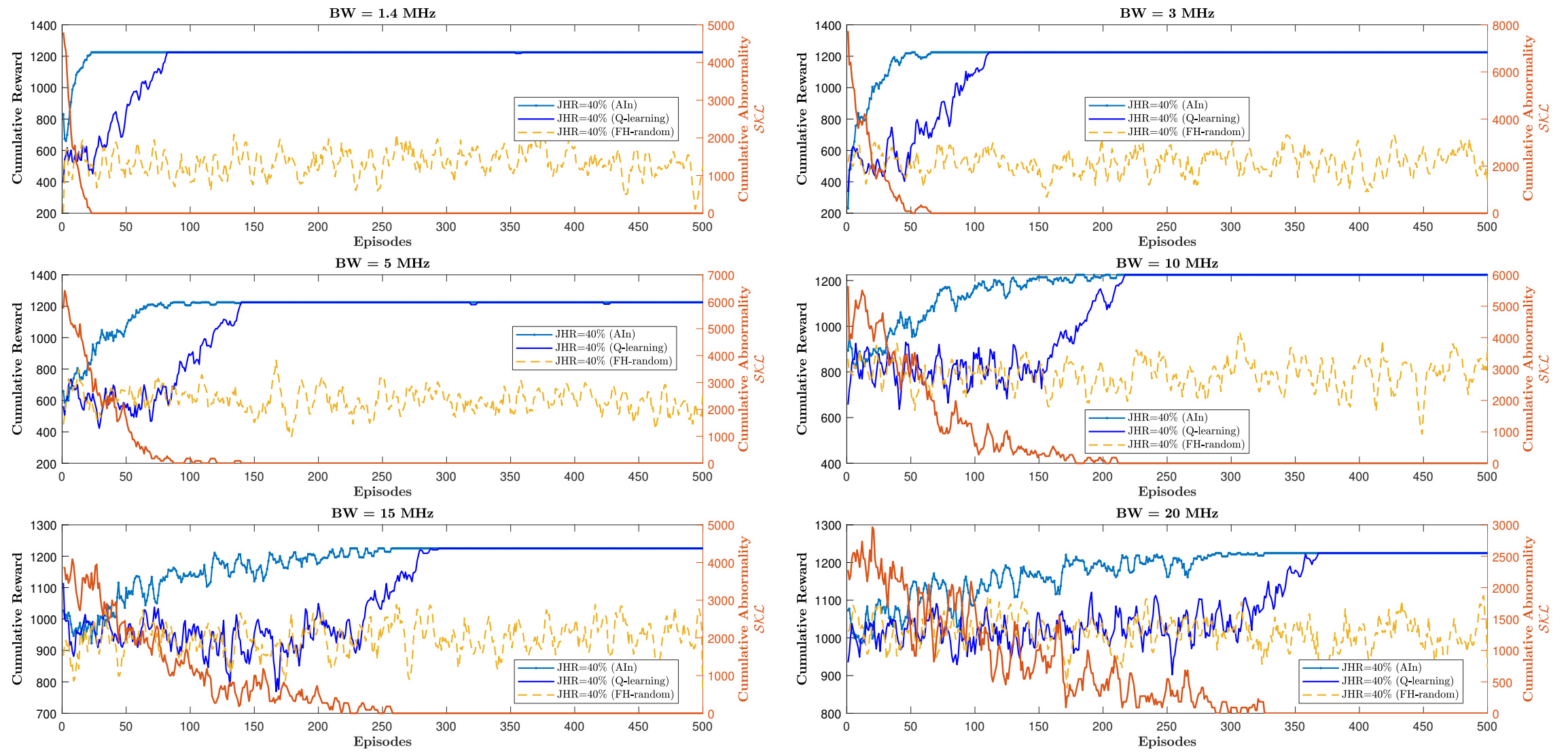

Fig. 6.7 Performance comparison of cumulative reward and abnormality with the proposed *AIn*, FH-random and QL under the CtU-AD mode, *Sweep* jamming strategy and different Bandwidth (BW).

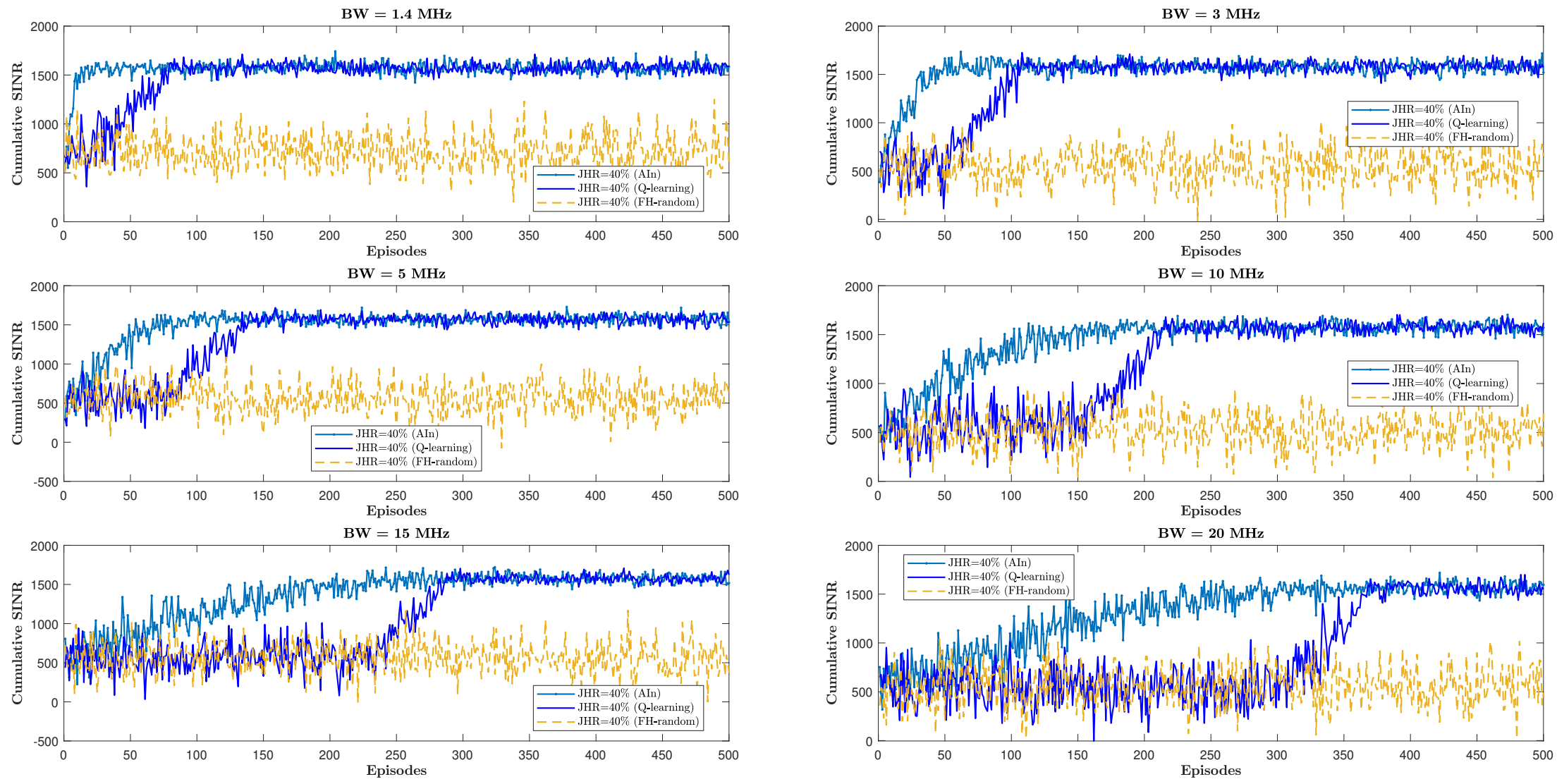

Fig. 6.8 Performance comparison of cumulative SINR with the proposed *AIn*, random FH and QL under the CtU-AD mode, *Constant* jamming strategy and different Bandwidth (BW).

Finally, Figs. 6.12-6.13-6.14 show some examples of how the active UAV resolved uncertainty and discovered the jamming strategy while interacting with the environment. An example of a constant jamming strategy is shown in Fig. 6.11 and the corresponding $\mathcal{P}^u$, $\mathcal{P}^j$ and $\Pi^{a^u}$ matrices are shown in Fig. 6.2-6.3-6.4 for constant, random and sweep jammers, respectively. Initially, these matrices contain equiprobable elements with probability $p = 1/N$ where $N = 6$ is the total number of available PRBs, and the agent adopts randomized actions as it is unfamiliar with the environment. While taking actions the agent starts to save its own transitions in $\mathcal{P}^u$, learn the jammers' transitions and the probabilistic state-action relationship by updating $\mathcal{P}^j_\tau$ and $\Pi^{a^u}_\tau$ once it got surprised by sensory signals. It can be seen in Fig. 6.12 for a constant jammer that after convergence, the active UAV succeeded to discover the jammer's strategy, which is attacking the $1^{st}$, $3^{rd}$ and $6^{th}$ PRBs (middle-bottom image of Fig. 6.12), and updating the $\Pi_{\tau}^{a^u}$ in an optimal way allowing to avoid the jammer which can be proved by observing the $\mathcal{P}^u$ matrix that encodes the user's transitions among the available PRBs. Similarly, Fig. 6.13 describes the learning procedure by facing a random jammer where we showed the time-varying matrices (i.e., $\mathcal{P}_{\tau}^u$, $\mathcal{P}_{\tau}^j$ and $\Pi_{\tau}^{a^u}$) for better understanding. The time-varying matrices allow inferring what action the agent needs to perform and when it should do it. The example shown in the figure is related to temporal matrices when the jammer attacks the $6^{th}$ PRB. It can be seen that the agent learnt to avoid selecting the $6^{th}$ PRB at a specific time since it is expected that the jammer will move there by relying on $\mathcal{P}_{\tau}^j$ (as shown in Fig. 6.13). Finally, Fig. 6.14 illustrates the learning procedure in a sweep jamming scenario. The time-varying matrices, in this case, encode the periodicity of a specific attack adopted by the sweep jammer, which repeats it strategy each $\tau$. The agent can learn this periodicity and use the proper matrices to update its transitions, jammers' transitions and the state-action probabilities. An example of the proper matrices that the agent can use when the jammer attacks the $5^{th}$ PRB is shown. After convergence, the probability of selecting the $5^{th}$ PRB being in any of the available PRBs is near zero (refer to $\Pi_{\tau}^{a^u}$), which is confirmed by $\mathcal{P}_{\tau}^j$ encoding high probabilities that the jammer will move to the 5-th PRB.

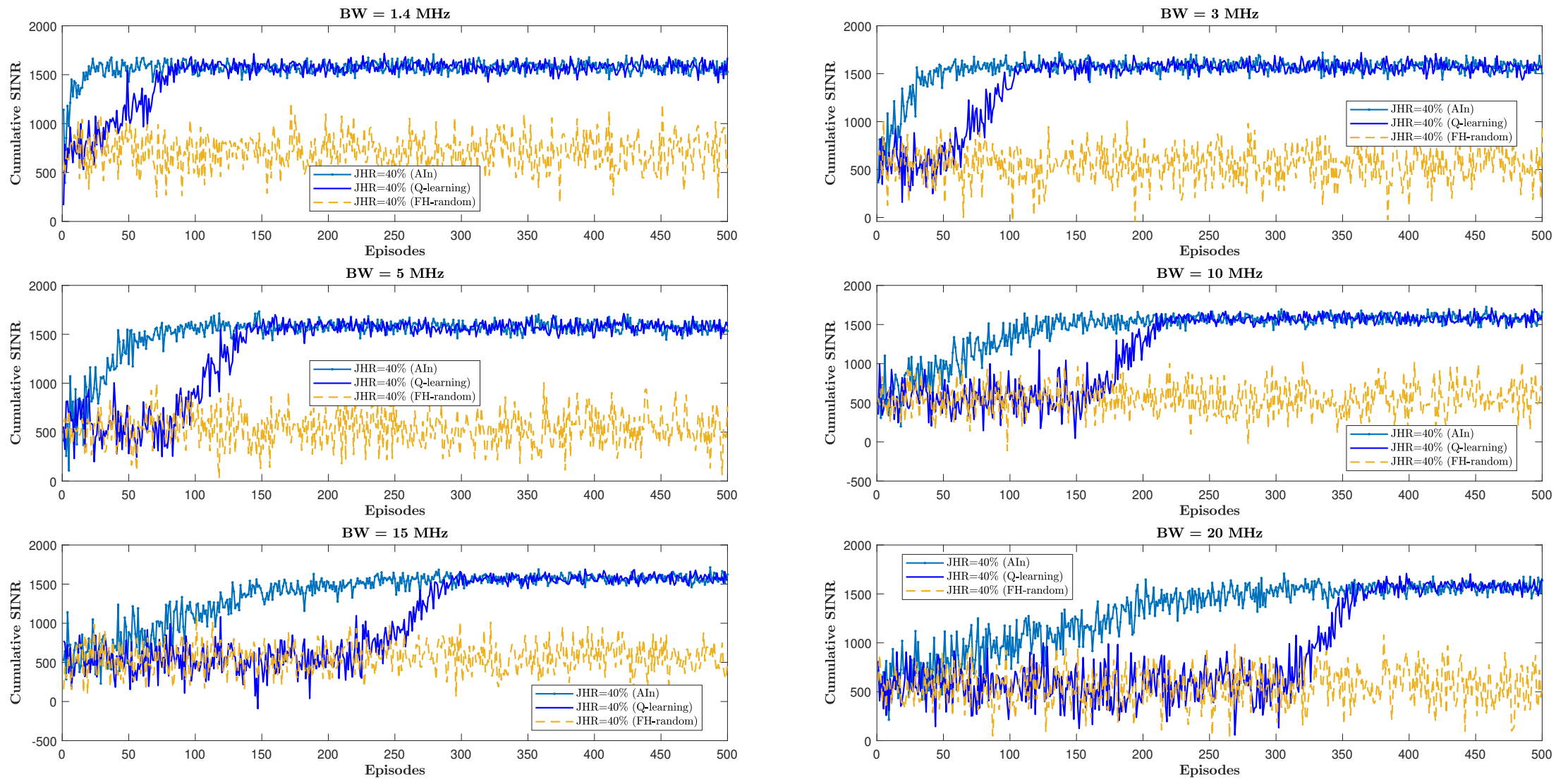

Fig. 6.9 Performance comparison of cumulative SINR with the proposed *AIn*, random FH and QL under the CtU-AD mode, *Random* jamming strategy and different Bandwidth (BW).

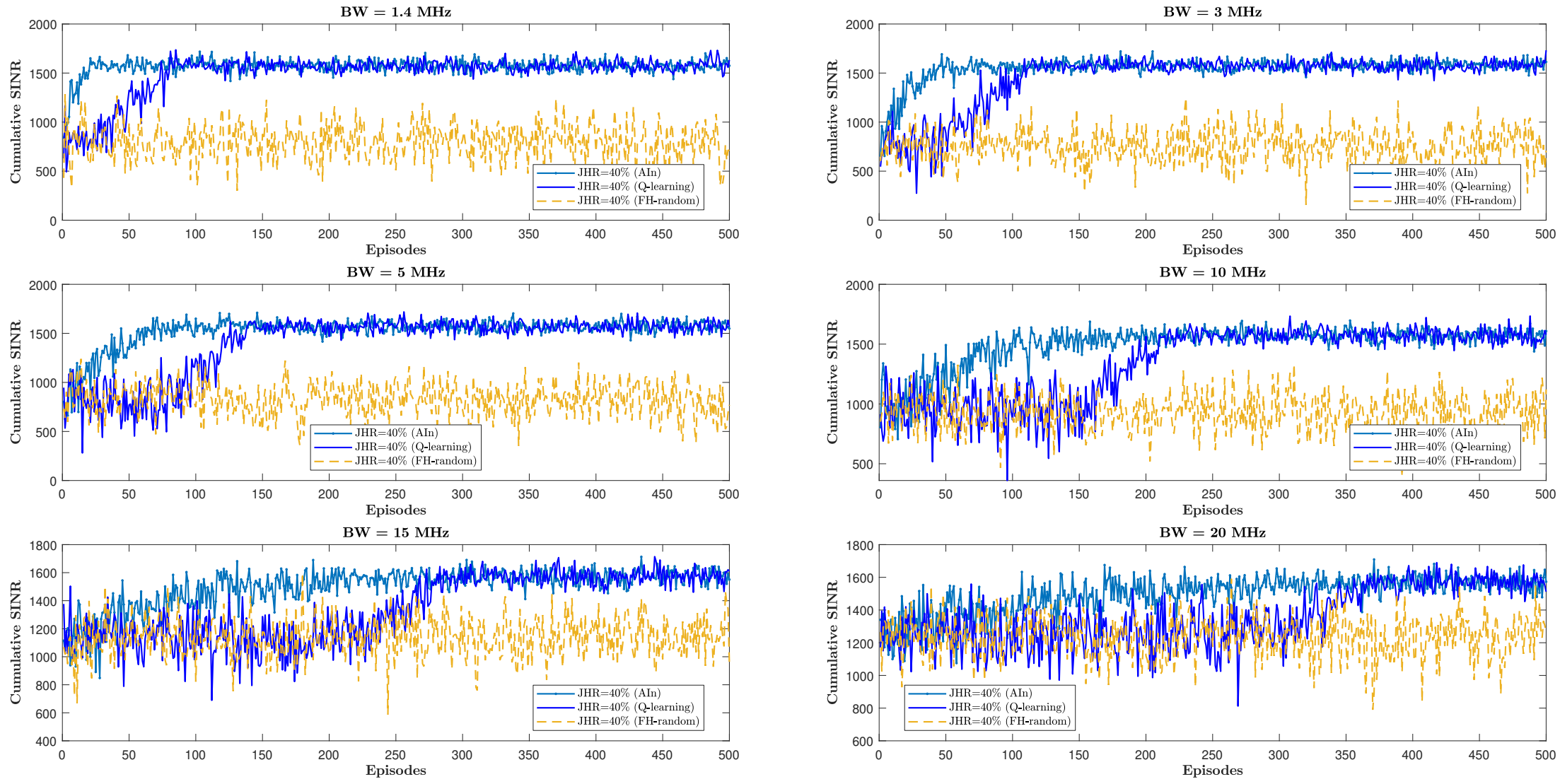

Fig. 6.10 Performance comparison of cumulative SINR with the proposed *AIn*, random FH and QL under the CtU-AD mode, *Sweep* jamming strategy and different Bandwidth (BW).

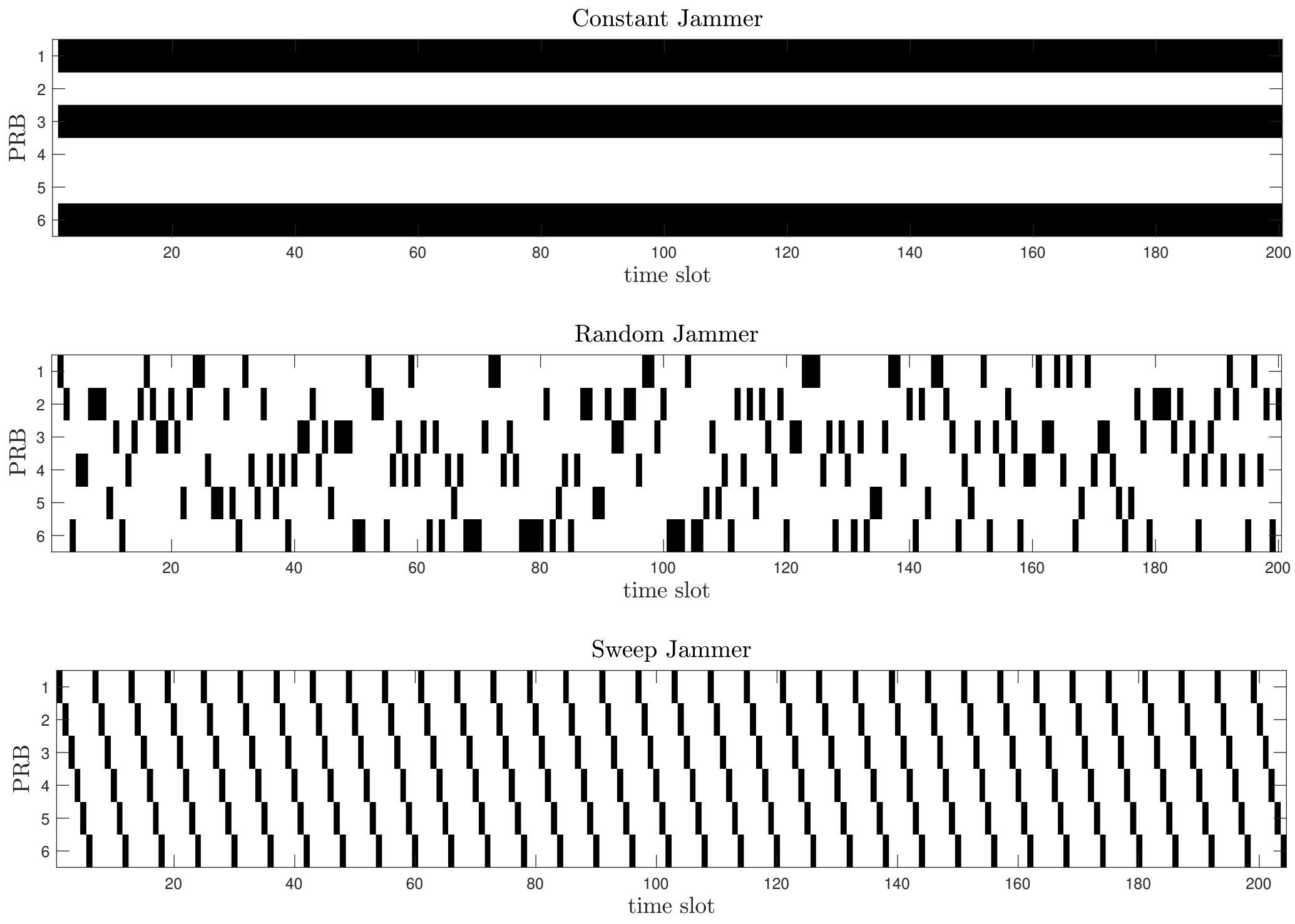

Fig. 6.11 An example of the Spectrum under different jamming strategies.

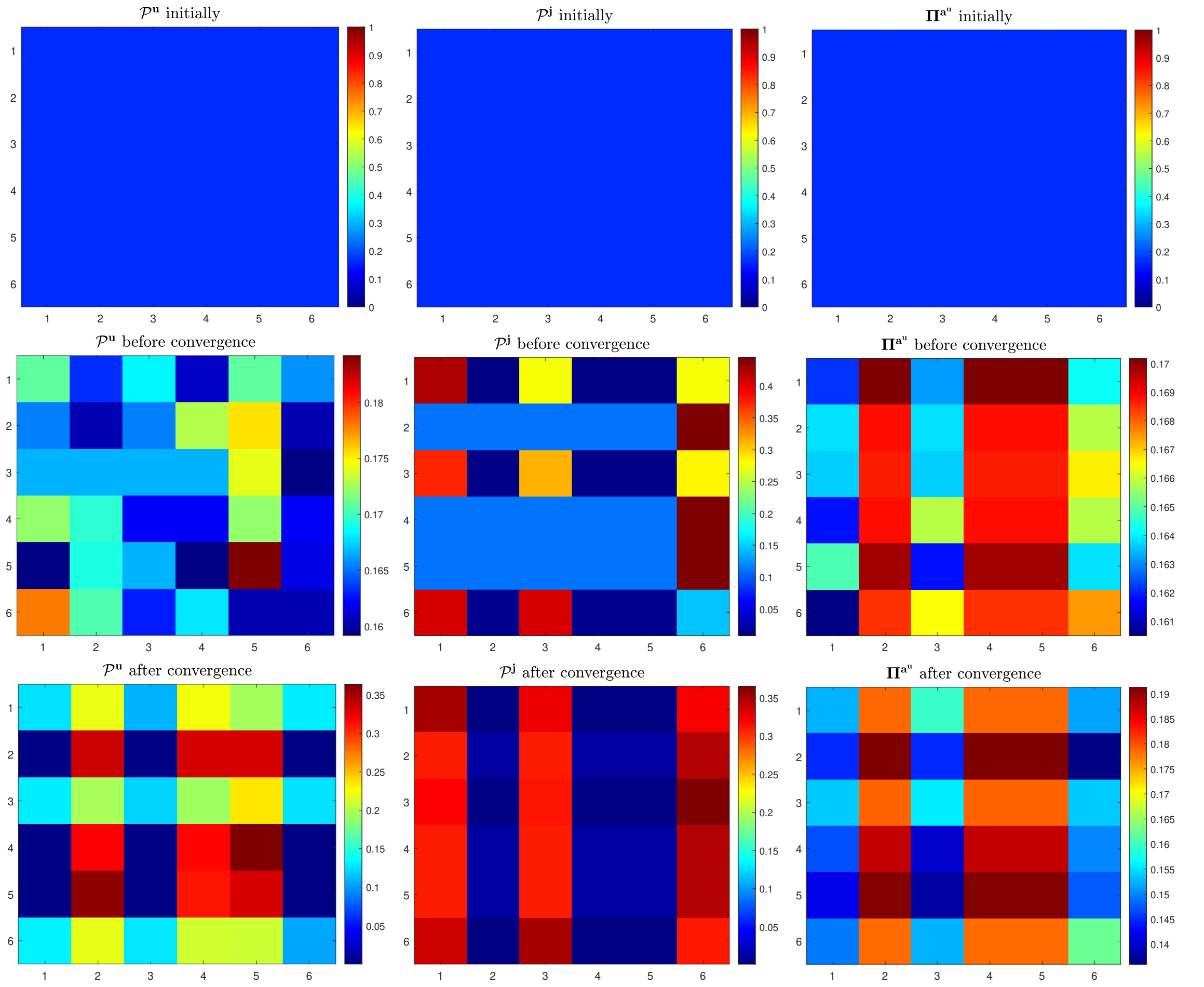

Fig. 6.12 Example of the active learning procedure from start to convergence under *Constant* jamming strategy.

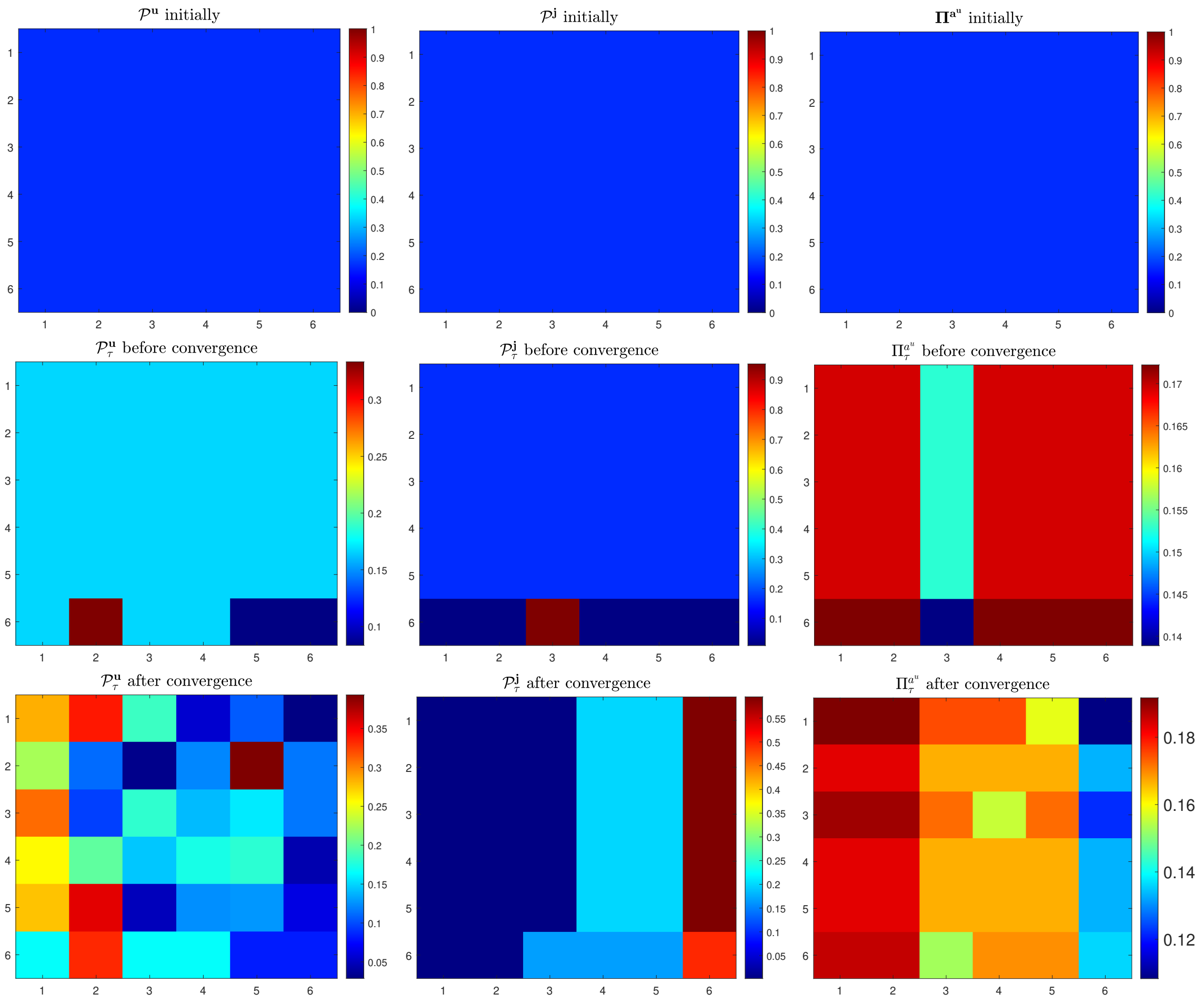

Fig. 6.13 Example of the active learning procedure from start to convergence under *Random* jamming strategy.

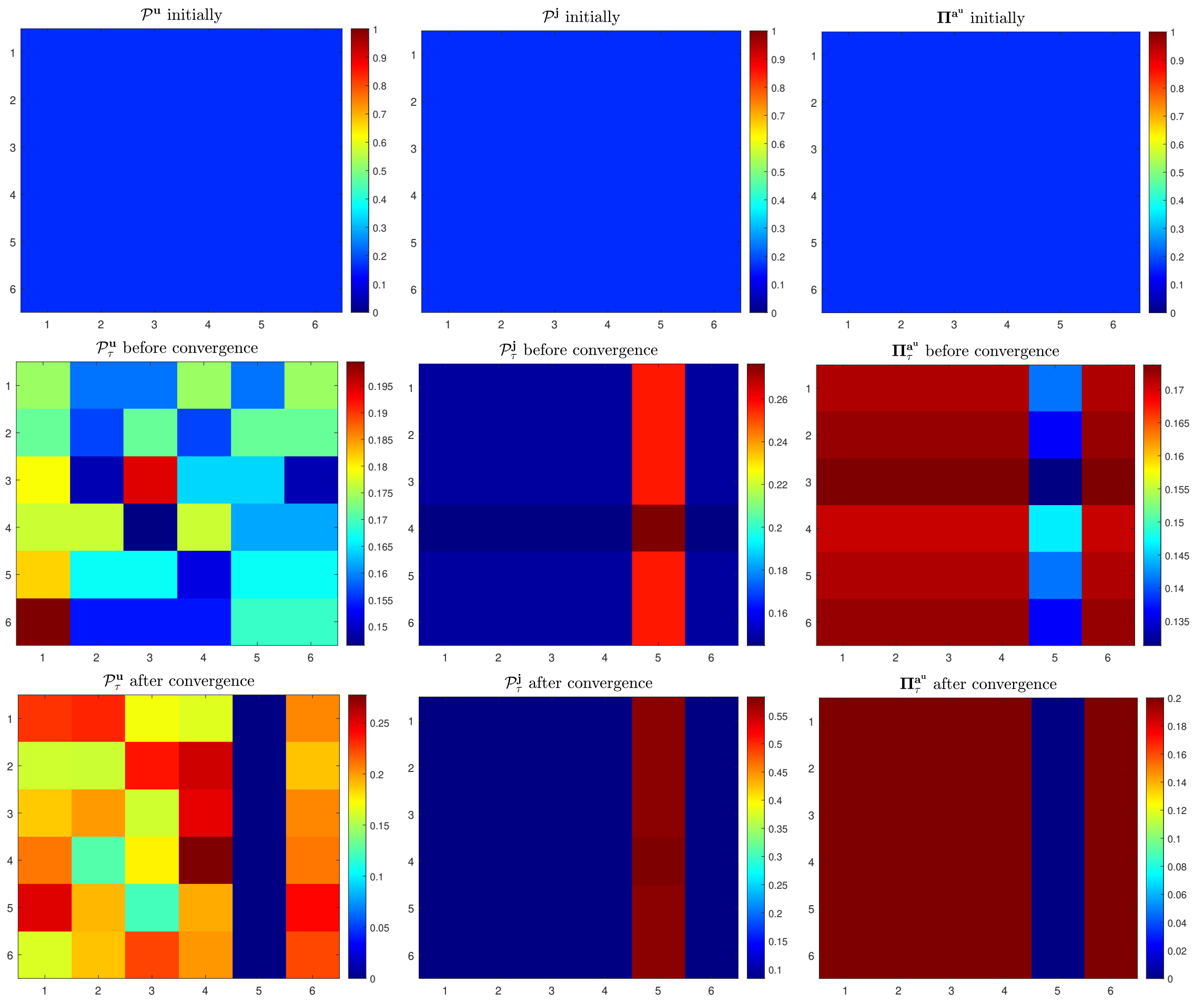

Fig. 6.14 Example of the active learning procedure from start to convergence under *Sweep* jamming strategy.

## 6.5 Conclusion

This chapter has proposed a novel resource allocation strategy using Active Inference (*AIn*) for anti-jamming where a Cognitive-UAV scenario is used as a case study. The Cognitive-UAV is endowed with a joint internal representation of the external environment representing the dynamics of the physical signals and the dynamic spectrum occupancy of both UAV and jammer. Under the *AIn* approach, UAV is capable of learning the best anti-jamming strategy by determining the actions that minimize abnormalities. Simulated results have indicated that the proposed method outperforms conventional Frequency Hopping and Q-Learning in terms of learning speed (convergence).

# Chapter 7

# Conclusion and Future directions

## 7.1 Conclusions

This thesis presents research work on the promising potentials of CR in empowering the future wireless communications with brain-powered solutions to solve several fundamental problems such as jamming attacks at the physical layer after being able to learn, think and understand the surrounding environment. It has been discussed that Artificial Intelligence (AI) can be seen as an effective tool to drive the development of CR to reach the required intelligence. However, most AI systems cannot understand and explain *why they do what they do, nor how*. Thus, CR empowered with AI (i.e., AI-enabled radios) cannot achieve its full potentials unless it reaches a high level of *Self-Awareness (SA)*. It has been shown that SA endows CR with the capability of maintaining a dynamic equilibrium with the external environment starting with null memory by capturing knowledge about itself and structural regularities from its external milieu variations through sensory signals and encoding them in its internal hierarchical Generative models. Furthermore, SA provides various probabilistic inference modes within the Generalized Bayesian Filtering involving predictive top-down messages propagating the belief in hidden variables from high levels (of the hierarchical models) towards the lower levels. In turn, bottom-up messages from lower levels report the evidence for expectations of beliefs generating predictions. Comparing predictive (top-down) messages with the sensory responses signalled via diagnostic (bottom-up) messages results in multi-level abnormality indicators and Generalized Errors (GE). Those errors are then fed back from bottom to up the hierarchy to update beliefs, incrementally encode new concepts, and finesse plans. Thus, improving future predictions, future actions while minimizing abnormalities and GEs (i.e., minimizing the free energy). In this perspective, CR exposed to random and unpredictable variations in its prevalent world can learn by minimizing the free energy (in terms of abnormality) to restrict itself from occupying surprising states.

**Chapter 3** proposed a coherent framework to learn different Self-Awareness representations for two CR applications. The proposed framework includes two methods that enrich CR with the capability to learn dynamic generative models and detect abnormal signals inside the wireless spectrum. The potential of the proposed SA framework is that it can be incorporated in CR systems and give a facility to adopt either of the abnormality detection methods depending on the application (e.g., data dimensionality, sampling rate). C-GAN, AC-GAN and VAE are employed in the high-dimensional data application due to its ability to deal with high data samples extracted from mmWave wide-band, while the DBN model is implemented in low dimensional data application due to its strength in representing in a probabilistic way low data samples extracted from sub-carriers of an OFDM signal at two levels. The dynamic models in both methods are generative models learnt from generalized state vectors. Generative models can generate synthesised data and provide a distance metric to measure the deviation of the predicted states from the observed one. Indeed, results reveal that both methods can effectively detect malicious or jammer attacks after learning the corresponding dynamic models and overcome conventional detection methods. Validation is performed on a real mmWave dataset and simulated OFDM data.

**Chapter 4** proposed introducing an emergent data-driven Self Awareness (SA) module to enhance the physical layer security in CR, where the cognitive-UAV scenario is used as a case study. It has been shown that the SA module allows the radio to build up its own memories incrementally by observing the stimulus received from the radio environment and learning with reasoning a hierarchical representation of such observation. The radio augmented with SA can predict the environment and identify any abnormality within a received signal. The SA module is also capable of characterizing the abnormal situation caused by a jammer while studying the rules of how (Power), when (Time) and where (Frequency) the jammer is attacking. Additionally, an incremental learning process is proposed to learn a new model representing the new situation. This chapter also showed that the radio could decide and act efficiently by suppressing the jamming signal. These tasks are performed by the radio itself in an incremental approach without any external supervision. The chapter showed that the proposed HDBN framework accurately characterizes the jammer's behaviours in different situations while the probability of detection is significantly high even at low Jamming-to-Signal-Ratio. The results also show that after learning the jammer's behaviours, the CR can correctly predict the future activities of the jammer, which can eventually help mitigate any future attacks.

**Chapter 5** proposed a novel method for joint detection and classification of jamming attacks in a Cognitive-UAV-based radio application. The method employs the acquired knowledge encoded in the GDBN model during the online phase to forecast what the cognitive-UAV is supposed to receive. While predicting and receiving actual stimuli, cognitive UAV can evaluate hierarchical abnormality measurements and generalised errors to explain the current situation by differentiating between normal and abnormal situations (i.e., jammer detection) and extract the jamming signal by exploiting the errors to encode it incrementally in a new GDBN model. In future jamming experiences, cognitive-UAV can perform multiple predictions in parallel using the M-MJPF and select the best model that explains the current situation to recognize the modulation scheme of the detected jammer. Simulation results showed that the proposed method achieves high probabilities in detecting jammer and high accuracy in classifying them and outperforms LSTM and CNN in classifying multiple jamming signals and different modulation schemes using a real dataset. In addition, the proposed approach provides interpretable results where multiple abnormality measurements and generalised errors can be used as self-information to keep learning incrementally. Furthermore, the chapter proposed applying transport planning concept for modulation conversion and classification, which is a prospective candidate technology in future wireless communications. The proposed automatic modulation conversion and classification (AMCC) framework is based data-driven approach following the inherent intelligent capability of Generalised filtering. Simulation results validate that the proposed approach allows transporting low-order to high-order modulation schemes without introducing distortions and performs modulation classification with high accuracy.

**Chapter 6** has proposed a novel resource allocation strategy through Active Inference (AIn) for anti-jamming in a Cognitive-UAV scenario. The Cognitive-UAV is endowed with a joint internal representation of the external environment representing the dynamics of the physical signals and the dynamic spectrum occupancy of both UAV and jammer. Under the AIn approach, UAV can learn the best anti-jamming strategy by determining the actions that minimize abnormalities. Simulated results have indicated that the proposed method outperforms conventional Frequency Hopping and Q-Learning in terms of learning speed (convergence).

## 7.2 Future Directions

The following research issues have been identified and are to be addressed in future work, for the application and implementation of SA in Cognitive and AI-enabled radios:

- It is worth noting that the functionalities of the SA module can help realize other goals targeted by the radio, such as channel characterization (in this thesis, the goal was to enhance the PHY security). The proposed SA module can largely help CR estimate the channel's dynamic changes (not only those related to the active transmissions in the spectrum) and discover its characteristics. In this way, CR can estimate the noise level (i.e., SNR value) and adapt to different channel conditions to ensure stable link communication and improve the decoding process. This task can be studied in future work.

- Another possible future direction is bridging high-dimensional and low-dimensional data through a Hierarchical Dynamic Bayesian Network model as shown in Fig. 3.41. The bridging task must be preceded by dimensionality reduction using Variational AutoEncoder to obtain low-dimensional probabilistic latent representations from high-dimensional data. The advantage of this approach is that representing useful features extracted from high-dimensional multi-signals in a probabilistic graphical model endows the radio with the ability to produce explainable and interpretable predictions allowing to understand surprising events and link them to the causal interpretation that provides the motivation supporting actions.
- A possible future direction regarding the classification task includes optimizing the proposed approach to achieve high classification accuracy at low SNR. Moreover, including more modulation schemes and addressing the transport planning in a fully unsupervised way, i.e., learning to synchronize without fixing the graph vertices of different schemes, realized a possible future direction to improve the modulation conversion task. In addition, investigating the effect of different channel conditions and channel impairments on the performance of the proposed framework for modulation conversion and studying the optimal source-to-target mapping between graph vertices leads to other interesting future work to demonstrate optimality for all modulation scenarios.
- Further research can be done to explore performance improvements by facing smart reactive jammers and study the extension of *Active Inference* to be applicable in fully-observable environments. Also, developing a joint frequency and power allocation is of great importance to deal with intelligent jammers. The proposed approach allows solving this task by exploiting the generalized errors at the continuous level, guiding the action selection process and beating the jammer even in the power domain under the NOMA framework. It is to note that the joint allocation strategy (frequency and power) involves discrete and continuous actions, which is still an open task in literature. In addition, integrating deep learning with *Active Inference* (i.e., *Deep Active Inference*) to substitute the state-action table and the users' transition tables is another possible future direction to speed up the action selection process when dealing with a high number of PRBs.

- The estimated transition matrices of both user and jammer as well as the Active inference (AIn) table encoding the optimal set of actions the user acquired after convergence (as we discussed in Chapter 6), can be used to transfer the knowledge to other CR users building by that an intelligent collaborative learning. Also, the estimated jammers matrix (from which the radio can infer jammers' actions) can be used to identify the jammer. In this context, the classification is based on the jammers' actions (not on jammers' modulation schemes).

- We believe that the proposed transport maps can be used to enhance the physical layer security in future semantic communications. The transmitted source can be multiplied with the transport map and then propagated through the channel. Only users holding a copy of that map can decode the signal and recover original information (it can be seen as an analogy to the CDMA method). Also, transport maps can be used to identify smart jammers trying to fool CR by sending the same information but with different modulation schemes. In this sense, CR using transport maps can be more intelligent by understanding that this jammer has been detected so far, but it changed the modulation scheme. In addition, the acquired transport plans (or maps) can significantly reduce complexity since the radio can save one dynamic model instead of multiple models that are substituted by multiple maps that are less complex and allow converting the single prediction into multiple ones.

- Finally, we believe that our approach plays a significant role in the Federated Learning (FL) Framework. Specifically, the learned transport maps (shown in Chapter 5) and the learned AIn tables, along with the matrices encoding best transitions of users and jammer strategies (shown in Chapter 6). Those maps and matrices can be shared with other CR devices in the cell, forming a Global Brain at the cloud (or Base station with a global brain). FL was proposed to overcome the issue of sending AI learning models to the remote cloud or centralized edge servers for training in large scale wireless networks, which requires high communication costs and poses severe threats to AI data privacy [323]. Thus, FL integrated with transport planning allows sharing the transport maps (that require low data rates and encodes encrypted information), which can be used by specific users improving by that complexity and privacy, which are the main concerns in FL.

# Appendix A

# Algorithms

## A.1 Modified Markov Jump Particle Filter (M-MJPF)

## A.2 Automatic Jamming Classification (AJC)

**Algorithm 1:** M-MJPF

**Input:** $\tilde{Z}_t$, $N$, $T$, $\Pi$, $\Pi_\tau$, $\tilde{S}_m : (\tilde{\mu}_{\tilde{S}_m}, \Sigma_{\tilde{S}_m})\ \forall\ \tilde{S}_m \in \mathbf{S}$

**for** *each* $t \in T$ **do**
  **for** *each* $n \in N$ **do**
    **Prediction at Discrete Level:**
    $W_n = \frac{1}{N} \leftarrow$ particle weight
    **if** $t = 1 \leftarrow$ *initial state* **then**
      *Sample:* $\tilde{X}_1$ from initial prior density $P(\tilde{X}_1)$
      $\tilde{X}_t = \tilde{X}_1 \leftarrow$ current state
      *Estimate:* $\tilde{S}_t^n$ from $P(\tilde{X}_t|\tilde{S}_t^n)$
    **else** $\leftarrow$ *remaining states*
      $\tilde{S}_t^n \sim \pi_\tau(\tilde{S}_{t-1})$
      **if** $\tilde{S}_t^n = \tilde{S}_{t-1}^n$ **then**
        $\tau^n = \tau^n + 1 \leftarrow$ time elapsed in $\tilde{S}_t^n$
      **else**
        $\tau^n = 1$
    **Prediction at Continuous Level:**
    $U_{\tilde{S}_t} = U_{\tilde{S}_t^n} \leftarrow$ Control Vector
    $P_{t-1|t-1} = \Sigma_{\tilde{S}_t^n} \leftarrow$ Covariance matrix
    $\tilde{X}_t = A\tilde{X}_{t-1} + BU_{\tilde{S}_t} \leftarrow$ state
    $P_{t|t-1} = AP_{t-1|t-1}A^\intercal + \Sigma_{\tilde{w}_t} \leftarrow$ covariance
    *Calculate:* $\pi(\tilde{X}_t)$ using (5.28)
    **Current Observation $\tilde{Z}_t$:**
    $\lambda(\tilde{X}_t) = P(\tilde{Z}_t|\tilde{X}_t) \leftarrow$ *diagnostic msg1*
    *Calculate:* $\lambda(\tilde{S}_t)$ using (5.31) $\leftarrow$ *diagnostic msg2*
    **Abnormality measurements:**
    *Calculate:* **KLDA** using (5.35)
    *Calculate:* **CLB** using (5.40)
    *Calculate:* **CLA** using (5.42)
    **Generalized Errors:**
    *Calculate:* $\tilde{\varepsilon}_{\tilde{Z}_t}^{[1]}$, $\tilde{\varepsilon}_{\tilde{Z}_t}^{[2]}$ using (5.16), (5.46)
    *Calculate:* $\tilde{\varepsilon}_{\tilde{X}_t}^{[1]}$, $\tilde{\varepsilon}_{\tilde{X}_t}^{[2]}$ using (5.17), (5.45)
    *Calculate:* $\tilde{\varepsilon}_{\tilde{S}_t}$ using (5.39)
    **Update Belief in hidden variables:**
    $K_t = P_{t|t-1}H^\intercal(HP_{t|t-1}H + \Sigma_{\tilde{Z}_t})^{-1} \leftarrow$ Kalman gain
    $\hat{\tilde{X}}_t = \tilde{X}_t + K_t\tilde{\varepsilon}_{\tilde{X}_t}^{[1]} \leftarrow$ updated state
    $\hat{P}_{t|t} = (1 - K_tHP_{t|t-1}) \leftarrow$ updated covariance
    $W_n = W_n \times \lambda(\tilde{S}_t) \leftarrow$ update particles weight
  *SIR resampling*

**Output:** *KLDA*, *CLA*, *CLB*, $\tilde{\varepsilon}_{\tilde{Z}_t}^{[1]}$, $\tilde{\varepsilon}_{\tilde{Z}_t}^{[2]}$, $\tilde{\varepsilon}_{\tilde{X}_t}^{[1]}$, $\tilde{\varepsilon}_{\tilde{X}_t}^{[2]}$, $\tilde{\varepsilon}_{\tilde{S}_t}$

**Algorithm 2:** AJC

**Input:** $\tilde{Z}_t$, $N$ $T$, $\Pi$, $\Pi_\tau$, $\tilde{S}_m : (\tilde{\mu}_{\tilde{S}_m}, \Sigma_{\tilde{S}_m}) \; \forall \; \tilde{S}_m \in \mathbf{S} \leftarrow$ Reference GDBN Model

$\mathcal{N}$, $\Pi^{(k)}$, $\Pi_t^{(k)}$, $\tilde{S}_m^{(k)} : (\tilde{\mu}_{\tilde{S}_m^{(k)}}, \Sigma_{\tilde{S}_m^{(k)}}) \; \forall \; \tilde{S}_m^{(k)} \in \mathcal{M}_k$ (where $\mathcal{M}_k \subset \mathcal{S}_{\mathcal{M}}$) $\leftarrow$ Jammers GDBN Models

**for** *each* $t \in T$ **do**

- **Predict normal commands and detect jammer using M-MJPF (Algorithm 1):**
- $[KLDA, CLA, CLB, \tilde{\varepsilon}_{\tilde{Z}_t}^{[1]}, \tilde{\varepsilon}_{\tilde{Z}_t}^{[2]}, \tilde{\varepsilon}_{\tilde{X}_t}^{[1]}, \tilde{\varepsilon}_{\tilde{X}_t}^{[2]}, \tilde{\varepsilon}_{\tilde{S}_t}]$ = M-MJPF($\tilde{Z}_t$, $N$, $T$, $\Pi$, $\tilde{S}_m$, $\tilde{\mu}_{\tilde{S}_m}$, $\Sigma_{\tilde{S}_m}$)
- **if** $KLDA > \psi \;||\; CLA > \eta \leftarrow$ *Jammer is present* **then**
  - **Predict jammer dynamics:**
  - **for** *each* $k \in \mathcal{S}_{\mathcal{M}} \leftarrow$ *available jammers' models* **do**
    - **for** *each* $n \in \mathcal{N}$ **do**
      - **Prediction at Discrete Level:**
      - $W_n = \frac{1}{N} \leftarrow$ particle weight
      - **if** $t = 1 \leftarrow$ *initial state* **then**
        - *Sample:* $\tilde{X}_1^{(k)}$ from initial prior density $P(\tilde{X}_1^{(k)})$
        - $\tilde{X}_t^{(k)} = \tilde{X}_1^{(k)} \leftarrow$ current state
        - *Estimate:* $\tilde{S}_{n,t}^{(k)}$ from $P(\tilde{X}_t^{(k)} | \tilde{S}_{n,t}^{(k)})$
      - **else** $\leftarrow$ *remaining states*
        - $\tilde{S}_{n,t}^{(k)} \sim \pi_\tau(\tilde{S}_{n,t-1}^{(k)})$
        - **if** $\tilde{S}_{n,t}^{(k)} = \tilde{S}_{n,t-1}^{(k)}$ **then**
          - $\tau^{n,k} = \tau^{n,k} + 1 \leftarrow$ time elapsed in $\tilde{S}_{n,t}^{(k)}$
        - **else**
          - $\tau^{n,k} = 1$
      - **Prediction at Continuous Level:**
      - $U_{\tilde{S}_t^{(k)}} = U_{\tilde{S}_{n,t}^{(k)} | \tilde{S}_{n,t-1}^{(k)}} \leftarrow$ *Conditional Control Vector* using (5.49)
      - $P_{t-1|t-1} = \Sigma_{\tilde{S}_{n,t}^{(k)} | \tilde{S}_{n,t-1}^{(k)}} \leftarrow$ *Conditional Covariance matrix* using (5.50)
      - $\tilde{X}_t^{(k)} = A\tilde{X}_{t-1}^{(k)} + BU_{\tilde{S}_{n,t}^{(k)}} \leftarrow$ state
      - $P_{t|t-1} = AP_{t-1|t-1}A^\intercal + \Sigma_{\tilde{w}_t} \leftarrow$ covariance
      - *Calculate:* $\pi(\tilde{X}_t^{(k)})$ using (5.28)
      - **Estimate the current observation** $\tilde{\varepsilon}_{\tilde{Z}_t}^{[2]}$ **using** (5.46)**:**
      - $\lambda(\tilde{X}_t^{(k)}) = P(\tilde{\varepsilon}_{\tilde{Z}_t}^{[2]} | \tilde{X}_t^{(k)}) \leftarrow$ *diagnostic msg1*
      - *Calculate:* $\lambda(\tilde{S}_t^{(k)})$ using (5.31) $\leftarrow$ *diagnostic msg2*
      - **Abnormality measurements:**
      - *Calculate:* $Abn_k$ using (5.51)
      - **Update Belief in hidden variables:**
      - $\hat{\tilde{\varepsilon}}_{\tilde{Z}_t}^{[2]} = (\tilde{\varepsilon}_{\tilde{Z}_t}^{[2]} - H\tilde{X}_t^{(k)}) \leftarrow$ Kalman innovation
      - $K_t = P_{t|t-1}H^\intercal(HP_{t|t-1}H + \Sigma_{\tilde{\varepsilon}_{\tilde{Z}_t}^{[2]}})^{-1} \leftarrow$ Kalman gain
      - $\hat{\tilde{X}}_t = \tilde{X}_t + K_t\hat{\tilde{\varepsilon}}_{\tilde{Z}_t}^{[2]} \leftarrow$ updated state
      - $\hat{P}_{t|t} = (1 - K_tHP_{t|t-1}) \leftarrow$ updated covariance
      - $W_n = W_n \times \lambda(\tilde{S}_t^{(k)}) \leftarrow$ update particles weight

**Output:** $\mathcal{S}_{Abn}$

# Appendix B

# List of Publications

## B.1 Journal Papers

1. **A. Krayani**, A. Alam, L. Marcenaro, A. Nallanthan, C. Regazzoni, "A Novel Resource Allocation for Anti-jamming in Cognitive-UAV Radios: an Active Inference Approach", *IEEE Communications Letters*, 2022 (Under Review).

2. **A. Krayani**, A. Alam, L. Marcenaro, A. Nallanthan, C. Regazzoni, "Automatic Jamming Signal Classification in Cognitive UAV Radios", *IEEE Transactions on Vehicular Technology*, 2022. (Under Review $2^{nd}$ Round).

3. **A. Krayani**, A. Alam, L. Marcenaro, A. Nallanathan, C. Regazzoni, "An Emergent Self-Awareness Module for Physical Layer Security in Cognitive UAV Radios," in *IEEE Transactions on Cognitive Communications and Networking*, March 2022, doi: 10.1109/TCCN.2022.3161937.

4. A. Toma, **A. Krayani**, M. Farrukh, H. Qi, L. Marcenaro, Y. Gao, and C. S. Regazzoni, "AI-Based Abnormality Detection at the PHY-Layer of Cognitive Radio by Learning Generative Models," in *IEEE Transactions on Cognitive Communications and Networking*, vol. 6, no. 1, pp. 21-34, March 2020, doi: 10.1109/TCCN.2020.2970693.

## B.2 Conference Papers

1. **A. Krayani**, N. J. William, A. S. Alam, L. Marcenaro, Z. Qin, A. Nallanathan, C. Regazzoni, "Generalized Filtering with Transport Planning for Joint Modulation Conversion and Classification in AI-enabled Radios", *2022 IEEE International Conference on Communications (ICC)*, 2022. (Accepted to appear).

## B.3 Book Chapters